\documentclass[8.5pt,twoside,twocolumn]{article}

\usepackage[super,sort&compress,comma]{natbib} 
\usepackage{sectsty}
\usepackage{balance} 

\usepackage[hang,splitrule]{footmisc}
\usepackage{graphicx} %eps figures can be used instead
\usepackage{lastpage}

\usepackage{geometry}
\makeatletter
\renewcommand*{\@biblabel}[1]{\hfill#1.}
\makeatother

\usepackage{amsmath}
\usepackage{amssymb}
\usepackage{amsfonts}
\usepackage{amsthm}
\usepackage{color}
\usepackage[T1]{fontenc}%needed in order to use Polish Characters !!!
\usepackage[normalem]{ulem}

\usepackage{fixltx2e}

\usepackage[section]{placeins}

\newcommand{\bfe}{{\bf e}}
\newcommand{\ctimes}{\!\times\!}
\newcommand{\cE}{\mathcal{E}}

\sectionfont{\large}
\subsectionfont{\normalsize}

\newcommand{\captionfonts}{\footnotesize}

\makeatletter  % Allow the use of @ in command names
\long\def\@makecaption#1#2{%
  \vskip\abovecaptionskip
  \sbox\@tempboxa{{\captionfonts #1: #2}}%
  \ifdim \wd\@tempboxa >\hsize
    {\captionfonts #1: #2\par}
  \else
    \hbox to\hsize{\hfil\box\@tempboxa\hfil}%
  \fi
  \vskip\belowcaptionskip}
\makeatother   % Cancel the effect of \makeatletter

\begin{document}

\twocolumn[
  \begin{@twocolumnfalse}
\noindent\LARGE{\textbf{Influence of material stretchability on the equilibrium shape of a M\"obius band}}
\vspace{0.3125cm}

\noindent\large{\textbf{David M.\ Kleiman,\textit{$^{a}$} Denis F.\ Hinz,\textit{$^{b}$} and Eliot Fried\textit{$^{c}$}}}\vspace{0.3125cm}

\linespread{0.9}
\small

\noindent 
\small{%The nonorientable topology of M\"obius bands is known to influence quantum effects, superconducting states, and electronic and thermal properties. {\color{red}
We use a discrete, lattice-based model for two-dimensional materials to show that M\"obius bands made with stretchable materials are less likely to crease or tear. This stems a delocalization of twisting strain that occurs if stretching is allowed. The associated low-energy configurations provide strategic target shapes for the guided assembly of nanometer and micron scale M\"obius bands. To predict macroscopic band shapes for a given material, we establish a connection between stretchability and relevant continuum moduli, leading to insight regarding the practical feasibility of synthesizing M\"obius bands from materials with continuum parameters that can be measured experimentally or estimated by upscale averaging. To take advantage of stretchability in the case of M\"obius bands made of graphene, DNA, and other effectively unstretchable materials, we develop and explore a novel architecture that uses the Chinese finger trap as a fundamental building block and imparts notable stretchability to otherwise unstretchable materials.
} %157 words!

\vspace{0.3125cm}
\end{@twocolumnfalse}
]

\normalsize

%\linespread{0.9}

\renewcommand{\baselinestretch}{0.9}\normalsize

\renewcommand{\thefootnote}{\alph{footnote}}

\makeatletter\renewcommand*\@makefnmark{{\footnotesize\@thefnmark$^a$}}
\footnotetext{\textit{Department of Mathematics and Statistics, McGill University, Montr\'eal, QC, Canada H3A 2K6. dave.kleiman2@gmail.com}}
\makeatother
\makeatletter\renewcommand*\@makefnmark{{\footnotesize\@thefnmark$^b$}}
\footnotetext{\textit{Kamstrup A/S, Industrivej 28, Stilling, 8660 Skanderborg, Denmark. dfhinz@gmail.com}}
\makeatother
\makeatletter\renewcommand*\@makefnmark{{\footnotesize\@thefnmark$^b$}}
\footnotetext{\textit{Mathematical Soft Matter Unit, Okinawa Institute of Science and Technology Graduate University, Onna, Okinawa, Japan 904-0495. eliot.fried@oist.jp}}
\makeatother

\noindent
Recent technical advances have made it increasingly clear that the properties of a material are determined not only by its composition but also by geometrical and, importantly, topological factors.\cite{Gupta2014} With this realization and breakthroughs in the ability to fabricate objects with molecular-scale precision, research into using the one-sided topology of the M\"obius band in scientific applications is burgeoning.  In chemical topology, for example, mechanically interlinked molecules, or catanenes, have been created using M\"obius molecules as intermediaries, which may set the stage for synthesizing programmable topological nanostructures.\cite{Fujita1996, Han2010} A branch of algebraic topology known as homotopy figures prominently in the description of defects in ordered media.\cite{Mermin1979} In that context, every defect is characterized by a topological charge, the conservation of which results in stability. In particular, the stability of topological solitons and singular defects is connected to the topological stability of M\"obius bands.\cite{Kleman2003} An additional connection between M\"obius topology and defects is provided by micron scale M\"obius crystals, which were first created just over a decade ago by spooling niobium triselenide ribbons onto selenium droplets.\cite{Tanda2002} Such objects can be considered as global disclinations.\cite{Osipov2006} The ability of a recently synthesized expanded porphyrinoid to switch between H\"uckel and M\"obius topologies presents the possibility of novel memory devices.\cite{Ajami2003,Stepien2007} M\"obius topology is also exhibited by cyclotides, macrocylic plant proteins involved in plant defense.\cite{Jennings2005} Due to their topologically derived structural stability, these proteins have the potential to serve as drug scaffolds and pharmaceutical templates.\cite{Wang2011a,Smith2011c,Poth2013}

This article focuses on determining energetically preferred equilibrium shapes of M\"obius bands. Work on this problem was initiated by Sadowsky,\cite{Hinz2013a,Hinz2014a,Hinz2013,Hinz2014,Hinz2013b,Hinz2014b} who focused exclusively on materials like paper which are easy to bend but essentially unstretchable and, thus, must adopt shapes that are very closely approximated by developable surfaces. Aside from proving that it is possible to construct a developable band from a rectangular strip of width sufficiently small relative to its length, Sadowsky established an upper bound for bending energy of a developable band and derived a dimensionally reduced expression for the bending energy of a band made from an infinitesimally thin rectangular strip. Wunderlich\cite{Wunderlich1961,Todres2014} later sharpened Sadowsky's bound and generalized Sadowsky's bending energy to incorporate the effect of finite width. The problem of constructing developable equilibrium configurations was first considered by Mahadevan and Keller,\cite{Mahadevan1993} whose numerically determined solutions led to a tighter upper bound on the bending energy but are inconsistent with results of Randrup and R{\o}gen,\cite{Randrup1996} who showed that the centerline of a M\"obius band must have an odd number of switching points at which its curvature and torsion both vanish. Using Wunderlich's energy, Starostin and van der Heijden\cite{Starostin2007,Starostin2007a} computed equilibria that meet these requirements and also found evidence to suggest that Sadowsky's energy is a singular limit that produces centerlines with discontinuous curvatures.

%for which the curvature of the centerline manifests discontinuities.

We depart from established tradition and explore the influence of material stretchability on the shape of an equilibrated M\"obius band. For bands with sufficiently large width-to-length ratios, Starostin and van der Heijden observed localized zones of concentrated bending-energy density. Their results point to the emergence of singularities indicative of the onset of failure. Even a slight degree of stretchability should alleviate such concentrations. We seek to quantify both this effect and accompanying shape variations that bands of a given aspect ratio exhibit with increasing stretchability. Additionally, a model that allows for small but discernible stretchability applies even to conventional paper, which stretches by a percent or two in the direction of loading without creasing or tearing.\cite{caulfield1988}

Our approach utilizes a discrete, lattice-based model that incorporates stretching but can also accurately approximate developable shapes for sufficiently small values of the stretchability. Regardless of the aspect ratio of the band, we find that the total energy decreases monotonically with stretchability. Bands made of stretchable materials are therefore easier to form than bands made of unstretchable materials.

For further insight regarding how stretchability influences equilibrium shape, we compute the mean and Gaussian curvatures for bands of various length-to-width aspect ratios, the latter of which is nonzero only for stretchable materials.\cite{Gauss1827} Consistent with the observation that unstretchable materials must adopt developable shapes, we find that the Gaussian curvature plays a key role in bands comprised of such materials. Except for cases involving combinations of the smallest aspect ratio and the two largest values of stretchability investigated, the mean and Gaussian curvatures become more evenly distributed across the band as stretchability increases. Bending is concomitantly transferred to stretching, thereby eliminating localized zones of concentrated bending-energy density and the associated possibility of creasing or tearing. 

We also show that our model is energetically consistent with a simple continuum theory and derive relationships between our material parameters and those of the continuum theory. These relationships could provide a basis for future material design and applications.

Finally, we describe a novel architecture consisting of a coiled Chinese finger trap. Our construction adopts a shape characteristic of those exhibited within the stretchable regime of our lattice model. Comparing the particular shape adopted by our construction with results from our model thus allows us to infer the effective stretchability of the underlying architecture.

\section{Discrete, lattice-based model}
\label{sec:model}

We first provide a brief description of our model. Details are provided in the Supplementary Information (SI).

%%%%%%%%%%%%%%%%%%%%%%%%%%%%%
%\vspace{-4pt}
\subsection{Kinematics}
%\vspace{-2pt}
%%%%%%%%%%%%%%%%%%%%%%%%%%%%%

The shape of an unstretchable M\"obius band is uniquely determined (up to a rigid transformation) by the curvature $\kappa$ and torsion $\tau$ of its centerline.\cite{Hinz2013a,Hinz2014a} The shapes adopted by such bands must be developable. That is, they must be ruled surfaces that can be continuously flattened onto planar regions without stretching/contracting or, equivalently, with the preservation of intrinsic lengths and angles. In contrast, a band made of a stretchable material can adopt a nondevelopable equilibrium configuration, in which case its shape is uniquely determined (up to a rigid transformation) by its first and second fundamental forms, as discussed in the SI. 

%%%%%%%%%%%%%%%%%%%%%%%%%%%%%%%%%%%%%	
%\vspace{-4pt}  	
\subsection{Linear and angular springs}
%\vspace{-2pt}
%%%%%%%%%%%%%%%%%%%%%%%%%%%%%%%%%%%%%	  

We approximate a rectangular strip of length $L$ and width $w$ by a lattice of equilateral triangles with $N$ points uniformly separated by a distance $r_0$. To incorporate resistance to stretching, we connect each pair of lattice points by a linear spring with stiffness $k_l$ and equilibrium length $r_e$. Further, to incorporate resistance to out-of-plane bending, we connect each triplet of lattice points with a torsional spring of stiffness $k_\theta>0$ and equilibrium angle $\theta_{e}=\pi$. The total energy $E$ of the band is then given by the sum
\begin{equation}
\label{eq:E_01}
E =\frac{k_{l}}{2}\sum_{i\in S_{l}}(r(i)-r_e)^{2}
+\frac{k_{\theta}}{2}\sum_{i\in S_{\theta}}(\theta(i)-\theta_{e})^{2},
%\label{eq:ESandEB}
\end{equation}
where $S_l$ and $S_\theta$ denote the sets of all linear and torsional springs, $r(i)$ is the current length of the $i$-th linear spring, and $\theta(i)$ is the angle between triplets of points associated with the $i$-th torsional spring.
	  
%%%%%%%%%%%%%%%%%%%%%%%%%%%%%%%%%%%%%
%\vspace{-4pt}	  
\subsection{Nondimensionalization}
\vspace{-2pt}
%\label{sec:nondim}
%%%%%%%%%%%%%%%%%%%%%%%%%%%%%%%%%%%%%	 
%
%
The width $w$, length $L$, linear spring constant $k_l$, and angular spring constant $k_\theta$ yield dimensionless parameters 
\begin{equation}
\label{eq:nondimParam_01}
a = \frac{L}{w} \qquad \text{and} \qquad k = \frac{k_\theta}{w^2k_l}.
\end{equation}
While the (length-to-width) aspect ratio $a$ embodies the geometry of the band, the stretchability $k$ characterizes the ratio of bending resistance to stretching resistance. Smaller values of $k$ describe materials like paper, graphene, and DNA which bend easily but are difficult to stretch. The limiting case $k\rightarrow0$ of vanishing stretchability embodies the idealized limit of a material which cannot be stretched and thus can adopt only developable shapes. 

Next, using $A= Lw$ as a reference area, $k_lA$ as a reference energy, and defining
\begin{equation}
\label{nrgND_param}
\Psi = \frac{E}{k_{l}A}, 
%\Psi_s = \frac{E_s}{k_{l}A}, 
%\qquad
%\Psi_b = \frac{E_b}{k_{l}A}, 
\quad\,\,
\tilde r(i) = \frac{r(i)}{w},
\quad\,\, \text{and} \quad\,\,
\tilde r_e = \frac{r_{e}}{w}
\end{equation}
leads to a dimensionless version 
\begin{equation}
\label{nrgND_03}
\Psi=\Psi_s+\Psi_b,
\end{equation}
of the total energy \eqref{eq:E_01}, with stretching and bending contributions
\begin{equation}
%\Psi_s=\frac{1}{2a} \sum_{i \in S_{l}}(\tilde r(i)-\tilde r_e)^2
\Psi_s=\sum_{i \in S_{l}}\frac{(\tilde r(i)-\tilde r_e)^2}{2a}
\quad\text{and}\quad
\Psi_b=\sum_{i \in S_{\theta}}
\frac{k(\theta(i) - \theta_{e})^2}{2a}.
\end{equation}
For brevity, the dimensionless quantities $\Psi$, $\Psi_s$, and $\Psi_b$ are hereafter referred to as energies.

%%%%%%%%%%%%%%%%%%%%%%%%%%%%%%%%%%%%%%
\section{Simulation results} %{Stretchability leads to different shapes} %{In-plane stretchability leads to different shapes}
\label{sec:eqshps}
%%%%%%%%%%%%%%%%%%%%%%%%%%%%%%%%%%%%%%

Our simulations indicate that stretchable M\"obius bands adopt three characteristic equilibrium shapes, depending on the combination of stretchability $k$ and aspect ratio $a$. For small values of $k$, the bands take shapes that resemble those of paper models and previous simulation results. Increasing $k$ leads to a loss of developability, and, for a certain combination of low aspect ratio and high stretchability, a self-intersecting achiral shape emerges.

%%%%%%%%%%%%%%%%%%%%%%%%%%%%%
%\vspace{-4pt}
\subsection{Equilibrium shapes}
\label{ss:shapes}
%\vspace{-2pt}
%%%%%%%%%%%%%%%%%%%%%%%%%%%%%

\begin{figure*}[!htb]
\begin{center}
\hspace{6mm}\includegraphics[width=.1\textwidth, trim=0cm 0.1cm 29cm 1.0cm, clip=true] {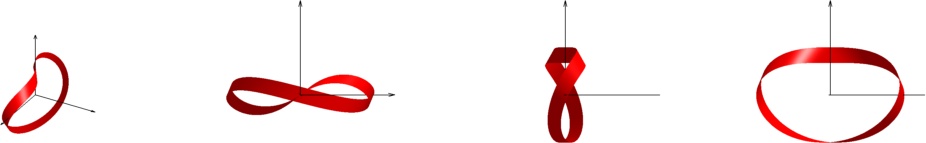}\hspace{3mm}
\includegraphics[width=.1\textwidth, trim=0cm 0.1cm 29cm 1.0cm, clip=true] {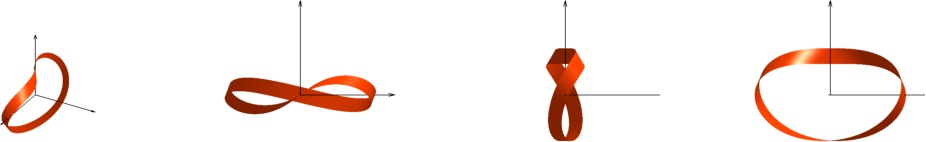}\hspace{3mm}
\includegraphics[width=.1\textwidth, trim=0cm 0.1cm 29cm 1.0cm, clip=true] {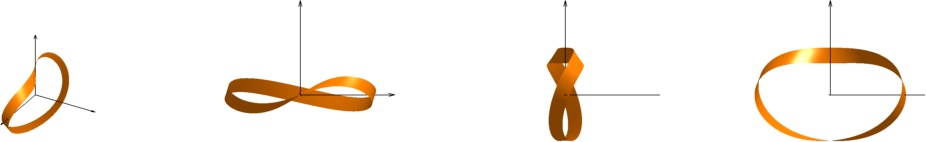}\hspace{3mm}
\includegraphics[width=.1\textwidth, trim=0cm 0.1cm 29cm 1.0cm, clip=true] {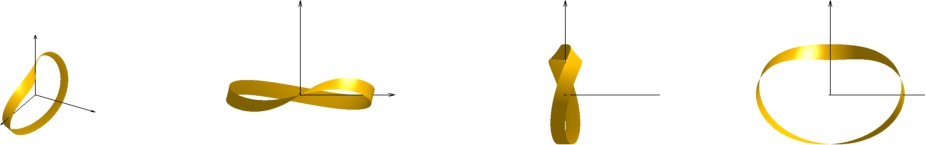}\hspace{3mm}
\includegraphics[width=.1\textwidth, trim=0cm 0.1cm 29cm 1.0cm, clip=true] {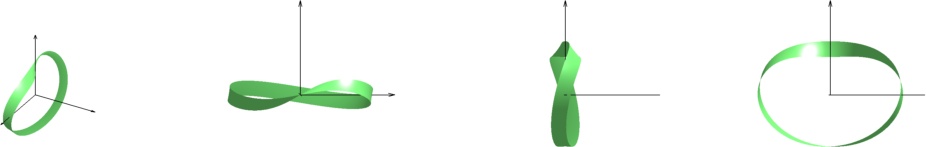}\hspace{3mm}
\includegraphics[width=.1\textwidth, trim=0cm 0.1cm 29cm 1.0cm, clip=true] {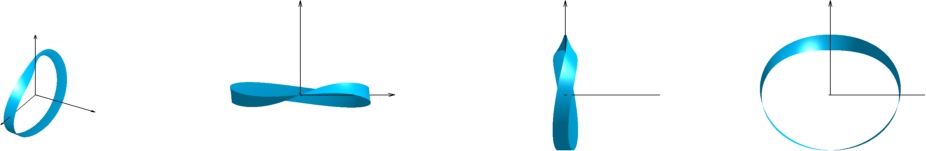}\hspace{3mm}
\includegraphics[width=.1\textwidth, trim=0cm 0.1cm 29cm 1.0cm, clip=true] {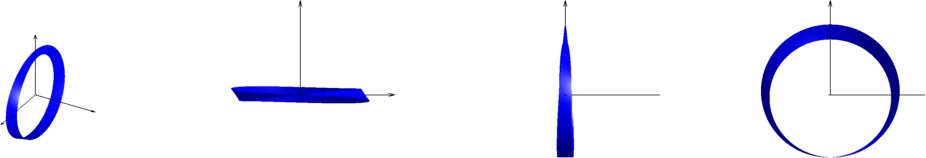}
\put(-458,20){\small$a= 8\pi$}

\hspace{6mm}\includegraphics[width=0.1\textwidth, trim=0cm 0.1cm 29cm 1.0cm, clip=true] {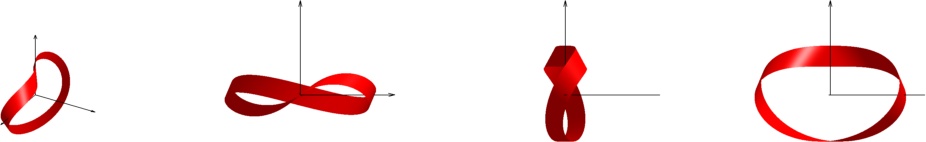}\hspace{3mm}
\includegraphics[width=0.1\textwidth, trim=0cm 0.1cm 29cm 1.0cm, clip=true] {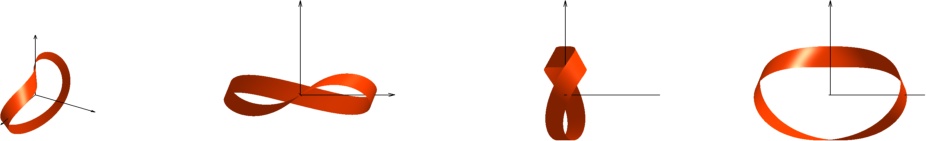}\hspace{3mm}
\includegraphics[width=0.1\textwidth, trim=0cm 0.1cm 29cm 1.0cm, clip=true] {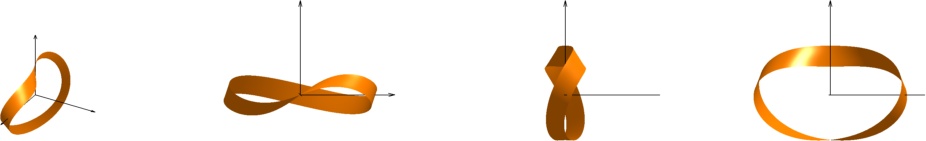}\hspace{3mm}
\includegraphics[width=0.1\textwidth, trim=0cm 0.1cm 29cm 1.0cm, clip=true] {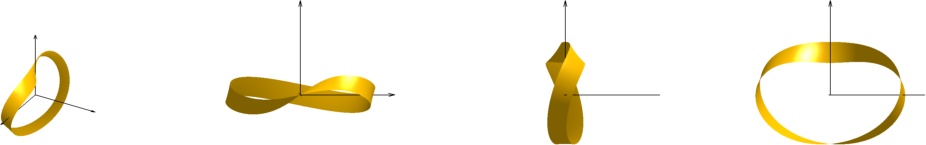}\hspace{3mm}
\includegraphics[width=.1\textwidth, trim=0cm 0.1cm 29cm 1.0cm, clip=true] {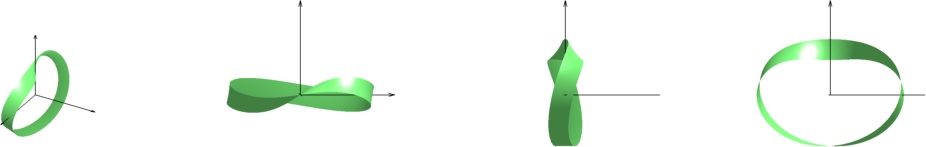}\hspace{3mm}
\includegraphics[width=.1\textwidth, trim=0cm 0.1cm 29cm 1.0cm, clip=true] {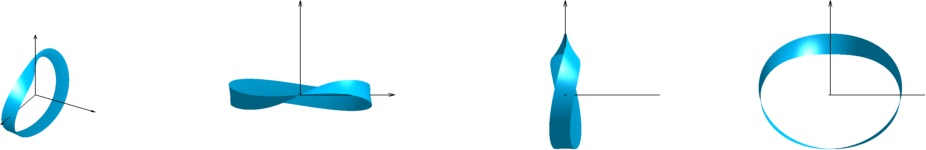}\hspace{3mm}
\includegraphics[width=.1\textwidth, trim=0cm 0.1cm 29cm 1.0cm, clip=true] {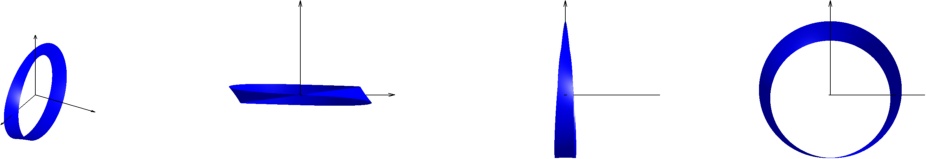}
\put(-458,20){\small$a= 6\pi$}

\hspace{6mm}\includegraphics[width=.1\textwidth, trim=0cm 0.1cm 29cm 1.0cm, clip=true] {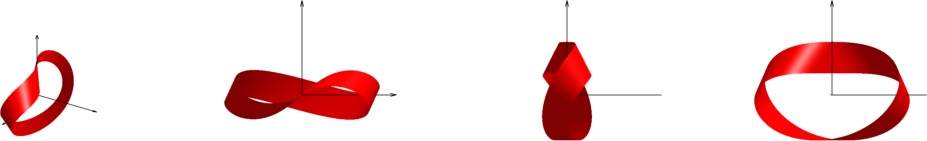}\hspace{3mm}
\includegraphics[width=.1\textwidth, trim=0cm 0.1cm 29cm 1.0cm, clip=true] {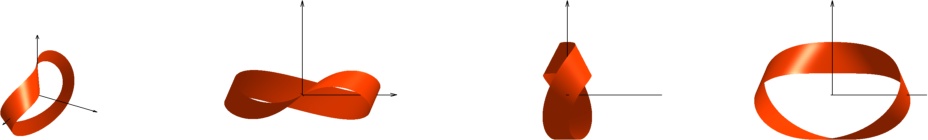}\hspace{3mm}
\includegraphics[width=.1\textwidth, trim=0cm 0.1cm 29cm 1.0cm, clip=true] {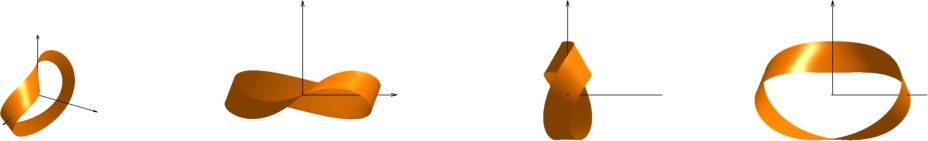}\hspace{3mm}
\includegraphics[width=.1\textwidth, trim=0cm 0.1cm 29cm 1.0cm, clip=true] {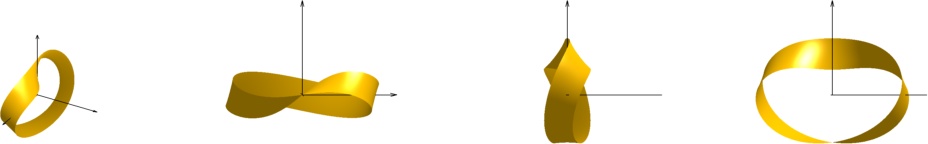}\hspace{3mm}
\includegraphics[width=.1\textwidth, trim=0cm 0.1cm 29cm 1.0cm, clip=true] {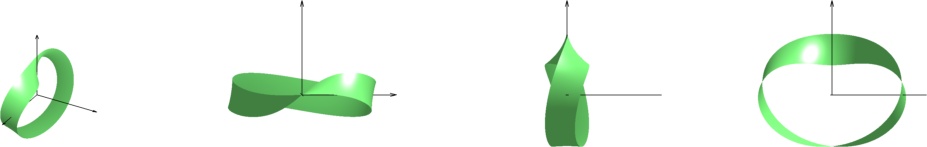}\hspace{3mm}
\includegraphics[width=.1\textwidth, trim=0cm 0.1cm 29cm 1.0cm, clip=true] {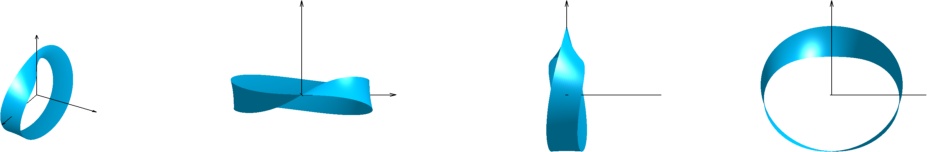}\hspace{3mm}
\includegraphics[width=.1\textwidth, trim=0cm 0.1cm 29cm 1.0cm, clip=true] {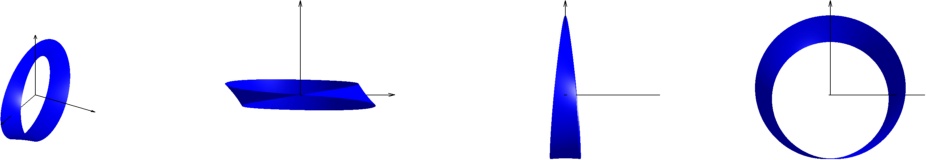}
\put(-458,20){\small$a= 4\pi$}

\hspace{6mm}\includegraphics[width=.1\textwidth, trim=0cm 0.1cm 29cm 1.0cm, clip=true] {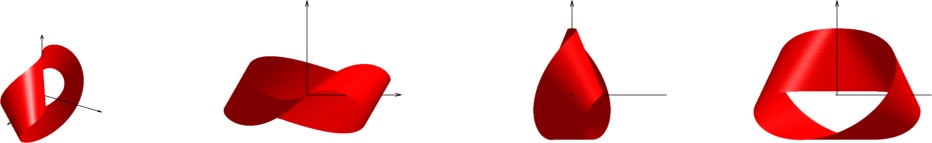}\hspace{3mm}
\includegraphics[width=.1\textwidth, trim=0cm 0.1cm 29cm 1.0cm, clip=true] {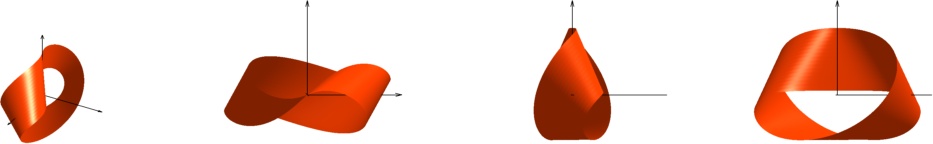}\hspace{3mm}
\includegraphics[width=.1\textwidth, trim=0cm 0.1cm 29cm 1.0cm, clip=true] {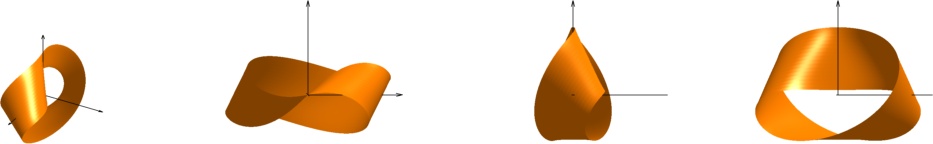}\hspace{3mm}
\includegraphics[width=.1\textwidth, trim=0cm 0.1cm 29cm 1.0cm, clip=true] {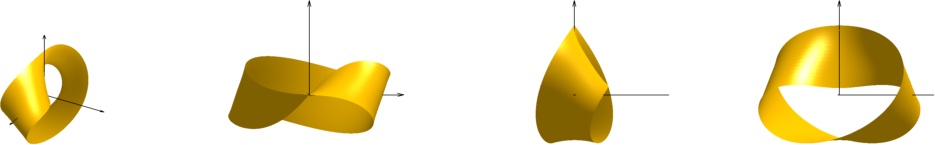}\hspace{3mm}
\includegraphics[width=.1\textwidth, trim=0cm 0.1cm 29cm 1.0cm, clip=true] {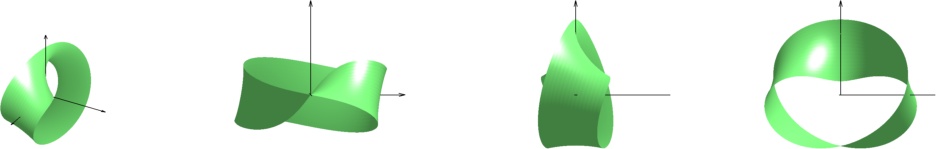}\hspace{3mm}
\includegraphics[width=.1\textwidth, trim=0cm 0.1cm 29cm 1.0cm, clip=true] {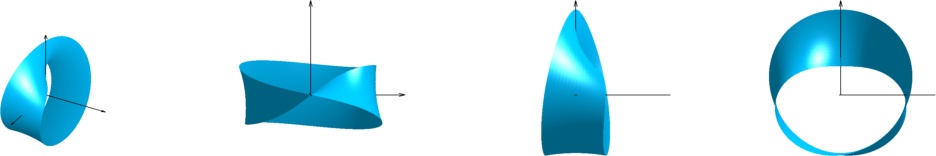}\hspace{3mm}
\includegraphics[width=.1\textwidth, trim=0cm 0.1cm 29cm 1.0cm, clip=true] {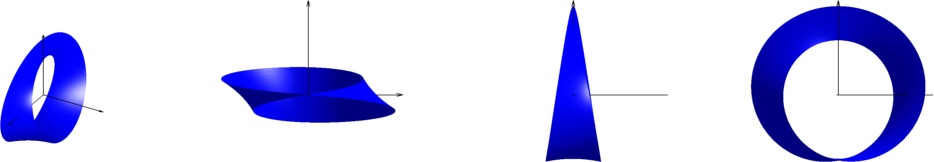}
\put(-458,20){\small$a= 2\pi$}

\hspace{6mm}\includegraphics[width=.1\textwidth, trim=0cm 0.1cm 29cm 1.0cm, clip=true] {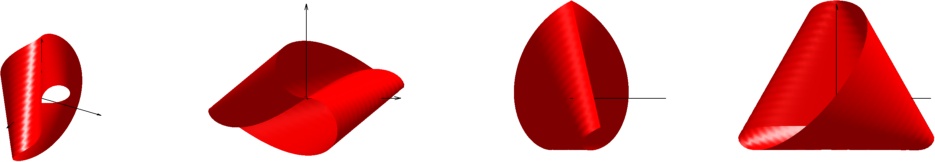}\put(-45,-14){\small$k= 10^{-6}$}\hspace{3mm}
\includegraphics[width=.1\textwidth, trim=0cm 0.1cm 29cm 1.0cm, clip=true] {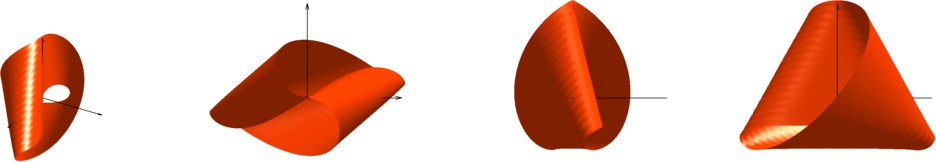}\put(-45,-14){\small$k= 10^{-5}$}\hspace{3mm}
\includegraphics[width=.1\textwidth, trim=0cm 0.1cm 29cm 1.0cm, clip=true] {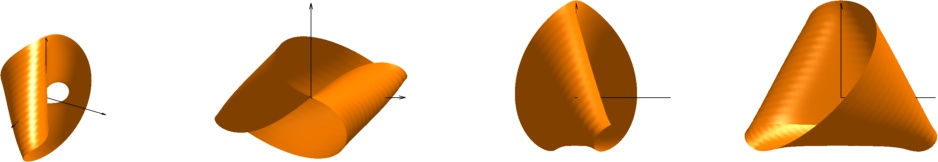}\put(-45,-14){\small$k= 10^{-4}$}\hspace{3mm}
\includegraphics[width=.1\textwidth, trim=0cm 0.1cm 29cm 1.0cm, clip=true] {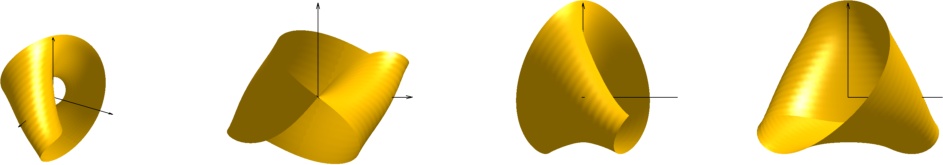}\put(-45,-14){\small$k= 10^{-3}$}\hspace{3mm}
\includegraphics[width=.1\textwidth, trim=0cm 0.1cm 29cm 1.0cm, clip=true] {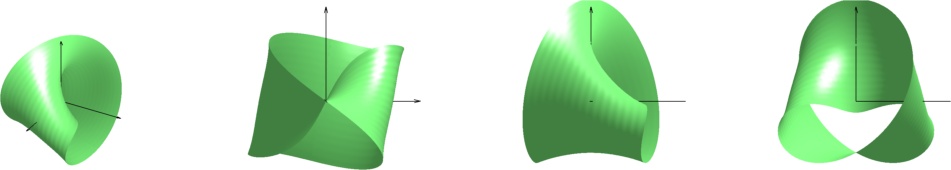}\put(-45,-14){\small$k= 10^{-2}$}\hspace{3mm}
\includegraphics[width=.1\textwidth, trim=0cm 0.1cm 29cm 1.0cm, clip=true] {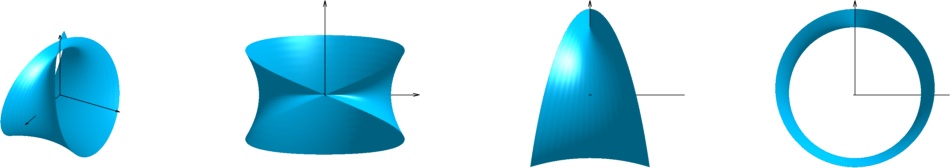}\put(-45,-14){\small$k= 10^{-1}$}\hspace{3mm}
\includegraphics[width=.1\textwidth, trim=0cm 0.1cm 29cm 1.0cm, clip=true] {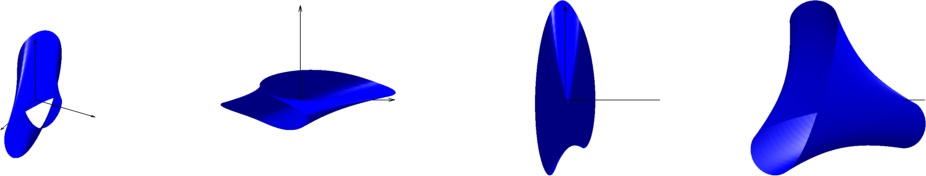}\put(-45,-12){\small$k=10^0$}
\put(-458,20){\small$a= \pi$}
%\put(-265,-8){\small$k= 10^0$}
%\put(-225,-8){\small$k= 10^{-1}$}
%\put(-185,-8){\small$k= 10^{-2}$}
%\put(-145,-8){\small$k= 10^{-3}$}
%\put(-105,-8){\small$k= 10^{-4}$}
%\put(-65,-8){\small$k= 10^{-5}$}
%\put(-25,-8){\small$k= 10^{-6}$}
\end{center}
\caption{M\"obius bands adopt characteristic equilibrium shapes depending on stretchability, $k$, and aspect ratio, $a$: Equilibrium shapes for different values of $a$ and ${k}$. Bands are rotated into their main axes. For views from other perspectives, see SI Figure \ref{fig:Moebius_shapeAll}.}
\label{fig:Moebius_shape}
\end{figure*}

\begin{figure}[!b]
\centering
\vspace{-8pt}
\includegraphics[width=.425\textwidth]{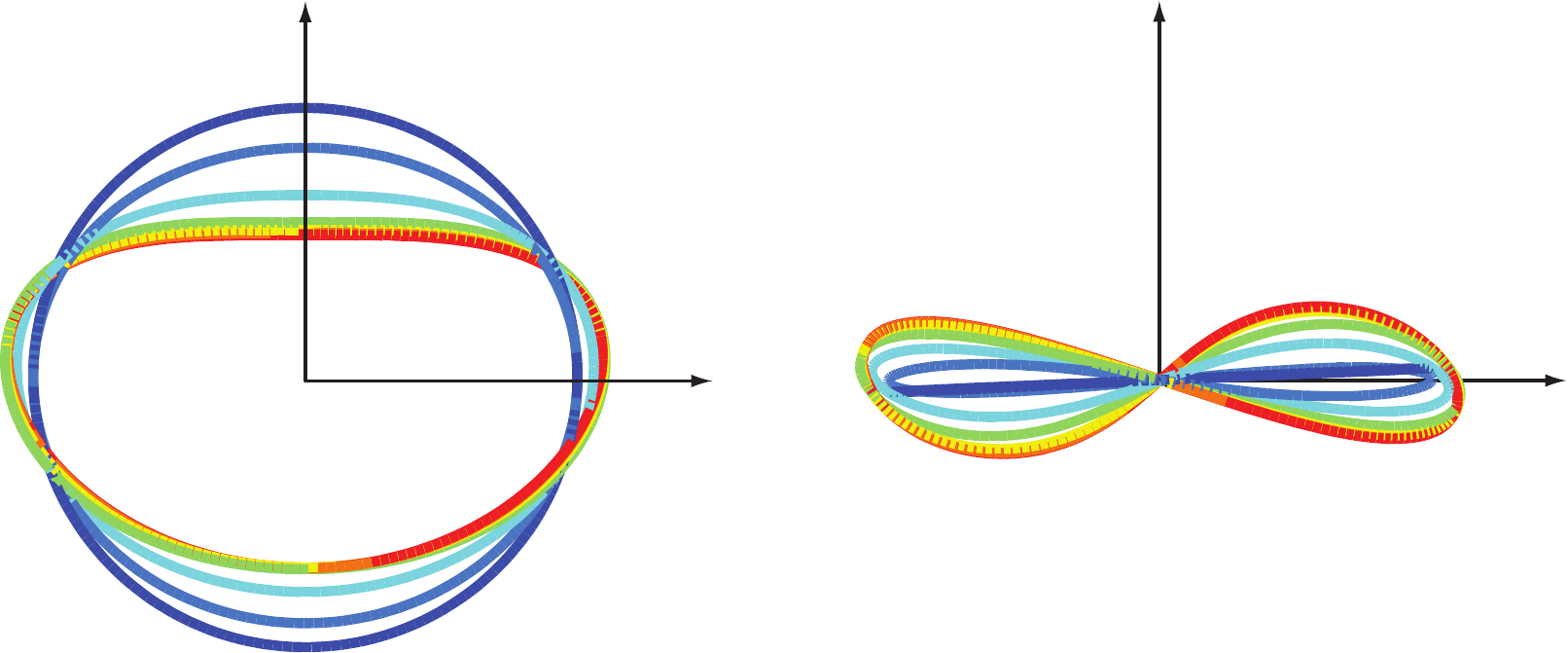}
%\put(-95,96){\scriptsize$k=10^{-6}$}
%\put(-47.75,100.5){$k=10^{-5}$}
%\put(-47.75,88.5){$k=10^{-4}$}
%\put(-47.75,76.25){$k=10^{-3}$}
%\put(-47.75,64.25){$k=10^{-2}$}
%\put(-47.75,52.125){$k=10^{-1}$}
%\put(-47.75,40){$k=10^{0}$}
%\put(-68,140.0){$k=2\pi$}
%\put(-68,130.5){$a=4\pi$}
%\put(-68,121.0){$a=6\pi$}
%\put(-68,111.5){$a=8\pi$}
\put(-116,35){\small$x$}
\put(-175.5,93){\small$z$}
\put(1,35){\small$x$}
\put(-58.5,93){\small$y$}
\put(-130,54){\includegraphics[width=.12\textwidth]{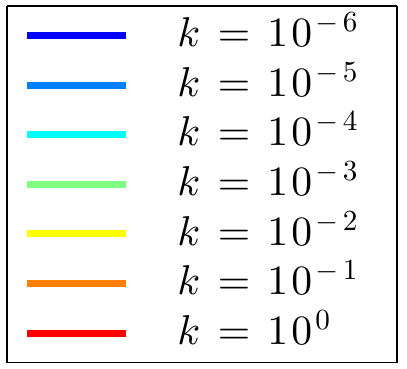}}
%\put(-182,186){$z$}
%\put(-161,21.25){$x$}
%\put(-210,73){$y$}
%\put(-69.5,21.25){$x$}
%\put(-118.5,73){$z$}
%\put(2,21.25){$y$}
%\put(-48,73){$z$}
\caption{Centerlines of the equilibrium shape of bands with aspect ratio $a=\pi$ made from materials of various stretchabilities $k$.}
\label{fig:centerlinesk}
\end{figure}

Equilibrium band shapes obtained by minimizing the total energy $\Psi$ are provided in Figure \ref{fig:Moebius_shape} for representative combinations of $k$ and $a$.
%
%%%%%%%%%%%%%%%
%%%%%%%%%%%%%%%
%
For sufficiently small values of $k$ and each value of $a$ considered, equilibrium shapes qualitatively resemble those of model bands made from rectangular strips of paper and thus appear to be nearly developable. Regardless of the value of $a$, reducing $k$ appears to have a negligible influence below $k=10^{-4}$. However, the influence of $k$ becomes progressively more evident above $k=10^{-4}$ and is increasingly obvious for smaller values of $a$. The centerlines of bands appear to be more circular and less out of plane for larger values of $k$, an impression that is confirmed by plots for bands of aspect ratio $a=2\pi$ that appear in Figure \ref{fig:centerlinesk}.

%%%%%%%%%%%%%%%
\begin{figure}[!t]
%\vspace{-12pt}
\centering
%\begin{picture}(500,210)
%\put(10,30){
\includegraphics[width=.975\linewidth]{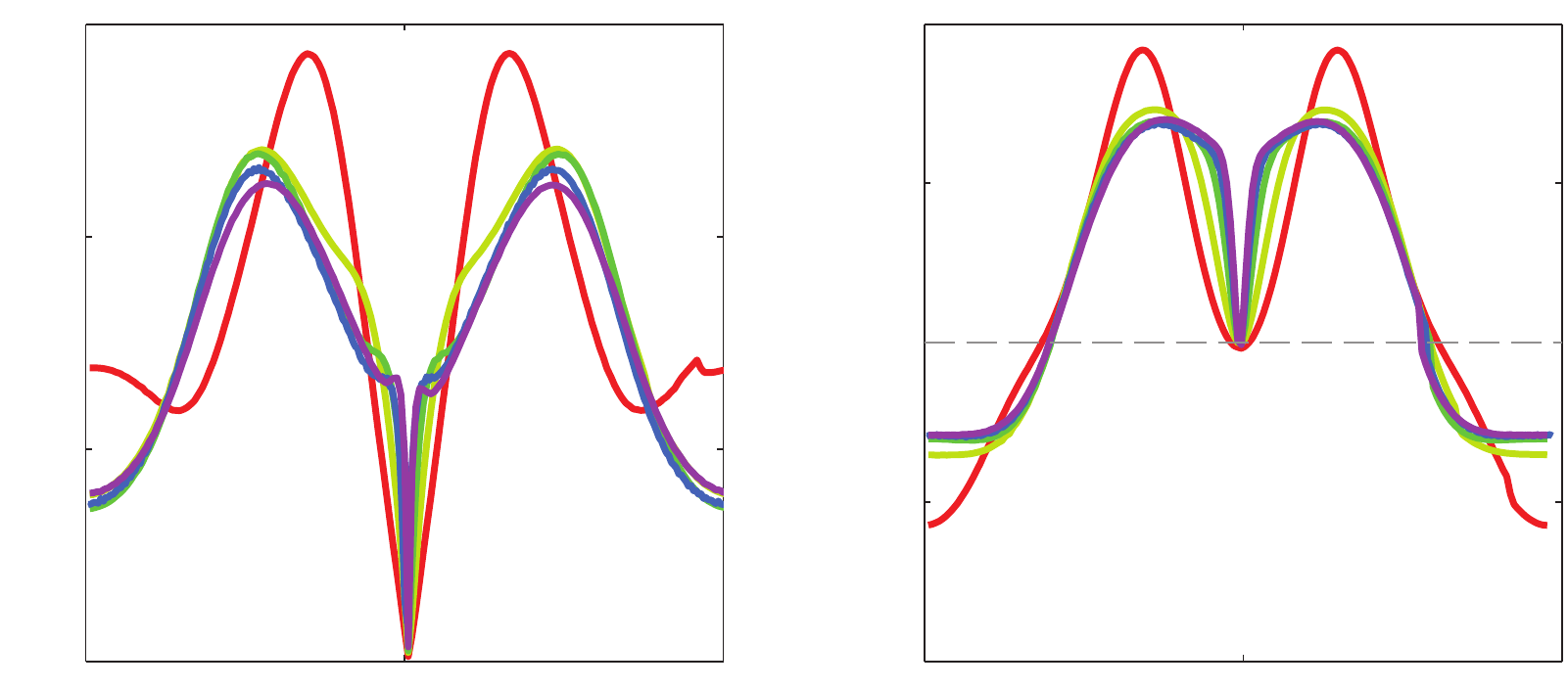}%}
\put(-229,-5){\small$1$}
\put(-189,-5){\small$N_l/2$}
\put(-135,-5){\small$N_l$}
\put(-102,-5){\small$1$}
\put(-62,-5){\small$N_l/2$}
\put(-8,-5){\small$N_l$}
%\put(105,25){\large$N_l/2$}
\put(-81.5,8){\includegraphics[width=.24\linewidth]{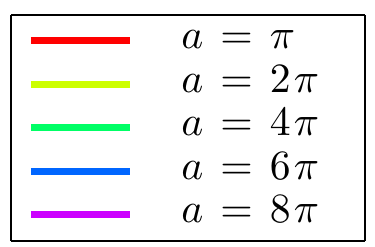}}
\put(-242,50.5){\rotatebox{90}{$\kappa$}}
\put(-116,50.5){\rotatebox{90}{$\tau$}}
\put(-234.5,2){\small$0$}
\put(-234.5,34){\small$5$}
\put(-239.3,67){\small$10$}
\put(-239.3,99.4){\small$15$}
\put(-117,2){\small$-10$}
\put(-112.5,26){\small$-5$}
\put(-112.5,51){\small$\phantom{-}0$}
\put(-112.5,75.25){\small$\phantom{-}5$}
\put(-117,99.4){\small$\phantom{-}10$}
%\put(113,10){\large$s$}
%\put(193,25){\large$N_l$}
%\put(250,25){\large$1$}
%\put(322,25){\large$N_l/2$}
%\put(330,10){\large$s$}
%\put(410,25){\large$N_l$}
%\end{picture}
\caption{In the unstretchable limit, the model recovers analytical predictions of Randrup and R{\o}gen~\cite{Randrup1996} and the characteristic shape obtained by Starostin and van der Heijden.\cite{Starostin2007,Starostin2007a} Curvature $\kappa$ and torsion $\tau$ of the centerline of M\"obius bands for stretchability $k=10^{-6}$ and various values of the aspect ratio $a$ versus the arclength along the centerline in terms of the number $N_l$ of points along the centerline.}
\label{fig:curvatureTwist}
\vspace{-14pt}
\end{figure}
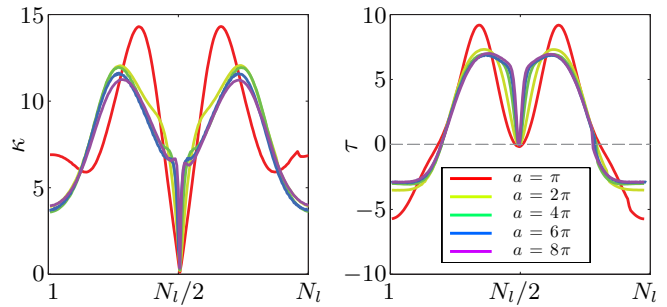
\begin{figure}[!t]
\centering
\includegraphics[width=.425\textwidth]{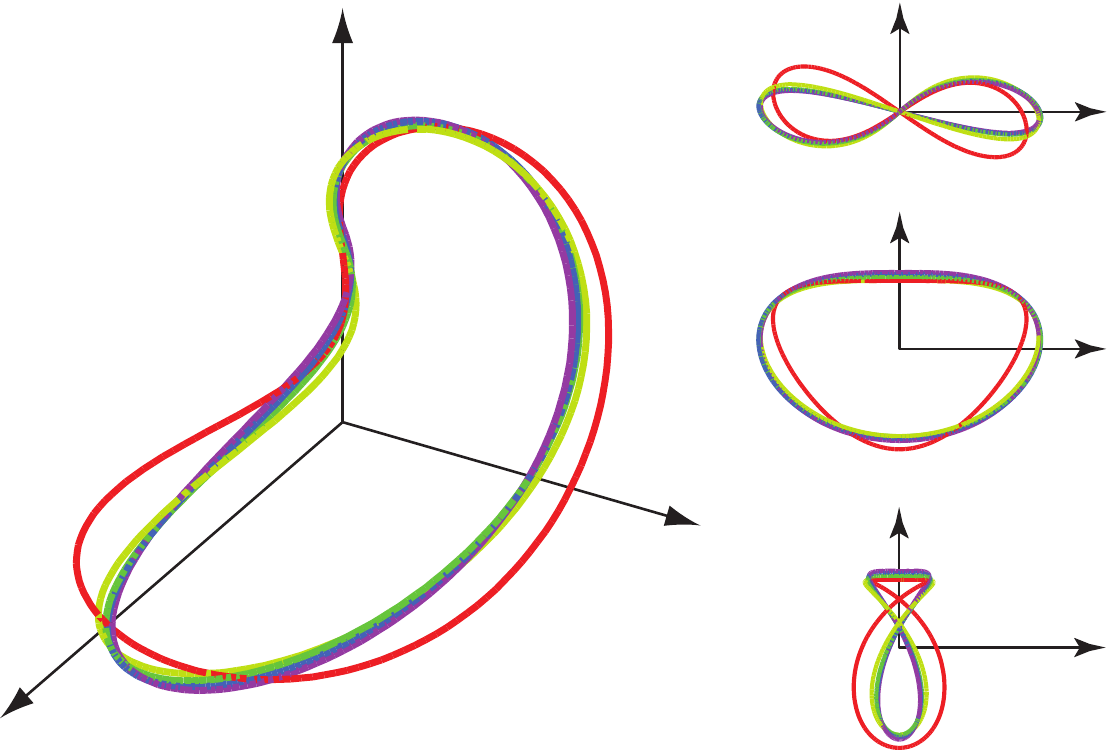}
%\put(-68,149.5){$a=\pi$}
%\put(-68,140.0){$a=2\pi$}
%\put(-68,130.5){$a=4\pi$}
%\put(-68,121.0){$a=6\pi$}
%\put(-68,111.5){$a=8\pi$}
\put(-222,2){\small$x$}
\put(-77,40.5){\small$y$}
\put(-151,146){\small$z$}
\put(2.75,121.5){\small$x$}
\put(-43,150){\small$y$}
\put(2.75,75.5){\small$x$}
\put(-43,107.5){\small$z$}
\put(2.75,18){\small$y$}
\put(-43,49){\small$z$}
\put(-224,86){\includegraphics[width=.13\textwidth]{figlegenda}}
\caption{The shape of the centerline of an effectively unstretchable M\"obius band depends on its aspect ratio $a$: Centerlines of equilibrium shapes of M\"obius bands for stretchability $k=10^{-6}$ and representative values of $a$. Bands are rotated into their main axes as in Figure \ref{fig:Moebius_shape}.}
\label{fig:centerlinesa}
\end{figure}

%%%%%%%%%%%%%%%
%
If $k$ is sufficiently small, the tangent vector to the centerline of each equilibrated band exhibits an odd number of switching points, at which its curvature, $\kappa$, and torsion, $\tau$, vanish simultaneously. While $\kappa$ has two peaks and one zero, $\tau$ has two peaks and three zeroes (Figure \ref{fig:curvatureTwist}). These findings are consistent with previous analytical and numerical results.\cite{Randrup1996, Starostin2007, Starostin2007a} Moreover, for smaller values of $a$, the peak values of $\kappa$ and $\tau$ are higher. In keeping with findings of Mahadevan and Keller\cite{Mahadevan1993} and Starostin and van der Heijden,\cite{Starostin2007, Starostin2007a} the centerlines of these bands are more out-of-plane than those of bands with larger aspect ratios (Figure \ref{fig:centerlinesa}). See the SI for comprehensive convergence and validation studies in the unstretchable limit $k\to0$. % and further discussion

For $a=\pi$, bands made of materials with stretchabilities $k=10^{-1}$ and $k=10^{-2/3}$ exhibit self-intersecting achiral equilibrium shapes. This degeneration, which is possible only if unpenalized self-intersections are allowed, resembles the collapse observed in M\"obius soap films with small throat distances.\cite{Goldstein2010} However, the constraint of lattice connectivity inherent to our model delivers shapes different from those of collapsed M\"obius soap films. Additionally, the choice $k=10^{-1/3}$ yields not a band but a flat annulus.

%%%%%%%%%%%%%%%%%%%%%%%%%%%%%
%\vspace{-4pt}
\subsection{Energy measures}
\label{ss:energy}
%\vspace{-2pt}
%%%%%%%%%%%%%%%%%%%%%%%%%%%%%

\begin{figure}[!b]
%\centering
%\begin{center}
%\includegraphics[width=.28\textwidth] {Psi_vs_a}
%\includegraphics[width=.265\textwidth] {Psi_S_vs_a}
%\includegraphics[width=.28\textwidth]{Psi_B_vs_a}

\hspace{8pt}
\includegraphics[width=.2125\textwidth] {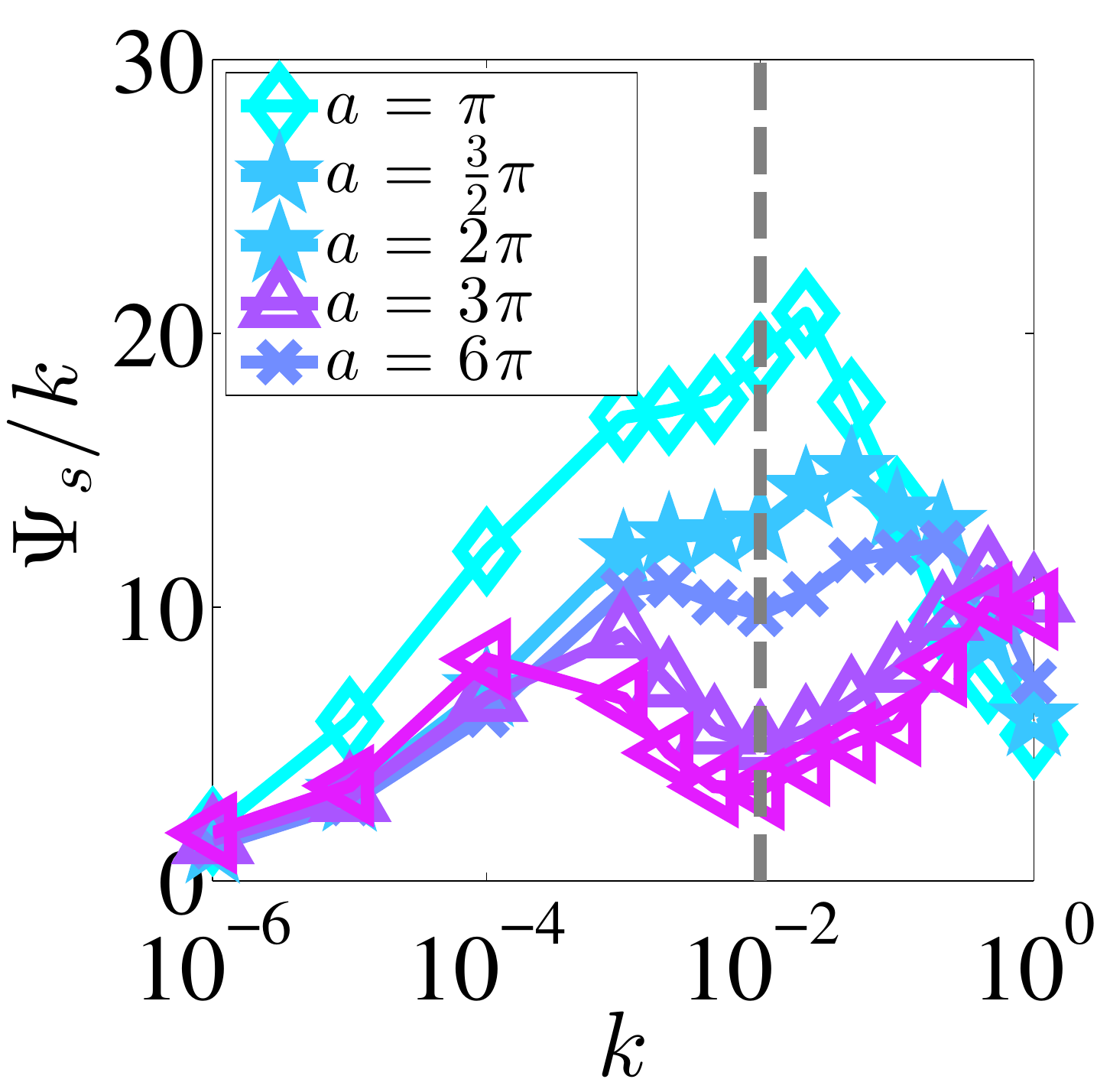}\put(-22,89){(a)}
\hspace{1pt}
\includegraphics[width=.2155\textwidth] {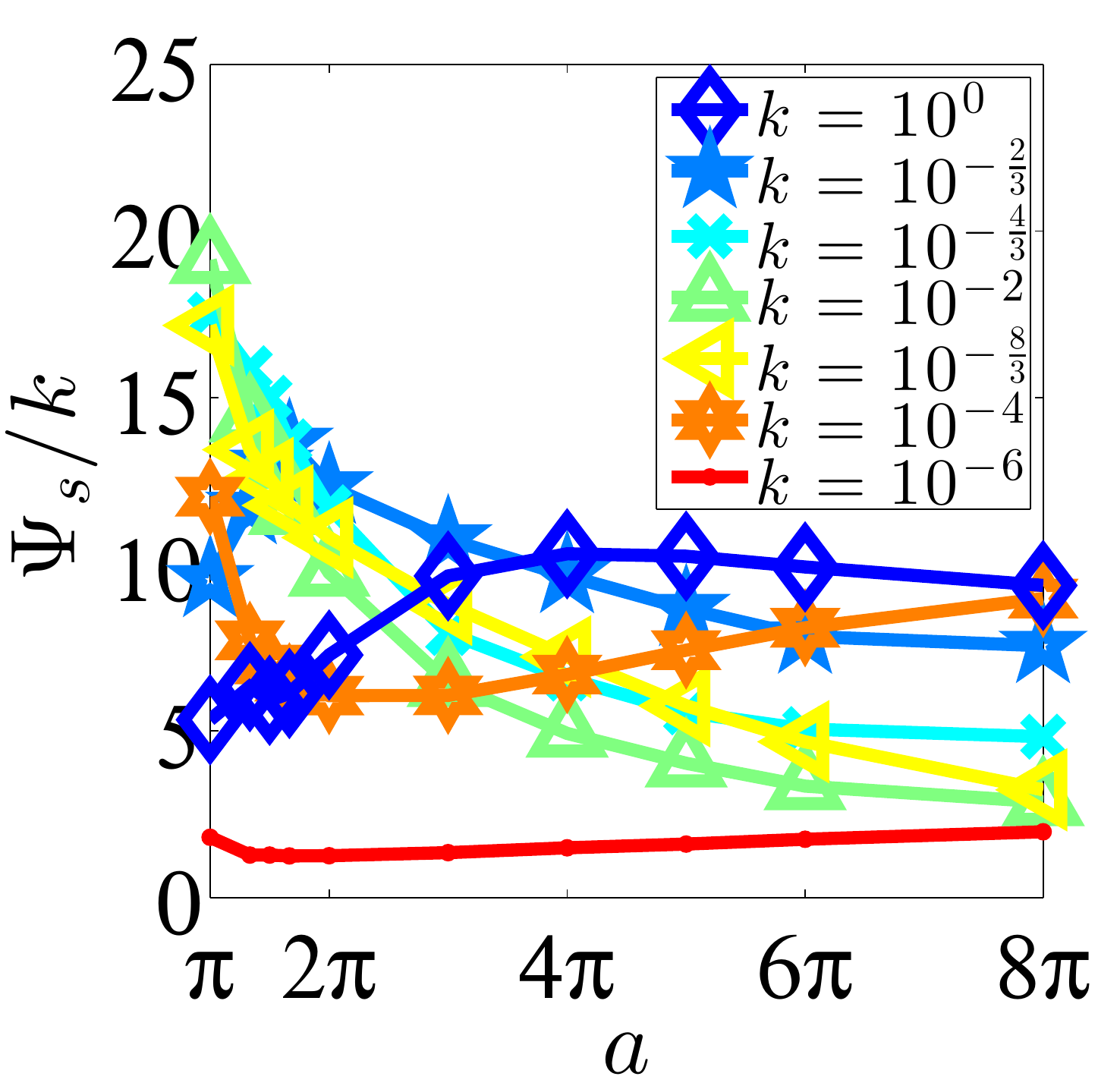}\put(-85,89){(d)}

\hspace{4pt}
\includegraphics[width=.2221\textwidth] {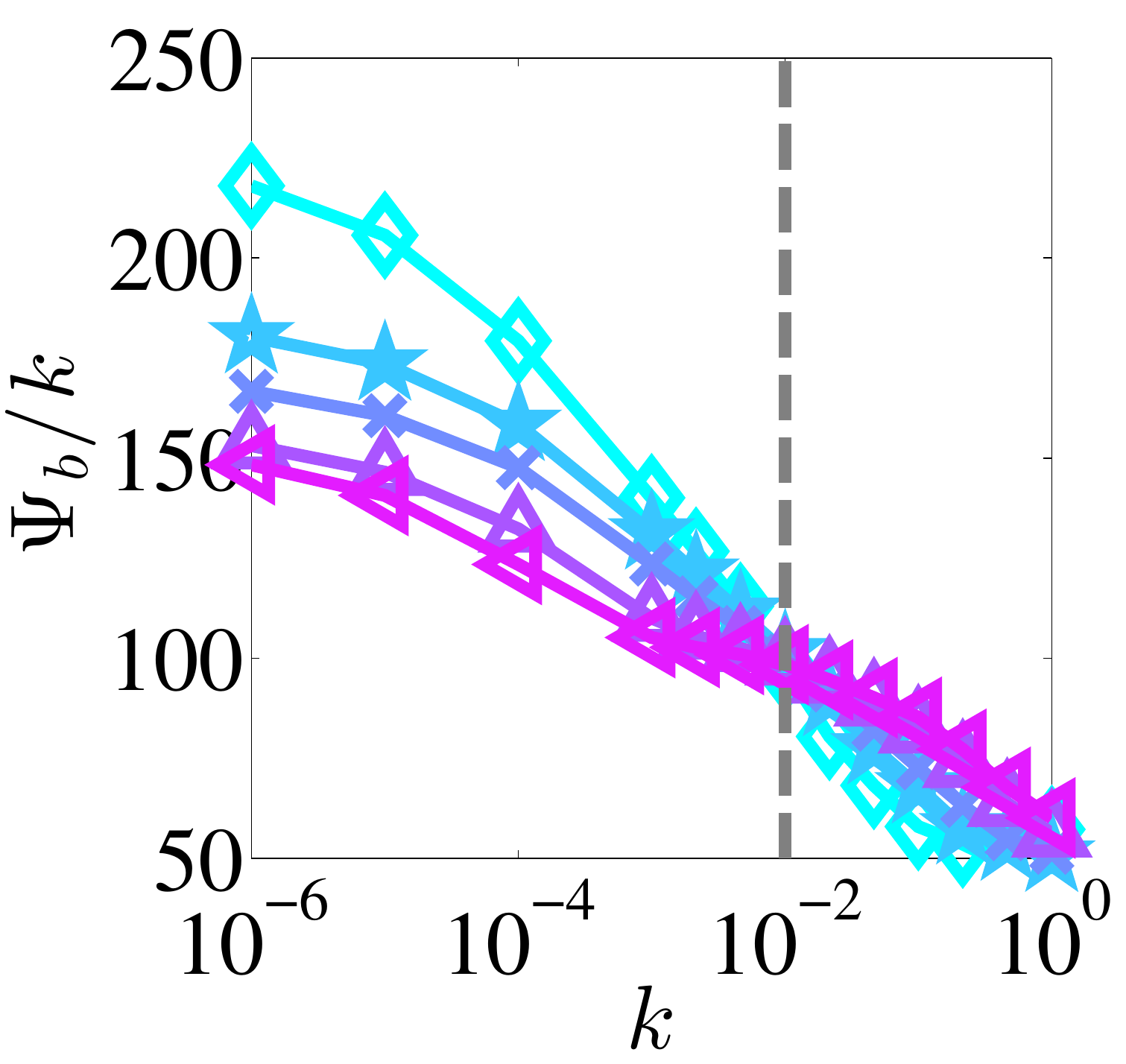}\put(-22,89){(b)}
%\vspace{12pt}
\includegraphics[width=.2252\textwidth] {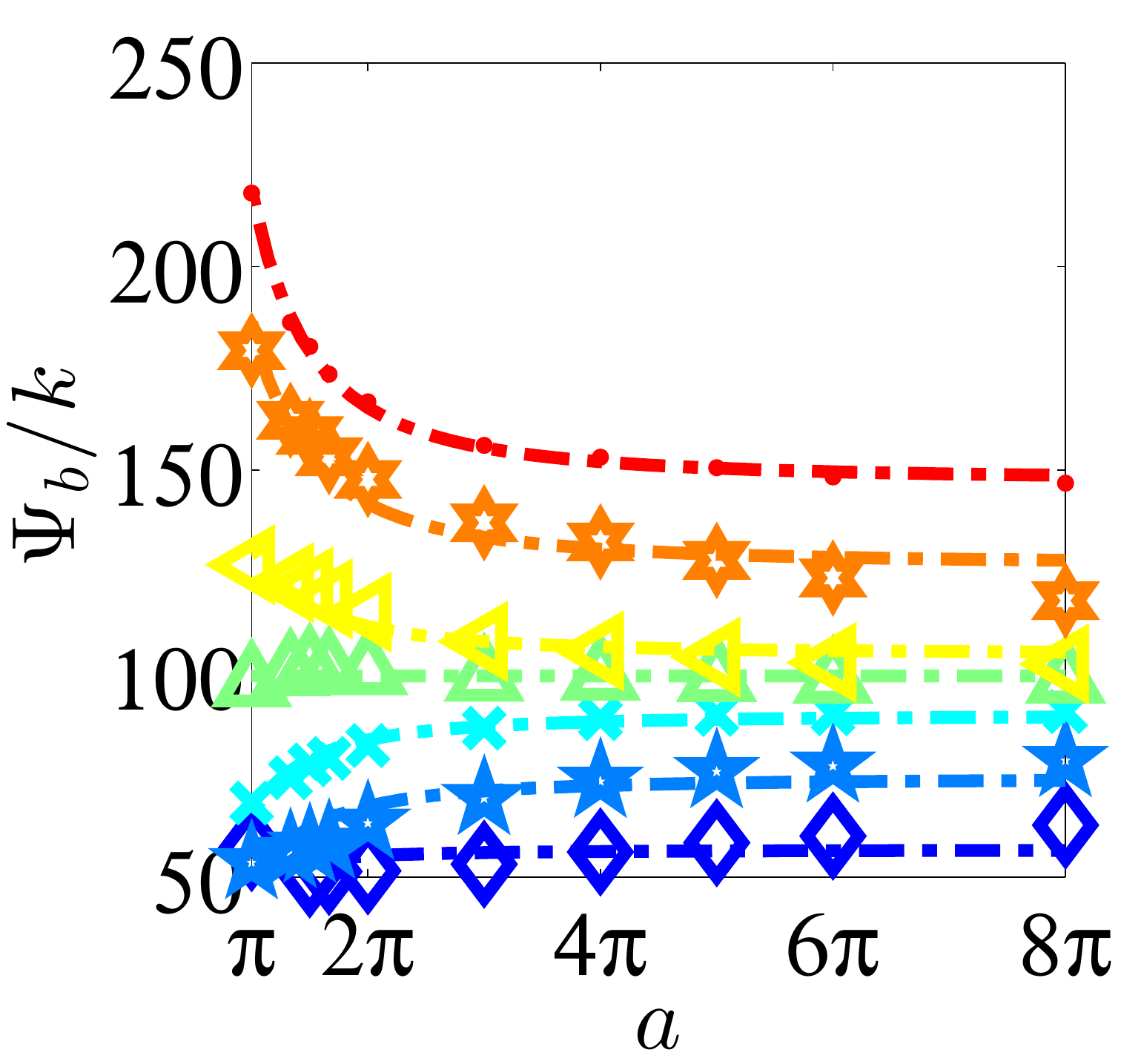}\put(-22,89){(e)}

\hspace{4pt}
\includegraphics[width=.2221\textwidth] {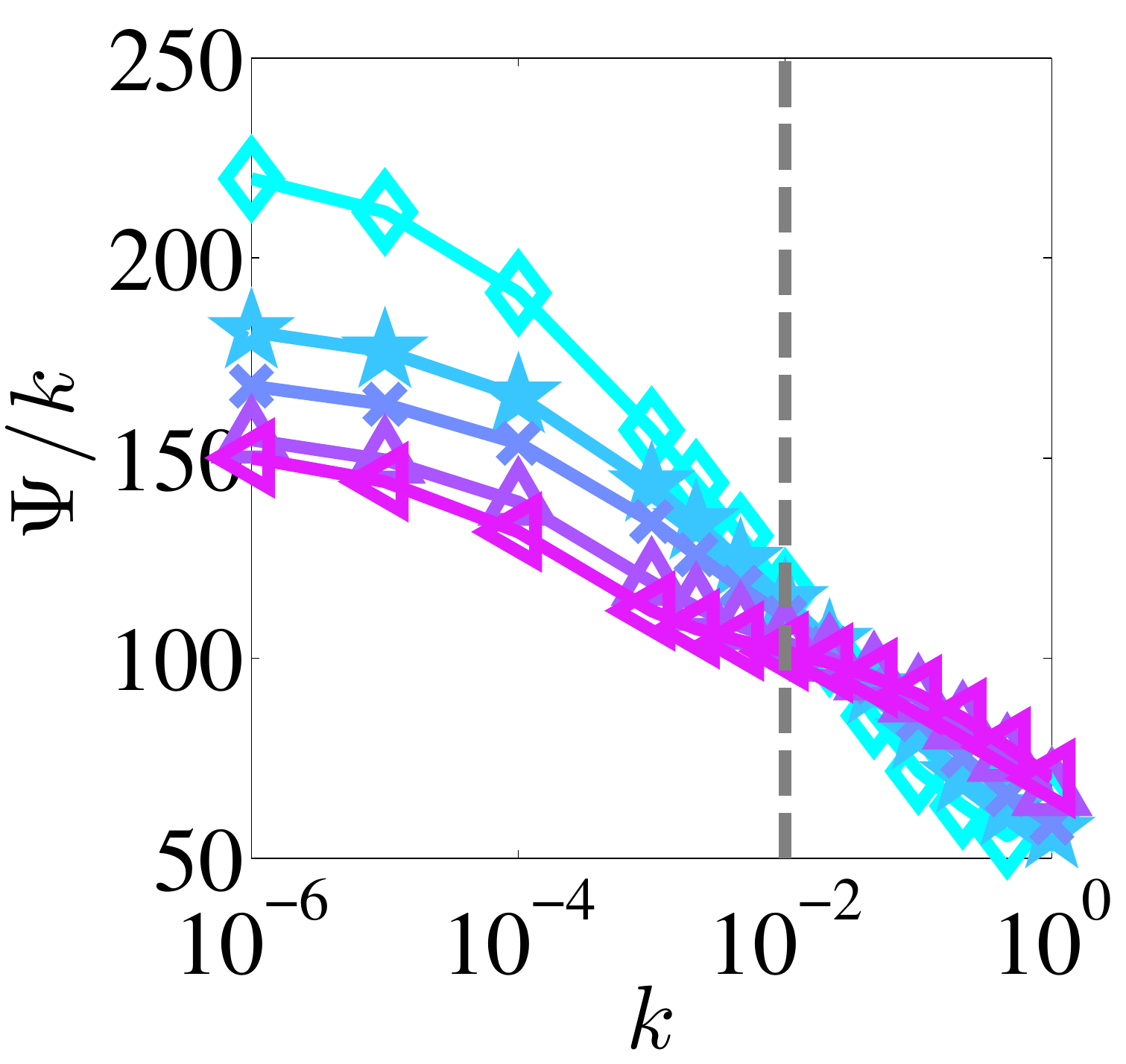}\put(-22,89){(c)}
%\vspace{12pt}
\includegraphics[width=.2252\textwidth]{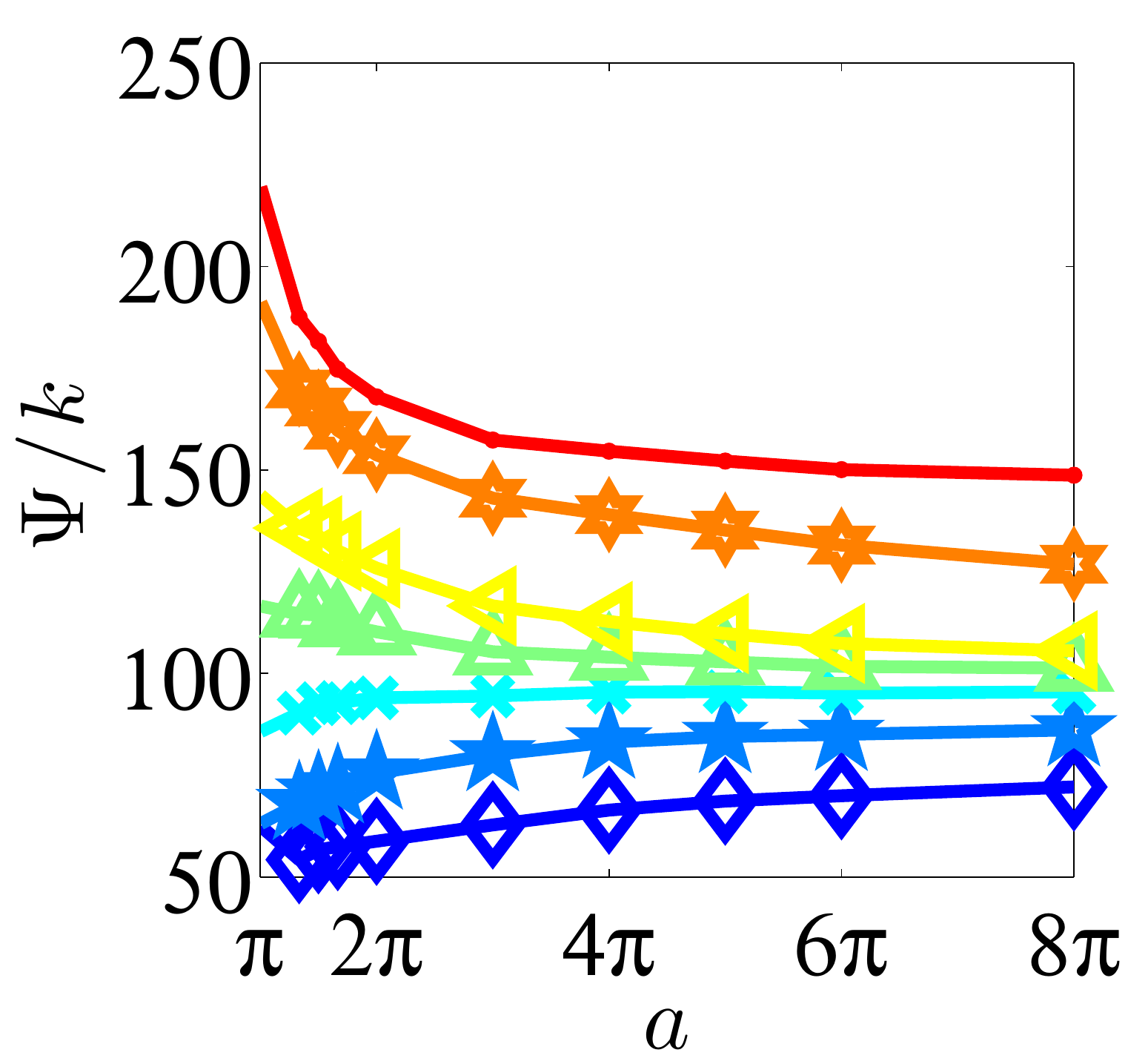}\put(-22,89){(f)}
%\includegraphics[width=.25\textwidth] {Psi_1_vs_k}\put(-92,98){(d)}

%\put(222,194){(g.i)}
%\put(265,194){(g.ii)}

%\includegraphics[width=.25\textwidth] {Psi_2_vs_k}\put(-92,100){(h)}
%\end{center}
%
\caption{Stretching energy $\Psi_{s}$, bending energy $\Psi_{b}$, and total energy $\Psi$ for representative combinations of stretchability, $k$, and aspect ratio, $a$. These quantities are normalized by $k$ to encompass changes in overall stiffness.}
\label{fig:Moebius_energy}
\end{figure}

Plots of the stretching energy $\Psi_{s}$, the bending energy $\Psi_{b}$, and the total energy $\Psi$, all normalized by $k$ to encompass changes in overall stiffness, are provided in Figure \ref{fig:Moebius_energy} for representative combinations of $k$ and $a$. For each choice of $a$, $\Psi_s$ first increases as $k$ increases but is dominated by the order-of-magnitude larger $\Psi_b$, which decreases. Hence, the sum $\Psi=\Psi_s+\Psi_b$ decreases monotonically as $k$ increases and the minimum value of $\Psi$ for any band is attained at the largest value of $k$ considered. From an energetic perspective, it therefore seems reasonable to infer that bands made of stretchable materials are significantly less costly to make than bands made of unstretchable materials.

%
%%%%%%%%%%%%%%%
%%%%%%%%%%%%%%%

The influence of $a$ is most evident for $k$ less than a critical value $k_c\sim10^{-2}$. For each $k$ less than $k_c$, $\Psi$ decreases monotonically with $a$. This effect is in keeping with everyday experience: the effort needed to twist a rectangular sheet of an essentially unstretchable material like paper into a M\"obius band increases notably as its length-to-width aspect ratio deceases. Moreover, performing the same task with a stretchable material, such as a thin sheet of rubber, is substantially easier. Interestingly, however, for $k>k_c$ and $a$ sufficiently small, $\Psi$ appears to increase, albeit weakly, with increasing $a$.

With reference to the previously noted observation that the centerlines of bands become more circular and less out-of-plane as $k$ increases, the observed dependence of $\Psi$ on $k$ indicates that more planar, less bent configurations are energetically favored above a certain critical stretchability. For the particular choice $k=10^{-6}$, $\Psi_s$ is negligibly small in comparison to $\Psi_b$ and, thus, $\Psi\sim\Psi_b$. The choice $k=10^{-6}$ therefore suffices to capture the strictly unstretchable limit $k=0$. The influence of $k$ is otherwise most significant near that limit. This provides potentially important insight regarding the guided assembly or synthesis of M\"obius bands, as it indicates that small increases in stretchability yield significant gains in the ease of engineering. The synthesis of M\"obius bands is challenging mainly due to their significantly higher energy states relative to untwisted rings. However, allowing for stretching diminishes this energy difference and could therefore reduce the difficulty of fabrication strategies that rely on bending rectangular strips.

Since $\Psi$ decreases as $k$ increases for each value of $a$ considered, the loss of chirality exhibited by bands of aspect ratio $a=\pi$ made from materials with stretchabilities $k=10^{-1}$ and $k = 10^0$ indicates that chirality is is accompanied by increased stored energy. To fabricate chiral objects, a system must therefore be sufficiently and properly constrained. For example, anisotropic particles can be used to guide the assembly of chiral objects.\cite{Fejer2007} However, we speculate that, for $k$ sufficiently large, non-collapsed bands are at best metastable, since $\Psi_s$ exhibits oscillations at $k = 10^0$ observed for small values of $a$. If a material is too stretchable, then small aspect ratio M\"obius bands are expected to be unstable. This provides a threshold for the stretchability below which the synthesis or guided assembly of small aspect ratio M\"obius bands should become feasible.

%%%%%%%%%%%%%%%
\begin{figure*}[!htb]

\hspace{1.5cm}
\includegraphics[height=.11\textwidth] {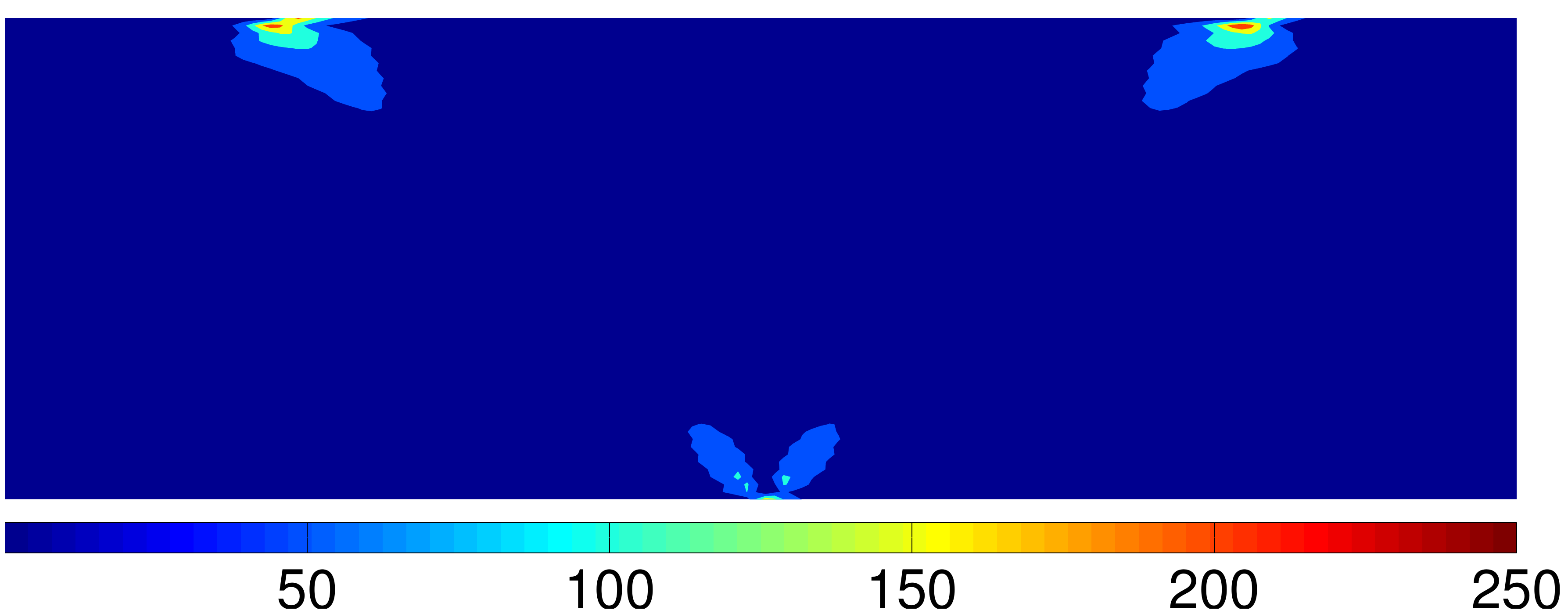}
\put(-174,20){\small$a = \pi$}
\put(-182,37){\small(a) $H^2$}
\put(-85,57){\small$k = 10^{-6}$}
\hspace{0.3mm}
\includegraphics[height=.11\textwidth] {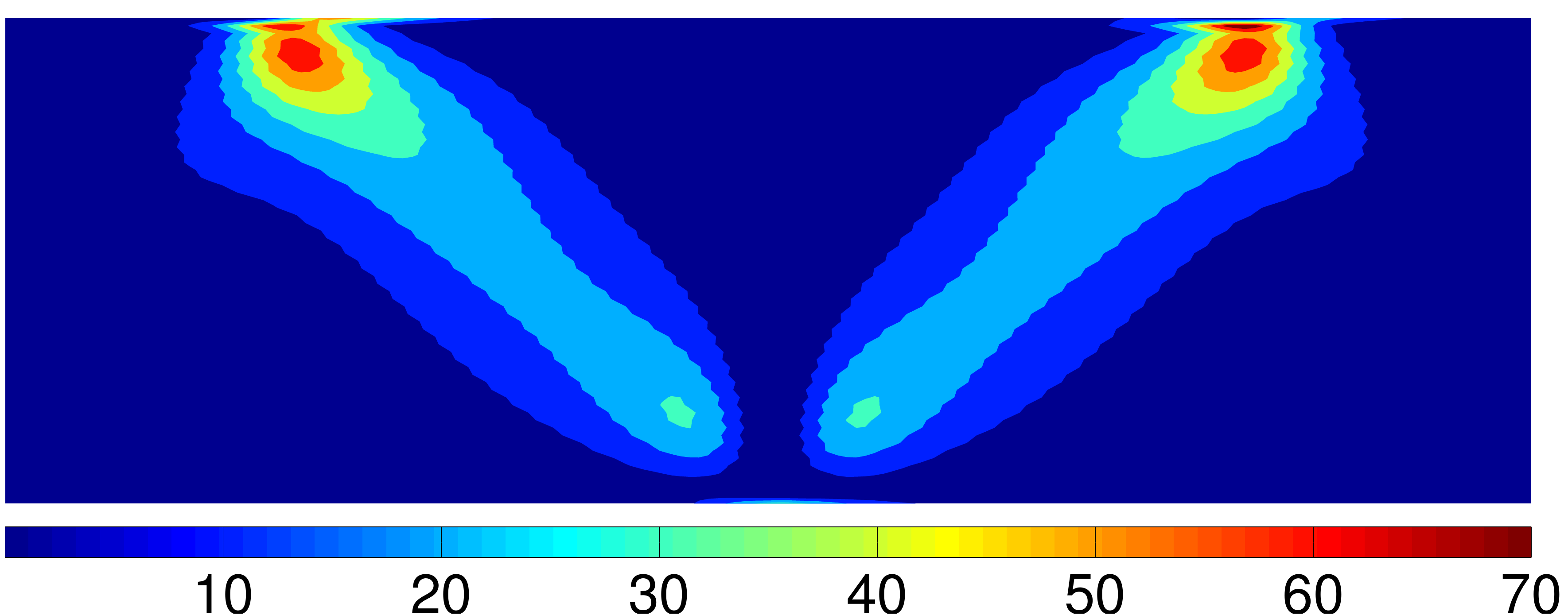}
\put(-85,57){\small$k = 10^{-4}$}
\hspace{1.4mm}
\includegraphics[height=.11\textwidth] {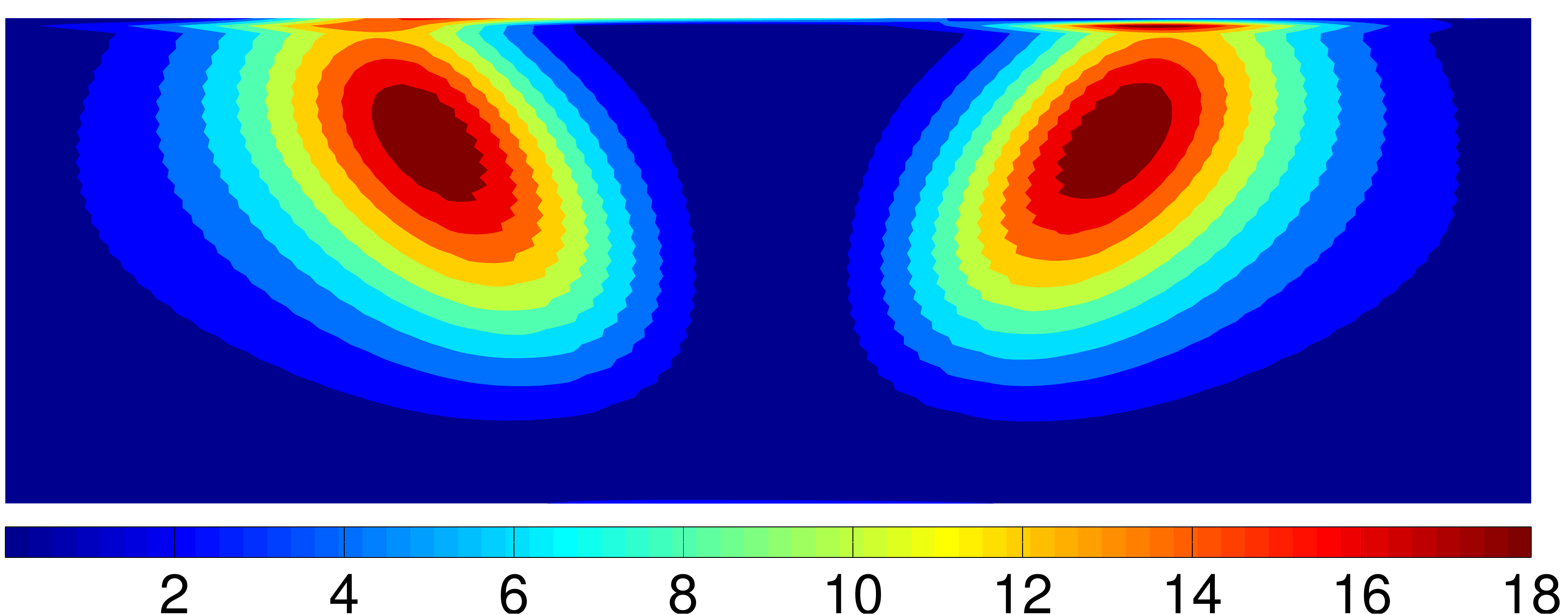}
\put(-85,57){\small$k = 10^{-2}$}

\hspace{1.5cm}
\includegraphics[height=.0665\textwidth] {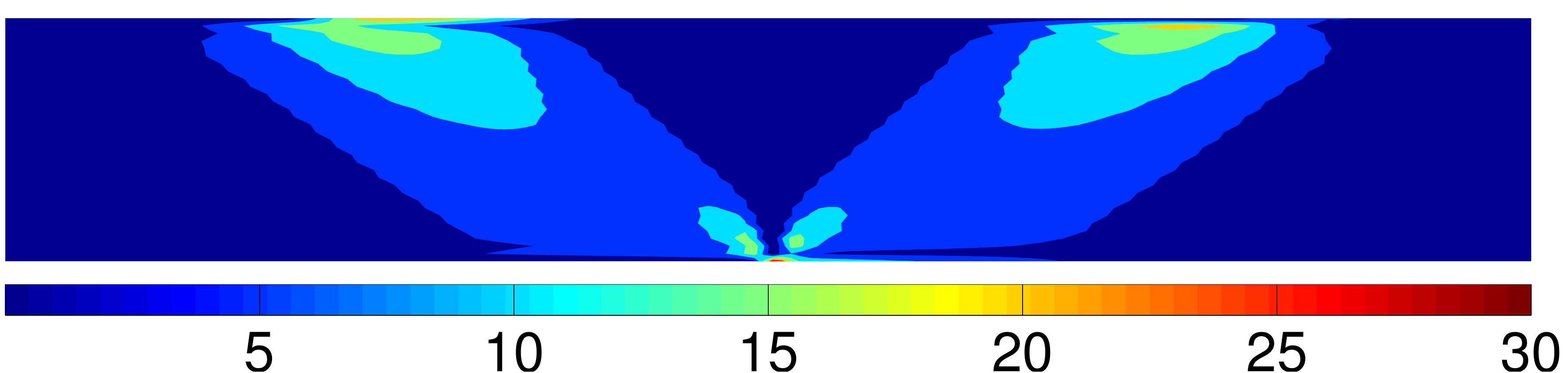}
\put(-173,12){\small$a = 2\pi$}
\hspace{0.7mm}
\includegraphics[height=.0665\textwidth] {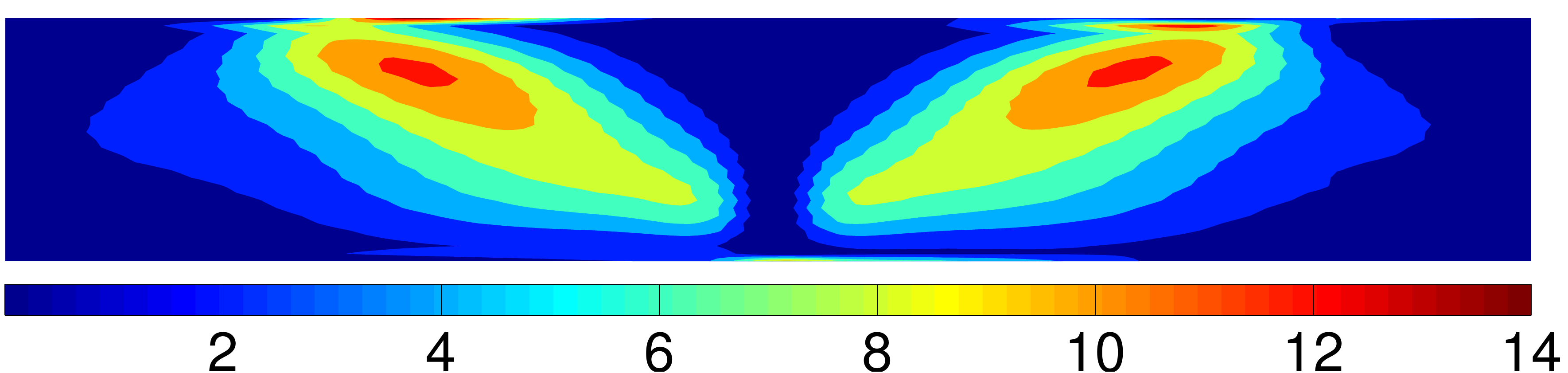}
\hspace{0.4mm}
\includegraphics[height=.0665\textwidth] {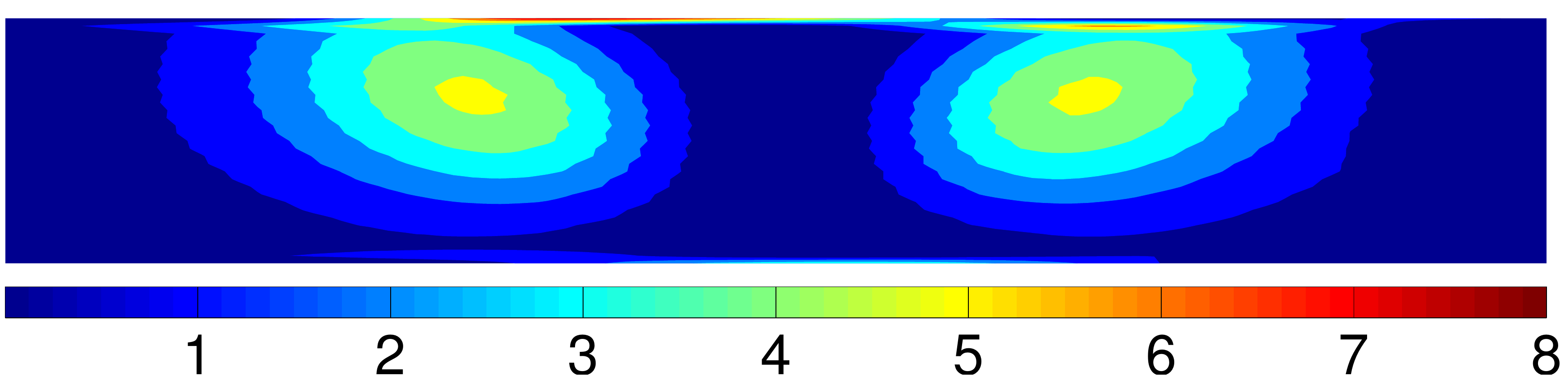}
%\put(5,10){$H^2$}

\hspace{1.5cm}
\includegraphics[height=.0424\textwidth] {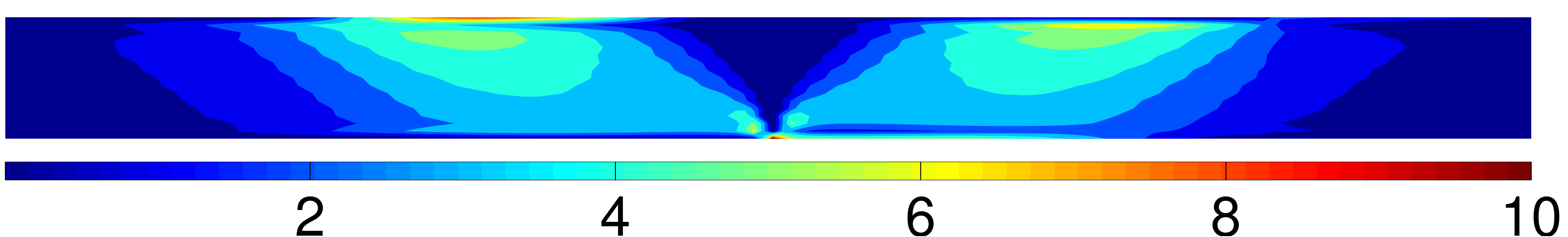}
\put(-173,8){\small$a = 4\pi$}
\hspace{0.75mm}
\includegraphics[height=.0424\textwidth] {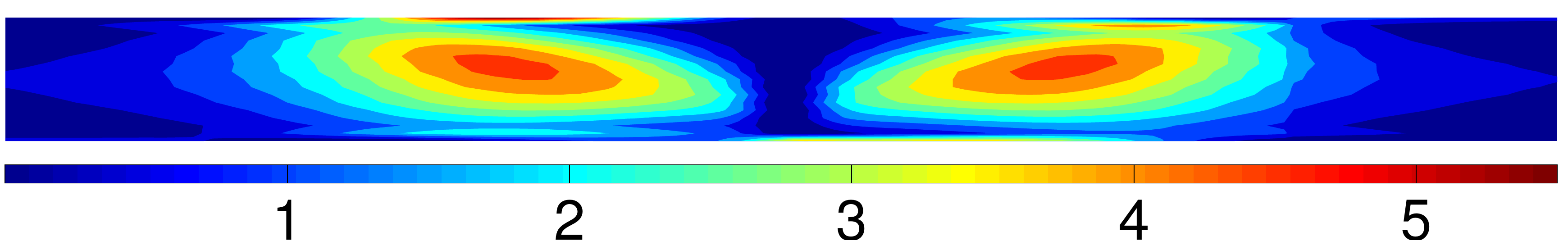}
\hspace{0.9mm}
\includegraphics[height=.0424\textwidth] {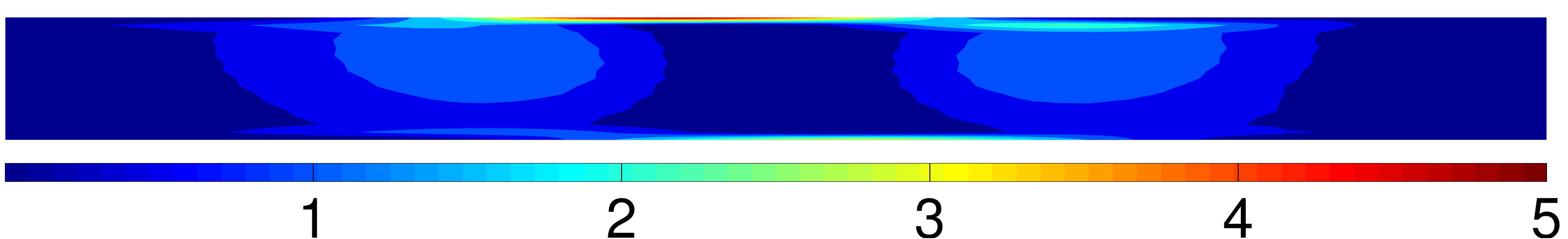}

\hspace{1.5cm}
\includegraphics[height=.0341\textwidth] {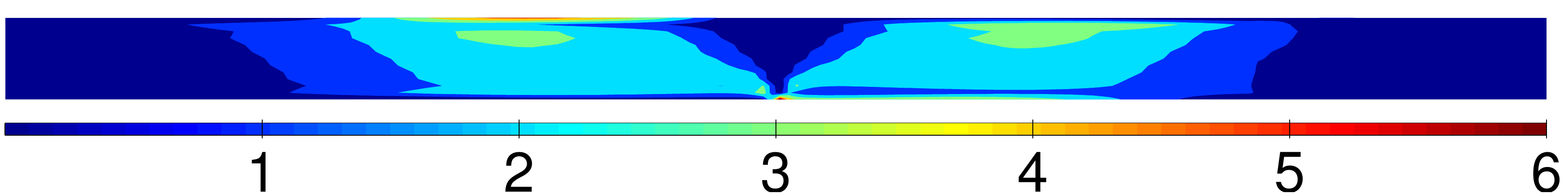}
\put(-171.5,7){\small$a = 6\pi$}
\hspace{1.1mm}
\includegraphics[height=.0341\textwidth] {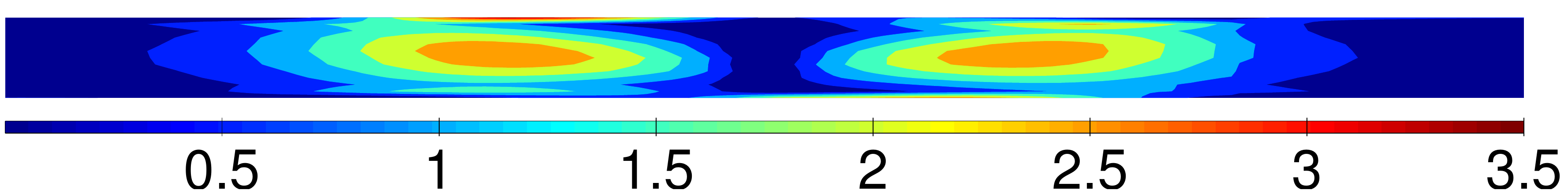}
\hspace{0.2mm}
\includegraphics[height=.0341\textwidth] {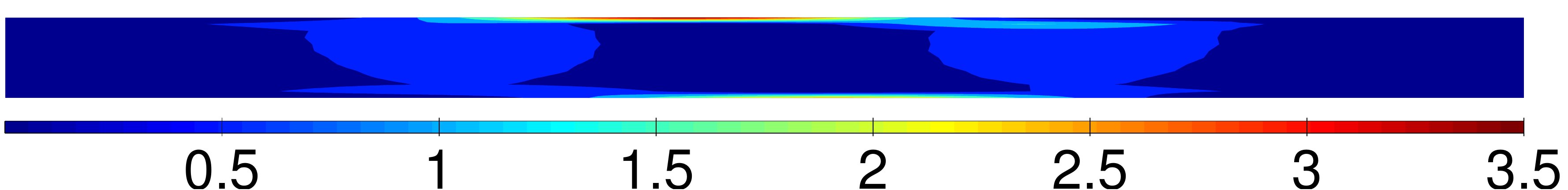}

\hspace{1.5cm}
\includegraphics[height=.0298\textwidth] {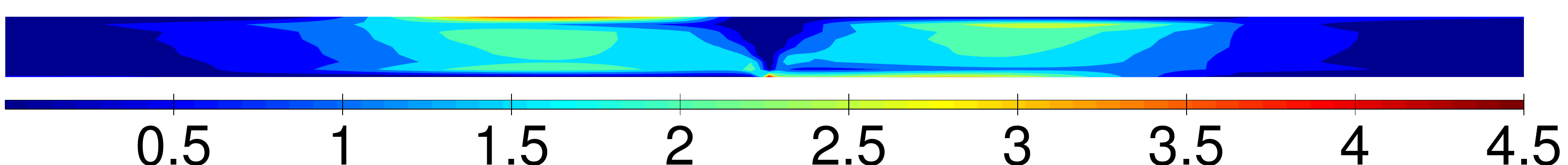}
\put(-173.5,6){\small$a = 8\pi$}
\hspace{0.1mm}
\includegraphics[height=.0298\textwidth] {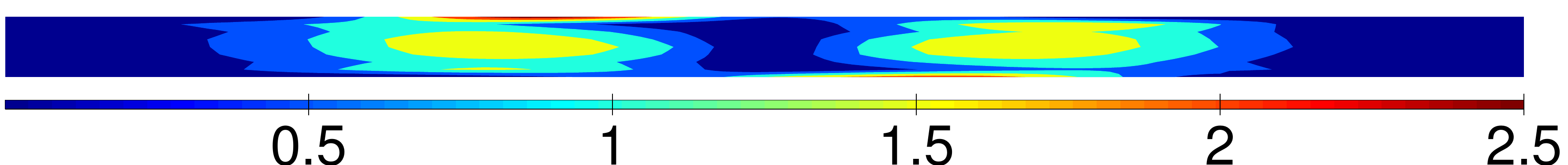}
\hspace{0.0mm}
\includegraphics[height=.0298\textwidth] {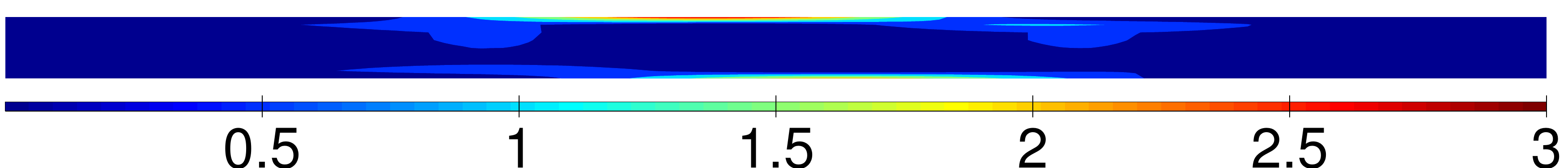}

\vspace{2mm}
\noindent
\hspace{1.501cm}
\includegraphics[height=.11\textwidth] {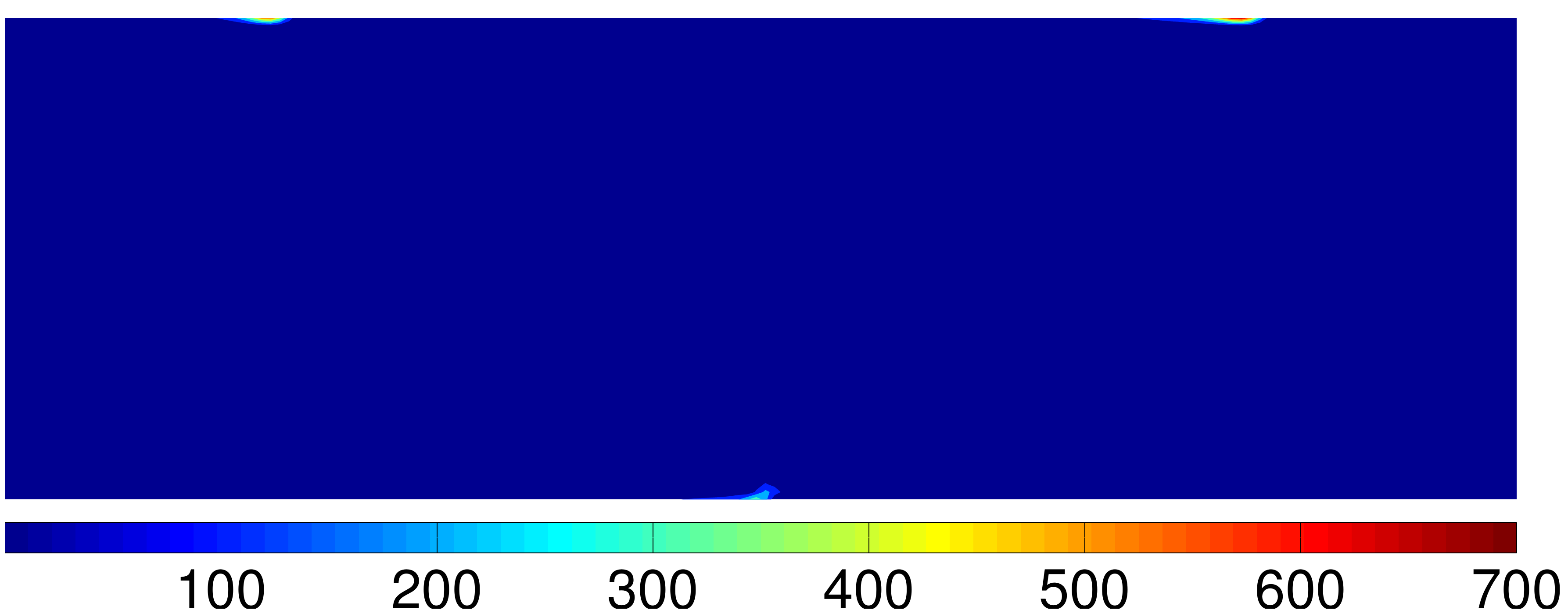}
\put(-173,20){\small$a = \pi$}
\put(-179,37){\small(b) $K$}
\hspace{0.0mm}
\includegraphics[height=.11\textwidth] {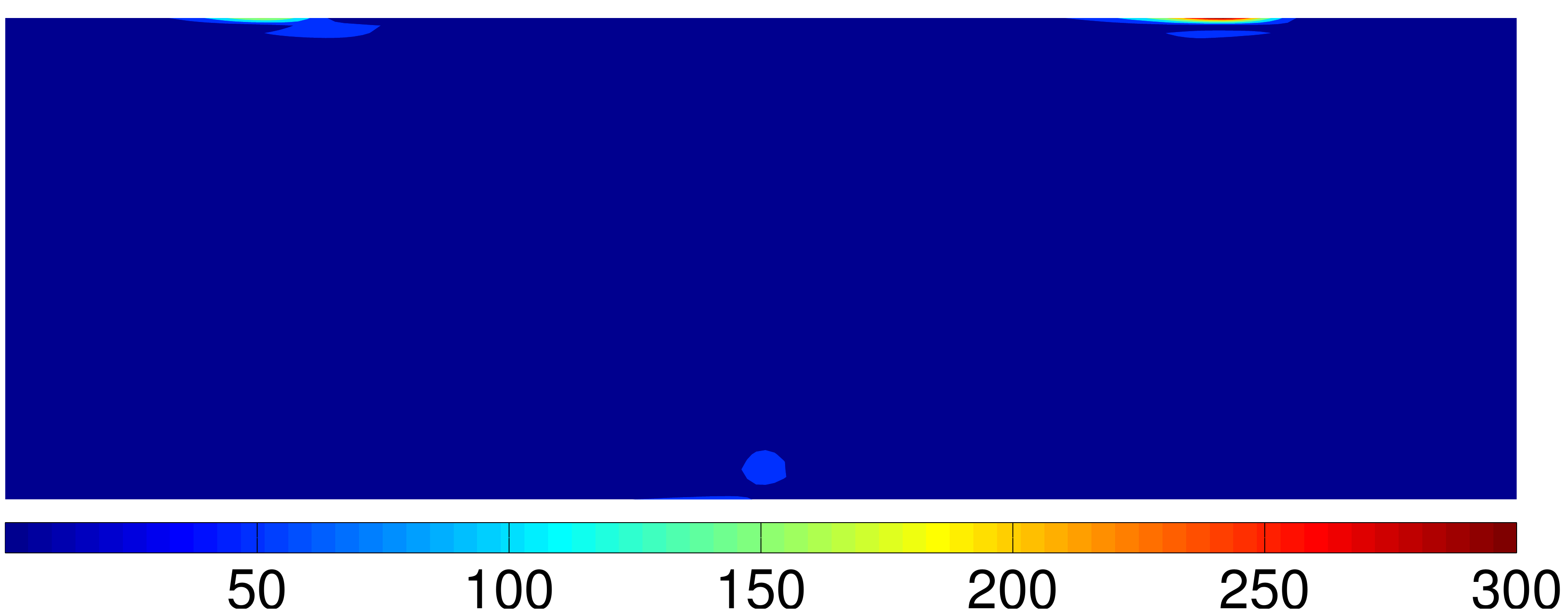}
\hspace{0.0mm}
\includegraphics[height=.11\textwidth] {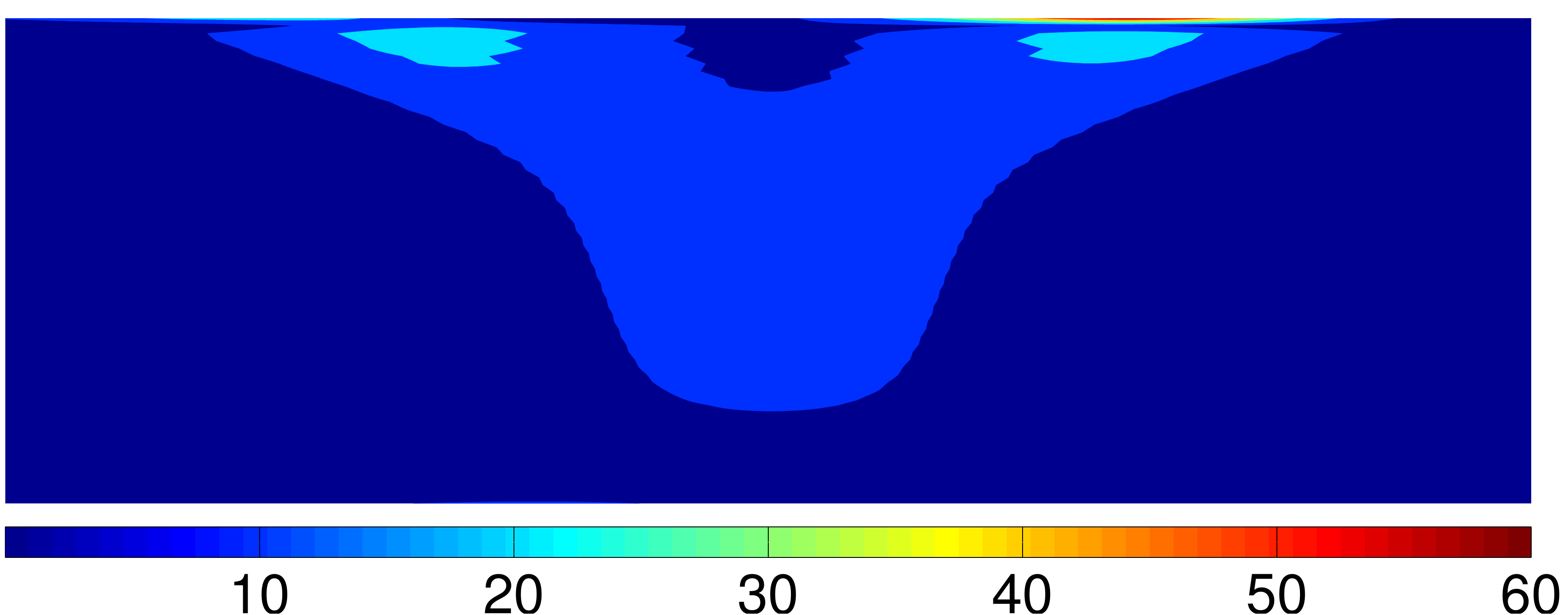}

\hspace{1.5cm}
\includegraphics[height=.0665\textwidth] {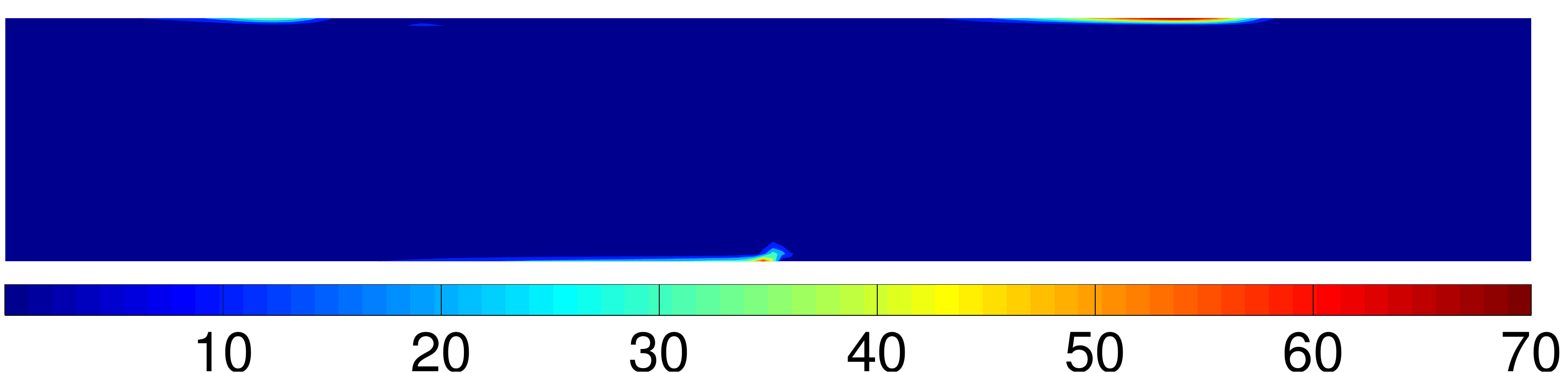}
\put(-173,12){\small$a = 2\pi$}
\hspace{0.6mm}
\includegraphics[height=.0665\textwidth] {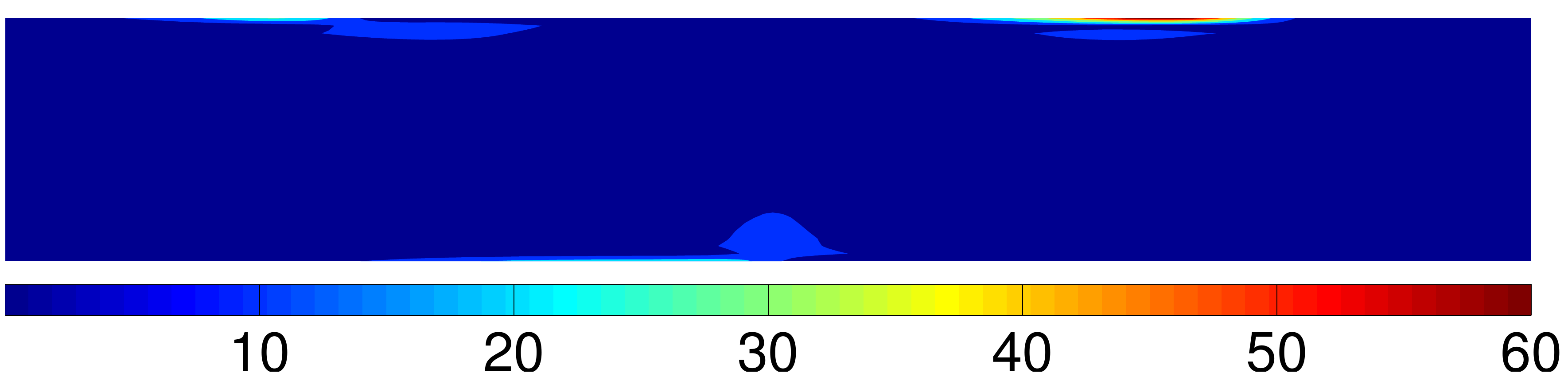}
\hspace{0.5mm}
\includegraphics[height=.0665\textwidth] {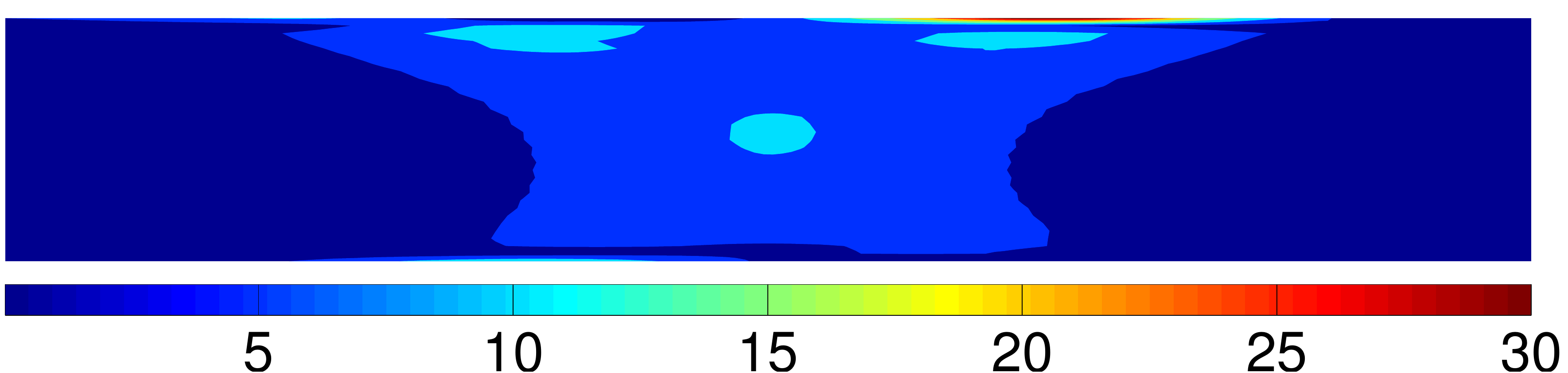}
%\put(5,10){$K$}

\hspace{1.5cm}
\includegraphics[height=.0424\textwidth] {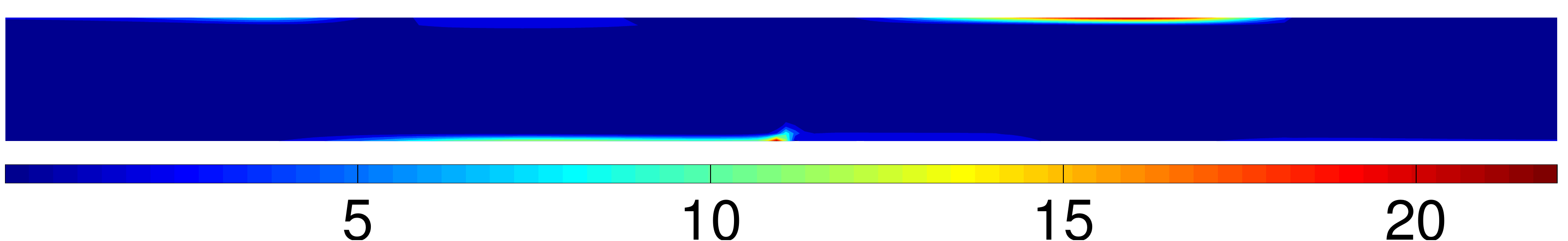}
\put(-173,8){\small$a = 4\pi$}
\hspace{1.0mm}
\includegraphics[height=.0424\textwidth] {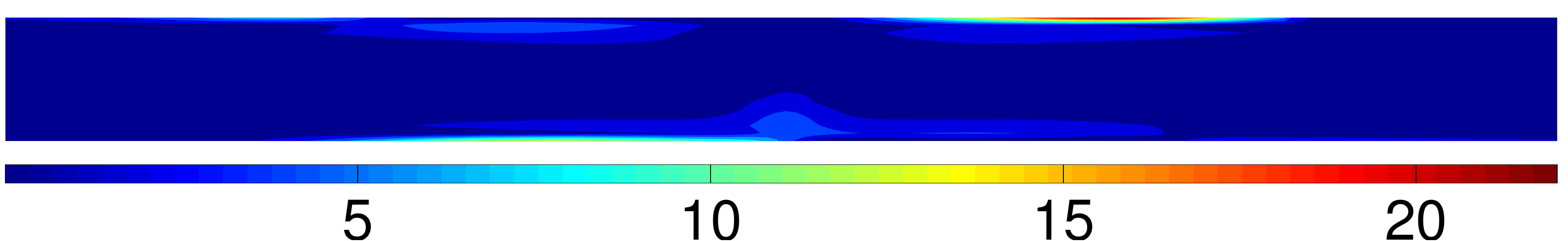}
\hspace{1.4mm}
\includegraphics[height=.0424\textwidth] {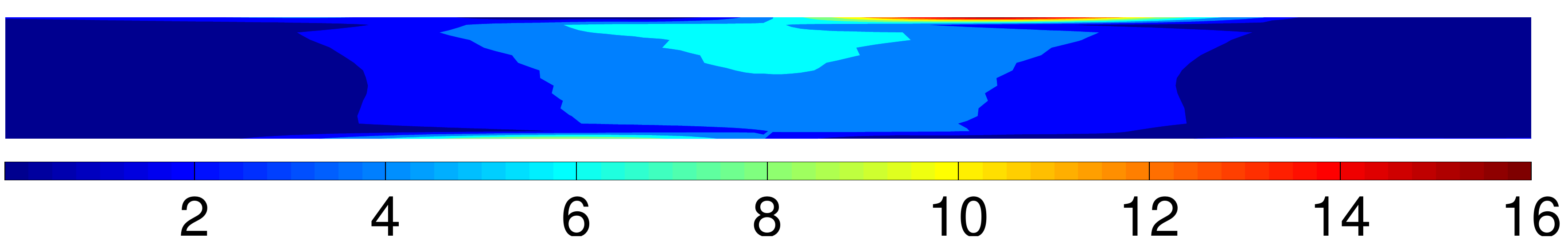}

\hspace{1.5cm}
\includegraphics[height=.0341\textwidth] {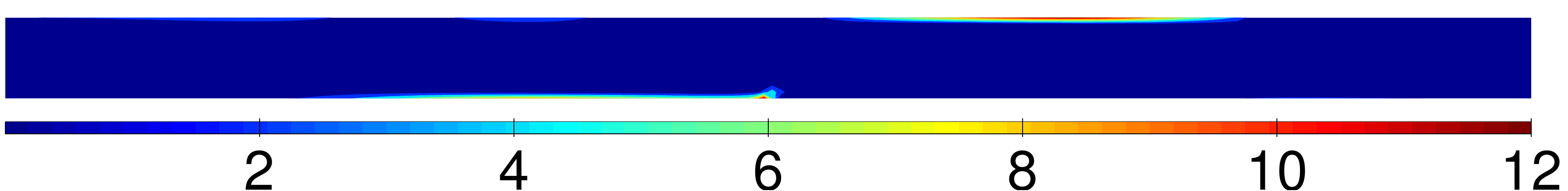}
\put(-173,7){\small$a = 6\pi$}
\hspace{0.3mm}
\includegraphics[height=.0341\textwidth] {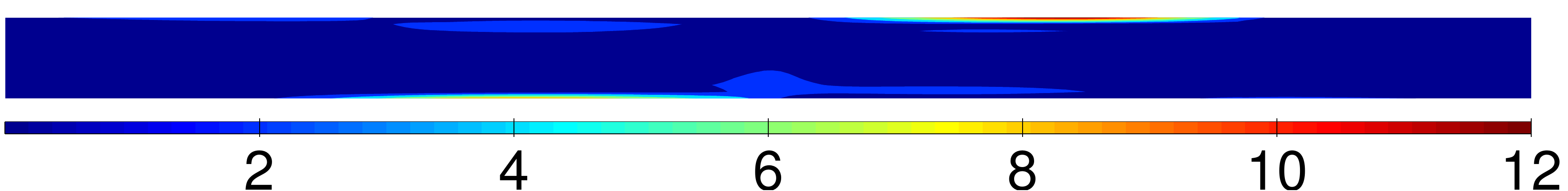}
\hspace{0.5mm}
\includegraphics[height=.0341\textwidth] {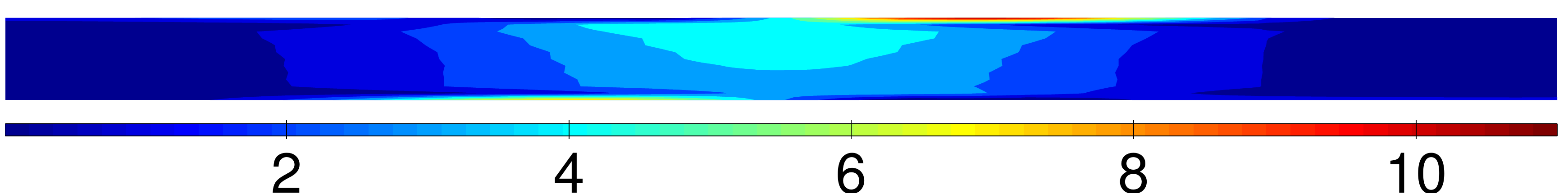}

\hspace{1.5cm}
\includegraphics[height=.0298\textwidth] {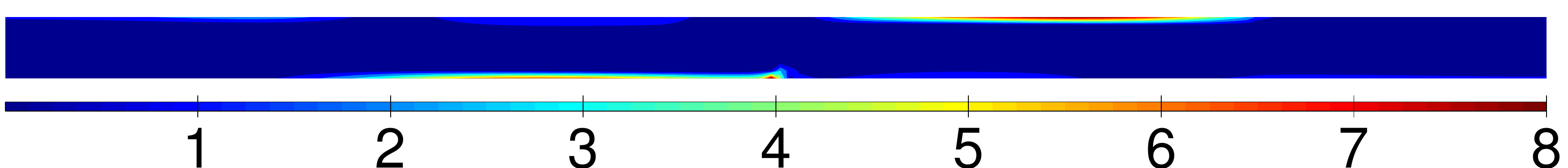}
\put(-172.5,6){\small$a = 8\pi$}
\hspace{0.8mm}
\includegraphics[height=.0298\textwidth] {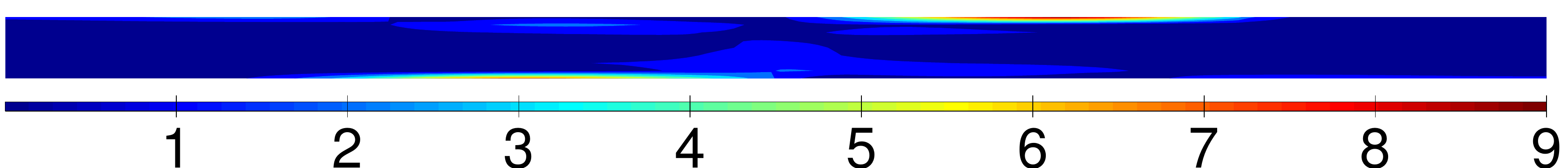}
\hspace{0.6mm}
\includegraphics[height=.0298\textwidth] {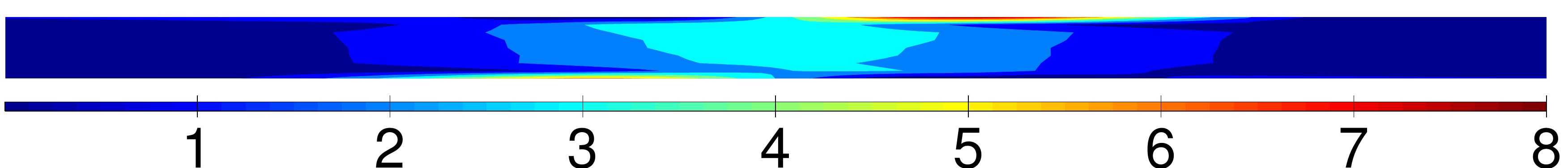}

\vspace{2mm}
\noindent
\hspace{1.5cm}
\includegraphics[height=.11\textwidth] {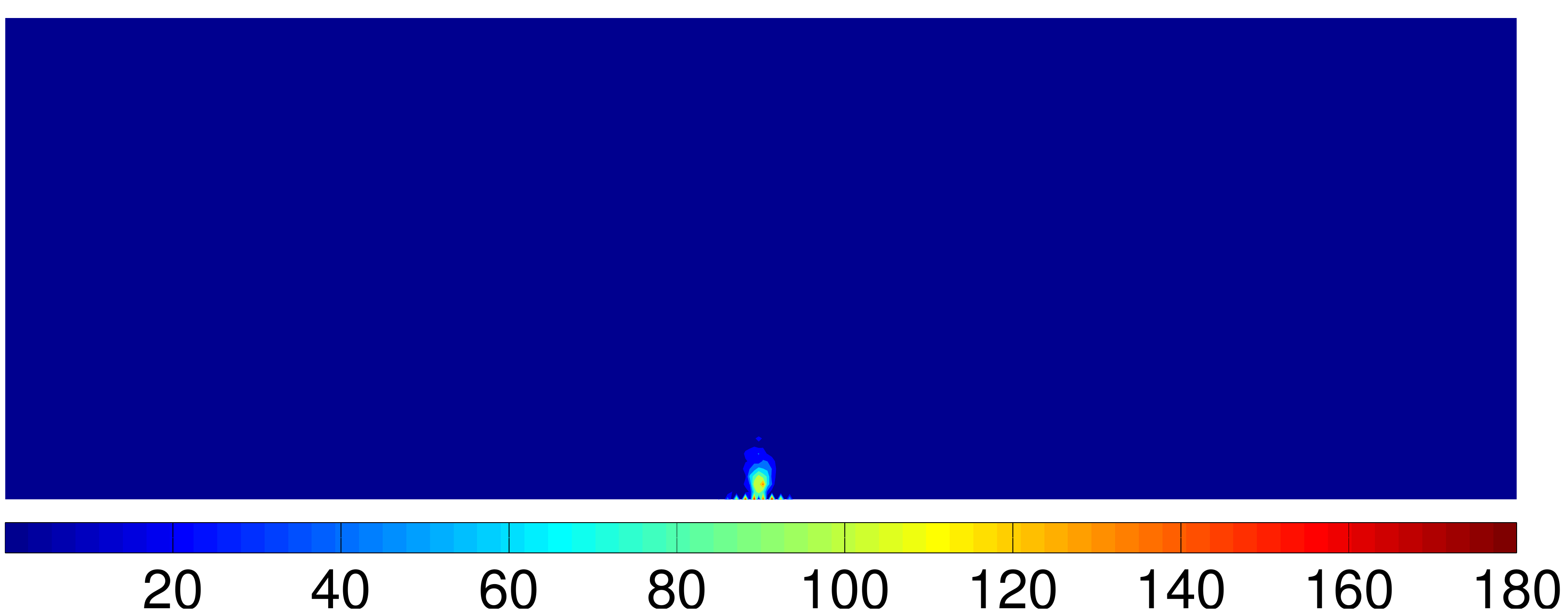}
\put(-175,20){\small$a = \pi$}
\put(-182,37){\small(c) $\epsilon^2$}
\hspace{0.0mm}
\includegraphics[height=.11\textwidth] {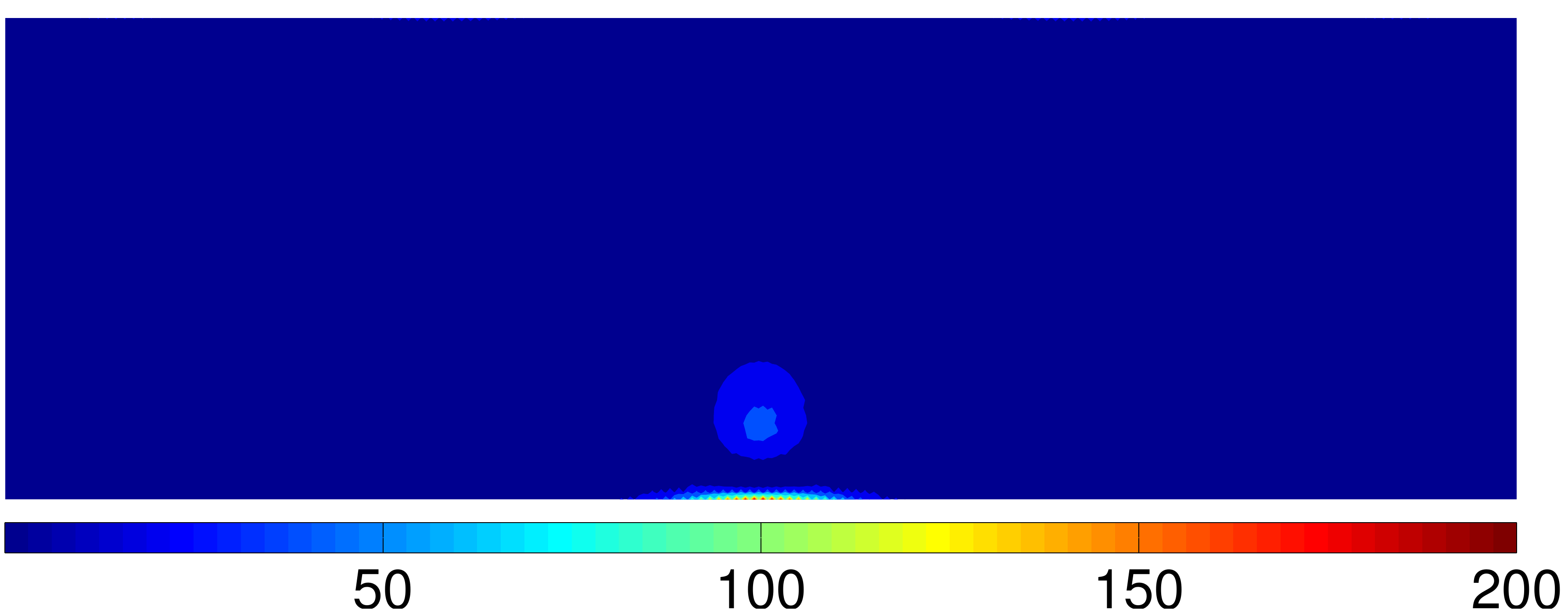}
\hspace{0.0mm}
\includegraphics[height=.11\textwidth] {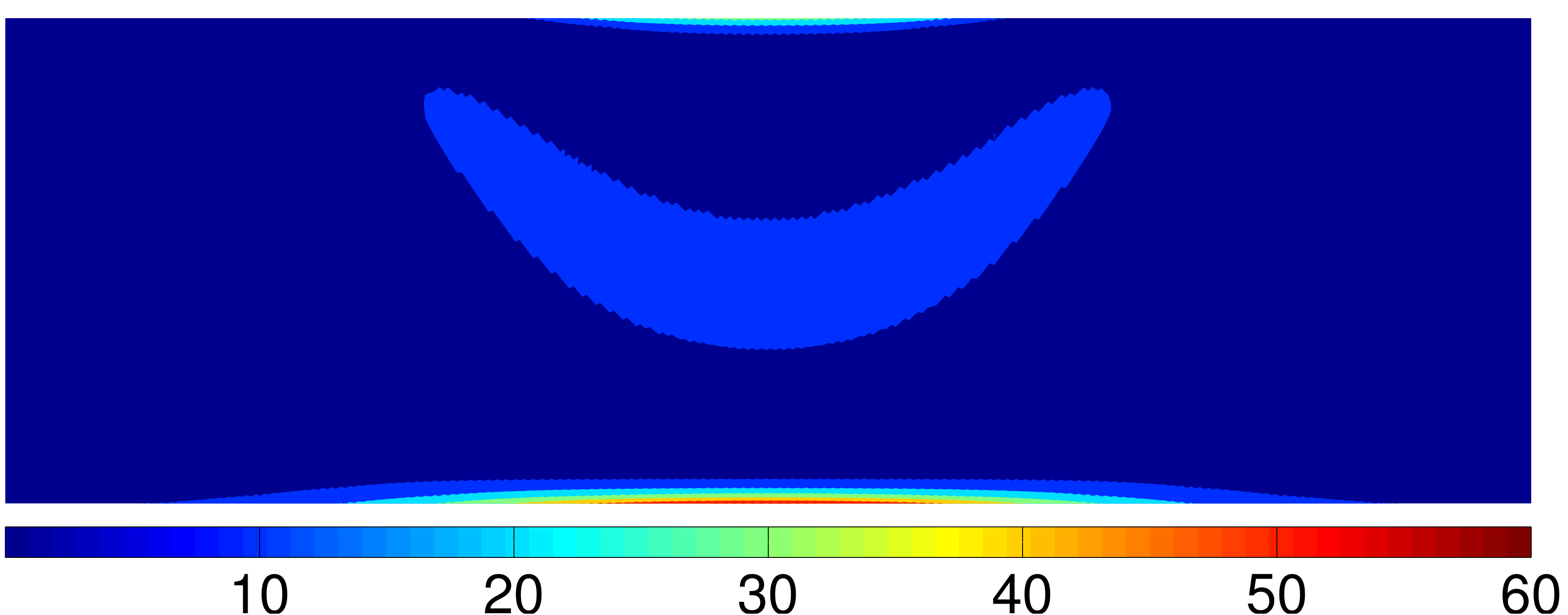}

\hspace{1.5cm}
\includegraphics[height=.0665\textwidth] {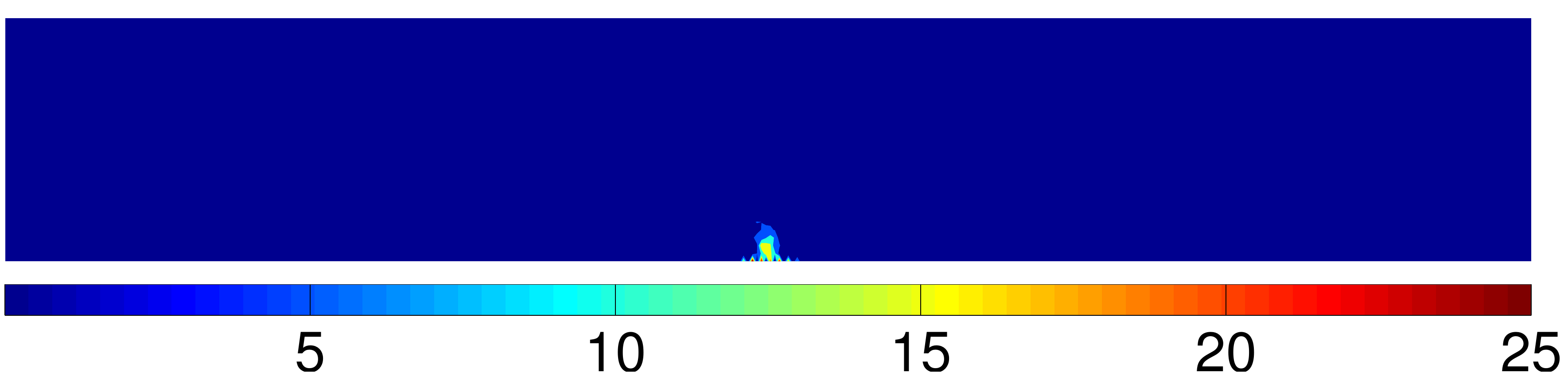}
\put(-174,12){\small$a = 2\pi$}
\hspace{0.7mm}
\includegraphics[height=.0665\textwidth] {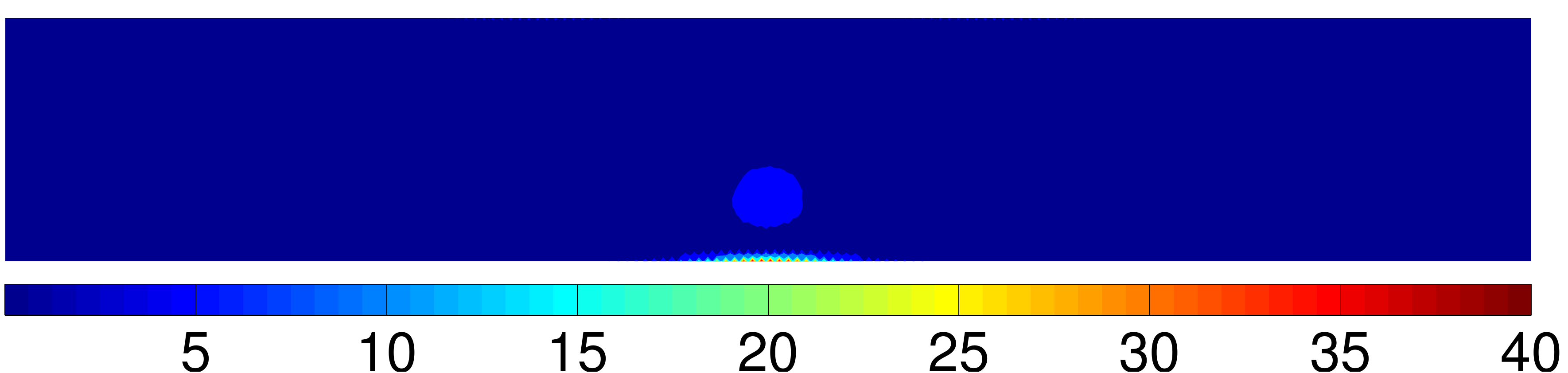}
\hspace{0.4mm}
\includegraphics[height=.0665\textwidth] {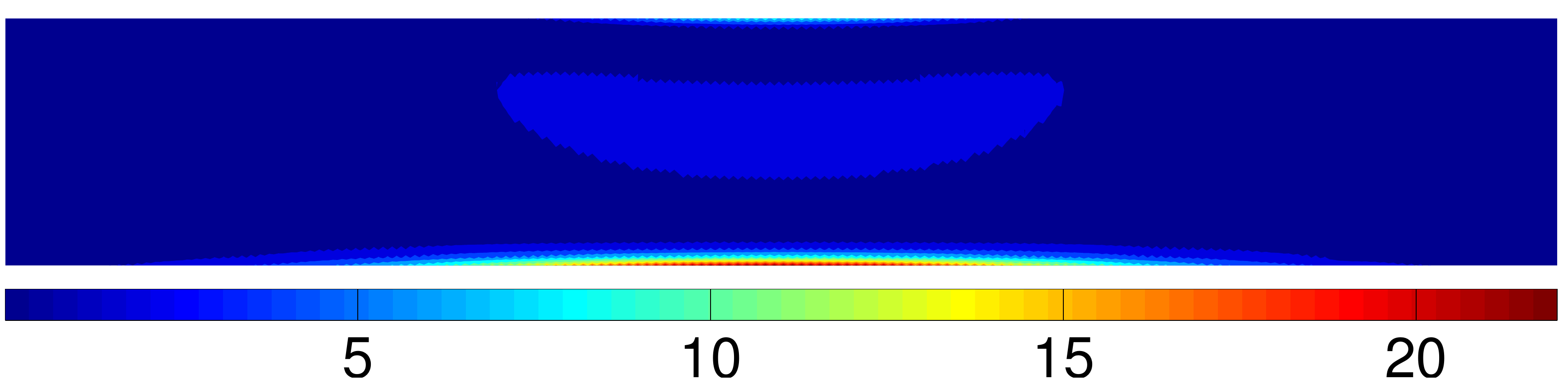}
%\put(5,10){$AF$}

\hspace{1.5cm}
\includegraphics[height=.0424\textwidth] {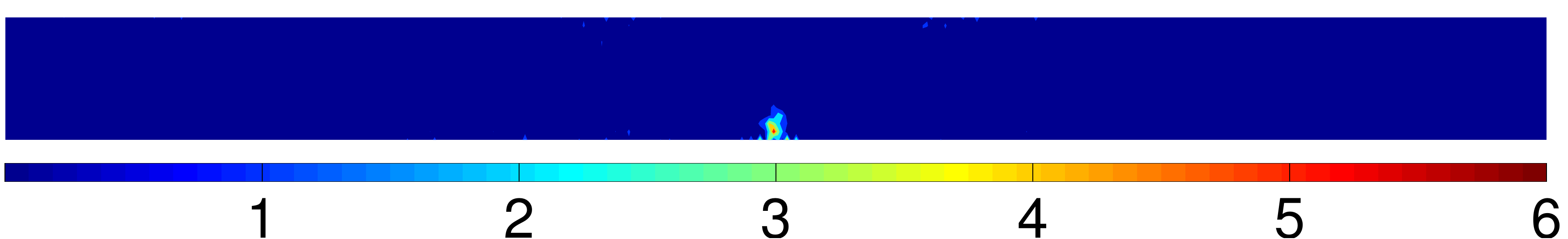}
\put(-173,8){\small$a = 4\pi$}
\hspace{1.0mm}
\includegraphics[height=.0424\textwidth] {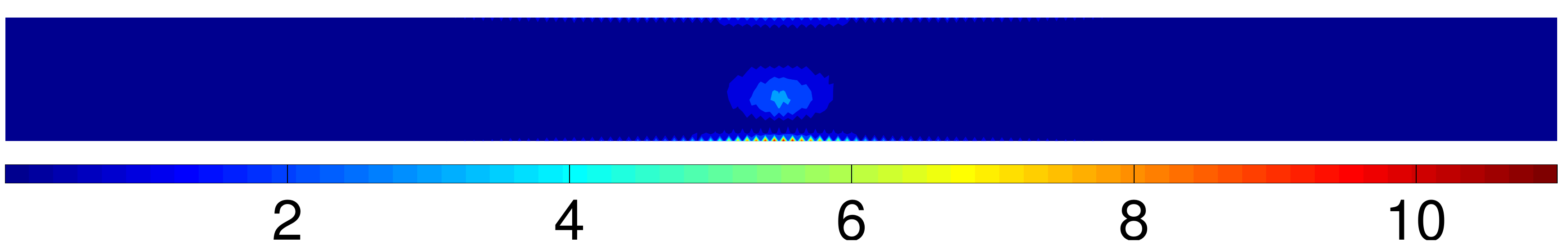}
\hspace{1.2mm}
\includegraphics[height=.0424\textwidth] {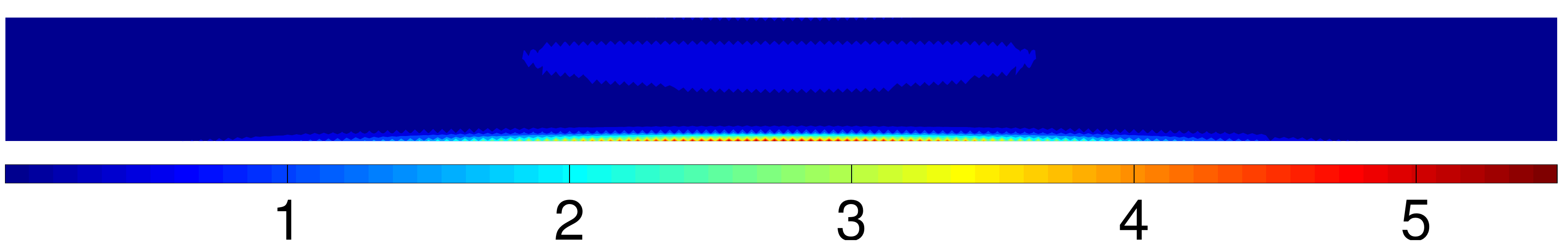}

\hspace{1.5cm}
\includegraphics[height=.0341\textwidth] {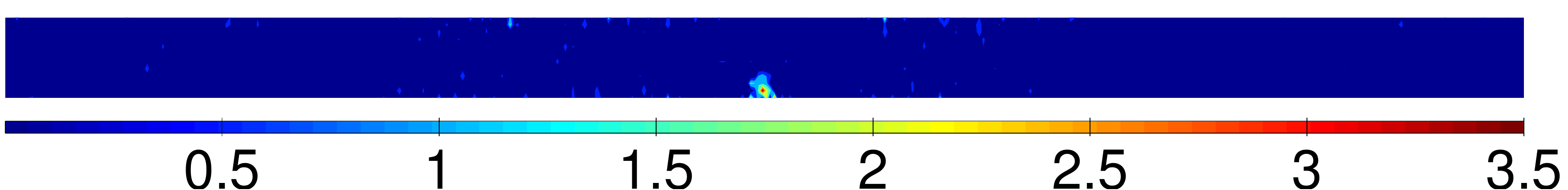}
\put(-175,7){\small$a = 6\pi$}
\hspace{0.3mm}
\includegraphics[height=.0341\textwidth] {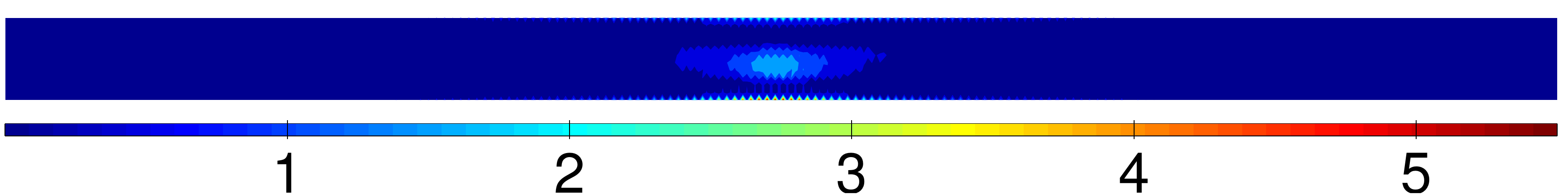}
\hspace{1.3mm}
\includegraphics[height=.0341\textwidth] {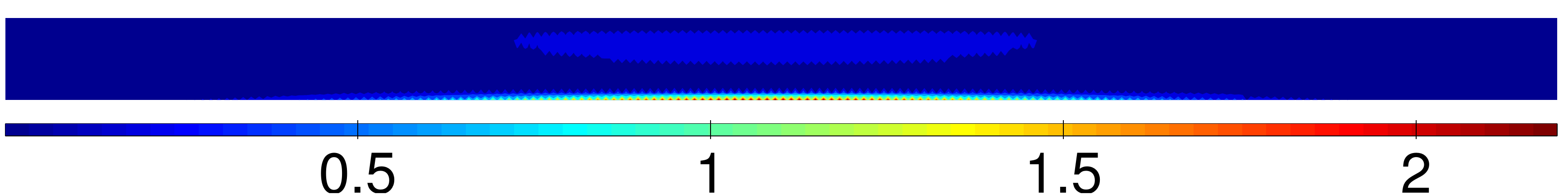}

\hspace{1.5cm}
\includegraphics[height=.0298\textwidth] {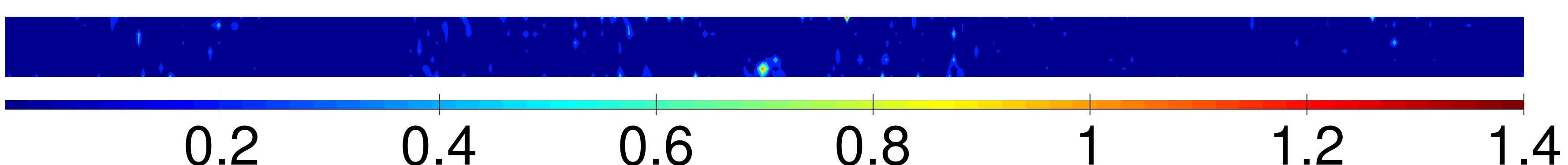}
\put(-175,6){\small$a = 8\pi$}
\hspace{0.1mm}
\includegraphics[height=.0298\textwidth] {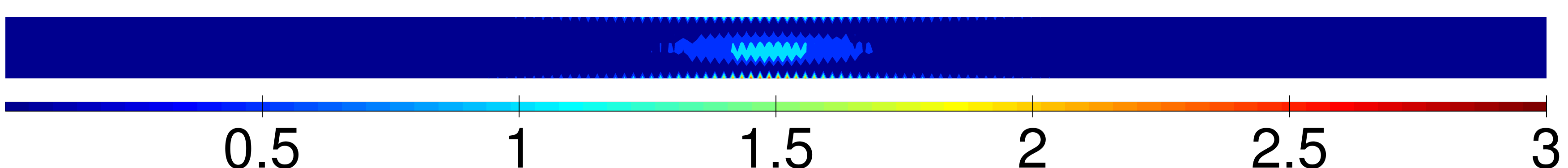}
\hspace{0.7mm}
\includegraphics[height=.0298\textwidth] {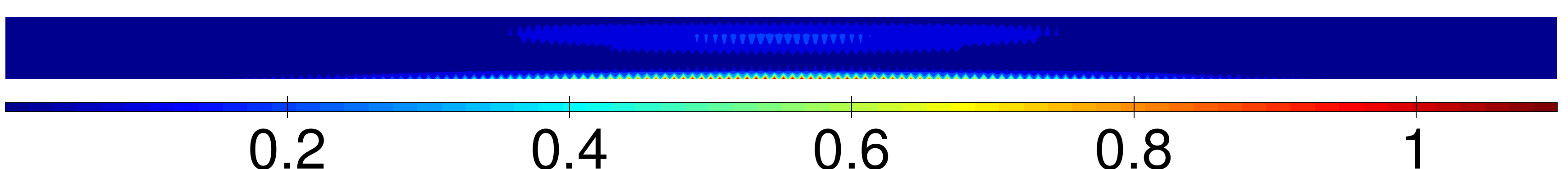}

%\end{center}
\caption{The mean and Gaussian curvatures are both more evenly distributed for increasing stretchability ${k}$: (a) Square of the mean curvature ${H}$, (b) magnitude $|K|$ of the Gaussian curvature $K$, and (c) square ${\epsilon^2}$ of the strain ${\epsilon}$ of M\"obius bands made of stretchable materials for different values of the aspect ratio ${a}$ and stretchability ${k}$. The aspect ratio of each contour plot is equal to the aspect ratio of the band it represents. Images of the bands corresponding to the contour plots appear in Figure~\ref{fig:Moebius_shape}. Note that different scales are adapted individually to capture the entire range of relevant values. Additional cases appear in Figure~\ref{fig:MeanCurvatureSqAll} of the SI. %\color{blue}(Please plot $|K|$ instead of $K$. This amounts to changing the scale on the color bars.)
}
\label{fig:MeanCurvatureSq}
\end{figure*}
%%%%%%%%%%%%%%%

%%%%%%%%%%%%%%%%%%%%%%%%%%%%%%%%%%%%%%
%\vspace{-4pt}
\subsection{Curvature and dilatation}
%\vspace{-2pt}
%%%%%%%%%%%%%%%%%%%%%%%%%%%%%%%%%%%%%%

In the SI, we show that the coarse-grained limit of our model corresponds to a continuum model for a surface $S$ that resists stretching and bending. Whereas stretching is characterized by an area modulus $\mu_a$ related to the linear spring stiffness $k_l$ by
\begin{equation}
\mu_a=\frac{\sqrt{3}k_l}{2},
\label{areamodulus}
\end{equation}
bending is characterized by splay and saddle-splay moduli $\mu$ and $\bar\mu$ related to the torsional spring stiffness $k_\theta$ by
\begin{equation}
\mu=\frac{3\sqrt{3}k_\theta}{2}
\qquad\text{and}\qquad
\bar\mu=-\frac{2\mu}{3}=\sqrt{3}k_\theta.
\label{splaymoduli}
\end{equation}
With \eqref{areamodulus} and \eqref{splaymoduli}, the total energy $\cE$ of $S$ takes the form
\begin{equation}
\cE=%\sqrt{3}k_l\int_S\epsilon^2\,\text{d}a
\sqrt{3}\int_S[k_l\epsilon^2+k_\theta(3H^2-K)]\,\text{d}a,
\label{limitE}
\end{equation}
where $\text{d}a$ denotes the area element on $S$ and where $\epsilon^2$, $H$, and $K$ denote the (two-dimensional) dilatation, mean curvature, and Gaussian curvature of $S$. The expression \eqref{areamodulus} for the area modulus $\mu_a$ is consistent with a result obtained by Seung and Nelson.\cite{Seung1988} The bending contribution to \eqref{limitE} arising from \eqref{splaymoduli}, which is identical to that of a Kichhoff plate with bending modulus $3\sqrt{3}k_\theta/2$ and Poisson's ratio $1/3$, coincides with an expression derived by Merchant and Keller.\cite{Merchant1991}

%
%%%%%%%%%%%%%%%
\begin{figure*}[!htb]
%\begin{center}
\centering
\begin{picture}(500,70)
\put(100,0){\includegraphics[width=.12\textwidth, trim=0cm 0.1cm 29cm 1.0cm, clip=true] {bandsar_1_spring_1}}
%\put(15,30){$a = \pi$}
\put(16,45){\small(a)}
\put(113,71){\small$k = 10^{-1}$}
\put(250,-3){\includegraphics[width=.11\textwidth, trim=0cm 0.1cm 29cm 1.0cm, clip=true] {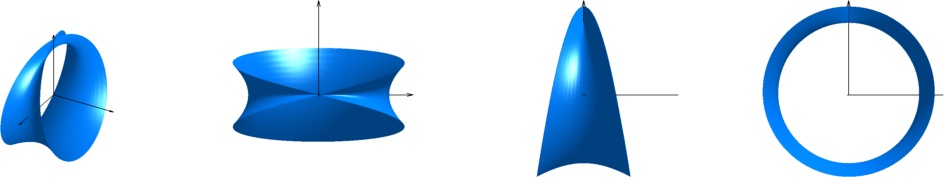}}
\put(260,71){\small$k = 10^{-2/3}$}
\put(400,-10){\includegraphics[width=.1\textwidth, trim=0cm 0.1cm 29cm 1.0cm, clip=true] {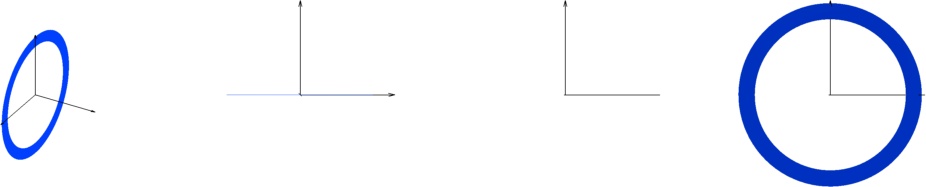}}
\put(400,71){\small$k = 10^{-1/3}$}
\end{picture}

\vspace{5mm}\noindent

\hspace{1.36cm}
\includegraphics[height=.123\textwidth] {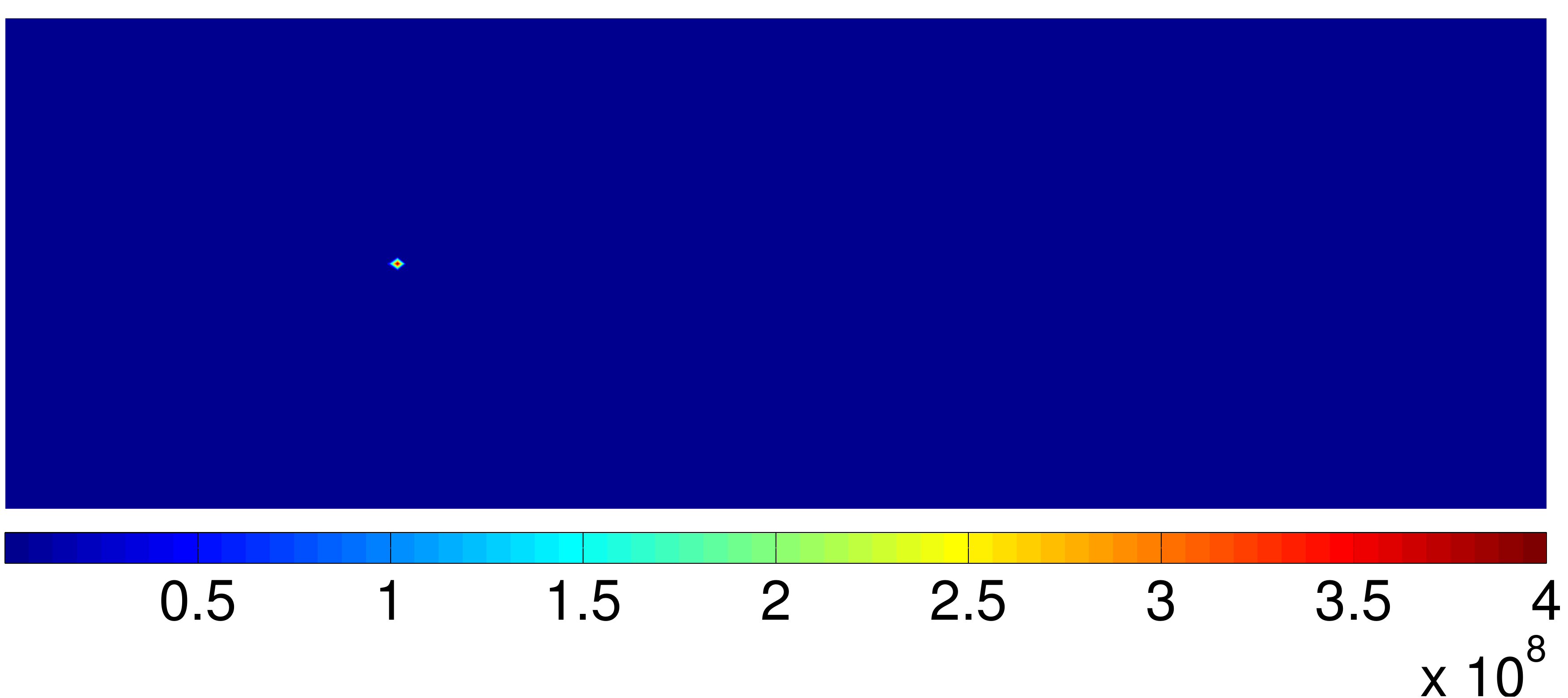}
\put(-178,45){\small(b) $H^2$}
\hspace{1.0mm}
\includegraphics[height=.123\textwidth] {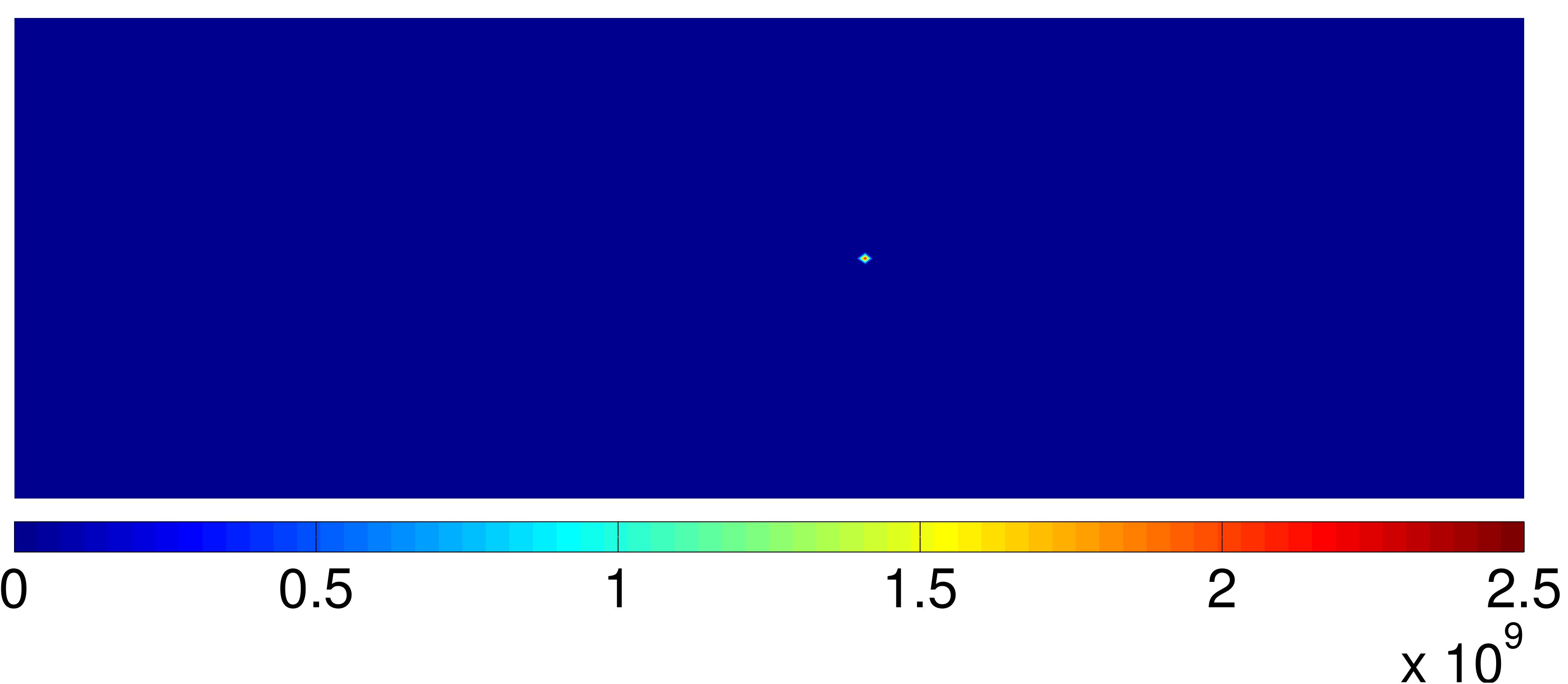}
\hspace{0.0mm}
\includegraphics[height=.123\textwidth] {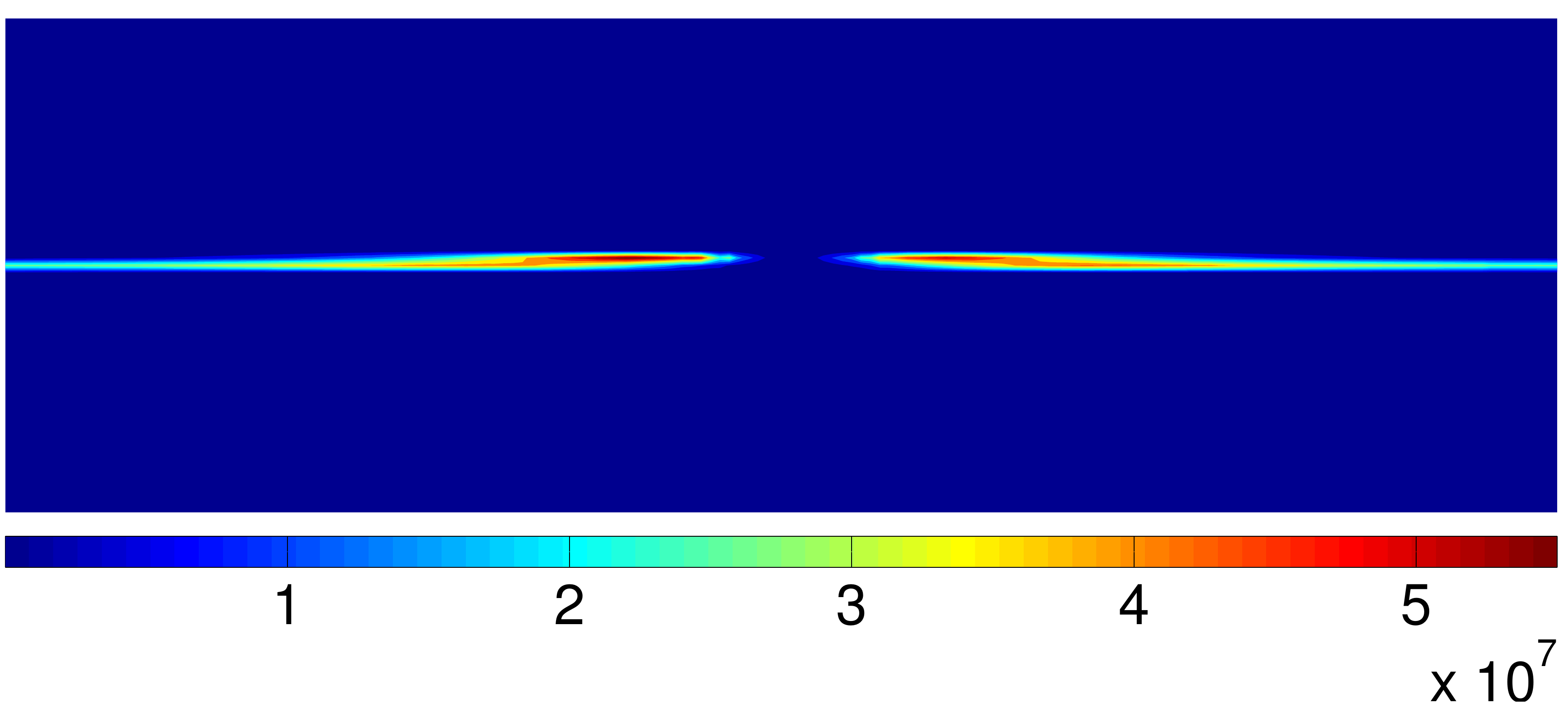}
%

%Gaussian curvature:
\hspace{1.36cm}
\includegraphics[height=.123\textwidth] {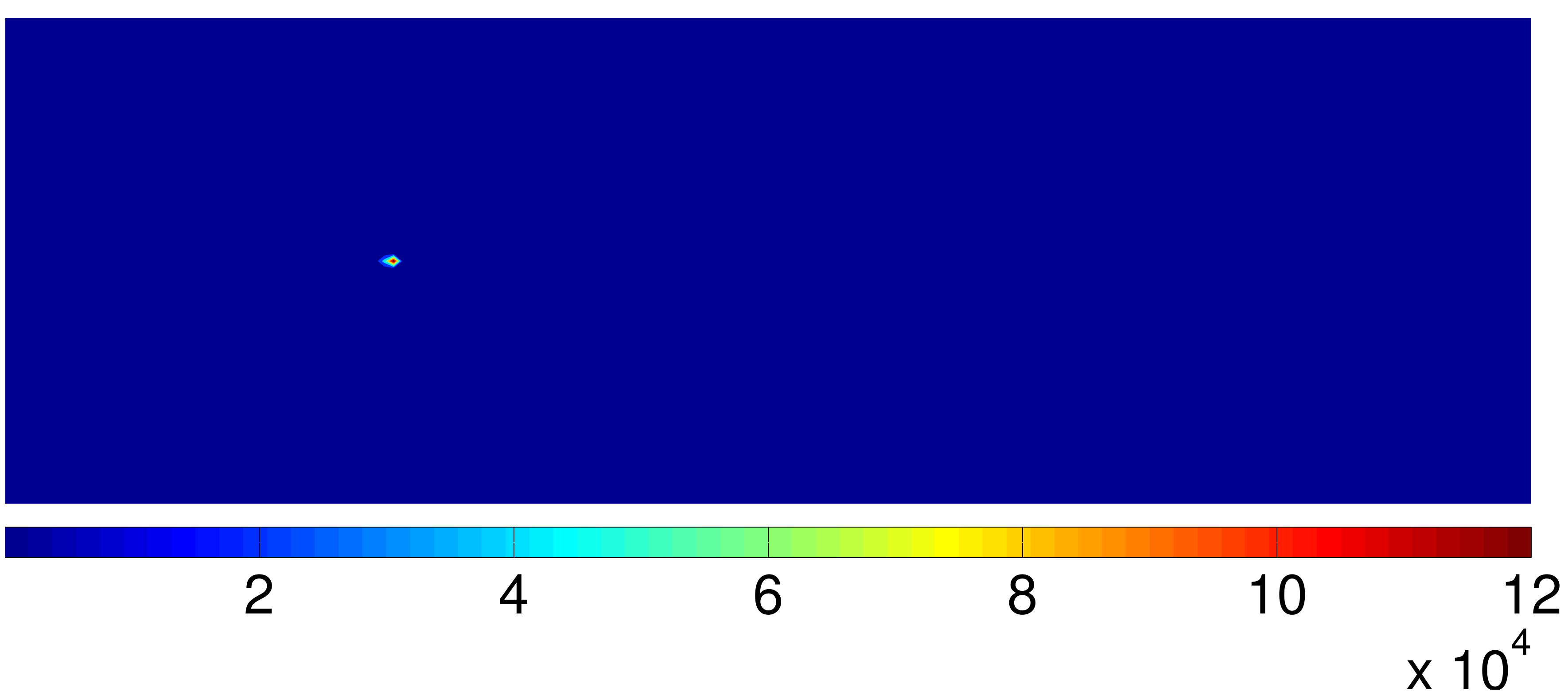}
\put(-178,45){\small(c) $K$}
\hspace{1.1mm}
\includegraphics[height=.123\textwidth] {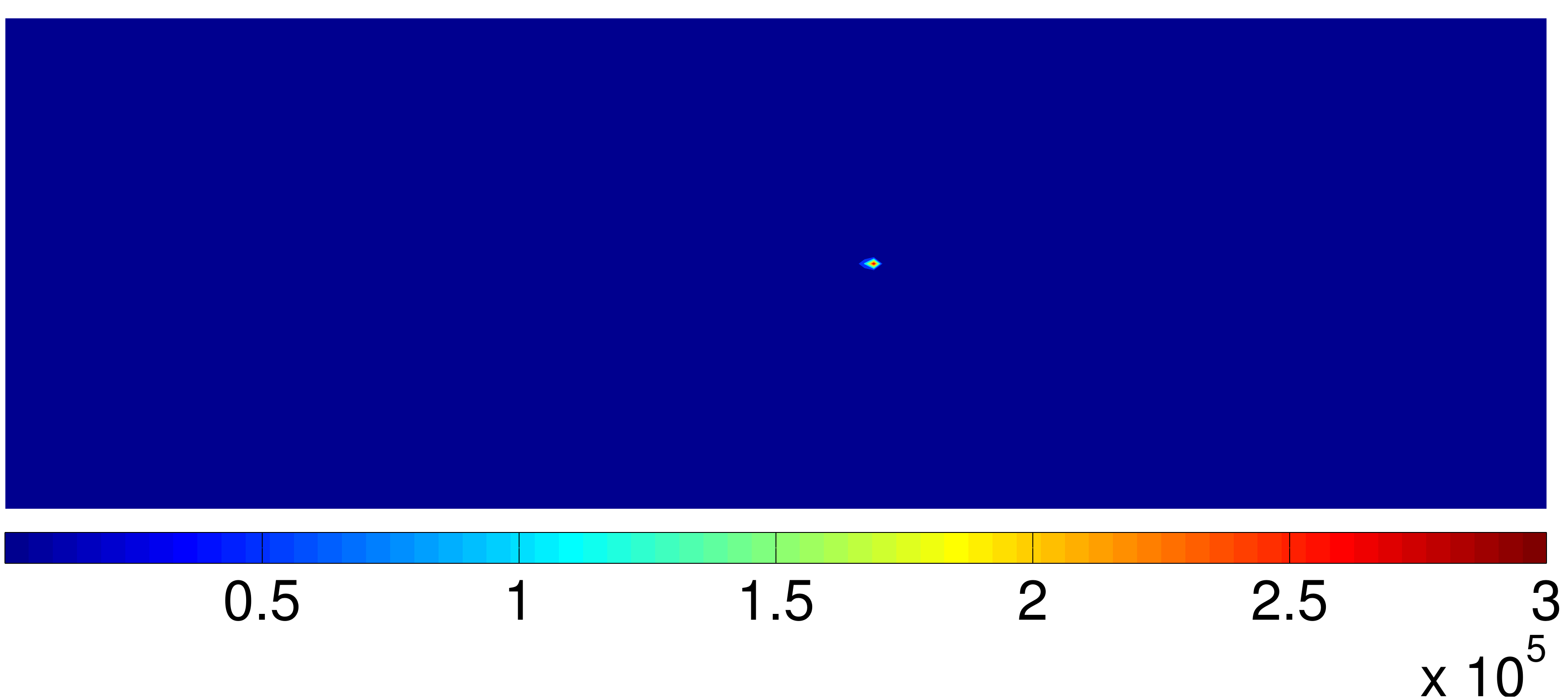}
\hspace{0.9mm}
\includegraphics[height=.123\textwidth] {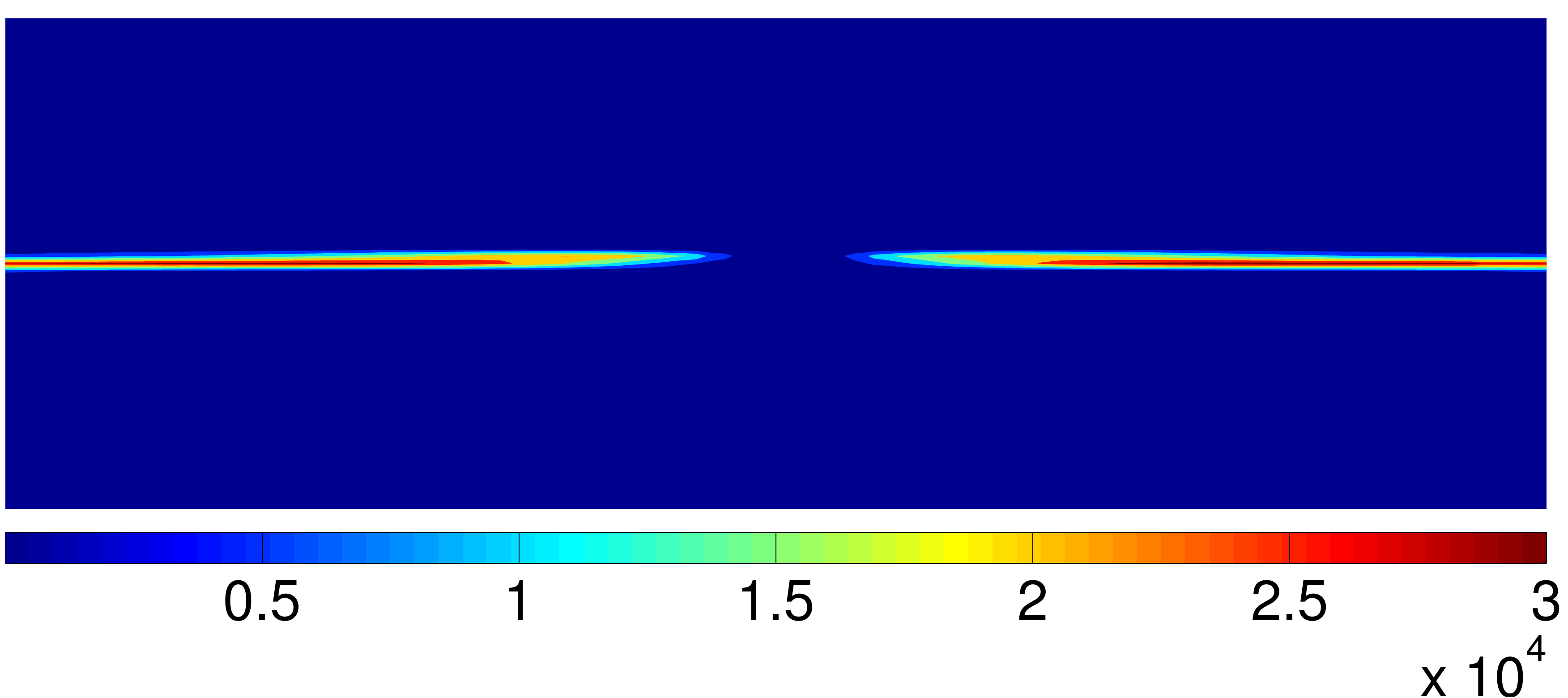}
%

%Areal Deformation:
\hspace{1.4cm}
\includegraphics[height=.11\textwidth] {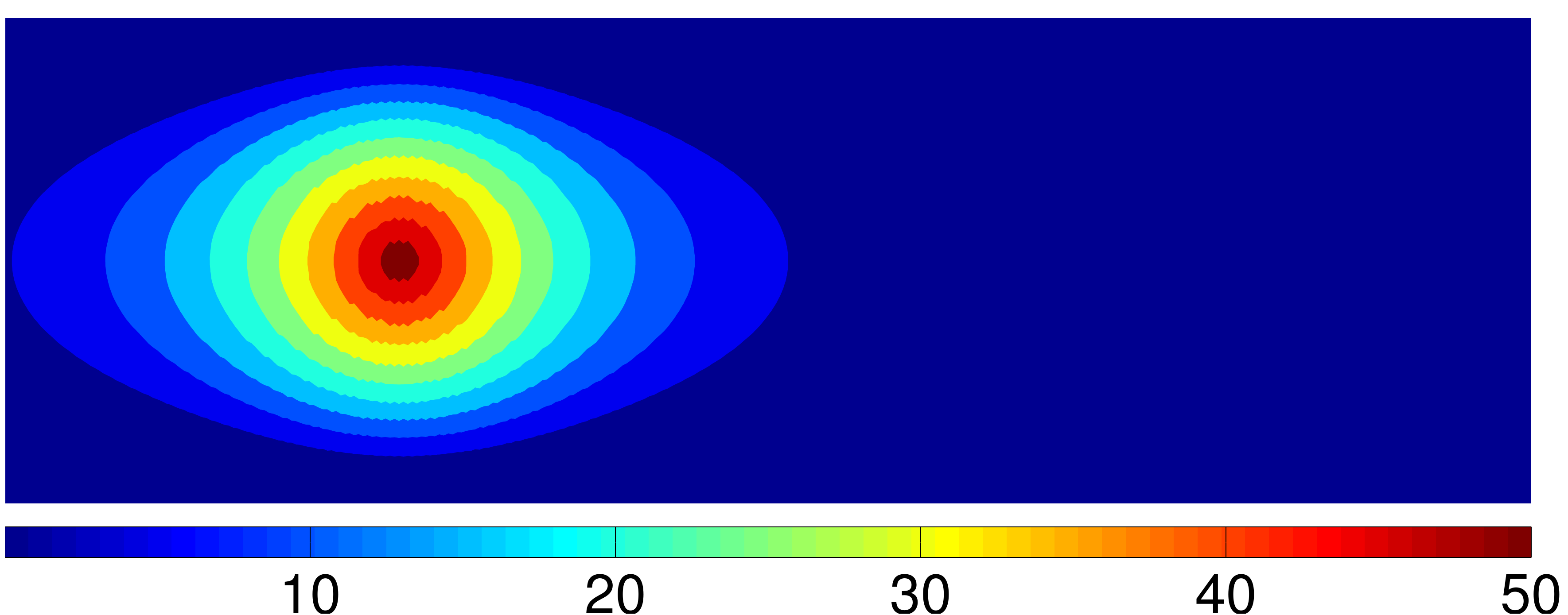}
\put(-180,40){\small(d) $\epsilon^2$}
\hspace{0.4mm}
\includegraphics[height=.11\textwidth] {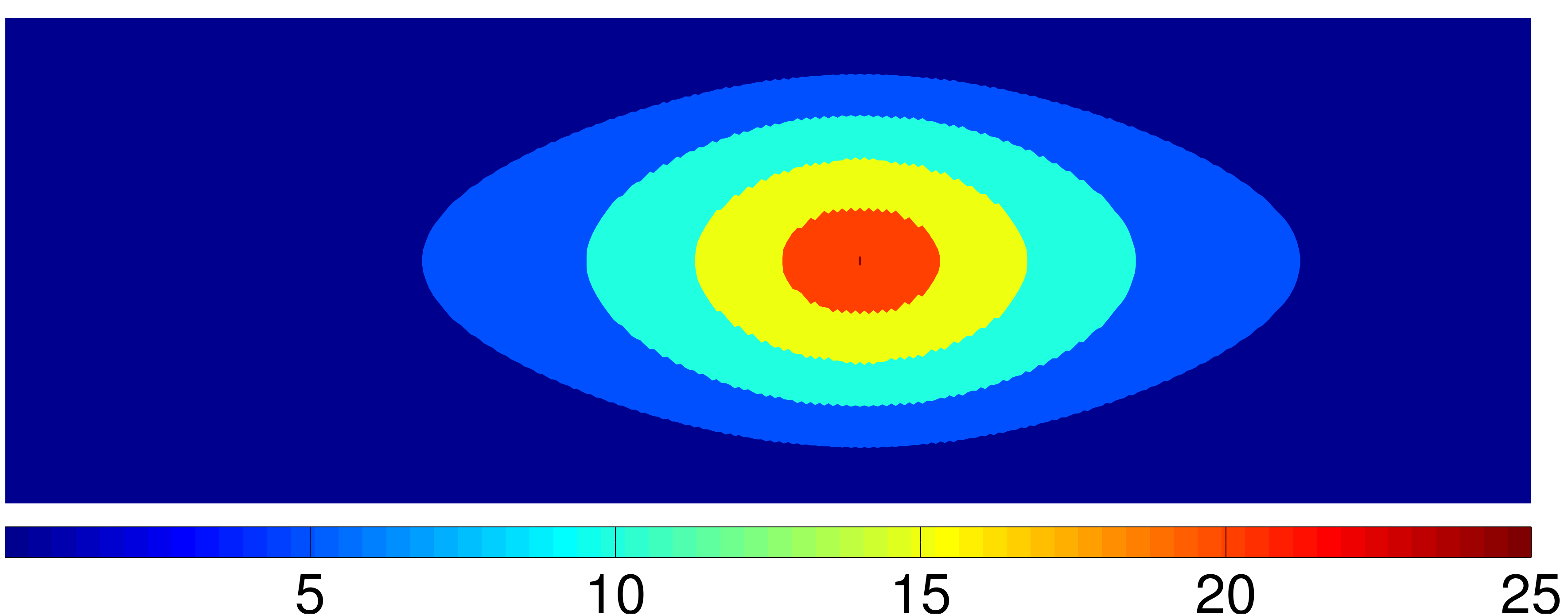}
\hspace{0.0mm}
\includegraphics[height=.11\textwidth] {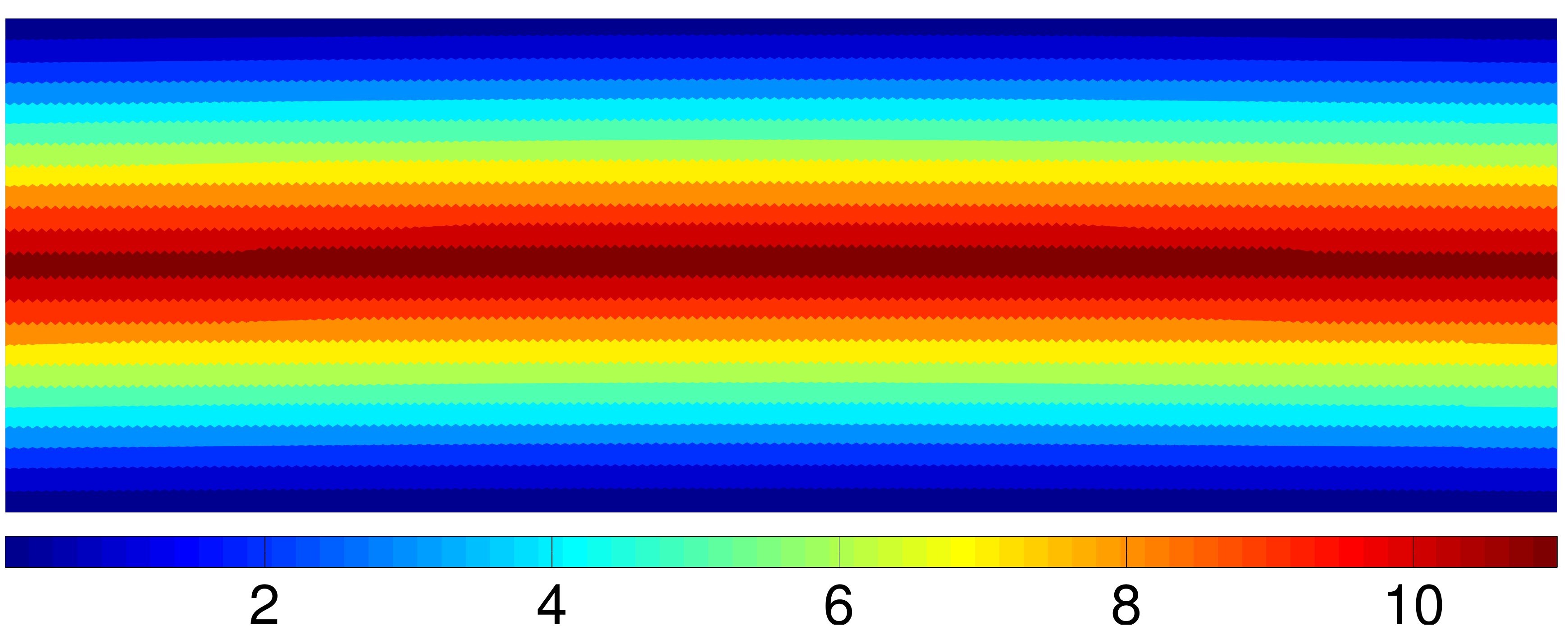}

%\end{center}
\caption{Both mean and Gaussian curvature are concentrated along the intersection of self-intersecting achiral bands: (a) Collapsed bands colored by stretchability as in Figure \ref{fig:Moebius_shape}, (b) Square $H^2$ of the mean curvature $H$, (c) magnitude $|K|$ of the Gaussian curvature $K$, and (d) square ${\epsilon^2}$ of the strain ${\epsilon}$ of collapsed M\"obius bands for different stretchabilities ${k}$. The aspect ratio of each contour plot is equal to the aspect ratio of the band it represents. Note that different scales are adapted individually to capture the entire range of relevant values. 
}
\label{fig:collapseCurv}
\end{figure*}
%%%%%%%%%%%%%%% 

This connection to continuum theory suggests that the pointwise distributions of suitable discrete approximations to $H^2$, $K$, and $\epsilon^2$ may provide further insight regarding how the stretchability $k$ and aspect ratio $a$ influence the shape of a M\"obius band. Plots of these distributions are provided in Figure \ref{fig:MeanCurvatureSq} for representative values of $k$ and $a$. In the effectively unstretchable case $k=10^{-6}$, each equilibrium shape exhibits a nearly flat triangular region bounded at its vertices by zones in which $H^2$ takes large values and is surrounded by a nearly flat trapezoidal region. These characterisics are consistent with previous findings,\cite{Hinz2013,Hinz2014,Hinz2013b,Hinz2014b} as is the increase in size of the zones in which $H^2$ exhibits large values with increasing $a$. However, as $k$ increases, $H^2$ becomes increasingly more evenly distributed over the band. Consistent with the observation that localized regions of concentrated bending-energy density indicate where creasing or tearing may occur,\cite{Starostin2007a, Starostin2007} our findings show that for bands of aspect ratio increasing the stretchability alleviates such concentrations.

All equilibrated bands exhibit non-vanishing Gaussian curvature $K$. However, for $k=10^{-6}$, $K$ is very close to zero almost everywhere. This confirms that the choice $k = 10^{-6}$ yields nearly developable shapes and thus provides a good approximation of the unstretchable limit, even though some degree of stretching is allowed for any value of $k>0$. For all larger values of $k$ considered, $H^2$ and $|K|$ are maximized at the same points. The maximum values of $|K|$ become more prominent as $a$ decreases. At the vertices of the previously discussed flat, triangular regions, where the contribution to the bending energy from $H^2$ is largest, the contribution to the bending energy from $|K|$ also attains its largest values. Thus, even in the approximately unstretchable limit, it is energetically favorable to locally transfer energy associated with $H^2$ to energy associated with $K$. Inspecting the local dilatation $\epsilon^2$ shows that bending is locally transferred to stretching unless $K\equiv0$, in which case the shape of the band is genuinely developable. Since most materials manifest some in-plane elasticity, our findings also indicate that an actual band should possess a region of local stretching. This also suggests that loci of large $H^2$ are likely to coincide with zones in which the bending contribution to the energy is concentrated. These zones are also most likely to be loci for inelastic deformation or failure. Conversely, allowing for local stretching may reduce the bending contribution to the energy density.

Both $H^2$ and $|K|$ become more evenly distributed over the band with increasing $k$, resulting in lower curvature gradients. Further, both $|K|$ and $\epsilon^2$ are transferred from the edge toward the centerline of the band, reducing the magnitude of the gradient of the continuum bending-energy density and decreasing the likelihood of creasing or tearing accordingly. Nonzero values of $|K|$ indicate that stretching contributes significantly to the overall shape of the band, as zones in which $|K|$ is largest coincide with zones of large $\epsilon^2$, to which the continuum stretching-energy density (see the SI) is directly related. This result confirms the previous observation that increasing $k$ causes the centerlines of bands to become more circular and less out-of-plane, thereby reducing bending. A zone of low bending energy is maintained as $k$ increases. The magnitude $|K|$ is largest in zones in which $H^2$ takes large values, resulting in large values of the continuum bending-energy density. However, this effect is mitigated by reduced gradients due to stretching.

The transition from bending to stretching occurs smoothly except for combinations of relatively large stretchability $k$ and very low aspect ratio $a$ leading to collapsed, achiral equilibrium shapes. Plots of $H^2$, $|K|$, and $\epsilon^2$ are provided in Figure \ref{fig:collapseCurv} for three illustrative choices of $k$ and $a=\pi$. For $k=10^{-1}$ and $k=10^{-2/3}$, $H^2$ and $|K|$ appear to diverge at a single point along the intersection and decay quickly with increasing distance from that point. The value of $\epsilon^2$ is also greatest at the point in question but does not decay as rapidly. However, for the choice $k=10^{-1/3}$ which results in collapse to an annulus, the values of $H^2$, $|K|$, and $\epsilon^2$ are largest along the centerline of the band, decaying toward the edge of the band.  

%%%%%%%%%%%%%%%%%%%%%%%%%
%\vspace{-4pt}
\subsection{Lines of curvature}
%\vspace{-2pt}
%%%%%%%%%%%%%%%%%%%%%%%%%

%
\begin{figure}[!t]
{\quad\includegraphics[width=.105\textwidth, trim=0cm 0cm 29cm 1.0cm, clip=true] {bandsar_2_spring_6}
\put(2,12){\includegraphics[width=.35\textwidth] {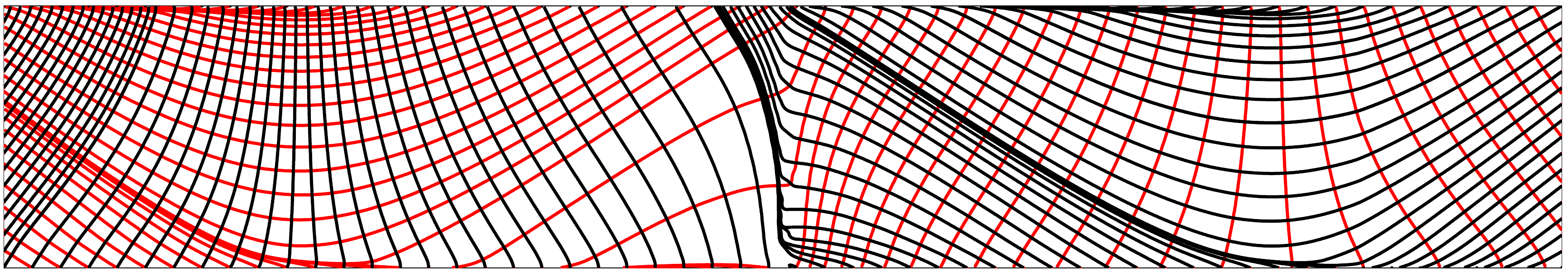}}
\put(-55,42){\small(a)}
\put(60,2){\small$a = 2\pi,$ $k = 10^{-6}$}
}

%\put(52,0){\small(a) $a = 2\pi,$ $k = 10^{-6}$}

\vspace{8pt}
{\quad\includegraphics[width=.095\textwidth, trim=0cm 0.1cm 29cm 1.0cm, clip=true] {bandsar_2_spring_0}
\put(7,14){\includegraphics[width=.35\textwidth] {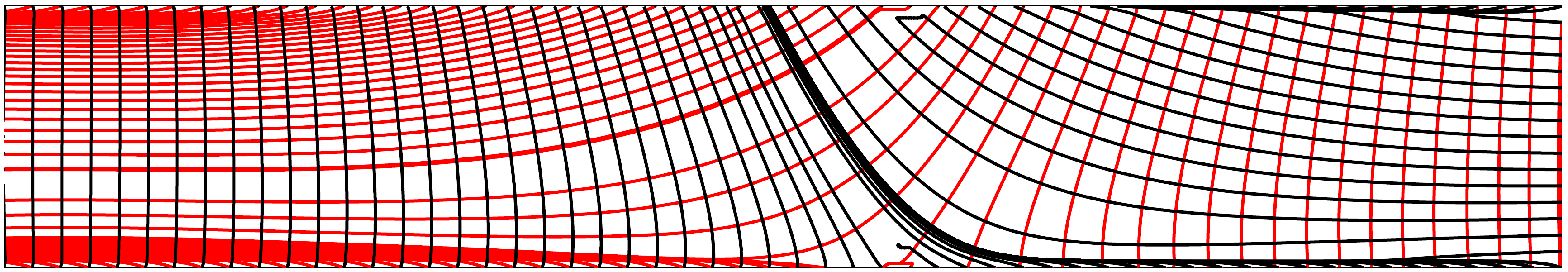}}
\put(-50,48){\small(b)}
\put(64,4){\small$a = 2\pi,$ $k = 10^0$}}

%\put(52,4){\small(b) $a = 2\pi,$ $k = 10^0$}

\vspace{4pt}

{\quad\includegraphics[width=.09\textwidth, trim=0cm 0.1cm 29cm 1.0cm, clip=true] {bandsar_1_spring_2o3}
\put(10,2){\includegraphics[width=.35\textwidth] {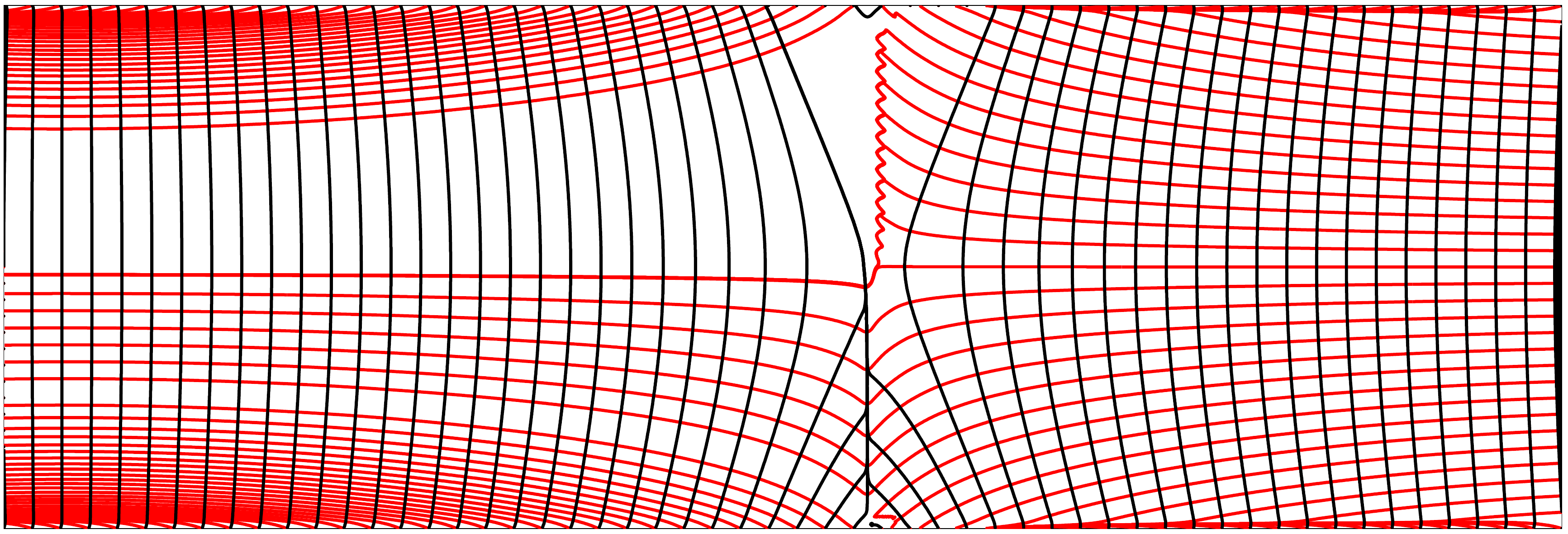}}
\put(-46,48){\small(c)}
\put(68,-8){\small$a = \pi,$ $k = 10^{-2/3}$}}

\caption{Lines of curvature are symmetric and trace out a nearly regular grid for bands made of stretchable materials, indicating reduced concentrations of bending-energy density: Lines of curvature for three representative characteristic shapes are shown on a rectangular strip with the associated equilibrium shape provided above. (a) Developable shapes have pronounced lines of curvature along the length of the band with nearly straight lines of curvature along their widths. (b) For bands made of stretchable materials, shapes with circular centerlines also have nearly straight lines of curvature, however any such band curves more gradually near its axis of symmetry. (c) Collapsed shapes have nearly straight lines of curvature with a sharp curve near the center of the intersection.}
\label{fig:linesOCurv}
\end{figure}
%%%%%%%%%%%%%%%

Differences in the distributions of curvature and stretch exhibited by the three characteristic shapes of M\"obius bands are also evident from the lines of curvature, which are curves with tangent vectors that align with a principal direction at every point. Plots of these curves for representative combinations of stretchability, $k$, and the aspect ratio, $a$, are provided in Figure \ref{fig:linesOCurv}. 

The least regular grid occurs for $k=10^{-6}$. The improvement in regularity that occurs as $k$ increases indicates that smaller gradients of the curvature correspond to smaller strain gradients. In agreement with the previous discussion, we thus infer that bands made of stretchable materials have reduced concentrations of bending. 

The lines of curvature of a collapsed band form a nearly regular grid but curve sharply near the intersection of the band (which coincides with the previously mentioned axis of symmetry). In the absence of collapse, for bands made of stretchable materials we otherwise observe similar behavior. However, lines of curvature bordering the axis of symmetry of such a band are smoother. 
%\vspace{-24pt}

%%%%%%%%%%%%%%%%%%%%%%%%%
\section{Applications}
%%%%%%%%%%%%%%%%%%%%%%%%%

%%%%%%%%%%%%%%%%%%%%%%%%%
%\vspace{-4pt}
\subsection{Graphene}
%\vspace{-2pt}
%%%%%%%%%%%%%%%%%%%%%%%%%

Based on available data for the (two-dimensional) Young's modulus $Y$ and (two-dimensional) Poisson's ratio $\nu$ of graphene, its area modulus $\mu_a=Y/2(1+\nu)$ should be no less than 130~N/m. 
%\cite{Sanchez1999,Kudin2001,Arroyo2004,Zhou2003,Liu2007,Zhou2007,Gui2008,Lee2008,Michel2008,Lu2009} 
\cite{Kudin2001,Lu2009} 
Using this lower bound in \eqref{areamodulus} shows that, to model graphene, the linear spring stiffness must satisfy
\begin{equation}
k_l\ge150~\text{N/m}.
\label{klgraphene}
\end{equation} 

In contrast to the classical Kirchhoff theory, the bending moduli of graphene are independent of $Y$ and $\nu$.\cite{Cadelano2010} For the reported value $22\cdot10^{-20}~\text{Nm}$ of the bending modulus of graphene,\cite{Kudin2001,Lu2009,Cadelano2010} the torsional spring stiffness determined by \eqref{splaymoduli}$_1$ is
\begin{equation}
k_\theta=85\cdot10^{-19}~\text{Nm}.
\label{kthetagraphene}
\end{equation} 

Granted that $k_l$ and $k_\theta$ satisfy \eqref{klgraphene} and \eqref{kthetagraphene}, we may use the definition \eqref{eq:nondimParam_01}$_2$ of the stretchability $k$ to determine upper bounds on the width $w$ along with associated lower bounds on the length, $L$, of a graphene band of given aspect ratio, $a\ge\pi$, needed to give rise to a targeted value of $k$. To achieve $k\le10^{-2}$, for instance, it would be necessary to take $w\le2.4~\text{nm}$ (corresponding to a width of approximately 10 unit cells) and $L\ge2.4\mskip1mua~\text{nm}$. Reducing the stretchability by another order of magnitude would require reducing the width to a few unit cells. Except for exceptionally large values of $a$, the equilibrium shape of a M\"obius nanoband made from graphene should therefore be nearly developable.
%
%%%%%%%%%%%%%%%
\begin{figure*}
\begin{center}
%\hspace{-0.5cm}
\includegraphics[width=.35\linewidth] {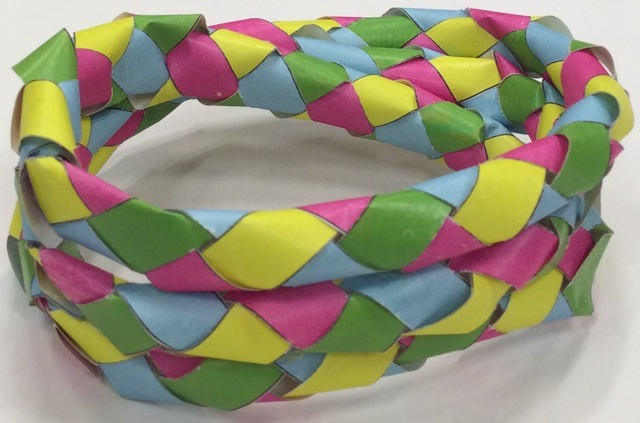}
\hspace{5.5mm}
\includegraphics[width=.114\linewidth] {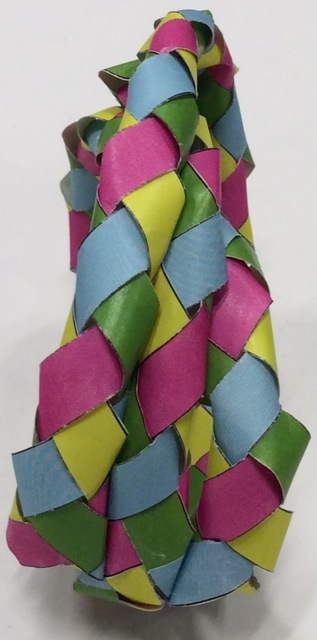}
\includegraphics[width=.12\textwidth,  trim=19.0cm 0.0cm 11.2cm 0cm, clip=true] {bandsar_2_spring_1}
\hspace{5.5mm}
\includegraphics[width=.112\linewidth] {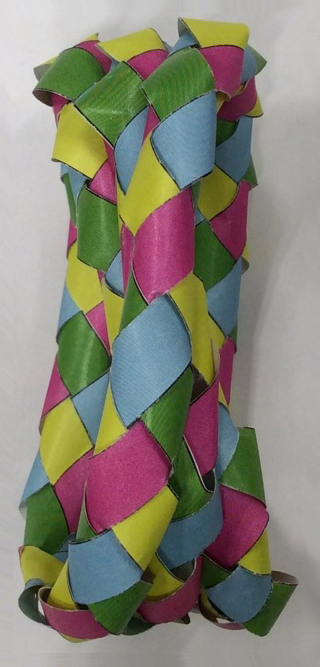}
%\hspace{3mm}
\includegraphics[width=.23\textwidth,  trim=8.41cm 0.7cm 19.5cm 2.1cm, clip=true, angle =90] {bandsar_2_spring_1}
\put(-461,125){\includegraphics[width=.44\linewidth] {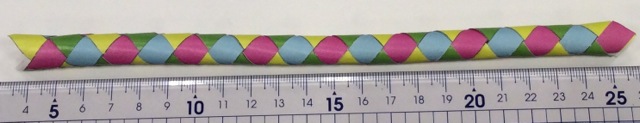}}
\put(-460,158){(a)}
\put(-230,125){\includegraphics[width=.45\linewidth] {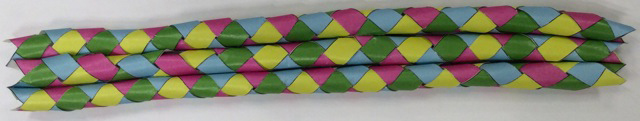}}
\put(-230,158){(b)}
\put(-463,108){(c)}
\put(-264,108){(d)}
\put(-200, 108){(e)}
\put(-122, 108){(f)}
\put(-58, 108){(g)}
\put(-54, 28){\color{white}$a=2\pi$}
\put(-54, 15){\color{white}$k=10^{-1}$}
\put(-193.75, 28){\color{white}$a=2\pi$}
\put(-193.75, 15){\color{white}$k=10^{-1}$}

%\vspace{-0.3cm}
\end{center}
\caption{A M\"obius band formed from finger trap structures adopts a characteristic stretchable shape: (a) One strand of the finger trap structure. (b) The three strands glued together. (c--g) Stretchable finger trap M\"obius band from different perspectives compared to simulation results for $a=2\pi$ and $k=10^{-1}$. }
\label{fig:Chin}
\end{figure*}

%%%%%%%%%%%%%%%%%%%%%%%%%
%\vspace{-4pt}
\subsection{A stretchable material architecture}
%\vspace{-2pt}
%%%%%%%%%%%%%%%%%%%%%%%%%

As a consequence of the structurally induced effective stretchability of a discrete lattice, there is a possibility of using a nanomaterial with novel architecture to fabricate M\"obius bands without regions of concentrated bending-energy density. As a particular example, effectively unstretchable graphene nanoribbons (GNRs) could be formed into a stretchable biaxial braid with the structure of a (Chinese) finger trap and then assembled side-by-side to form a ribbon. 

Such ribbons could then be twisted into a M\"obius band. Since our calculations as well as the work of Qi et al.~\cite{Qi2014} show that the behavior of strained graphene at the nanoscale is qualitatively similar to that of strained paper at the macroscale, it is instructive to realize this construction with a paper model. The macroscale M\"obius band shown in panels (c), (d), and (f) of Figure \ref{fig:Chin} adopts a shape close to that characteristic of bands with stretchability $k=10^{-1}$ and aspect ratios satisfying $a>\pi$. On this basis, we estimate that the stretchability of our proposed architecture is roughly equal to $k\approx10^{-1}$. More generally, this example consequently suggests that our model can be used to infer the effective stretchability of a material from the shape it adopts when twisted into a M\"obius band. 

%Importantly, 

DNA origami techniques pioneered by Han and coworkers\cite{Han2010} might be used to fabricate nanometer or micron scale versions of the macroscopic finger trap M\"obius bands considered here.

%%%%%%%%%%%%%%%%%%%%%%%%%%
\section{Conclusions and discussion}
%%%%%%%%%%%%%%%%%%%%%%%%%%

Using a discrete, lattice-based model, we confirm analytical results for the distributions of curvature and torsion on the centerlines of M\"obius bands\cite{Randrup1996} along with results qualitatively similar to those obtained for continuum models.\cite{Mahadevan1993, Starostin2007a, Starostin2007} In analogy to the effect of writhe,\cite{Herges2006, Schaller2014} we find that twisting strain becomes delocalized if stretching is allowed. To predict macroscopic band shapes for a given material, we establish a connection between stretchability and relevant continuum moduli. This affords insight regarding the practical feasibility of synthesizing M\"obius bands from materials with continuum parameters that can be measured experimentally or estimated by upscale averaging.

We find that M\"obius bands adopt three characteristic shapes depending on stretchability, $k$, and aspect ratio, $a$. These provide reasonable target shapes and strategies for the guided assembly and synthesis of M\"obius bands. In particular, we find that highly stretchable bands have lower energies, indicating that it would be easier to make M\"obius bands from more stretchable materials. Hence, M\"obius molecules might well correspond to these stretchable shapes. We also find that for bands with $k<10^{-2}$, the smaller $a$ becomes, the higher the energy, so that wider bands should be more difficult to make than narrower ones. Conversely, we find that for $k>10^{-2}$ bands with lower $a$ have lower energies, but for $a = \pi$ and either $k = 10^{-1}$ or $k = 10^{-2/3}$ they collapse into self-intersecting achiral shapes, which are degenerate M\"obius bands. This suggests that highly stretchable bands with small aspect ratios might be unstable and thus impossible to make. For $10^{-2}\lesssim k \lesssim 10^{-1}$, bands with smaller values of $a$ should, however, be easier to make than those with larger ones. 

Thus, $k$ and $a$ influence not only the equilibrium shape of a M\"obius band but also its ease of guided assembly. Due to the energetically advantageous nature of bands made of stretchable materials, the ability to design and fabricate M\"obius bands with nondevelopable shapes might be valuable. These findings could be helpful in a variety of applications and also serve as an archetype for other twisted topologies. 

In particular, our results indicate that bands made of stretchable materials are distinguished from bands with developable shapes by lower curvature gradients. Consequently, the stretchability $k$ is positively related to the extent to which the curvature of a band is homogeneous. This connection between stretchability, curvature gradients, and shape has implications for the electronic ground state of quantum M\"obius bands.\cite{Fomin2012} In particular, the ground and excited states of a M\"obius band determined by minimizing Wunderlich's functional over a restricted family of centerlines were found to exhibit wave functions that are significantly altered by curvature effects.\cite{Gravesen2005} Similarly, curvature has been found to influence electron localization, where deep potential wells arise in connection with singularities of the bending-energy density.\cite{Korte2009} Since stretchability and curvature are inextricably linked,\cite{Gauss1827} our results suggest that the electron localization of a quantum M\"obius band is directly related to its stretchability and aspect ratio. For bands made of stretchable materials, the combined influences of nontrivial Gaussian curvature and mean curvature may yield unprecedented effects.

M\"obius bands comprised of graphene with architectures that impart stretchability might allow for advances based on M\"obius topology. Importantly, the use of graphene M\"obius bands for topological insulators has already been investigated.~\cite{Guo2009}  Here, we presented a simple but realistic finger trap paper model as macroscopic analog for stretchable graphene M\"obius structures that allow determination of the underlying stretchability. Such analog models could thus be used as a straightforward means to test new architectures and gain further insight into their properties. Further, comparing the shape adopted by our finger trap M\"obius bands with numerical results generated by our model yields a quantitative estimate for the effective stretchability of the novel woven construction proposed here.

To suppress the fragility and fast decoherence of quantum states and yield topological stability, a notion resembling that underlying our finger trap model has been proposed for the formation of a topological qubit by braiding Marjoranas.~\cite{You2014, Ho2014,Boer2014} Also in the quantum realm and taking advantage of M\"obius topology, a method to enhance spiral intramolecular charge transfer in M\"obius cyclacene for use in novel optical and photoelectric devices has been proposed.~\cite{Zhong2012}

Given the ever-increasing spectrum of applications, the treatment of stretchability and aspect ratio and their implications for the M\"obius band presented here may help guide further developments as well as to increase the theoretical understanding of these fascinating objects.

\section{Methods}
\label{sec:methods}

%%%%%%%%%%%%%%%%%%%%%%%%%%%%%%%%%%%%%	  
%\vspace{-4pt}
\subsection{Energy minimization}
\label{sec:minim}	
%\vspace{-2pt}
%%%%%%%%%%%%%%%%%%%%%%%%%%%%%%%%%%%%%	  

\begin{figure}[!b]
\centering
\includegraphics[width=0.975\linewidth]{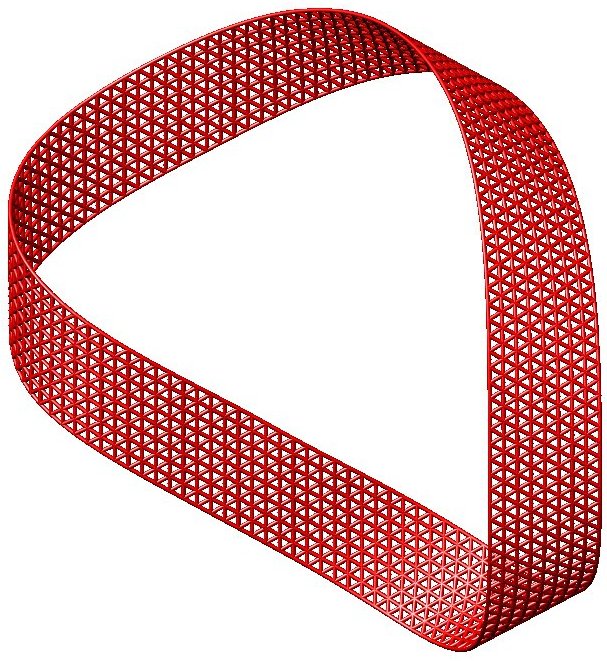}
\caption{Schematic of a trial configuration provided by Sadowsky's\cite{Hinz2013a,Hinz2014a} developable M\"obius band and used for computing the energetically preferred shape of a band of aspect ratio $a=6\pi$ discretized by $N=1170$ points separated by uniform dimensionless distance $r_0/L=\sqrt{3}/72\pi$.% 
}
%\vspace{1cm}
\label{trialband}
\end{figure}

To minimize the total energy $\Psi$ appearing in \eqref{nrgND_03}, we use a conjugate-gradient method within the molecular dynamics code LAMMPS (Large-scale Atomic/Molecular Massively Parallel Simulator).\cite{Plimpton1995} For all values of the aspect ratio $a$ considered, trial configurations for computing energy minima are provided by the piecewise isometric construction (Figure \ref{trialband}) described by Sadowsky.\cite{Hinz2013a,Hinz2014a} 
%
%%%%%%%%%%%%%%%%%%%%%%%%%%
Since each trial shape is developable, minimizing the energy of a band made of a material with stretchability $k$ very close to zero constrains stretching and thus maintains approximate developability. With the linear springs already at their equilibrium distance, this minimization amounts to reducing the curvature of the band without altering the  distances between points. Conversely, for non-negligible values of $k$, the spacing between points is unconstrained. The model and energy minimization strategy were subjected to extensive validation procedures, as described in the SI.

%%%%%%%%%%%%%%%%%%%%%%%%%%
\vspace{-4pt}
\subsection{Curvature approximations}
\vspace{-2pt}
%%%%%%%%%%%%%%%%%%%%%%%%%%

At each discrete point on the surface of the band, we compute discrete versions of the mean curvature $H$, the Gaussian curvature $K$, and the principal curvatures $\kappa_1$ and $\kappa_2$ from discrete versions of the first and second fundamental forms. To calculate the latter quantities, we proceed by analogy to the continuous case,\cite{doCarmo1976} where tangent vectors are approximated by vectors between neighboring points on the surface.

%%%%%%%%%%%%%%%%%%%%%%%%%%
%\vspace{-4pt}
\subsection{Lines of curvature}
%\vspace{-2pt}
%%%%%%%%%%%%%%%%%%%%%%%%%%

We determine the vector field of principal directions and approximate the lines of curvature by computing the relevant streamlines. All computed lines of curvature are symmetric about an axis perpendicular to the length of the band.

%%%%%%%%%%%%%%%%%%%%%%%%%%%%%%%%%%%%%	 
%\vspace{-4pt} 
\subsection{Finger trap models}
\label{sec:paper_models}	
%\vspace{-2pt}
%%%%%%%%%%%%%%%%%%%%%%%%%%%%%%%%%%%%%

Paper models for the finger trap M\"obius band were made using standard office paper ($80~\rm{g/m^2}$) and liquid craft glue.

\footnotesize
\setlength{\bibsep}{2pt}

%\end{document}

%%%%%%%%%%%%%%%%%%%%
% SI starts here
%%%%%%%%%%%%%%%%%%%%

\newcommand{\bft}{{\bf t}}
\newcommand{\bfb}{{\bf b}}
\newcommand{\bfn}{{\bf n}}
\newcommand{\bfI}{{\bf I}}%
\newcommand{\half}{{\textstyle{\frac{1}{2}}}}
\newcommand{\bfalpha}{\boldsymbol{\alpha}}%
\newcommand{\trace}{\ensuremath{\text{tr}\,}}

%\begin{document}

%\thispagestyle{plain}
%\fancypagestyle{plain}{
%\fancyhead[L]{\includegraphics[height=8pt]{headers/LH}}
%\fancyhead[C]{\hspace{-1cm}\includegraphics[height=20pt]{headers/CH}}
%\fancyhead[R]{\includegraphics[height=10pt]{headers/RH}\vspace{-0.2cm}}
%\renewcommand{\headrulewidth}{1pt}}
%\renewcommand{\thefootnote}{\fnsymbol{footnote}}
%\renewcommand\footnoterule{\vspace*{1pt}% 
%\hrule width 3.4in height 0.4pt \vspace*{5pt}} 
%\setcounter{secnumdepth}{5}

\renewcommand{\theequation}{SI.\arabic{equation}}
\renewcommand{\thefigure}{SI.\arabic{figure}}
\setcounter{equation}{0}    
\setcounter{figure}{0}
\clearpage
\setcounter{page}{1}

\makeatletter 
\def\subsubsection{\@startsection{subsubsection}{3}{10pt}{-1.25ex plus -1ex minus -.1ex}{0ex plus 0ex}{\normalsize\bf}} 
\def\paragraph{\@startsection{paragraph}{4}{10pt}{-1.25ex plus -1ex minus -.1ex}{0ex plus 0ex}{\normalsize\textit}} 
\renewcommand\@biblabel[1]{#1}            
\renewcommand\@makefntext[1]% 
{\noindent\makebox[0pt][r]{\@thefnmark\,}#1}
\makeatother 
\renewcommand{\figurename}{\small{Fig.}~}
\sectionfont{\large}
\subsectionfont{\normalsize} 

\setlength{\columnsep}{6.5mm}
\setlength\bibsep{1pt}

\onecolumn

%%%%%%%%%%%
\title{Supplementary Information for:\\[4pt]
Influence of material stretchability on the equilibrium\\ shape of a M\"obius band}

\author{
David M.\ Kleiman,\text{$^{a}$} Denis F.\ Hinz,\textit{$^{b}$} and Eliot Fried$^{\ast}$\textit{$^{c}$}
}
\date{}

\maketitle

\vspace{-18pt}

%\noindent
%\large
\begin{center}
\textit{$^{a}$Department of Matthematics and Statistics, McGill University\\ Montr\'eal, Qu\'ebec, Canada H3A 2K6. E-mail: dave.kleiman2@gmail.com}

\vspace{5pt}
\textit{$^{b}$Kamstrup A/S, Industrivej 28, Stilling, 8660 Skanderborg, Denmark. E-mail: dfhinz@gmail.com}
 
 \vspace{5pt}
 \textit{$^{c}$Mathematical Soft Matter Unit, Okinawa Institute of Science and Technology\\ Onna, Okinawa, Japan 904-0495. E-mail: eliot.fried@oist.jp}

\end{center}

\normalsize
%\onecolumn

%\appendix

%\section{Supplementary information}

%%%%%%%%%%%%%%%%%%%%%%%%%%%%%%%%%%%%%	  
%\section{Supporting information}
%%%%%%%%%%%%%%%%%%%%%%%%%%%%%%%%%%%%%	  
%\newpage

%%%%%%%%%%%%%%%%%%%%%%%%%%%%%
\section{Discrete approximations of various curvatures}%\section{Kinematics}
\label{sec:Discr}%\label{sec:kinem}
%%%%%%%%%%%%%%%%%%%%%%%%%%%%%
%

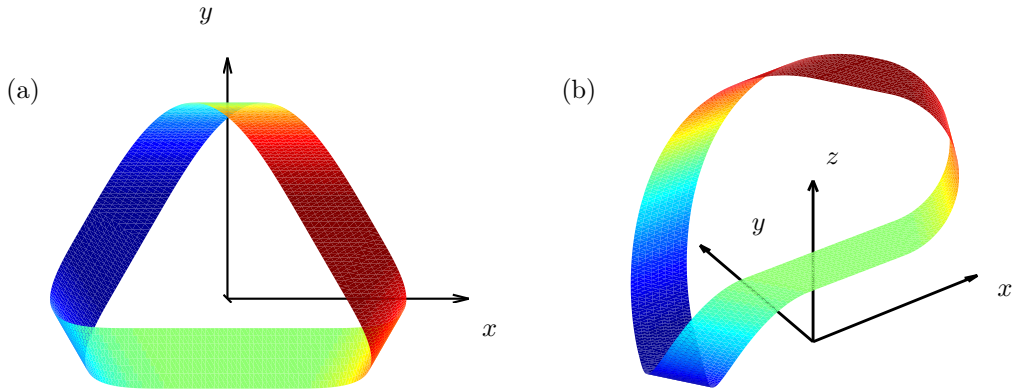
\begin{figure*}[b]
\begin{center}
\begin{picture}(500,150)
\put(100,30){\includegraphics[width=0.3\linewidth]{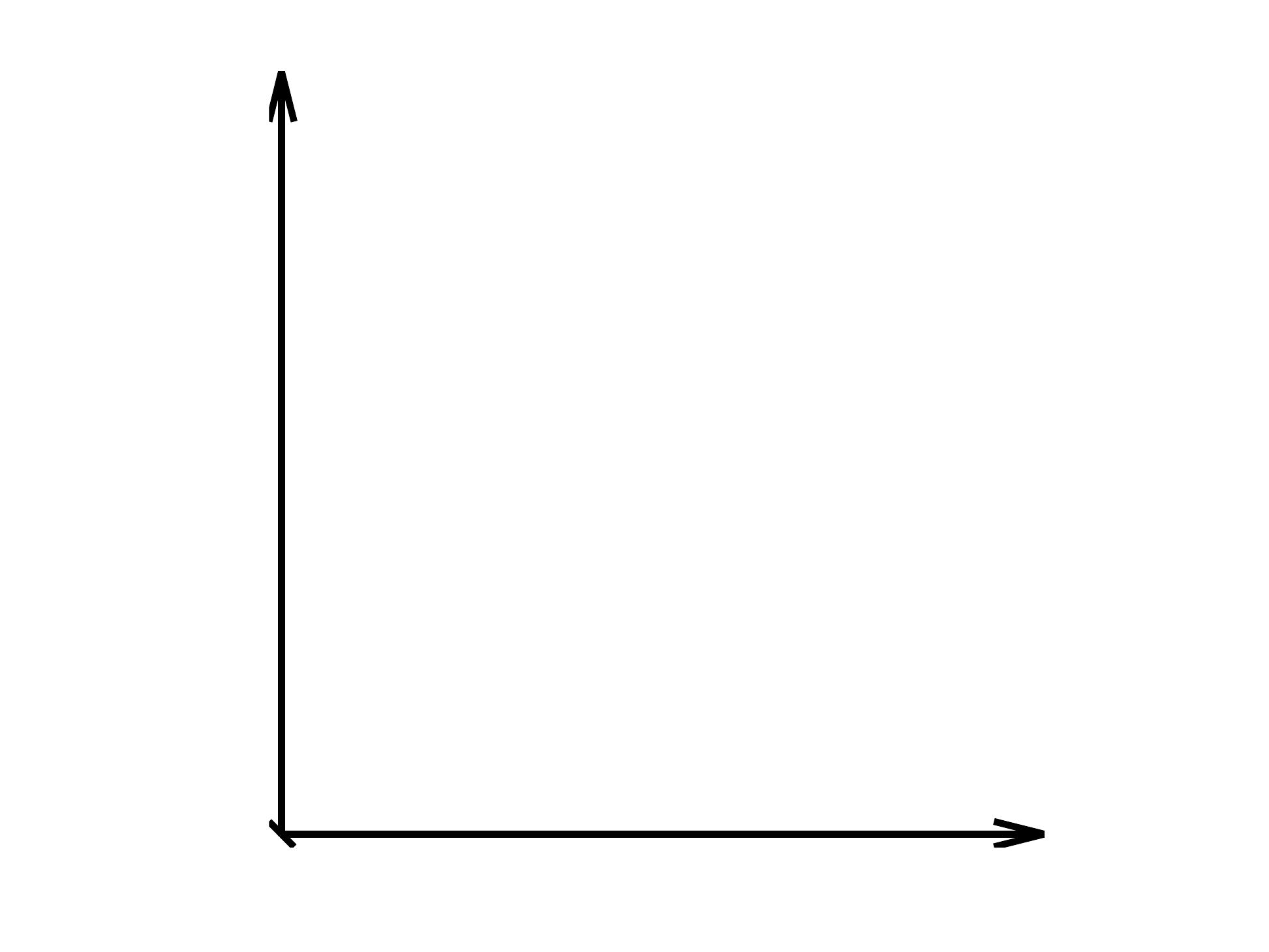}}
\put(123,150){$y$}
\put(230,30){$x$}
\put(50,10){\includegraphics[width=0.3\linewidth]{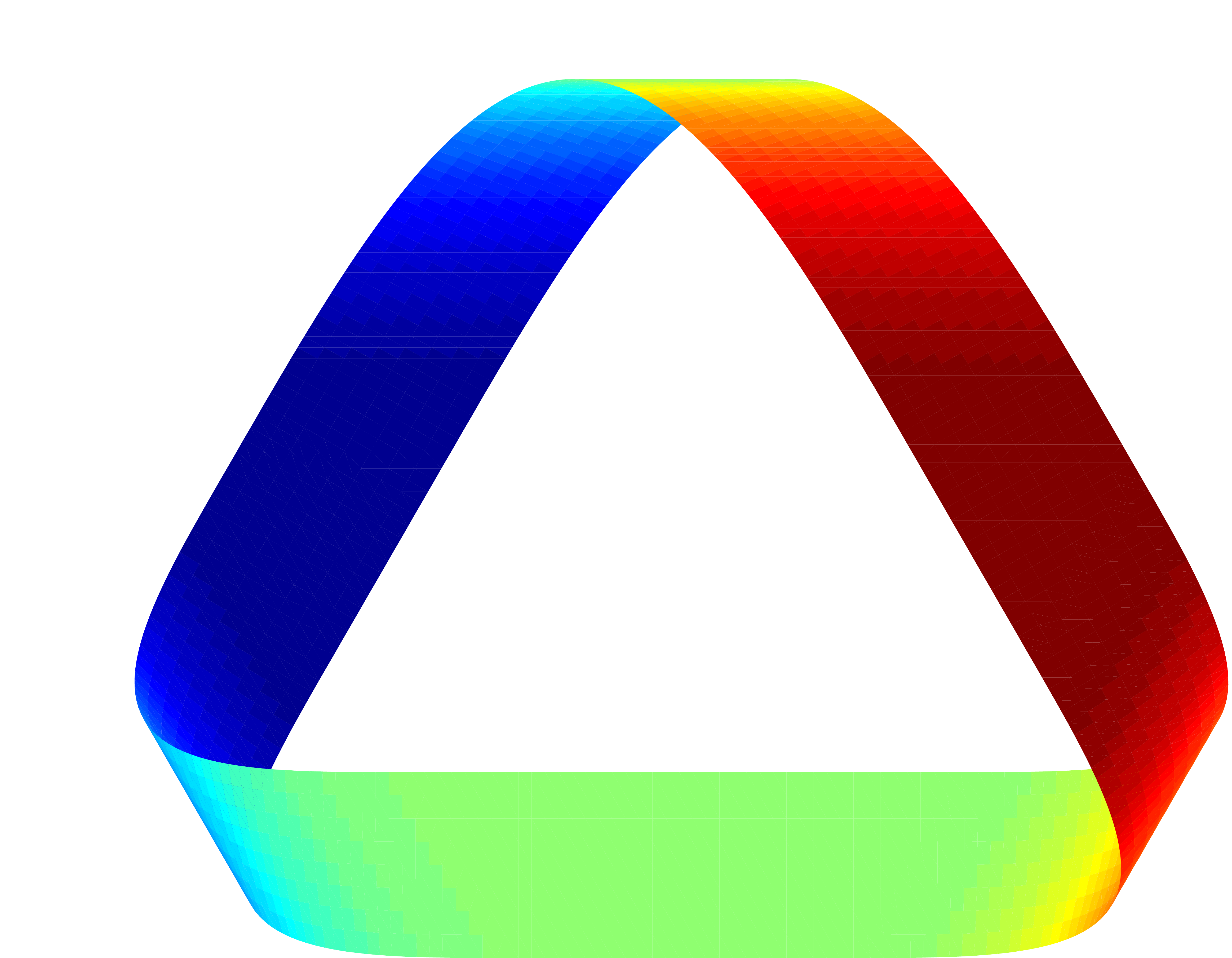}}
%\put(255,5){\includegraphics[width=50mm, height=16mm]{./latex/trapdiagram.pdf}}
%\put(285,26){\tiny{$30^{\circ}$}}
%\put(302,27){\tiny{$x$}} 
%\put(274,50){\tiny{$\frac{\sqrt{3}}{3}x$}}
\put(260,10){\includegraphics[width=0.4\linewidth]{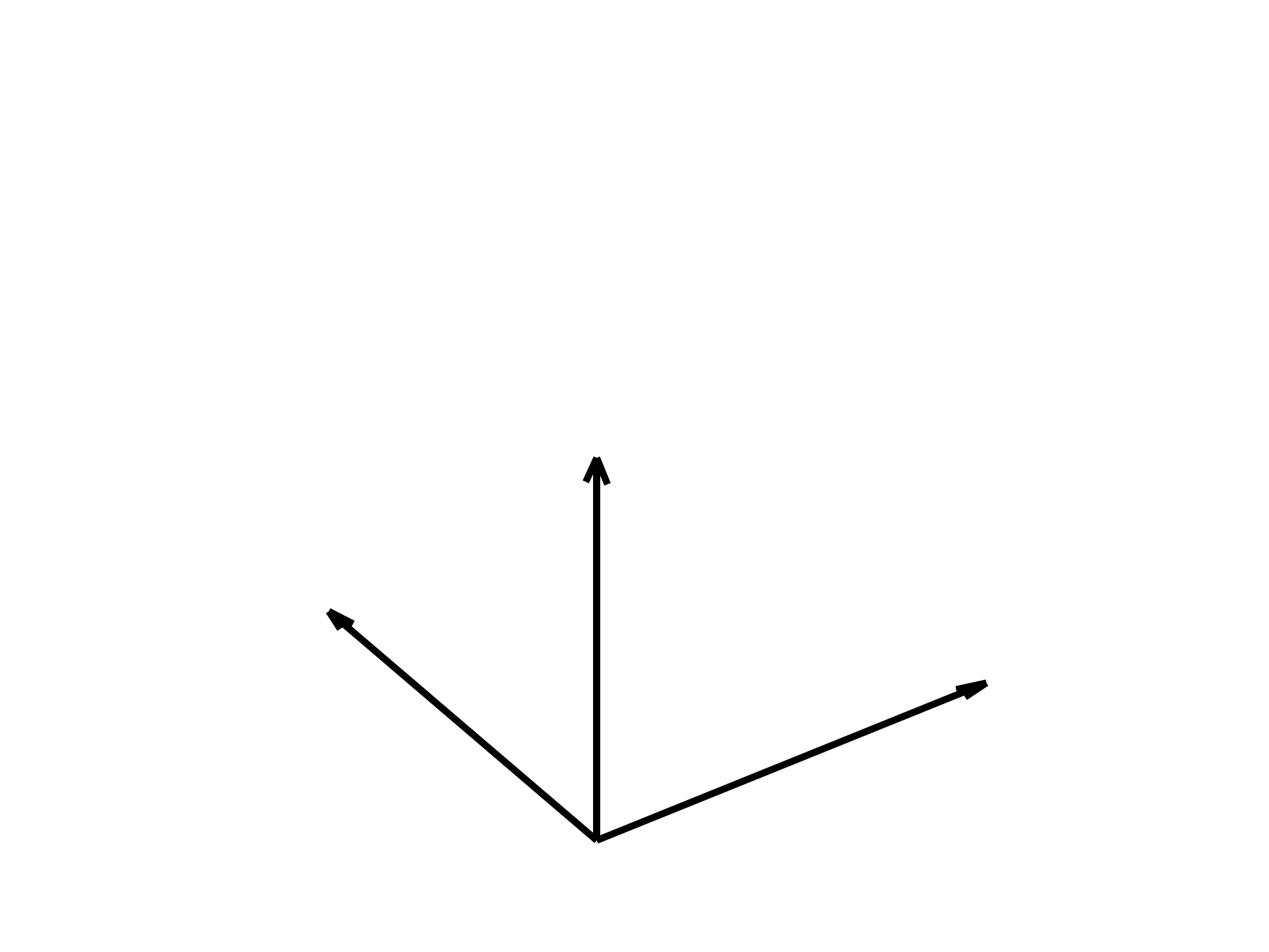}}
\put(332,70){$y$}
\put(425,45){$x$}
\put(360,95){$z$}
\put(210,10){\includegraphics[width=0.4\linewidth]{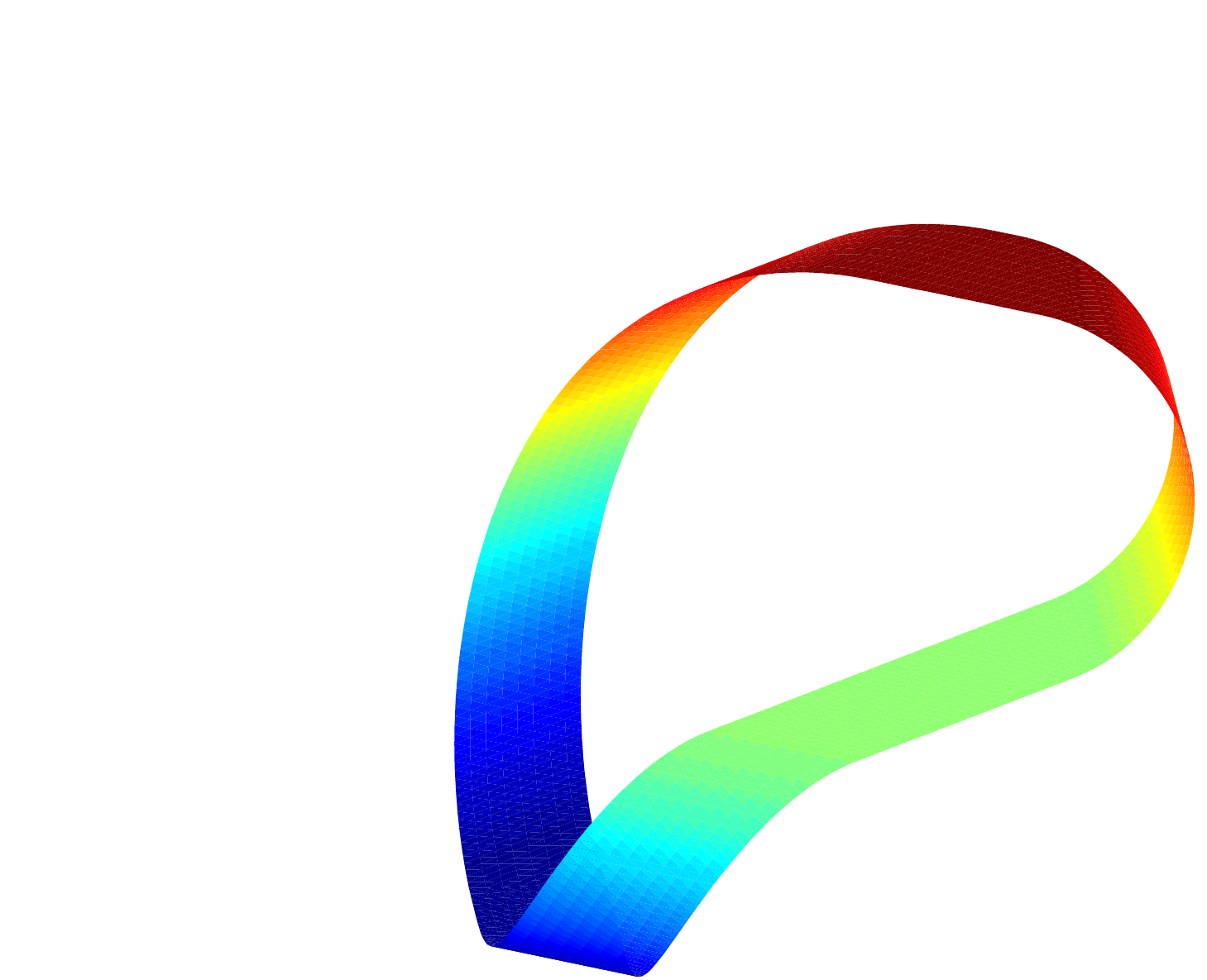}}
\put(50,120){(a)}
\put(260,120){(b)}
%\put(100,50){\includegraphics[width=20mm]{./latex/tripod.pdf}}
%\put(100,65){\bft}
%\put(140,50){\bfb}
%\put(123,100){\bfn}
\end{picture}
\end{center}
\caption{Sadowsky's parametrization of a M\"obius band with a developable shape in a Cartesian coordinate system colored by the $z$-coordinate of each point.}
 %\vspace{1cm}
 \label{SMB}
\end{figure*}

\begin{figure*}[t]
\begin{center}
%\begin{picture}(500,130)
\includegraphics[width=0.6\linewidth, trim = 1cm 5cm 1cm 3cm, clip=true]{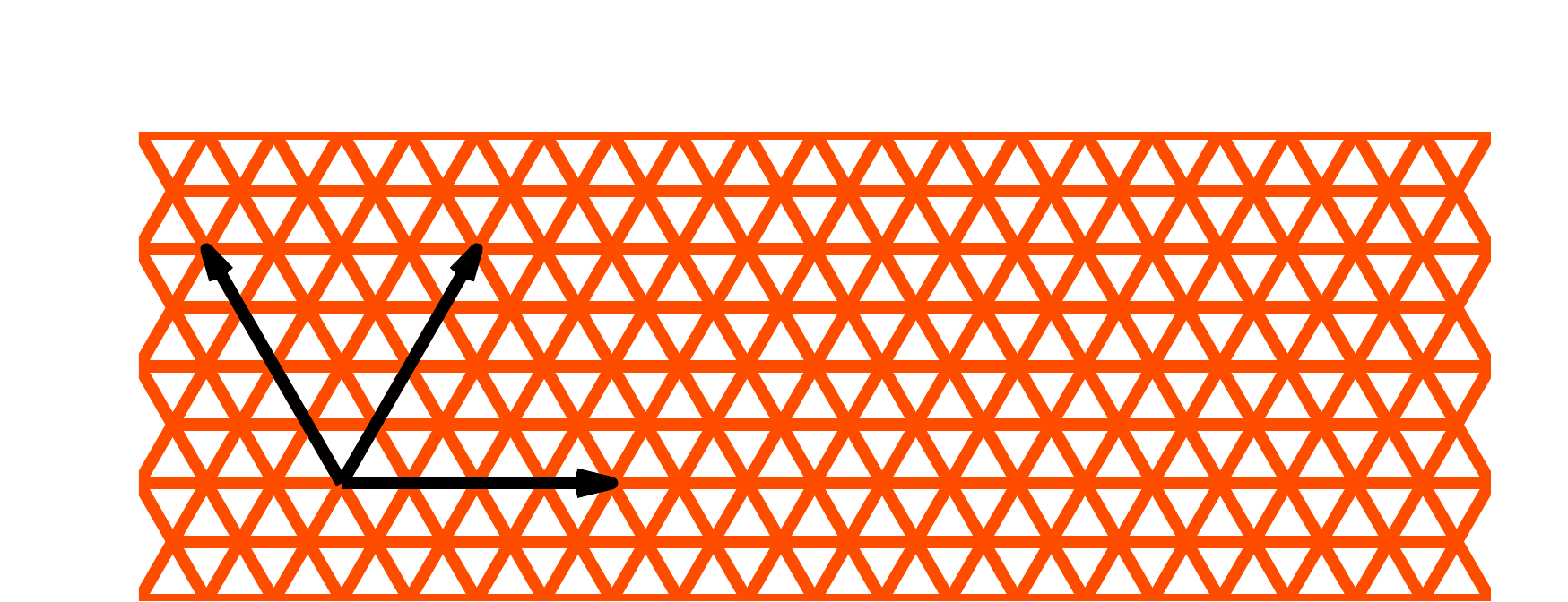}
%\put(194,34){$1$}
%\put(154,86){$2$}
%\put(90,86){$3$}
%\end{picture}
\end{center}
\caption{A lattice of equilateral triangles with the three directions of angular and linear springs illustrated.}
\label{triLat}
\end{figure*}

We approximate a rectangular strip of length $L$ and width $w$ by a lattice of equilateral triangles with $N$ points uniformly separated by a distance $r_0$. The piecewise developable mapping described by Sadowsky~\cite{Hinz2013a} is then used to bend the strip into a M\"obius band (Figure \ref{SMB}), which we then employ as the trial configuration for each numerically determined band with aspect ratio $a$. Each plane strip is discretized with a lattice of equilateral triangles (SI Figure \ref{triLat}). The variable number of points used to discretize the lattice is denoted by $N$. Further, the variable number of points along the centerline is denoted by $N_l$.
%%%%%%%%%%%%%%%%%%%%%%%%%%

For a discretized curve with point spacing $r_0$, the discrete curvature $\kappa(i)$ at the $i$-th point along the curve is given by
\begin{equation}
\kappa(i) = \frac{\pi-\angle(\bft(i),\bft(i+1))}{r_0},
\label{disCurv}
\end{equation}
where $\angle(\bft(i),\bft(i+1))$ denotes the angle between discrete tangent vectors at the point $i$---i.e., the angle at point $i$ between points $i-1,$ $i$, and $i+1$ (see, for example, Belyaev~\cite{Belyaev1999}). Similarly the discrete torsion $\tau(i)$ at the $i$-th point along the curve is given by
\begin{equation}
\tau(i) = \frac{\angle(\bfb(i),\bfb(i+1))}{r_0},
\end{equation}
where $\angle(\bfb(i),\bfb(i+1))$ denotes the angle between the discrete binormals to the curve at points $i$ and $i+1$. 

To calculate the first and second fundamental forms of surface represented by a discrete triangular lattice, we proceed by analogy to the continuous case (see, for example, do Carmo~\cite{doCarmo1976}), but we use vectors between neighboring points on the surface as tangent coordinate vectors. At each point $i$ of the lattice, we determine the discrete mean curvature $H(i)$, the discrete Gaussian curvature $K(i)$, and the discrete principal curvatures $\kappa_1(i)$ and $\kappa_2(i)$ from the first and second fundamental forms. 

Let $\alpha$ be the angle between the tangent vector to a curve and the principal direction associated with $\kappa_1$. The discrete normal curvature $\kappa_n(i)$ at the $i$-th point along a curve can then be determined pointwise from the continuous relation
\begin{equation}
\label{eq:kappa_n02}
\kappa_n = \kappa_1\cos^2\alpha + \kappa_2\sin^2\alpha.
\end{equation}
Further, having determined $\kappa(i)$ and $\kappa_n(i)$ from \eqref{disCurv} and \eqref{eq:kappa_n02}, the magnitude $|\kappa_g(i)|$ of the discrete geodesic curvature $\kappa_g(i)$ can be determined from the continuous relation
\begin{equation}
\label{eq:kappa_g02}
|\kappa_g|=\sqrt{\kappa^2-\kappa_n^2}.
\end{equation}

%%%%%%%%%%%%%%%%%%%%%%%%%
%

%%%%%%%%%%%%%%%
\begin{figure*}[t]
\begin{center}
\begin{picture}(500,300)
\put(20,0){\includegraphics[width=.8\linewidth] {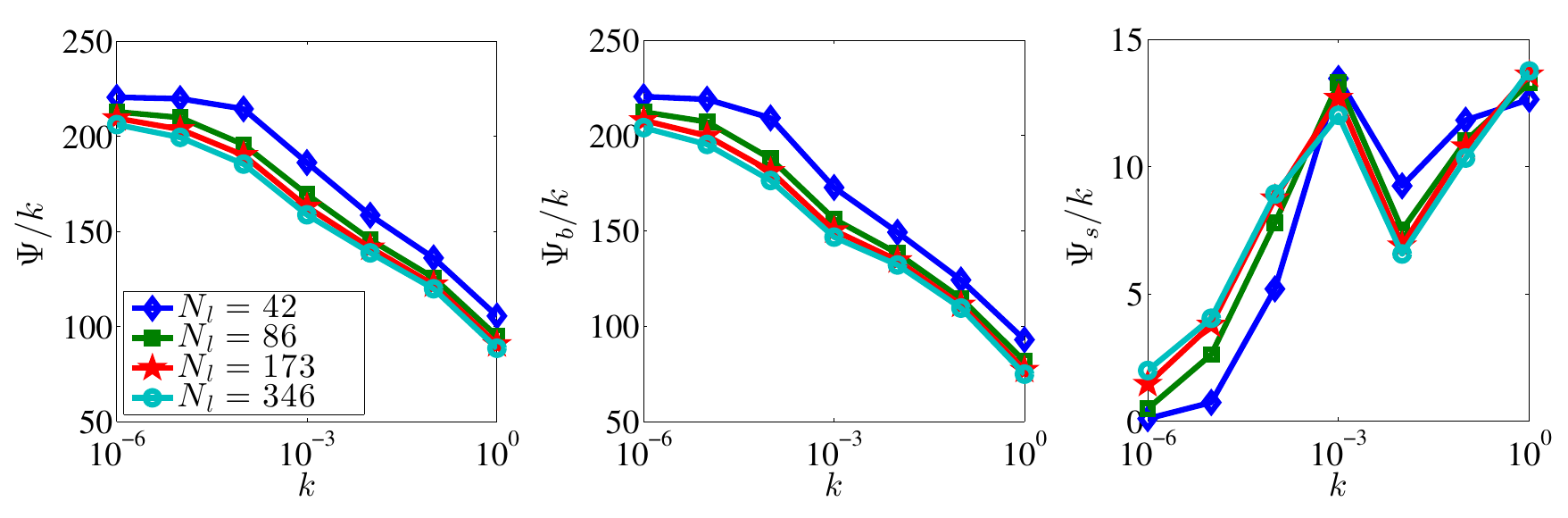}}
\put(108,185){\includegraphics[width=.47\linewidth] {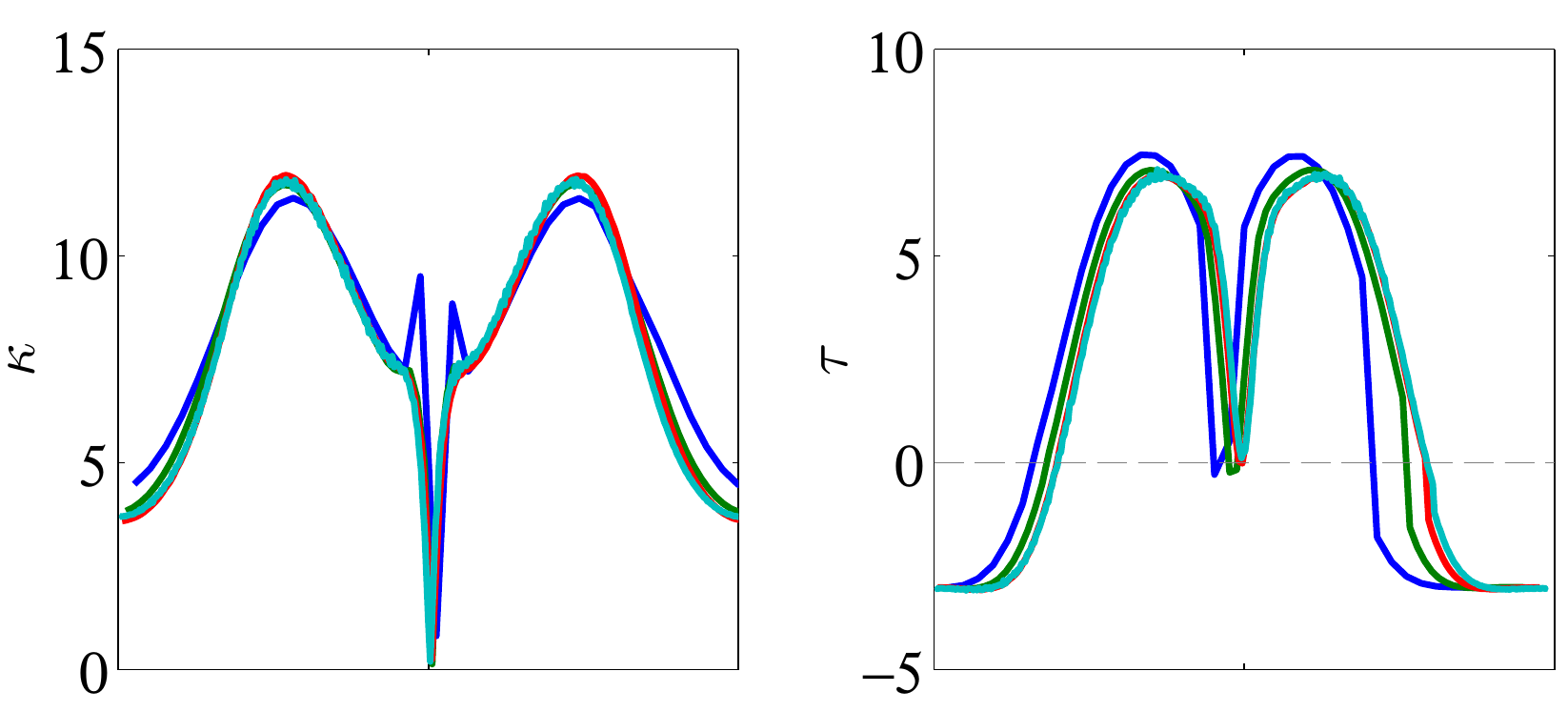}}
\put(125,180){$1$}
\put(164,180){$N_l/2$}
\put(172,170){$s$}
\put(218,180){$N_l$}
\put(252,180){$1$}
\put(290,180){$N_l/2$}
\put(298,170){$s$}
\put(345,180){$N_l$}
\put(20,135){(c)}
\put(160,135){(d)}
\put(300,135){(e)}
\put(105,300){(a)}
\put(240,300){(b)}
\end{picture}
\end{center}
\caption{ The model shows convergence for increasing number of points along the midline: The curvature ${\kappa}$ (a) and torsion ${\tau}$ (b) of the centerline for approximately unstretchable bands (${k=10^{-6}}$) approach a limiting value for a large number of points along the centerline ${173<N_l}$. Notice that, as ${N_l}$ is further increased beyond ${N_l=346}$, slight oscillations in ${\kappa}$ and ${\tau}$ are observed, indicating an ideal range of ${173<N_l<346}$. The total energy ${\Psi}$ (c), stretching energy ${\Psi_s}$ (d), and ${\Psi_b}$ (e) converge to a limiting shape corresponding to the continuum limit. Results are shown for ${a= 4\pi}$.}
\label{fig:convergence}
\end{figure*}
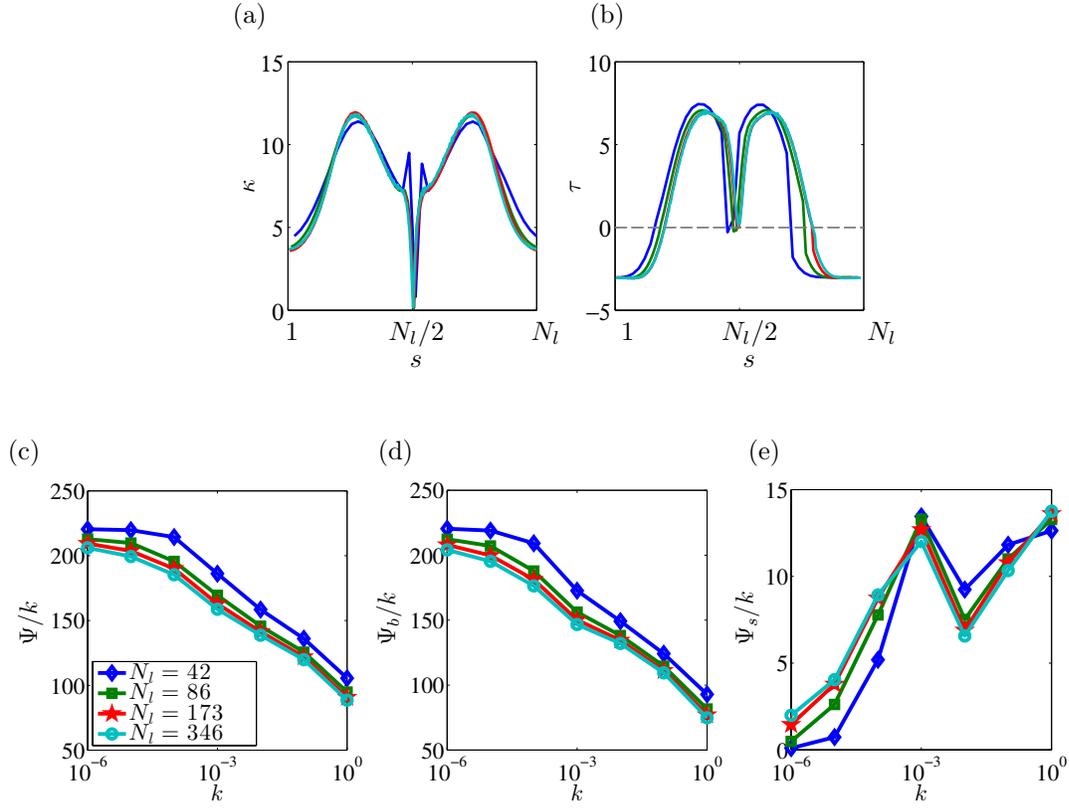
%%%%%%%%%%%%%%%

%%%%%%%%%%%%%%%
\begin{figure*}[t]
\begin{center}
\includegraphics[width=.3\textwidth, trim = 9cm 0cm 0cm 0cm, clip=true] {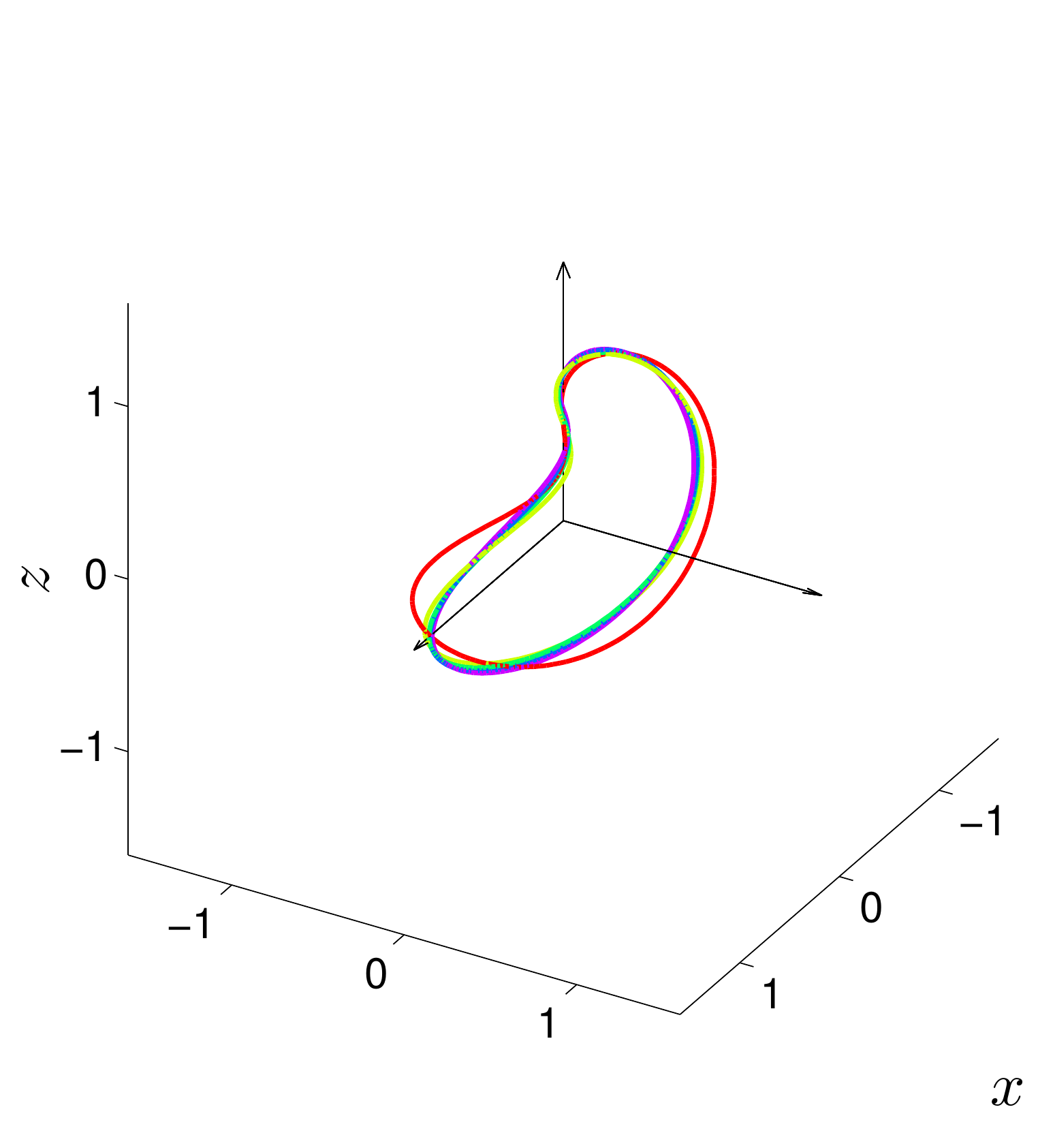}
\put(-125,17){$y$}
\includegraphics[width=.3\textwidth] {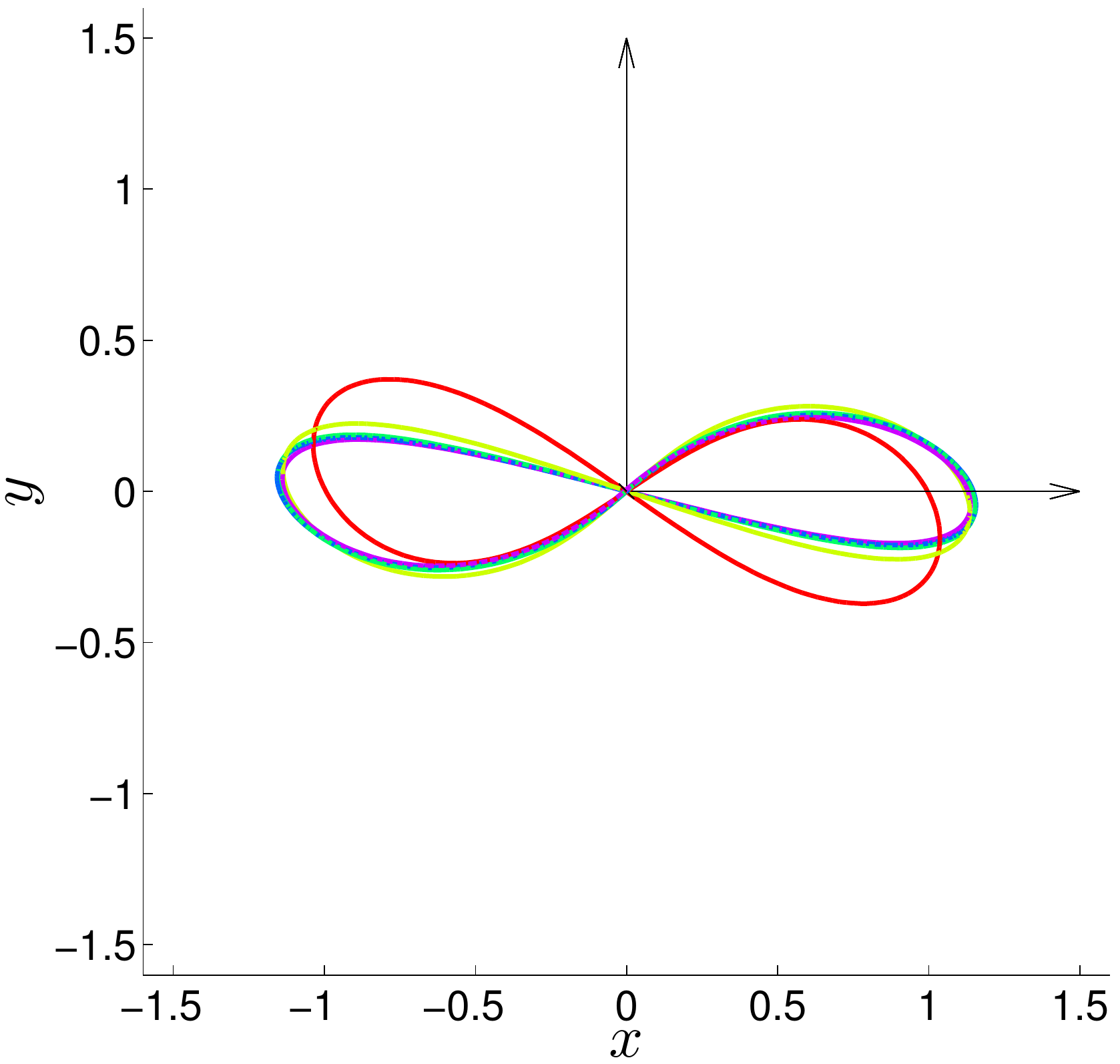}
\put(-120,20){\includegraphics[width=0.08\textwidth]{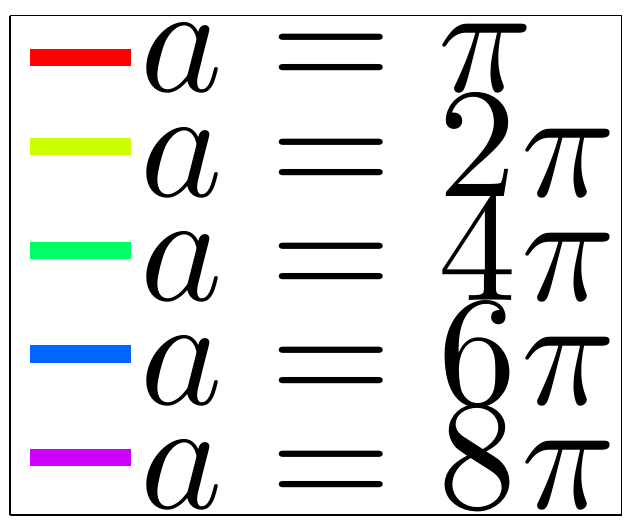}}

\vspace{5mm}
\includegraphics[width=.3\textwidth] {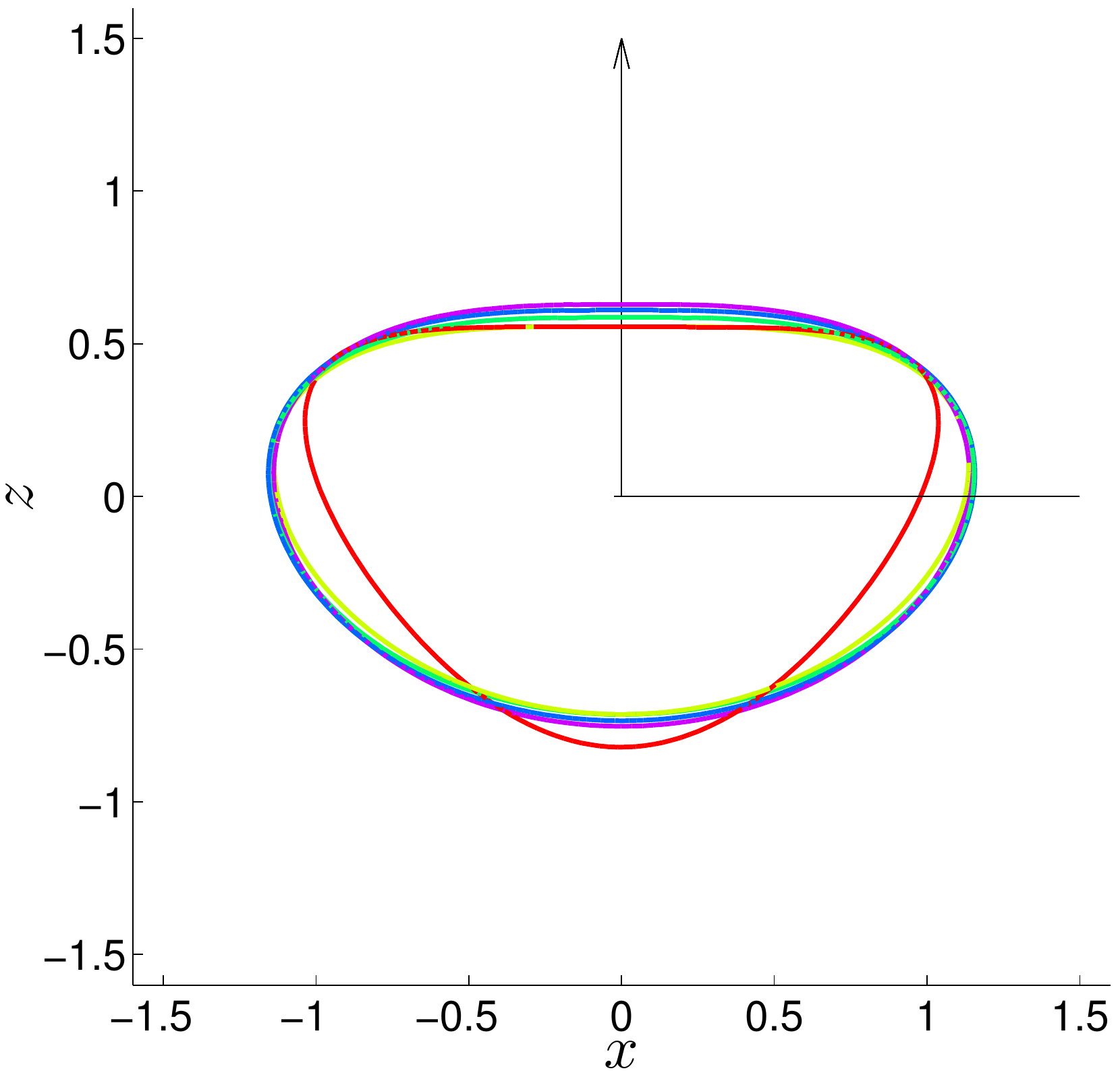}
\includegraphics[width=.3\textwidth] {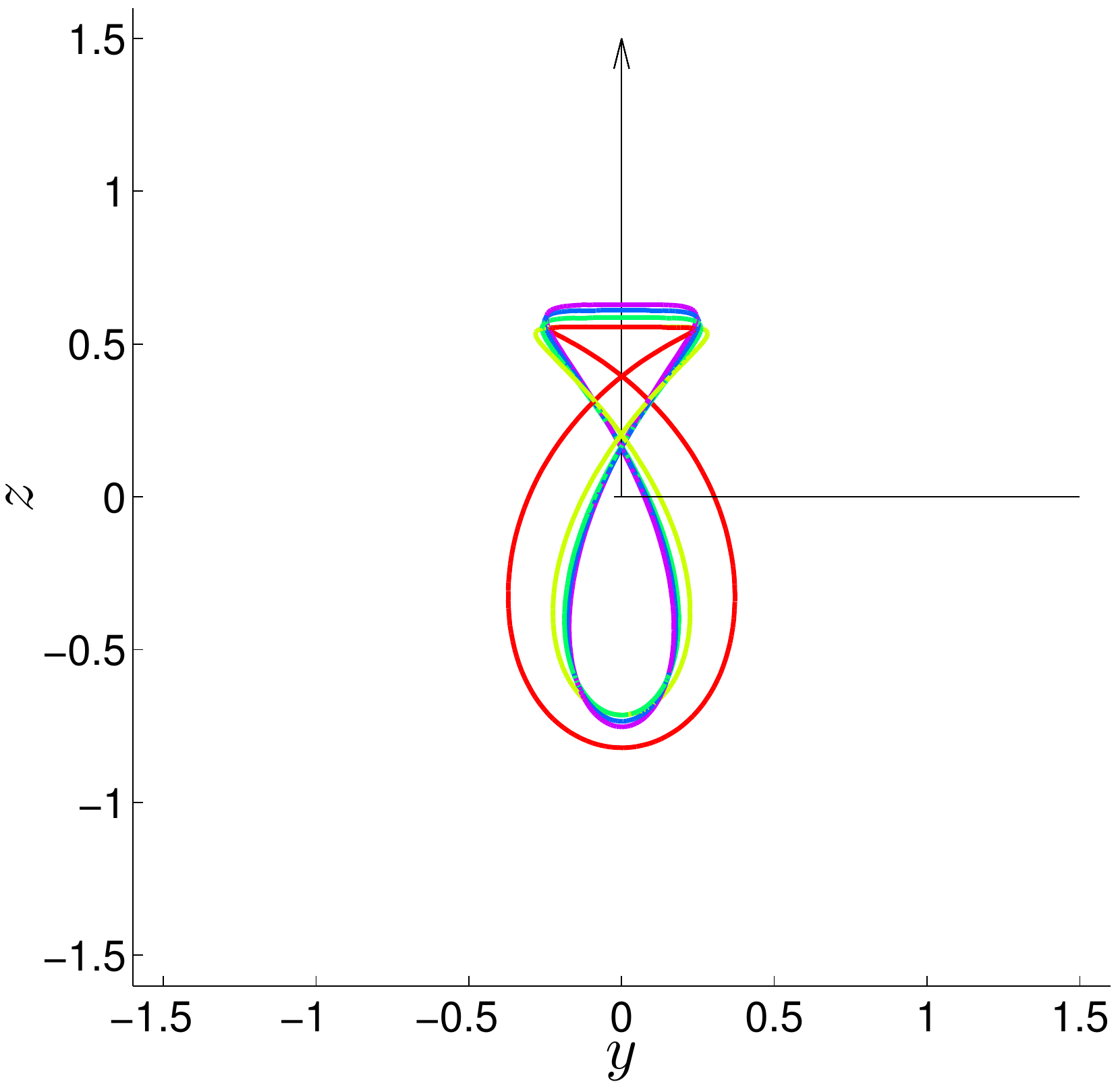}
\end{center}
\caption{The centerline of a approximately unstretchable M\"obius band depends on its aspect ratio: Centerlines of equilibrium shapes of general M\"obius bands for different values of the aspect ratio ${a}$ and stretchability ${k}$. The band is rotated into its main axes. }
\label{fig:Moebius_Centerline_shape}
\end{figure*}
%%%%%%%%%%%%%%%

%%%%%%%%%%%%%%%
\begin{figure*}[t!]
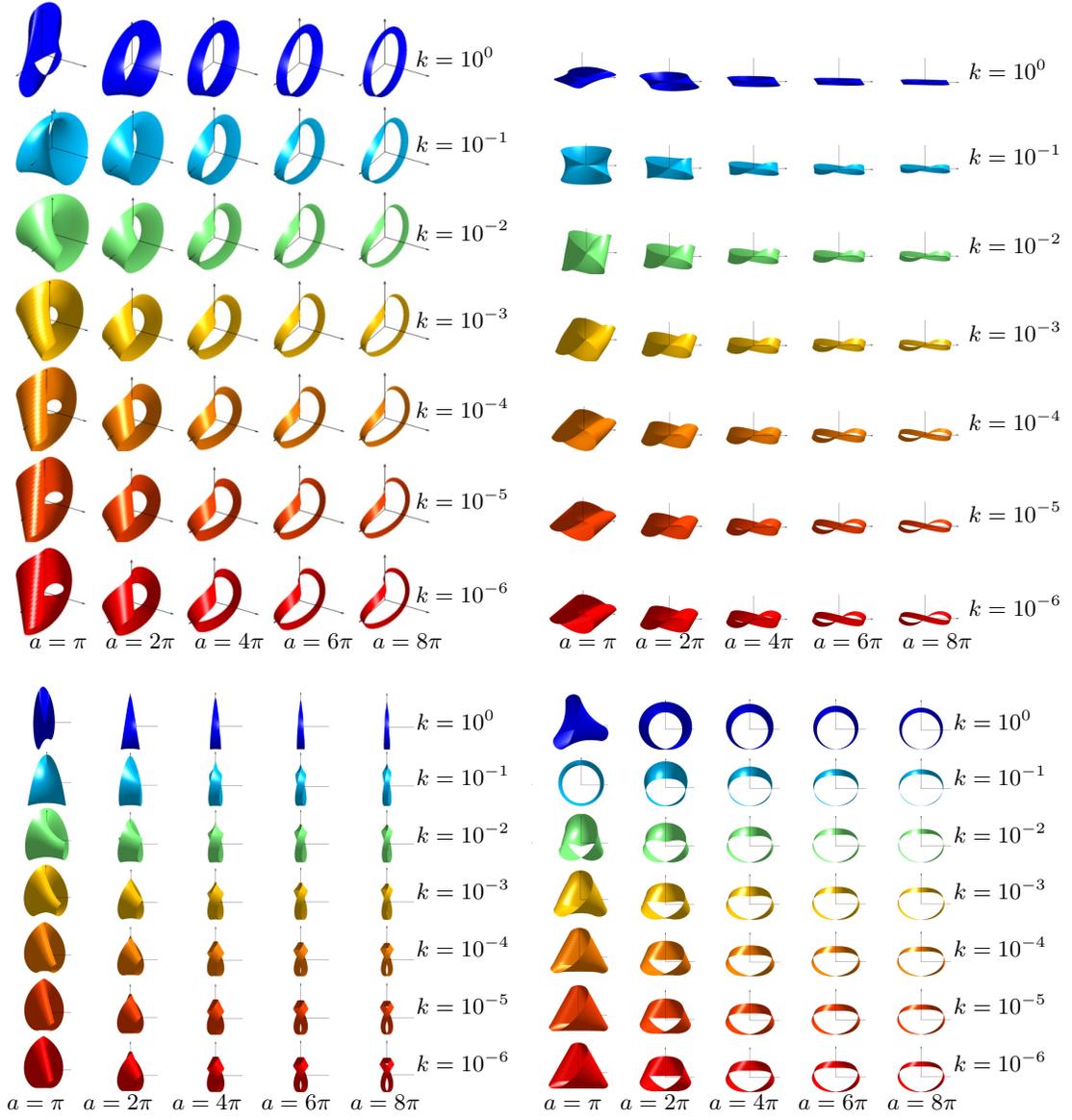

\begin{center}
\hspace{4mm}
\includegraphics[width=.06\textwidth, trim=0cm 0.5cm 29cm 1.0cm, clip=true] {bandsar_1_spring_0}
\includegraphics[width=.06\textwidth, trim=0cm 0.5cm 29cm 1.0cm, clip=true] {bandsar_2_spring_0}
\includegraphics[width=.06\textwidth, trim=0cm 0.5cm 29cm 1.0cm, clip=true] {bandsar_4_spring_0}
\includegraphics[width=.06\textwidth, trim=0cm 0.5cm 29cm 1.0cm, clip=true] {bandsar_6_spring_0}
\includegraphics[width=.06\textwidth, trim=0cm 0.5cm 29cm 1.0cm, clip=true] {bandsar_8_spring_0}
\put(-8,13){\small$k=10^0$}
\hspace{14mm}
\includegraphics[width=.06\textwidth, trim=7cm 0.5cm 18cm 1.0cm, clip=true] {bandsar_1_spring_0}
\includegraphics[width=.06\textwidth, trim=7cm 0.5cm 18cm 1.0cm, clip=true] {bandsar_2_spring_0}
\includegraphics[width=.06\textwidth, trim=7cm 0.5cm 18cm 1.0cm, clip=true] {bandsar_4_spring_0}
\includegraphics[width=.06\textwidth, trim=7cm 0.5cm 18cm 1.0cm, clip=true] {bandsar_6_spring_0}
\includegraphics[width=.06\textwidth, trim=7cm 0.5cm 18cm 1.0cm, clip=true] {bandsar_8_spring_0}
\put(1,8){\small$k=10^0$}

\hspace{4mm}
\includegraphics[width=.06\textwidth, trim=0cm 0.4cm 29cm 1.0cm, clip=true] {bandsar_1_spring_1}
\includegraphics[width=.06\textwidth, trim=0cm 0.4cm 29cm 1.0cm, clip=true] {bandsar_2_spring_1}
\includegraphics[width=.06\textwidth, trim=0cm 0.4cm 29cm 1.0cm, clip=true] {bandsar_4_spring_1}
\includegraphics[width=.06\textwidth, trim=0cm 0.4cm 29cm 1.0cm, clip=true] {bandsar_6_spring_1}
\includegraphics[width=.06\textwidth, trim=0cm 0.4cm 29cm 1.0cm, clip=true] {bandsar_8_spring_1}
\put(-8,13){\small$k=10^{-1}$}
\hspace{14mm}
\includegraphics[width=.06\textwidth, trim=7cm 0.4cm 18cm 1.0cm, clip=true] {bandsar_1_spring_1}
\includegraphics[width=.06\textwidth, trim=7cm 0.4cm 18cm 1.0cm, clip=true] {bandsar_2_spring_1}
\includegraphics[width=.06\textwidth, trim=7cm 0.4cm 18cm 1.0cm, clip=true] {bandsar_4_spring_1}
\includegraphics[width=.06\textwidth, trim=7cm 0.4cm 18cm 1.0cm, clip=true] {bandsar_6_spring_1}
\includegraphics[width=.06\textwidth, trim=7cm 0.4cm 18cm 1.0cm, clip=true] {bandsar_8_spring_1}
\put(1,8){\small$k=10^{-1}$}

\hspace{4mm}
\includegraphics[width=.06\textwidth, trim=0cm 0.1cm 29cm 1.0cm, clip=true] {bandsar_1_spring_2}
\includegraphics[width=.06\textwidth, trim=0cm 0.1cm 29cm 1.0cm, clip=true] {bandsar_2_spring_2}
\includegraphics[width=.06\textwidth, trim=0cm 0.1cm 29cm 1.0cm, clip=true] {bandsar_4_spring_2}
\includegraphics[width=.06\textwidth, trim=0cm 0.1cm 29cm 1.0cm, clip=true] {bandsar_6_spring_2}
\includegraphics[width=.06\textwidth, trim=0cm 0.1cm 29cm 1.0cm, clip=true] {bandsar_8_spring_2}
\put(-8,13){\small$k=10^{-2}$}
\hspace{14mm}
\includegraphics[width=.06\textwidth, trim=7cm 0.1cm 18cm 1.0cm, clip=true] {bandsar_1_spring_2}
\includegraphics[width=.06\textwidth, trim=7cm 0.1cm 18cm 1.0cm, clip=true] {bandsar_2_spring_2}
\includegraphics[width=.06\textwidth, trim=7cm 0.1cm 18cm 1.0cm, clip=true] {bandsar_4_spring_2}
\includegraphics[width=.06\textwidth, trim=7cm 0.1cm 18cm 1.0cm, clip=true] {bandsar_6_spring_2}
\includegraphics[width=.06\textwidth, trim=7cm 0.1cm 18cm 1.0cm, clip=true] {bandsar_8_spring_2}
\put(1,8){\small$k=10^{-2}$}

\hspace{4mm}
\includegraphics[width=.06\textwidth, trim=0cm 0.1cm 29cm 1.0cm, clip=true] {bandsar_1_spring_3}
\includegraphics[width=.06\textwidth, trim=0cm 0.1cm 29cm 1.0cm, clip=true] {bandsar_2_spring_3}
\includegraphics[width=.06\textwidth, trim=0cm 0.1cm 29cm 1.0cm, clip=true] {bandsar_4_spring_3}
\includegraphics[width=.06\textwidth, trim=0cm 0.1cm 29cm 1.0cm, clip=true] {bandsar_6_spring_3}
\includegraphics[width=.06\textwidth, trim=0cm 0.1cm 29cm 1.0cm, clip=true] {bandsar_8_spring_3}
\put(-8,13){\small$k=10^{-3}$}
\hspace{14mm}
\includegraphics[width=.06\textwidth, trim=7cm 0.1cm 18cm 1.0cm, clip=true] {bandsar_1_spring_3}
\includegraphics[width=.06\textwidth, trim=7cm 0.1cm 18cm 1.0cm, clip=true] {bandsar_2_spring_3}
\includegraphics[width=.06\textwidth, trim=7cm 0.1cm 18cm 1.0cm, clip=true] {bandsar_4_spring_3}
\includegraphics[width=.06\textwidth, trim=7cm 0.1cm 18cm 1.0cm, clip=true] {bandsar_6_spring_3}
\includegraphics[width=.06\textwidth, trim=7cm 0.1cm 18cm 1.0cm, clip=true] {bandsar_8_spring_3}
\put(1,8){\small$k=10^{-3}$}

\hspace{4mm}
\includegraphics[width=.06\textwidth, trim=0cm 0.1cm 29cm 1.0cm, clip=true] {bandsar_1_spring_4}
\includegraphics[width=.06\textwidth, trim=0cm 0.1cm 29cm 1.0cm, clip=true] {bandsar_2_spring_4}
\includegraphics[width=.06\textwidth, trim=0cm 0.1cm 29cm 1.0cm, clip=true] {bandsar_4_spring_4}
\includegraphics[width=.06\textwidth, trim=0cm 0.1cm 29cm 1.0cm, clip=true] {bandsar_6_spring_4}
\includegraphics[width=.06\textwidth, trim=0cm 0.1cm 29cm 1.0cm, clip=true] {bandsar_8_spring_4}
\put(-8,13){\small$k=10^{-4}$}
\hspace{14mm}
\includegraphics[width=.06\textwidth, trim=7cm 0.1cm 18cm 1.0cm, clip=true] {bandsar_1_spring_4}
\includegraphics[width=.06\textwidth, trim=7cm 0.1cm 18cm 1.0cm, clip=true] {bandsar_2_spring_4}
\includegraphics[width=.06\textwidth, trim=7cm 0.1cm 18cm 1.0cm, clip=true] {bandsar_4_spring_4}
\includegraphics[width=.06\textwidth, trim=7cm 0.1cm 18cm 1.0cm, clip=true] {bandsar_6_spring_4}
\includegraphics[width=.06\textwidth, trim=7cm 0.1cm 18cm 1.0cm, clip=true] {bandsar_8_spring_4}
\put(1,8){\small$k=10^{-4}$}

\hspace{4mm}
\includegraphics[width=.06\textwidth, trim=0cm 0.1cm 29cm 1.0cm, clip=true] {bandsar_1_spring_5}
\includegraphics[width=.06\textwidth, trim=0cm 0.1cm 29cm 1.0cm, clip=true] {bandsar_2_spring_5}
\includegraphics[width=.06\textwidth, trim=0cm 0.1cm 29cm 1.0cm, clip=true] {bandsar_4_spring_5}
\includegraphics[width=.06\textwidth, trim=0cm 0.1cm 29cm 1.0cm, clip=true] {bandsar_6_spring_5}
\includegraphics[width=.06\textwidth, trim=0cm 0.1cm 29cm 1.0cm, clip=true] {bandsar_8_spring_5}
\put(-8,13){\small$k=10^{-5}$}
\hspace{14mm}
\includegraphics[width=.06\textwidth, trim=7cm 0.1cm 18cm 1.0cm, clip=true] {bandsar_1_spring_5}
\includegraphics[width=.06\textwidth, trim=7cm 0.1cm 18cm 1.0cm, clip=true] {bandsar_2_spring_5}
\includegraphics[width=.06\textwidth, trim=7cm 0.1cm 18cm 1.0cm, clip=true] {bandsar_4_spring_5}
\includegraphics[width=.06\textwidth, trim=7cm 0.1cm 18cm 1.0cm, clip=true] {bandsar_6_spring_5}
\includegraphics[width=.06\textwidth, trim=7cm 0.1cm 18cm 1.0cm, clip=true] {bandsar_8_spring_5}
\put(1,8){\small$k=10^{-5}$}

\hspace{4mm}
\includegraphics[width=.06\textwidth, trim=0cm 0.0cm 29cm 1.0cm, clip=true] {bandsar_1_spring_6}
\includegraphics[width=.06\textwidth, trim=0cm 0.0cm 29cm 1.0cm, clip=true] {bandsar_2_spring_6}
\includegraphics[width=.06\textwidth, trim=0cm 0.0cm 29cm 1.0cm, clip=true] {bandsar_4_spring_6}
\includegraphics[width=.06\textwidth, trim=0cm 0.0cm 29cm 1.0cm, clip=true] {bandsar_6_spring_6}
\includegraphics[width=.06\textwidth, trim=0cm 0.0cm 29cm 1.0cm, clip=true] {bandsar_8_spring_6}
\put(-8,13){\small$k=10^{-6}$}
\put(-160,-5){\small$a= \pi$}
\put(-130,-5){\small$a= 2\pi$}
\put(-95,-5){\small$a= 4\pi$}
\put(-60,-5){\small$a= 6\pi$}
\put(-25,-5){\small$a= 8\pi$}
\hspace{14mm}
\includegraphics[width=.06\textwidth, trim=7cm 0.0cm 18cm 1.0cm, clip=true] {bandsar_1_spring_6}
\includegraphics[width=.06\textwidth, trim=7cm 0.0cm 18cm 1.0cm, clip=true] {bandsar_2_spring_6}
\includegraphics[width=.06\textwidth, trim=7cm 0.0cm 18cm 1.0cm, clip=true] {bandsar_4_spring_6}
\includegraphics[width=.06\textwidth, trim=7cm 0.0cm 18cm 1.0cm, clip=true] {bandsar_6_spring_6}
\includegraphics[width=.06\textwidth, trim=7cm 0.0cm 18cm 1.0cm, clip=true] {bandsar_8_spring_6}
\put(1,8){\small$k=10^{-6}$}
\put(-160,-5){\small$a= \pi$}
\put(-130,-5){\small$a= 2\pi$}
\put(-95,-5){\small$a= 4\pi$}
\put(-60,-5){\small$a= 6\pi$}
\put(-25,-5){\small$a= 8\pi$}
\vspace{5mm}

\includegraphics[width=.06\textwidth, trim=15cm 0cm 10cm 0cm, clip=true] {bandsar_1_spring_0}
\includegraphics[width=.06\textwidth, trim=15cm 0cm 10cm 0cm, clip=true] {bandsar_2_spring_0}
\includegraphics[width=.06\textwidth, trim=15cm 0cm 10cm 0cm, clip=true] {bandsar_4_spring_0}
\includegraphics[width=.06\textwidth, trim=15cm 0cm 10cm 0cm, clip=true] {bandsar_6_spring_0}
\includegraphics[width=.06\textwidth, trim=15cm 0cm 10cm 0cm, clip=true] {bandsar_8_spring_0}
\put(1,8){\small$k=10^0$}
\hspace{15mm}
\includegraphics[width=.06\textwidth, trim=24cm 0cm 0cm 0cm, clip=true] {bandsar_1_spring_0}
\includegraphics[width=.06\textwidth, trim=24cm 0cm 0cm 0cm, clip=true] {bandsar_2_spring_0}
\includegraphics[width=.06\textwidth, trim=24cm 0cm 0cm 0cm, clip=true] {bandsar_4_spring_0}
\includegraphics[width=.06\textwidth, trim=24cm 0cm 0cm 0cm, clip=true] {bandsar_6_spring_0}
\includegraphics[width=.06\textwidth, trim=24cm 0cm 0cm 0cm, clip=true] {bandsar_8_spring_0}
\put(1,8){\small$k=10^0$}

\includegraphics[width=.06\textwidth, trim=15cm 0cm 10cm 0cm, clip=true] {bandsar_1_spring_1}
\includegraphics[width=.06\textwidth, trim=15cm 0cm 10cm 0cm, clip=true] {bandsar_2_spring_1}
\includegraphics[width=.06\textwidth, trim=15cm 0cm 10cm 0cm, clip=true] {bandsar_4_spring_1}
\includegraphics[width=.06\textwidth, trim=15cm 0cm 10cm 0cm, clip=true] {bandsar_6_spring_1}
\includegraphics[width=.06\textwidth, trim=15cm 0cm 10cm 0cm, clip=true] {bandsar_8_spring_1}
\put(1,8){\small$k=10^{-1}$}
\hspace{15mm}
\includegraphics[width=.06\textwidth, trim=24cm 0cm 0cm 0cm, clip=true] {bandsar_1_spring_1}
\includegraphics[width=.06\textwidth, trim=24cm 0cm 0cm 0cm, clip=true] {bandsar_2_spring_1}
\includegraphics[width=.06\textwidth, trim=24cm 0cm 0cm 0cm, clip=true] {bandsar_4_spring_1}
\includegraphics[width=.06\textwidth, trim=24cm 0cm 0cm 0cm, clip=true] {bandsar_6_spring_1}
\includegraphics[width=.06\textwidth, trim=24cm 0cm 0cm 0cm, clip=true] {bandsar_8_spring_1}
\put(1,8){\small$k=10^{-1}$}

\includegraphics[width=.06\textwidth, trim=15cm 0cm 10cm 0cm, clip=true] {bandsar_1_spring_2}
\includegraphics[width=.06\textwidth, trim=15cm 0cm 10cm 0cm, clip=true] {bandsar_2_spring_2}
\includegraphics[width=.06\textwidth, trim=15cm 0cm 10cm 0cm, clip=true] {bandsar_4_spring_2}
\includegraphics[width=.06\textwidth, trim=15cm 0cm 10cm 0cm, clip=true] {bandsar_6_spring_2}
\includegraphics[width=.06\textwidth, trim=15cm 0cm 10cm 0cm, clip=true] {bandsar_8_spring_2}
\put(1,8){\small$k=10^{-2}$}
\hspace{15mm}
\includegraphics[width=.06\textwidth, trim=24cm 0cm 0cm 0cm, clip=true] {bandsar_1_spring_2}
\includegraphics[width=.06\textwidth, trim=24cm 0cm 0cm 0cm, clip=true] {bandsar_2_spring_2}
\includegraphics[width=.06\textwidth, trim=24cm 0cm 0cm 0cm, clip=true] {bandsar_4_spring_2}
\includegraphics[width=.06\textwidth, trim=24cm 0cm 0cm 0cm, clip=true] {bandsar_6_spring_2}
\includegraphics[width=.06\textwidth, trim=24cm 0cm 0cm 0cm, clip=true] {bandsar_8_spring_2}
\put(1,8){\small$k=10^{-2}$}

\includegraphics[width=.06\textwidth, trim=15cm 0cm 10cm 0cm, clip=true] {bandsar_1_spring_3}
\includegraphics[width=.06\textwidth, trim=15cm 0cm 10cm 0cm, clip=true] {bandsar_2_spring_3}
\includegraphics[width=.06\textwidth, trim=15cm 0cm 10cm 0cm, clip=true] {bandsar_4_spring_3}
\includegraphics[width=.06\textwidth, trim=15cm 0cm 10cm 0cm, clip=true] {bandsar_6_spring_3}
\includegraphics[width=.06\textwidth, trim=15cm 0cm 10cm 0cm, clip=true] {bandsar_8_spring_3}
\put(1,8){\small$k=10^{-3}$}
\hspace{15mm}
\includegraphics[width=.06\textwidth, trim=24cm 0cm 0cm 0cm, clip=true] {bandsar_1_spring_3}
\includegraphics[width=.06\textwidth, trim=24cm 0cm 0cm 0cm, clip=true] {bandsar_2_spring_3}
\includegraphics[width=.06\textwidth, trim=24cm 0cm 0cm 0cm, clip=true] {bandsar_4_spring_3}
\includegraphics[width=.06\textwidth, trim=24cm 0cm 0cm 0cm, clip=true] {bandsar_6_spring_3}
\includegraphics[width=.06\textwidth, trim=24cm 0cm 0cm 0cm, clip=true] {bandsar_8_spring_3}
\put(1,8){\small$k=10^{-3}$}

\includegraphics[width=.06\textwidth, trim=15cm 0cm 10cm 0cm, clip=true] {bandsar_1_spring_4}
\includegraphics[width=.06\textwidth, trim=15cm 0cm 10cm 0cm, clip=true] {bandsar_2_spring_4}
\includegraphics[width=.06\textwidth, trim=15cm 0cm 10cm 0cm, clip=true] {bandsar_4_spring_4}
\includegraphics[width=.06\textwidth, trim=15cm 0cm 10cm 0cm, clip=true] {bandsar_6_spring_4}
\includegraphics[width=.06\textwidth, trim=15cm 0cm 10cm 0cm, clip=true] {bandsar_8_spring_4}
\put(1,8){\small$k=10^{-4}$}
\hspace{15mm}
\includegraphics[width=.06\textwidth, trim=24cm 0cm 0cm 0cm, clip=true] {bandsar_1_spring_4}
\includegraphics[width=.06\textwidth, trim=24cm 0cm 0cm 0cm, clip=true] {bandsar_2_spring_4}
\includegraphics[width=.06\textwidth, trim=24cm 0cm 0cm 0cm, clip=true] {bandsar_4_spring_4}
\includegraphics[width=.06\textwidth, trim=24cm 0cm 0cm 0cm, clip=true] {bandsar_6_spring_4}
\includegraphics[width=.06\textwidth, trim=24cm 0cm 0cm 0cm, clip=true] {bandsar_8_spring_4}
\put(1,8){\small$k=10^{-4}$}

\includegraphics[width=.06\textwidth, trim=15cm 0cm 10cm 0cm, clip=true] {bandsar_1_spring_5}
\includegraphics[width=.06\textwidth, trim=15cm 0cm 10cm 0cm, clip=true] {bandsar_2_spring_5}
\includegraphics[width=.06\textwidth, trim=15cm 0cm 10cm 0.0cm, clip=true] {bandsar_4_spring_5}
\includegraphics[width=.06\textwidth, trim=15cm 0cm 10cm 0.0cm, clip=true] {bandsar_6_spring_5}
\includegraphics[width=.06\textwidth, trim=15cm 0cm 10cm 0.0cm, clip=true] {bandsar_8_spring_5}
\put(1,8){\small$k=10^{-5}$}
\hspace{15mm}
\includegraphics[width=.06\textwidth, trim=24cm 0cm 0cm 0cm, clip=true] {bandsar_1_spring_5}
\includegraphics[width=.06\textwidth, trim=24cm 0cm 0cm 0cm, clip=true] {bandsar_2_spring_5}
\includegraphics[width=.06\textwidth, trim=24cm 0cm 0cm 0cm, clip=true] {bandsar_4_spring_5}
\includegraphics[width=.06\textwidth, trim=24cm 0cm 0cm 0cm, clip=true] {bandsar_6_spring_5}
\includegraphics[width=.06\textwidth, trim=24cm 0cm 0cm 0cm, clip=true] {bandsar_8_spring_5}
\put(1,8){\small$k=10^{-5}$}

\includegraphics[width=.06\textwidth, trim=15cm 0.0cm 10cm 0.0cm, clip=true] {bandsar_1_spring_6}
\includegraphics[width=.06\textwidth, trim=15cm 0.0cm 10cm 0.0cm, clip=true] {bandsar_2_spring_6}
\includegraphics[width=.06\textwidth, trim=15cm 0.0cm 10cm 0.0cm, clip=true] {bandsar_4_spring_6}
\includegraphics[width=.06\textwidth, trim=15cm 0.0cm 10cm 0.0cm, clip=true] {bandsar_6_spring_6}
\includegraphics[width=.06\textwidth, trim=15cm 0.0cm 10cm 0.0cm, clip=true] {bandsar_8_spring_6}
\put(1,8){\small$k=10^{-6}$}
\put(-160,-7){\small$a= \pi$}
\put(-130,-7){\small$a= 2\pi$}
\put(-95,-7){\small$a= 4\pi$}
\put(-60,-7){\small$a= 6\pi$}
\put(-25,-7){\small$a= 8\pi$}
\hspace{15mm}
\includegraphics[width=.06\textwidth, trim=24cm 0cm 0cm 0cm, clip=true] {bandsar_1_spring_6}
\includegraphics[width=.06\textwidth, trim=24cm 0cm 0cm 0cm, clip=true] {bandsar_2_spring_6}
\includegraphics[width=.06\textwidth, trim=24cm 0cm 0cm 0cm, clip=true] {bandsar_4_spring_6}
\includegraphics[width=.06\textwidth, trim=24cm 0cm 0cm 0cm, clip=true] {bandsar_6_spring_6}
\includegraphics[width=.06\textwidth, trim=24cm 0cm 0cm 0cm, clip=true] {bandsar_8_spring_6}
\put(1,8){\small$k=10^{-6}$}
\put(-160,-7){\small$a= \pi$}
\put(-130,-7){\small$a= 2\pi$}
\put(-95,-7){\small$a= 4\pi$}
\put(-60,-7){\small$a= 6\pi$}
\put(-25,-7){\small$a= 8\pi$}

\end{center}
\caption{Generalized M\"obius bands adopt characteristic equilibrium shapes depending on the in-plane stretchability of the surface and the aspect ratio: Equilibrium shapes of generalized M\"obius bands for different aspect ratio ${a}$ and stretchability ${k}$. The band is rotated into its main axes. }
\label{fig:Moebius_shapeAll}
\end{figure*}
%%%%%%%%%%%%%%%

%%%%%%%%%%%%%%%
\begin{figure*}[t!]
%\begin{center}
\noindent
\put(-4,-3.6){
\includegraphics[width=.138\textwidth] {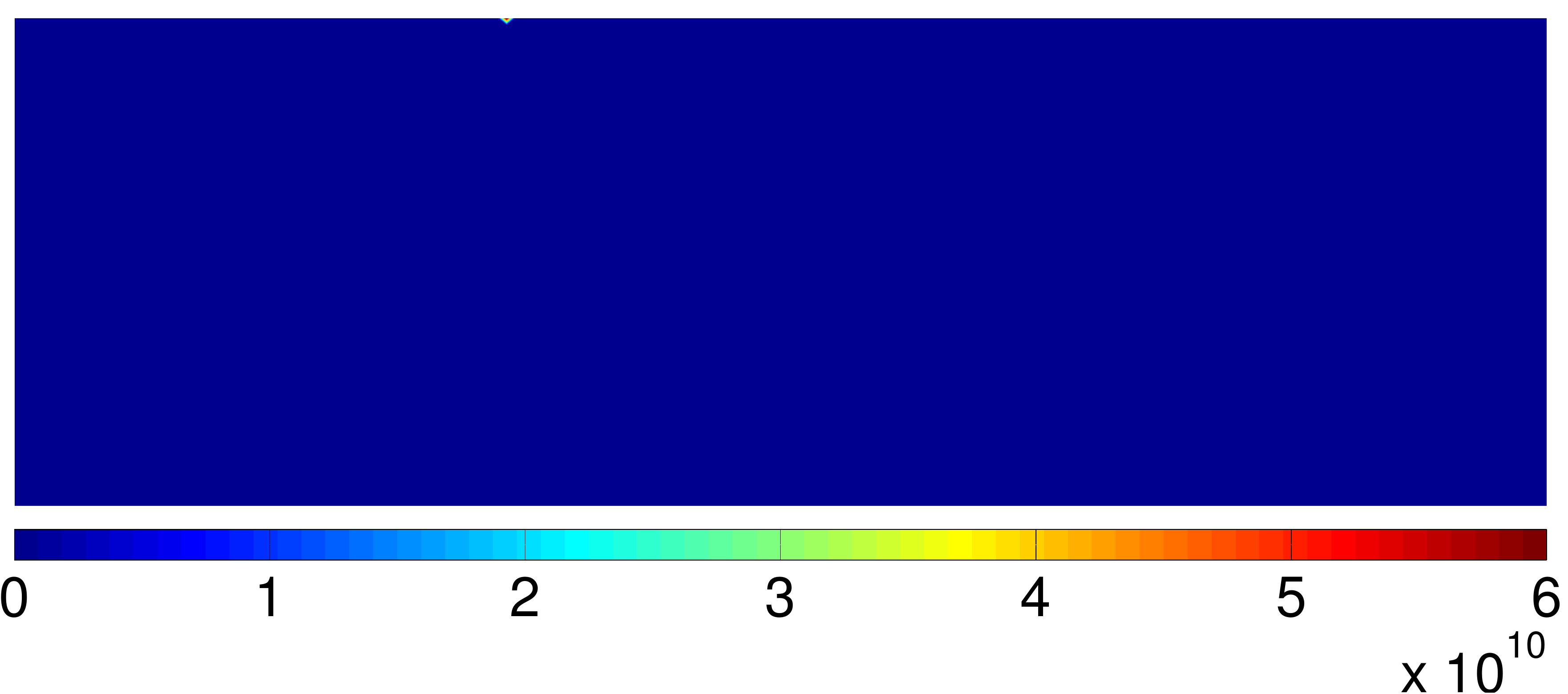}
\put(-70,43.6){(a)}
\put(-53,43.6){\small$k=10^0$}
}
\put(66,-3.6){
\includegraphics[width=.138\textwidth] {H_2_ar_1_spring_1}
\put(-53,43.6){\small$k=10^{-1}$}
}
\hspace{4.8cm}
\includegraphics[width=.138\textwidth] {H_2_ar_1_spring_2}
\put(-53,40){\small$k=10^{-2}$}
\includegraphics[width=.138\textwidth] {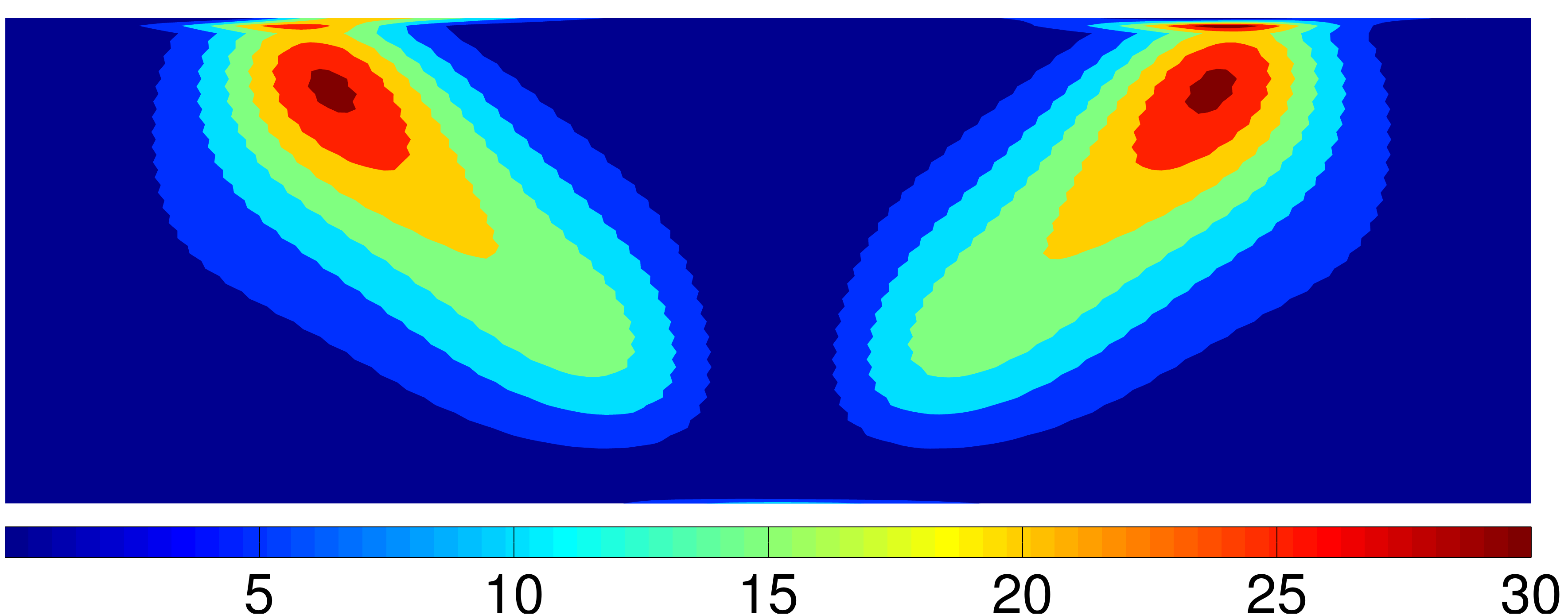}
\put(-53,40){\small$k=10^{-3}$}
\includegraphics[width=.138\textwidth] {H_2_ar_1_spring_4}
\put(-53,40){\small$k=10^{-4}$}
\includegraphics[width=.138\textwidth] {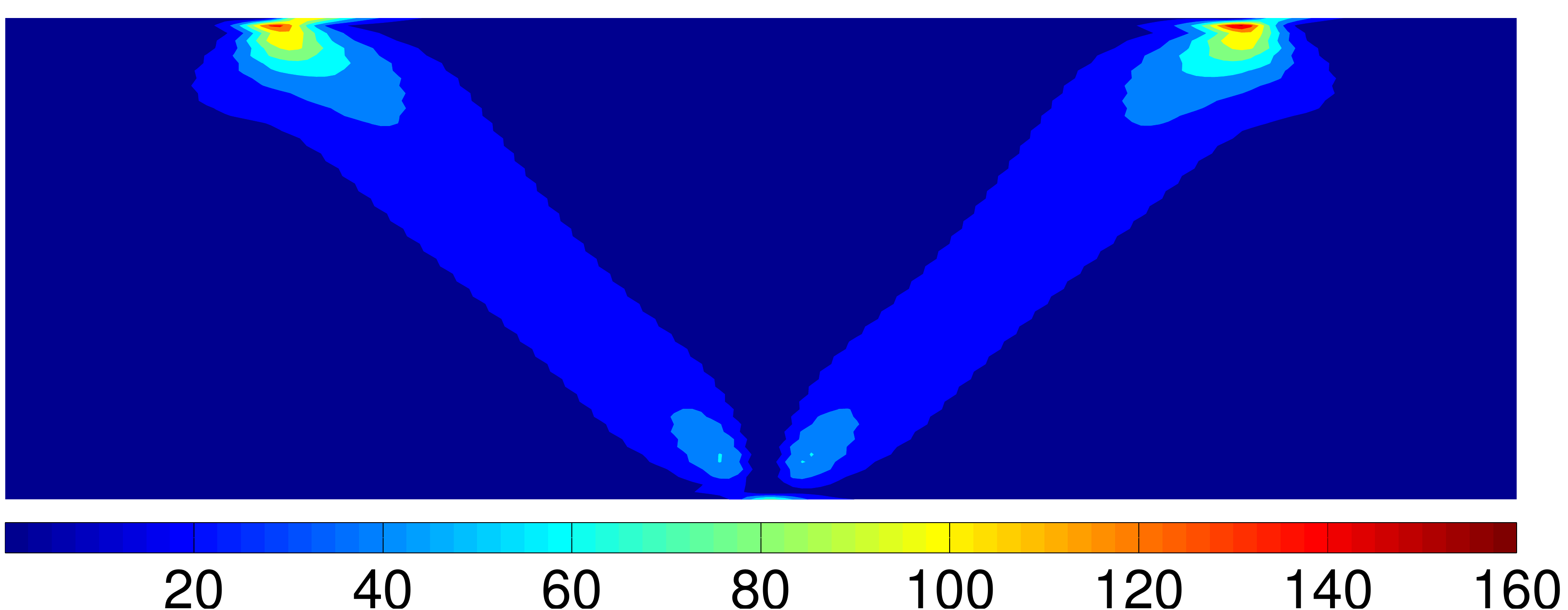}
\put(-53,40){\small$k=10^{-5}$}
\includegraphics[width=.138\textwidth] {H_2_ar_1_spring_6}
\put(-53,40){\small$k=10^{-6}$}
\vspace{0.0cm}

\includegraphics[width=.138\textwidth] {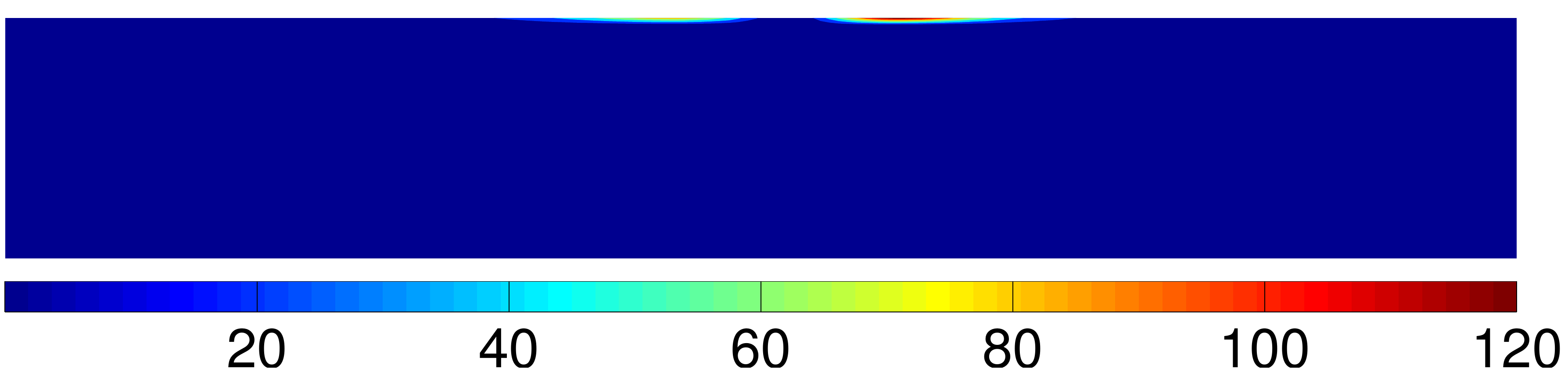}
\put(0,0){}
\includegraphics[width=.138\textwidth] {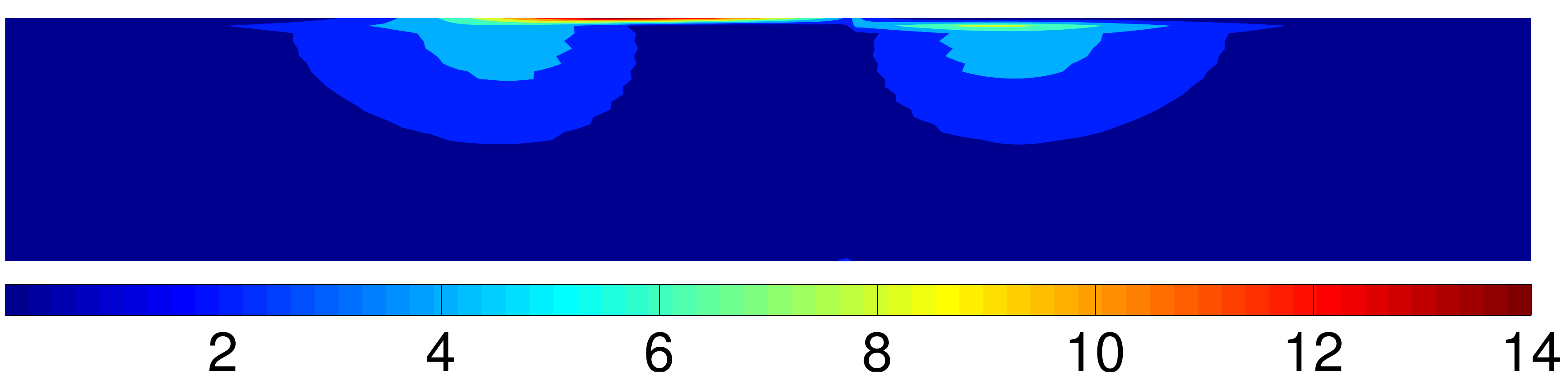}
\put(0,0){}
\includegraphics[width=.138\textwidth] {H_2_ar_2_spring_2}
\put(0,0){}
\includegraphics[width=.138\textwidth] {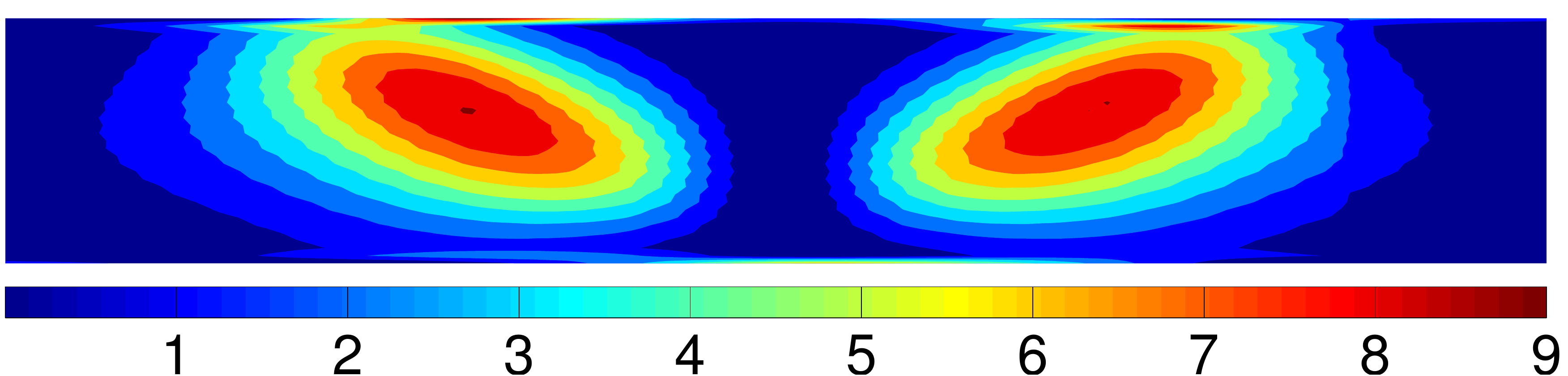}
\put(0,0){}
\includegraphics[width=.138\textwidth] {H_2_ar_2_spring_4}
\put(0,0){}
\includegraphics[width=.138\textwidth] {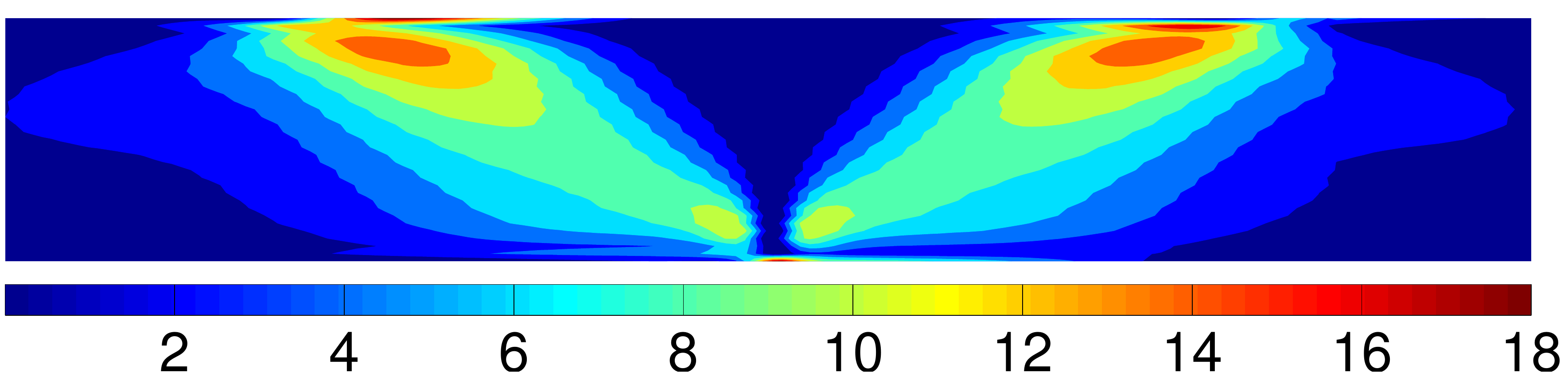}
\put(0,0){}
\includegraphics[width=.138\textwidth] {H_2_ar_2_spring_6}

\includegraphics[width=.138\textwidth] {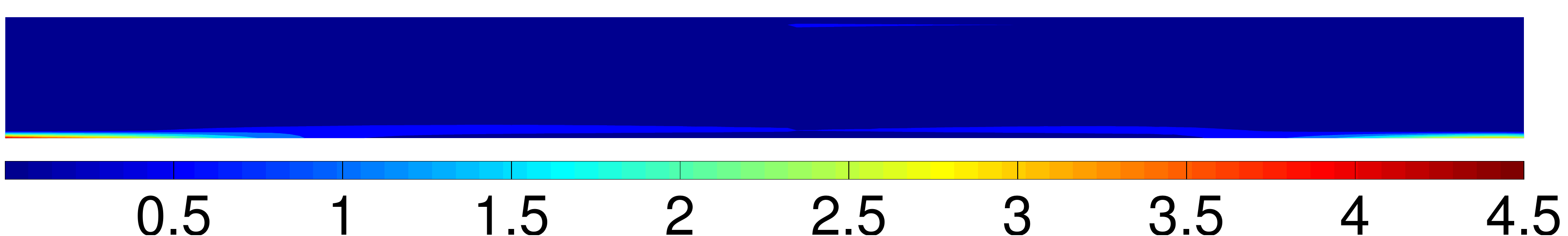}
\put(0,0){}
\includegraphics[width=.138\textwidth] {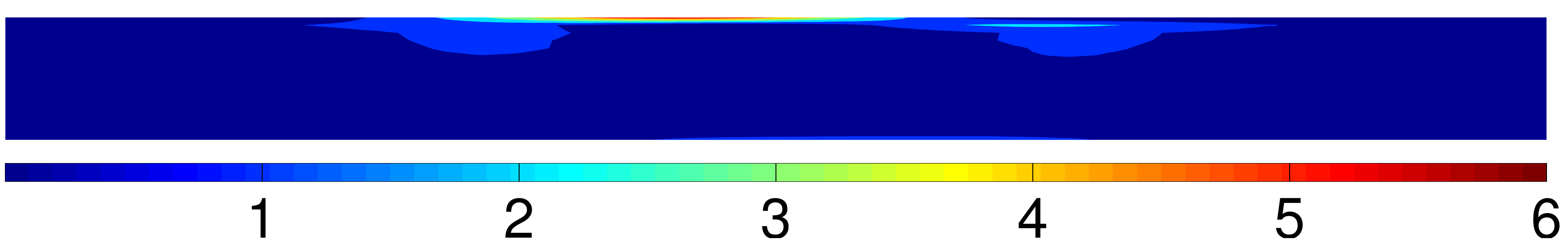}
\put(0,0){}
\includegraphics[width=.138\textwidth] {H_2_ar_4_spring_2}
\put(0,0){}
\includegraphics[width=.138\textwidth] {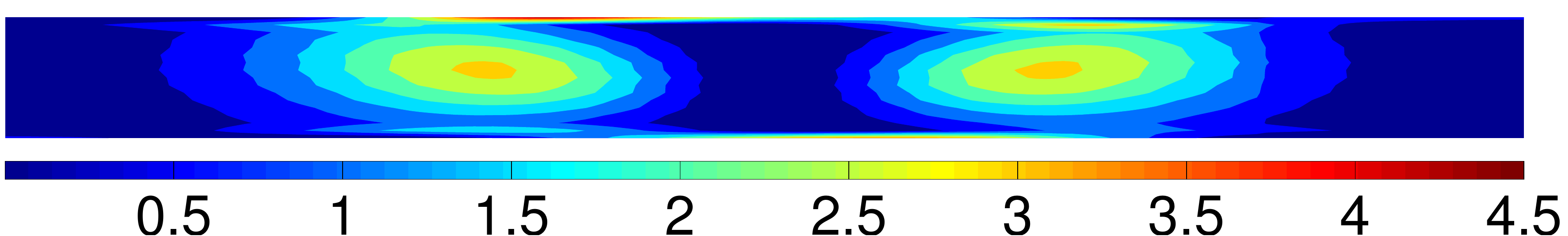}
\put(0,0){}
\includegraphics[width=.138\textwidth] {H_2_ar_4_spring_4}
\put(0,0){}
\includegraphics[width=.138\textwidth] {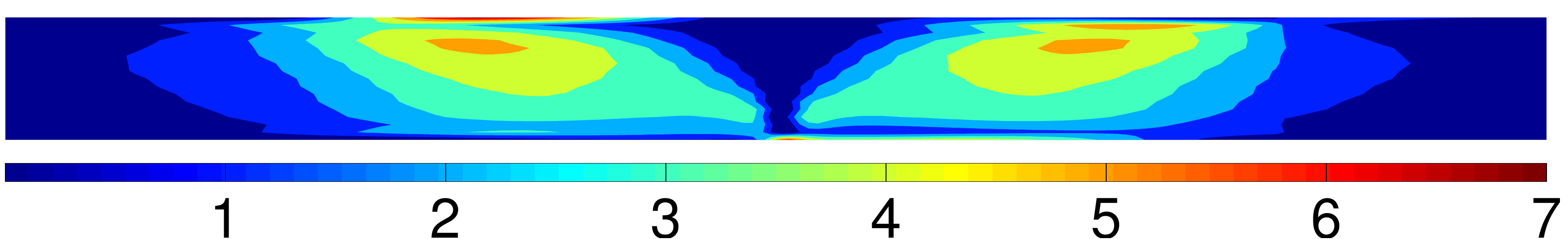}
\put(0,0){}
\includegraphics[width=.138\textwidth] {H_2_ar_4_spring_6}
\put(0,0){}

\includegraphics[width=.138\textwidth] {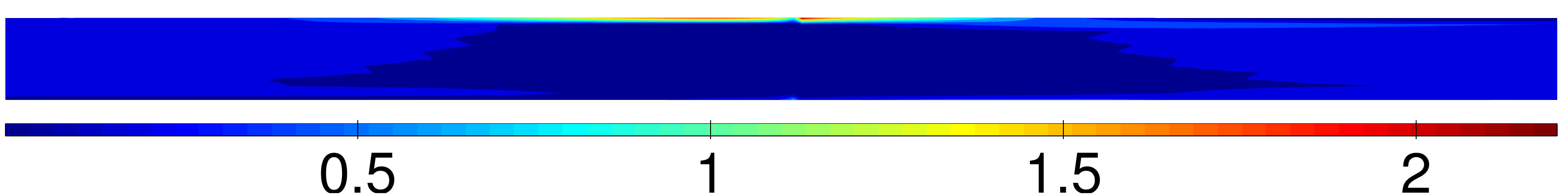}
\put(0,0){}
\includegraphics[width=.138\textwidth] {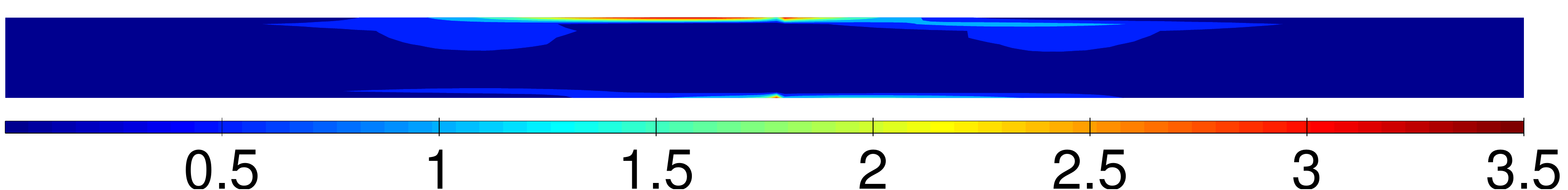}
\put(0,0){}
\includegraphics[width=.138\textwidth] {H_2_ar_6_spring_2}
\put(0,0){}
\includegraphics[width=.138\textwidth] {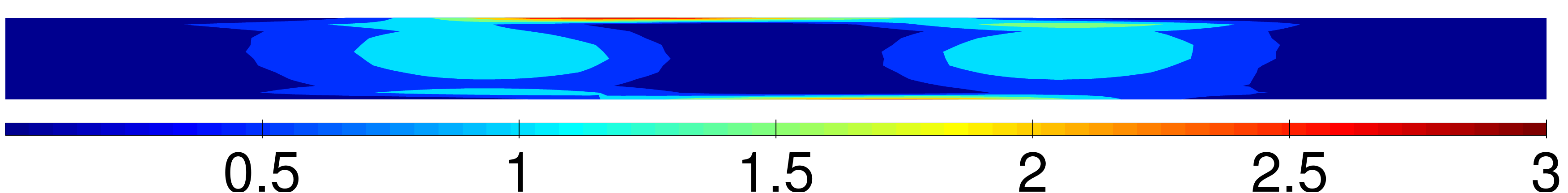}
\put(0,0){}
\includegraphics[width=.138\textwidth] {H_2_ar_6_spring_4}
\put(0,0){}
\includegraphics[width=.138\textwidth] {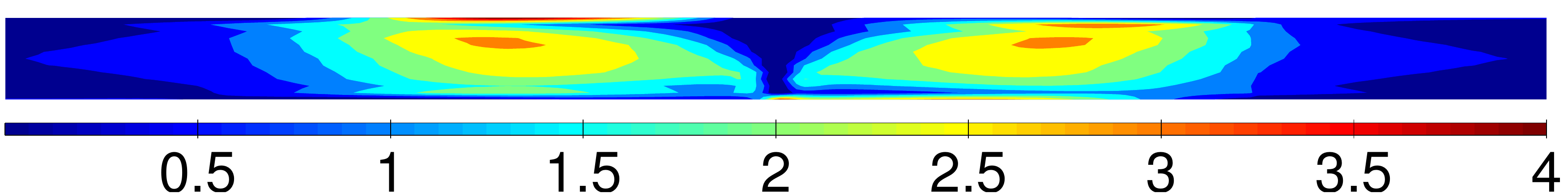}
\put(0,0){}
\includegraphics[width=.138\textwidth] {H_2_ar_6_spring_6}
\put(0,0){}

\includegraphics[width=.138\textwidth] {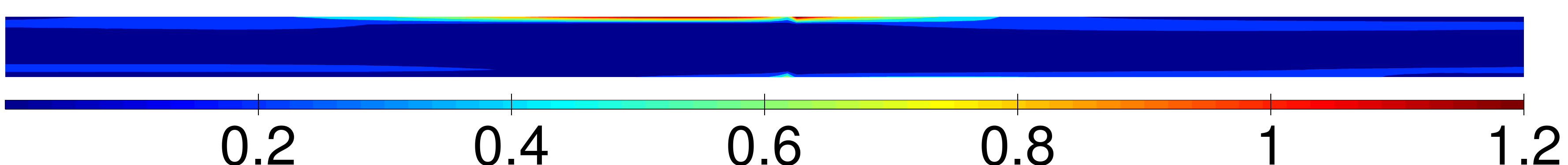}
\put(0,0){}
\includegraphics[width=.138\textwidth] {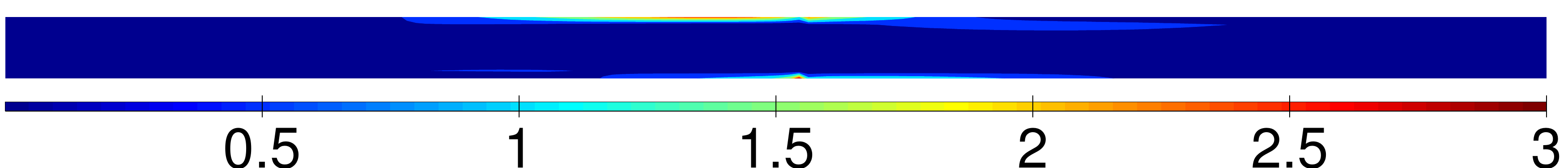}
\put(0,0){}
\includegraphics[width=.138\textwidth] {H_2_ar_8_spring_2}
\put(0,0){}
\includegraphics[width=.138\textwidth] {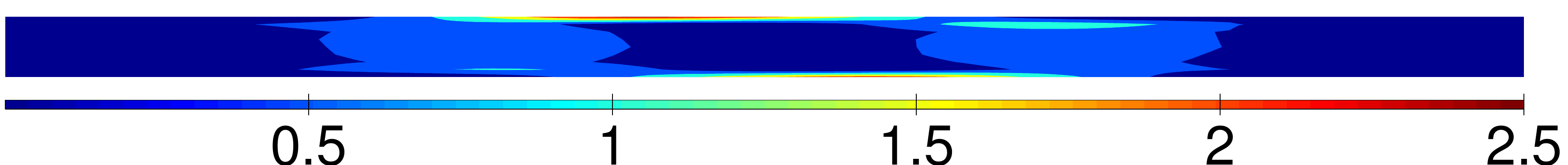}
\put(0,0){}
\includegraphics[width=.138\textwidth] {H_2_ar_8_spring_4}
\put(0,0){}
\includegraphics[width=.138\textwidth] {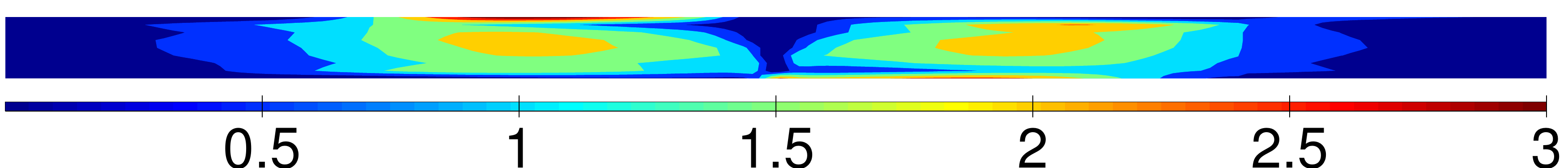}
\put(0,0){}
\includegraphics[width=.138\textwidth] {H_2_ar_8_spring_6}
\put(0,0){}

%
%Gaussian curvature:

\vspace{5mm}\noindent
\put(-4,-3.6){
\includegraphics[width=.138\textwidth] {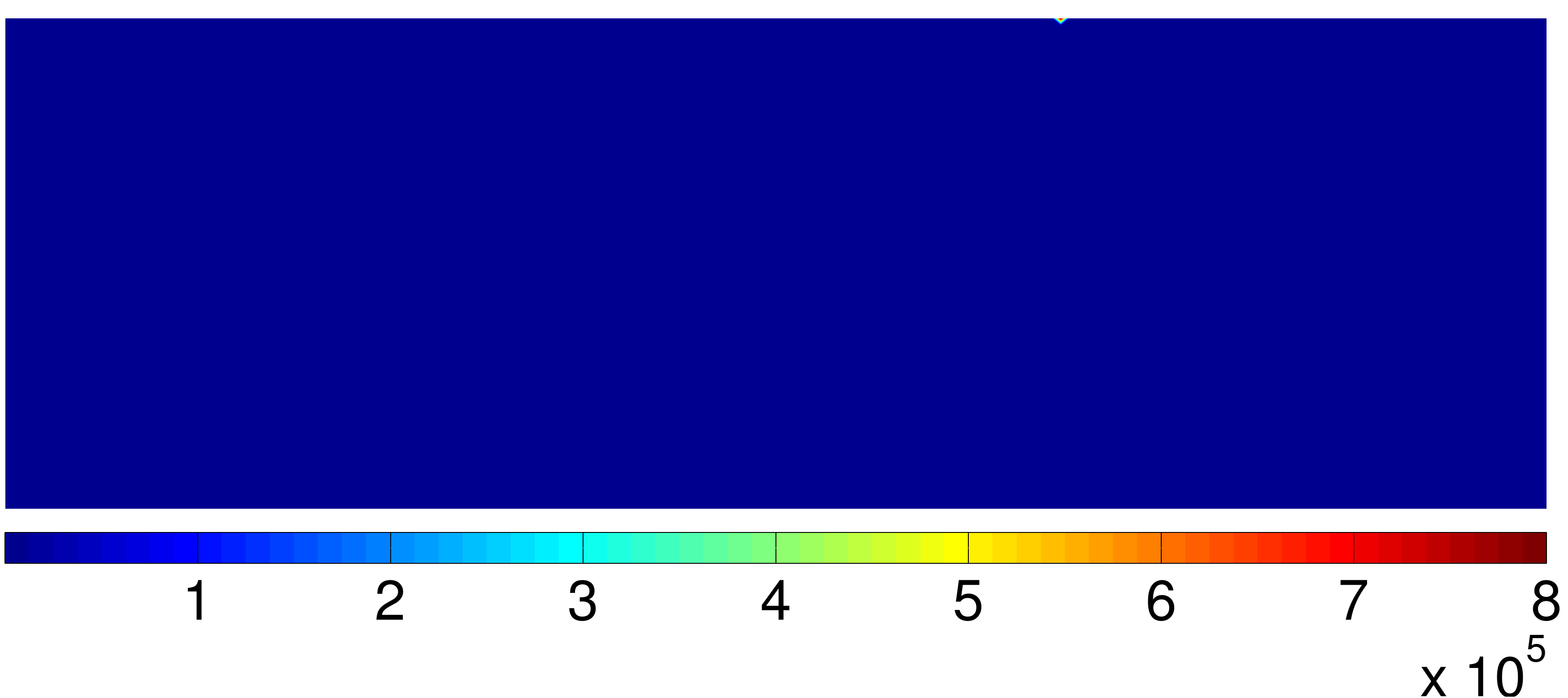}
\put(-70,43.6){(b)}
\put(-53,43.6){\small$k=10^0$}
}
\put(66,-3.6){
\includegraphics[width=.138\textwidth] {K_ar_1_spring_1}
\put(-53,43.6){\small$k=10^{-1}$}
}
\hspace{4.8cm}
%\includegraphics[width=.138\textwidth] {K_ar_1_spring_0}
%\put(-70,40){(b)}
%\put(-53,40){\small$k=10^{0}$}
%\includegraphics[width=.138\textwidth] {K_ar_1_spring_1}
%\put(-53,40){\small$k=10^{-1}$}
\includegraphics[width=.138\textwidth] {K_ar_1_spring_2}
\put(-53,40){\small$k=10^{-2}$}
\includegraphics[width=.138\textwidth] {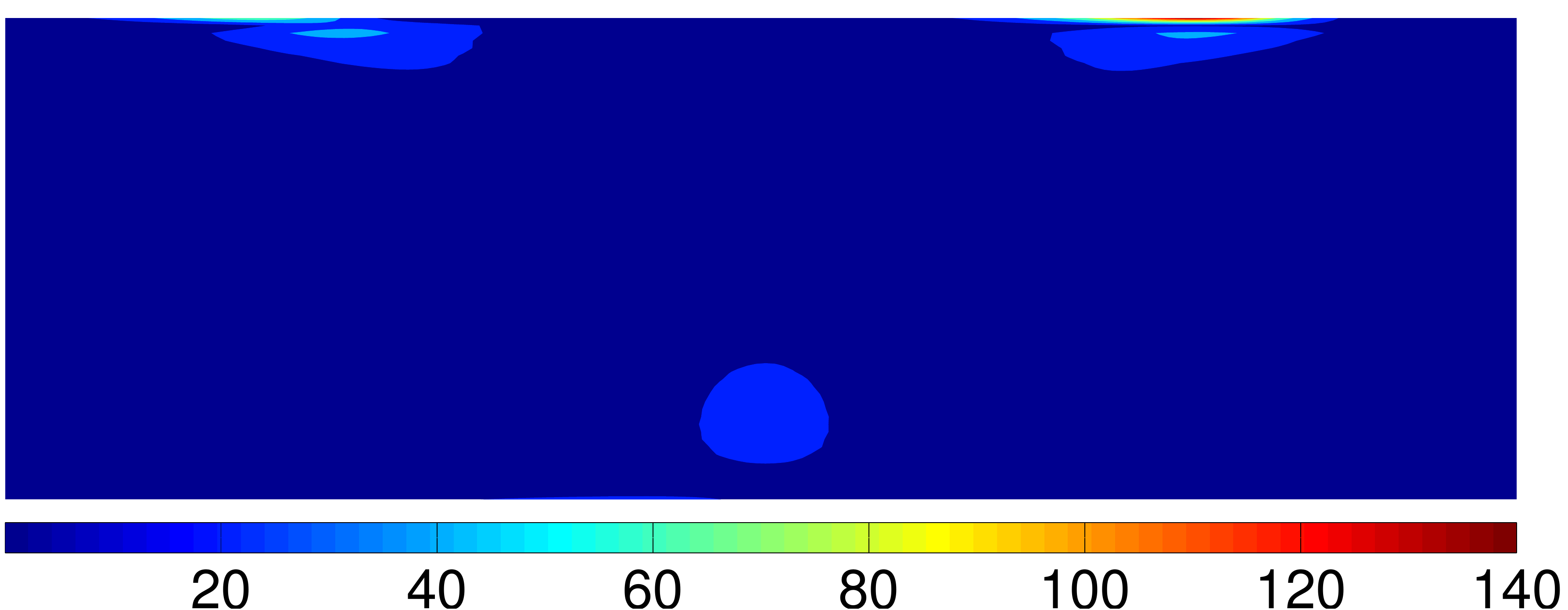}
\put(-53,40){\small$k=10^{-3}$}
\includegraphics[width=.138\textwidth] {K_ar_1_spring_4}
\put(-53,40){\small$k=10^{-4}$}
\includegraphics[width=.138\textwidth] {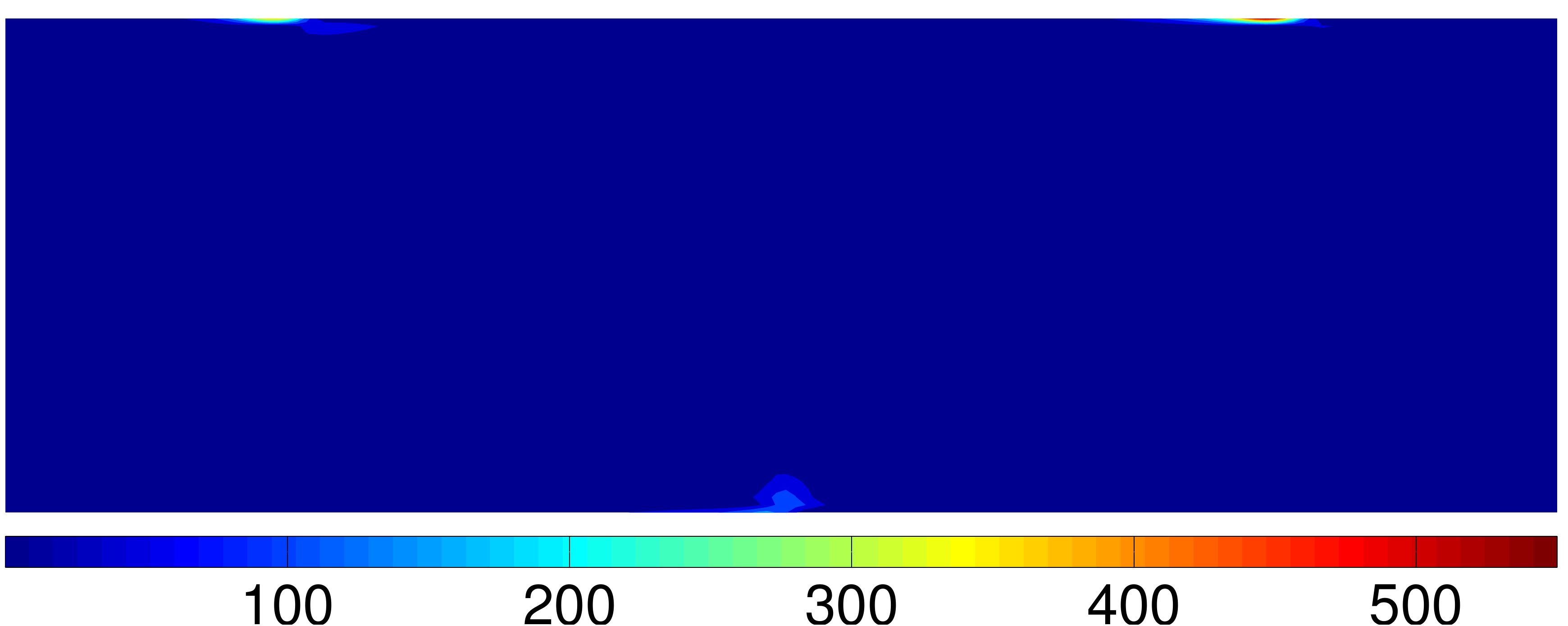}
\put(-53,40){\small$k=10^{-5}$}
\includegraphics[width=.138\textwidth] {K_ar_1_spring_6}
\put(-53,40){\small$k=10^{-6}$}

\includegraphics[width=.138\textwidth] {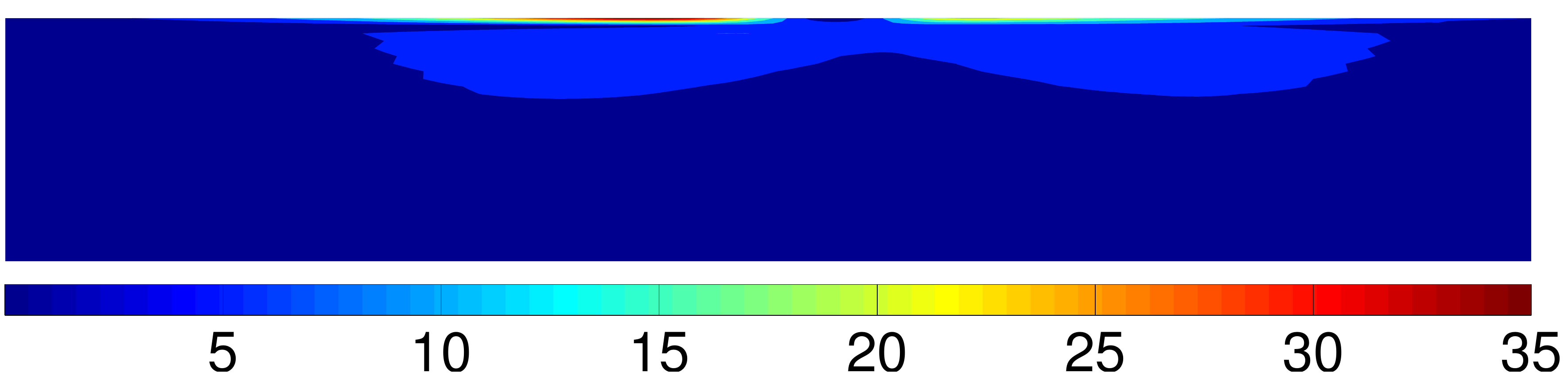}
\put(0,0){}
\includegraphics[width=.138\textwidth] {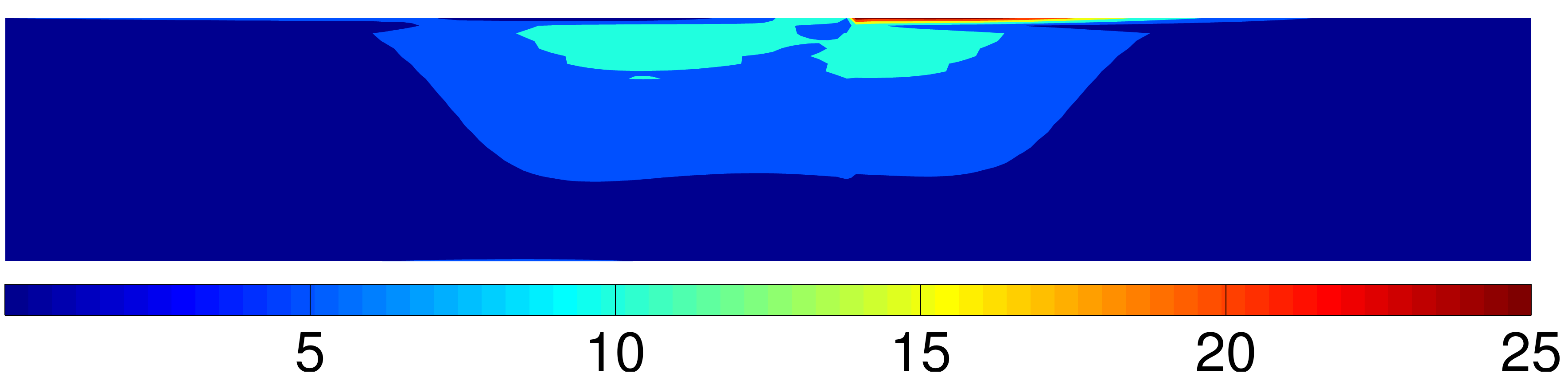}
\put(0,0){}
\includegraphics[width=.138\textwidth] {K_ar_2_spring_2}
\put(0,0){}
\includegraphics[width=.138\textwidth] {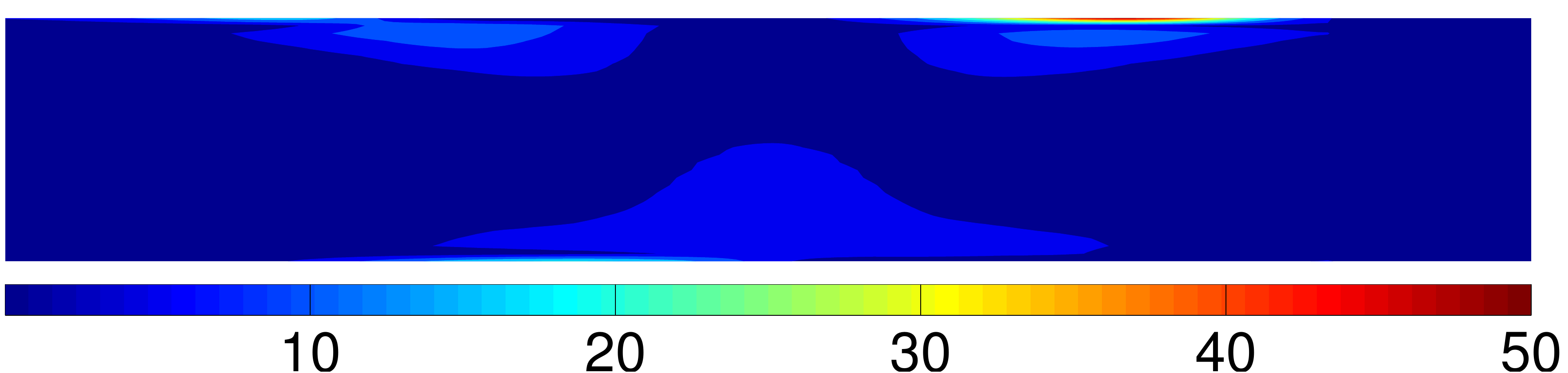}
\put(0,0){}
\includegraphics[width=.138\textwidth] {K_ar_2_spring_4}
\put(0,0){}
\includegraphics[width=.138\textwidth] {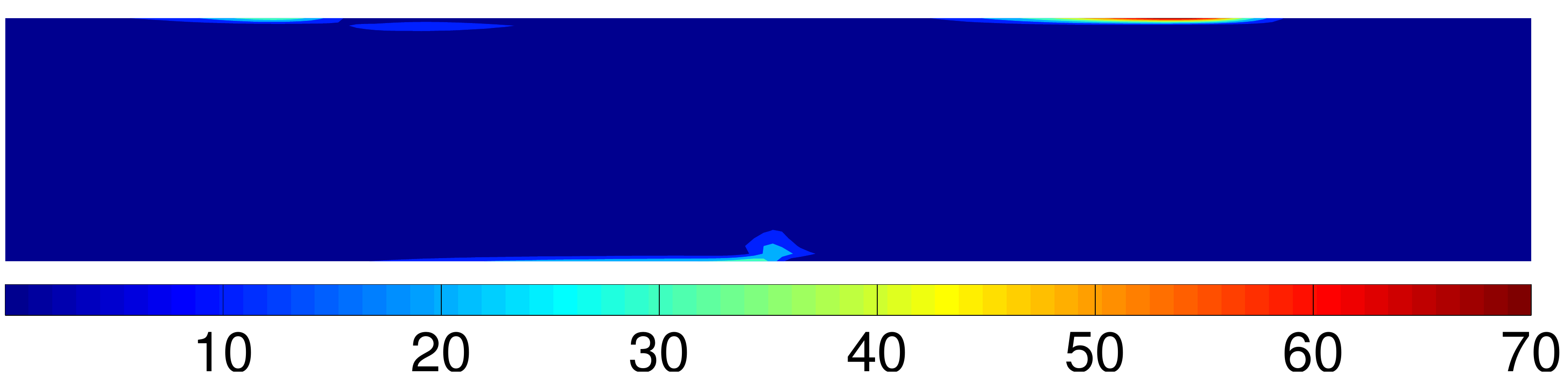}
\put(0,0){}
\includegraphics[width=.138\textwidth] {K_ar_2_spring_6}
\put(0,0){}

\includegraphics[width=.138\textwidth] {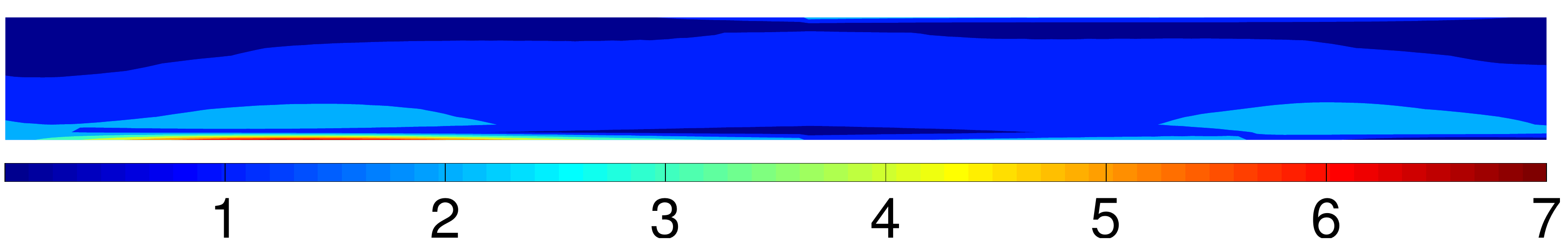}\put(0,0){}
\includegraphics[width=.138\textwidth] {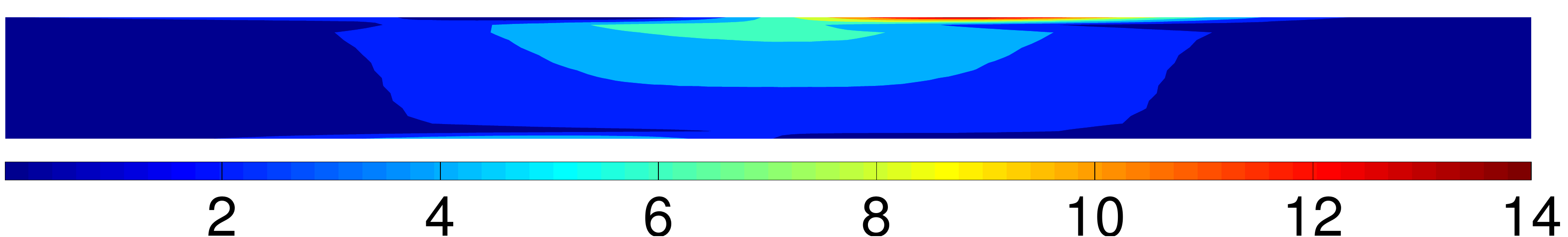}\put(0,0){}
\includegraphics[width=.138\textwidth] {K_ar_4_spring_2}\put(0,0){}
\includegraphics[width=.138\textwidth] {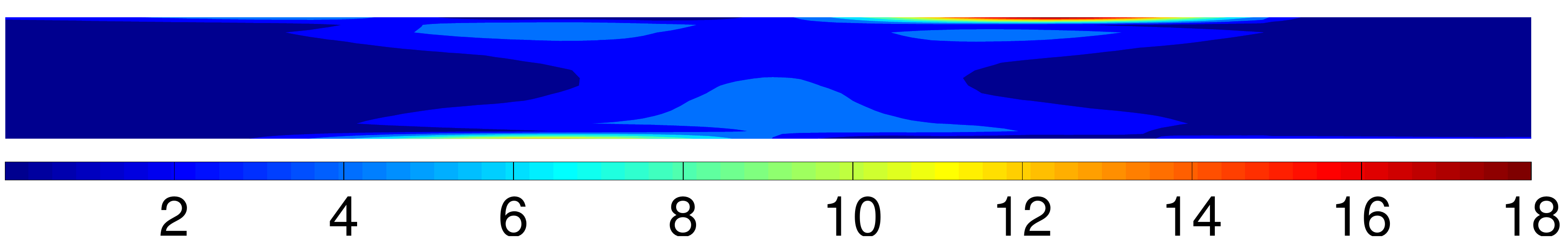}\put(0,0){}
\includegraphics[width=.138\textwidth] {K_ar_4_spring_4}\put(0,0){}
\includegraphics[width=.138\textwidth] {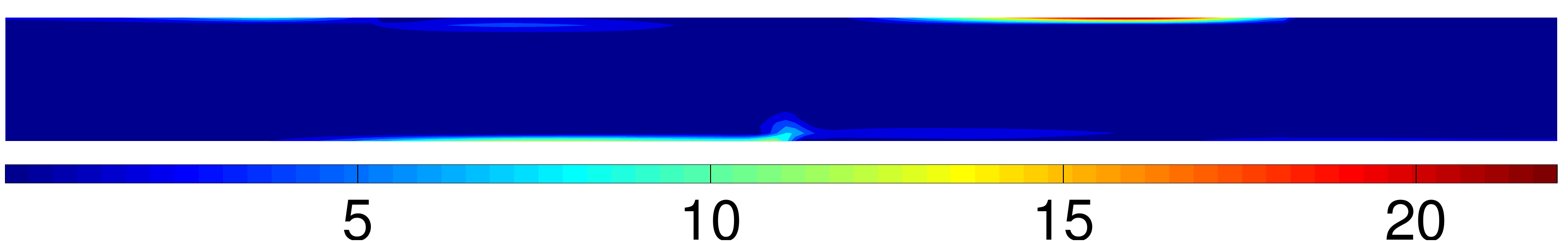}\put(0,0){}
\includegraphics[width=.138\textwidth] {K_ar_4_spring_6}\put(0,0){}

\includegraphics[width=.138\textwidth] {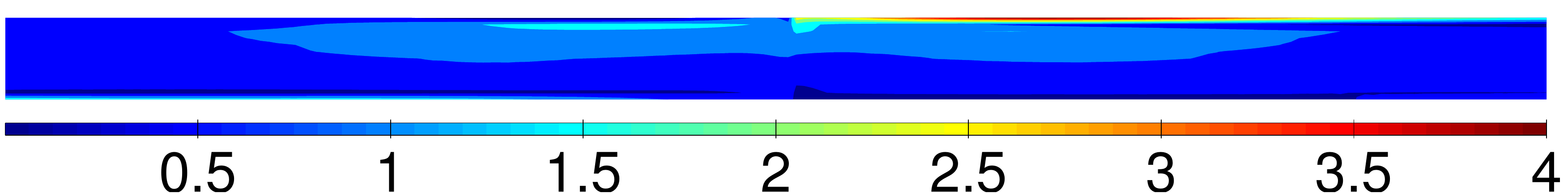}\put(0,0){}
\includegraphics[width=.138\textwidth] {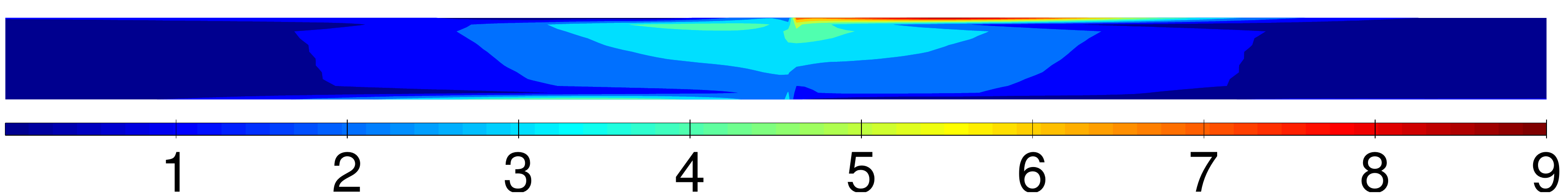}\put(0,0){}
\includegraphics[width=.138\textwidth] {K_ar_6_spring_2}\put(0,0){}
\includegraphics[width=.138\textwidth] {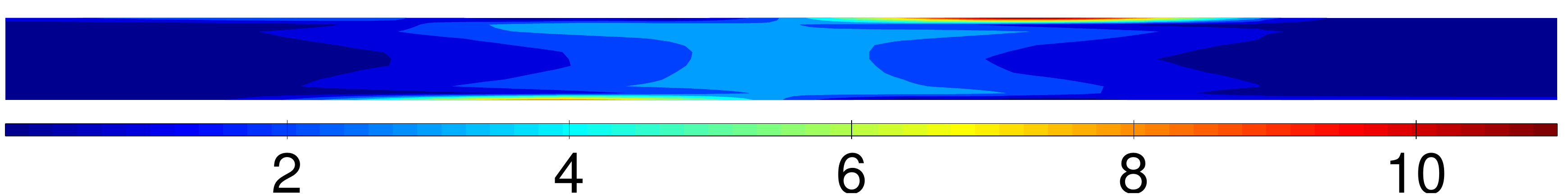}\put(0,0){}
\includegraphics[width=.138\textwidth] {K_ar_6_spring_4}\put(0,0){}
\includegraphics[width=.138\textwidth] {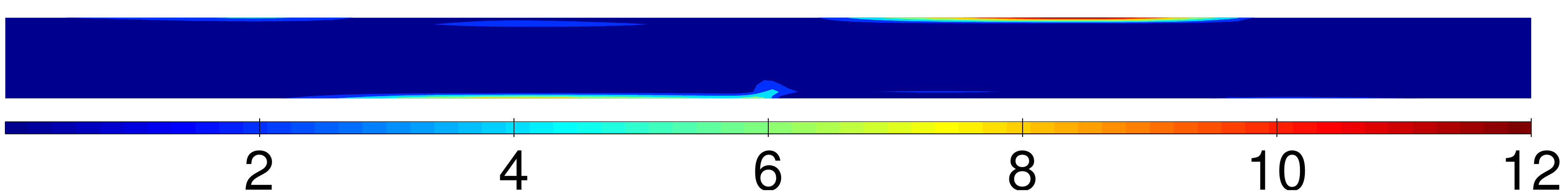}\put(0,0){}
\includegraphics[width=.138\textwidth] {K_ar_6_spring_6}\put(0,0){}

\includegraphics[width=.138\textwidth] {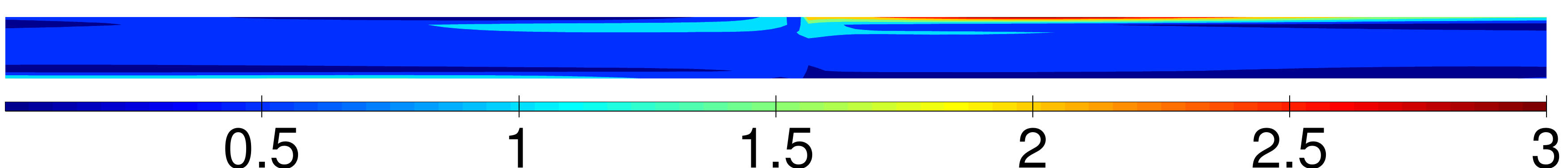}\put(0,0){}
\includegraphics[width=.138\textwidth] {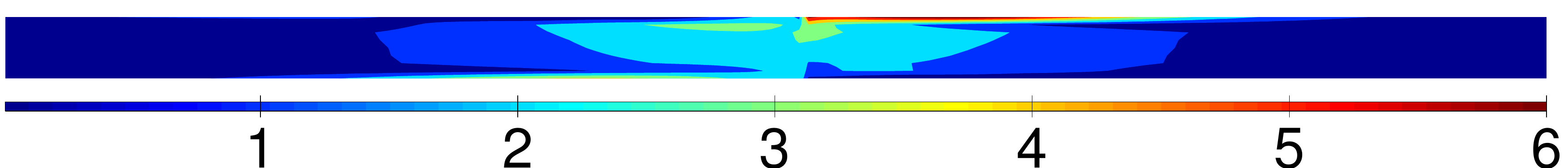}\put(0,0){}
\includegraphics[width=.138\textwidth] {K_ar_8_spring_2}\put(0,0){}
\includegraphics[width=.138\textwidth] {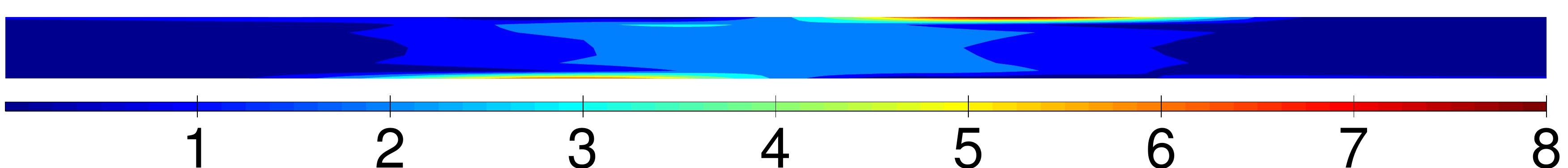}\put(0,0){}
\includegraphics[width=.138\textwidth] {K_ar_8_spring_4}\put(0,0){}
\includegraphics[width=.138\textwidth] {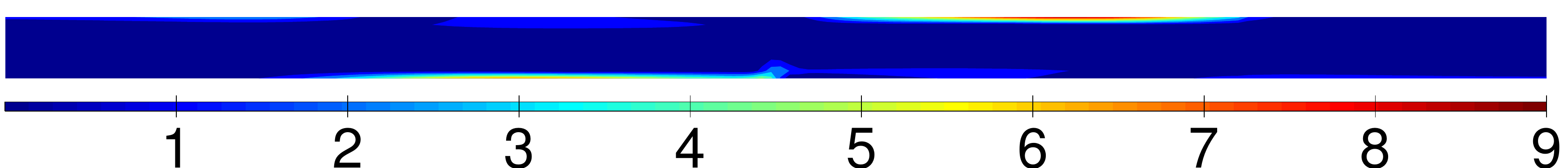}\put(0,0){}
\includegraphics[width=.138\textwidth] {K_ar_8_spring_6}\put(0,0){}

%Dilatation:

\vspace{5mm}\noindent
\includegraphics[width=.138\textwidth] {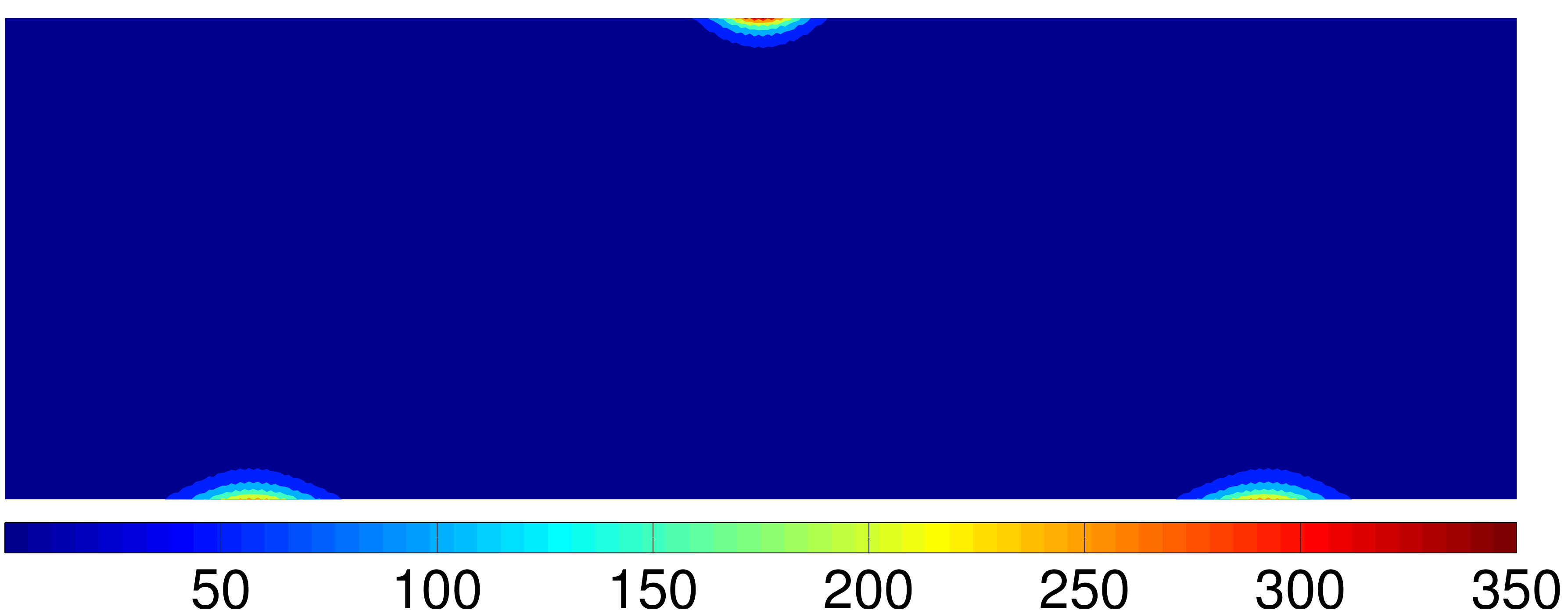}
\put(-70,40){(c)}
\put(-53,40){\small$k=10^{0}$}
\includegraphics[width=.138\textwidth] {AF_ar_1_spring_1}\put(-53,40){\small$k=10^{-1}$}
\includegraphics[width=.138\textwidth] {AF_ar_1_spring_2}\put(-53,40){\small$k=10^{-2}$}
\includegraphics[width=.138\textwidth] {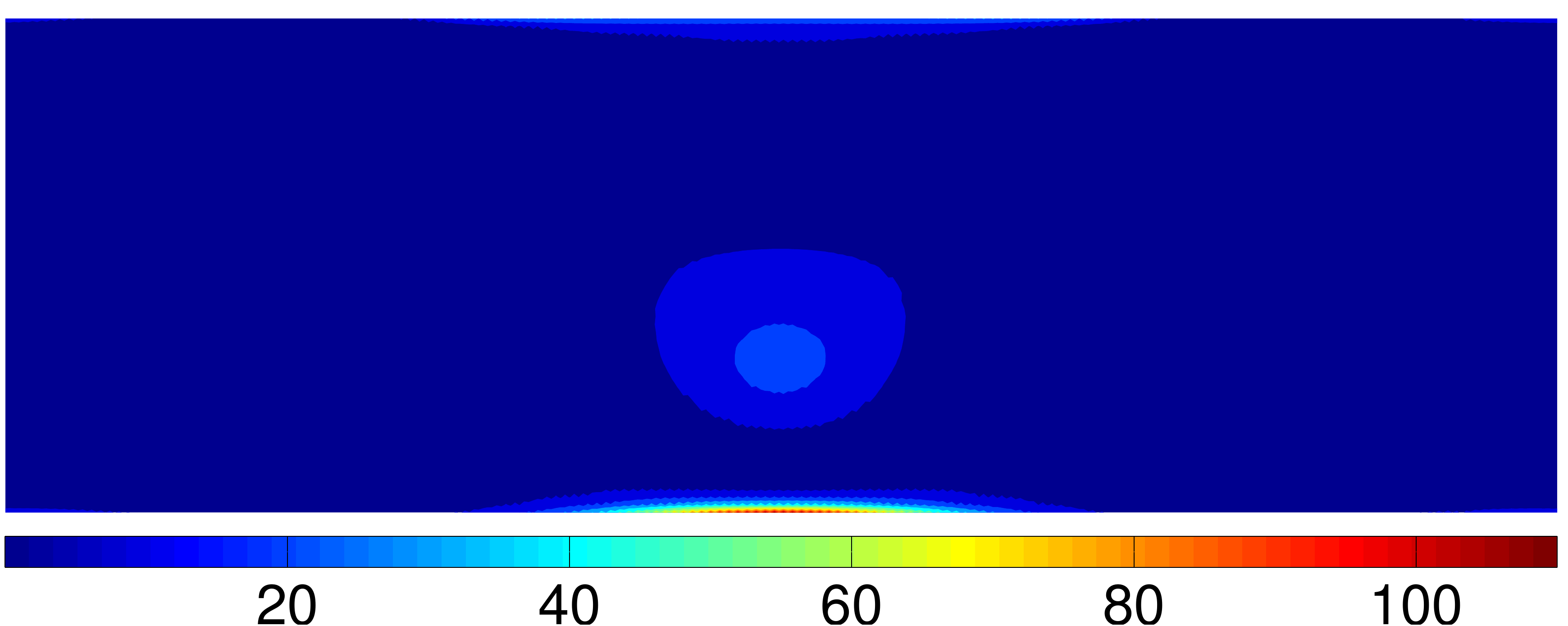}\put(-53,40){\small$k=10^{-3}$}
\includegraphics[width=.138\textwidth] {AF_ar_1_spring_4}\put(-53,40){\small$k=10^{-4}$}
\includegraphics[width=.138\textwidth] {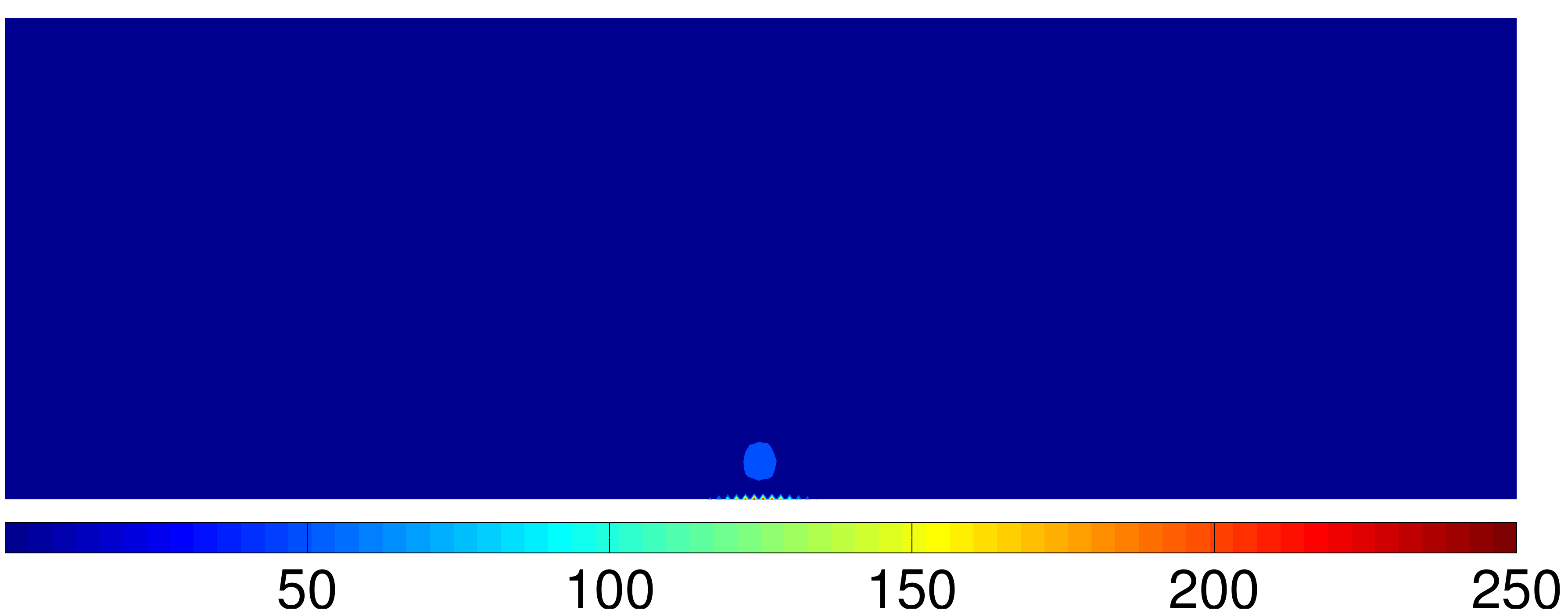}\put(-53,40){\small$k=10^{-5}$}
\includegraphics[width=.138\textwidth] {AF_ar_1_spring_6}\put(-53,40){\small$k=10^{-6}$}

\includegraphics[width=.138\textwidth] {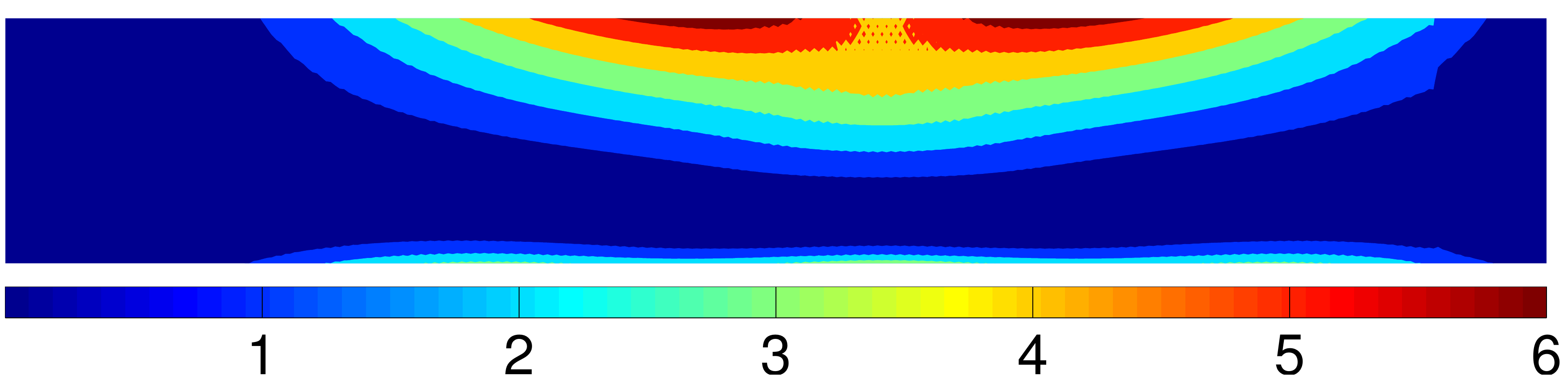}\put(0,0){}
\includegraphics[width=.138\textwidth] {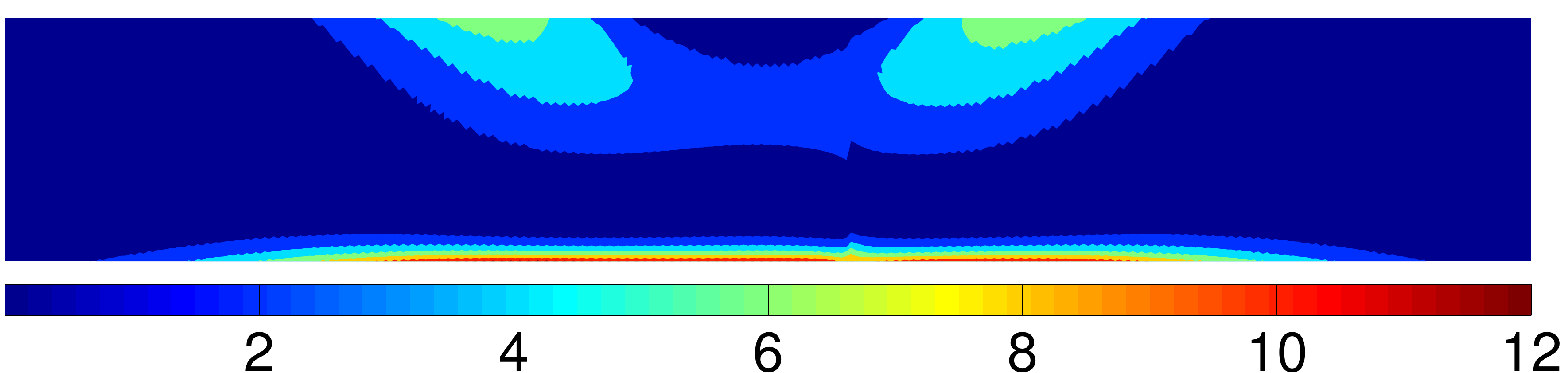}\put(0,0){}
\includegraphics[width=.138\textwidth] {AF_ar_2_spring_2}\put(0,0){}
\includegraphics[width=.138\textwidth] {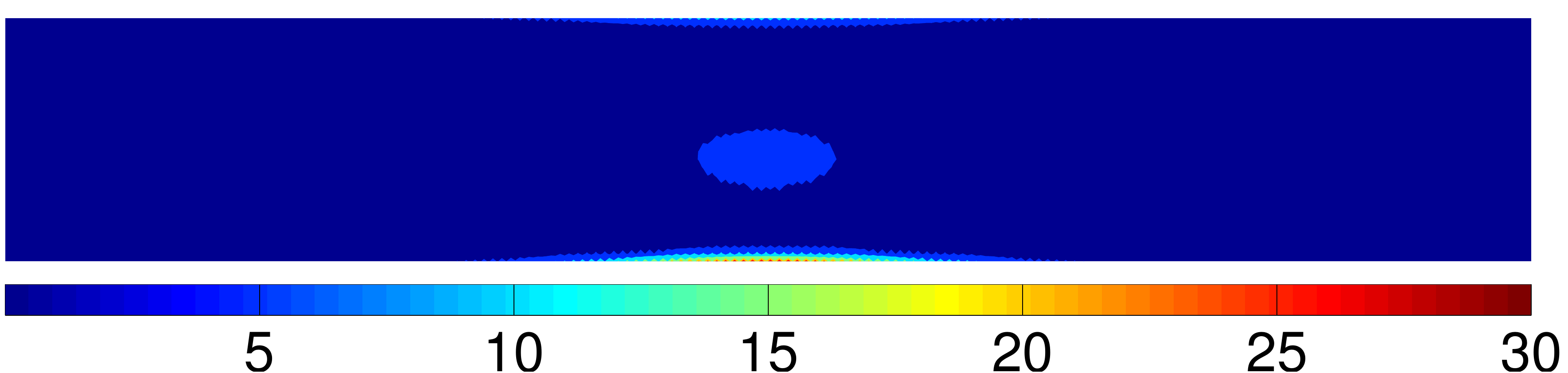}\put(0,0){}
\includegraphics[width=.138\textwidth] {AF_ar_2_spring_4}\put(0,0){}
\includegraphics[width=.138\textwidth] {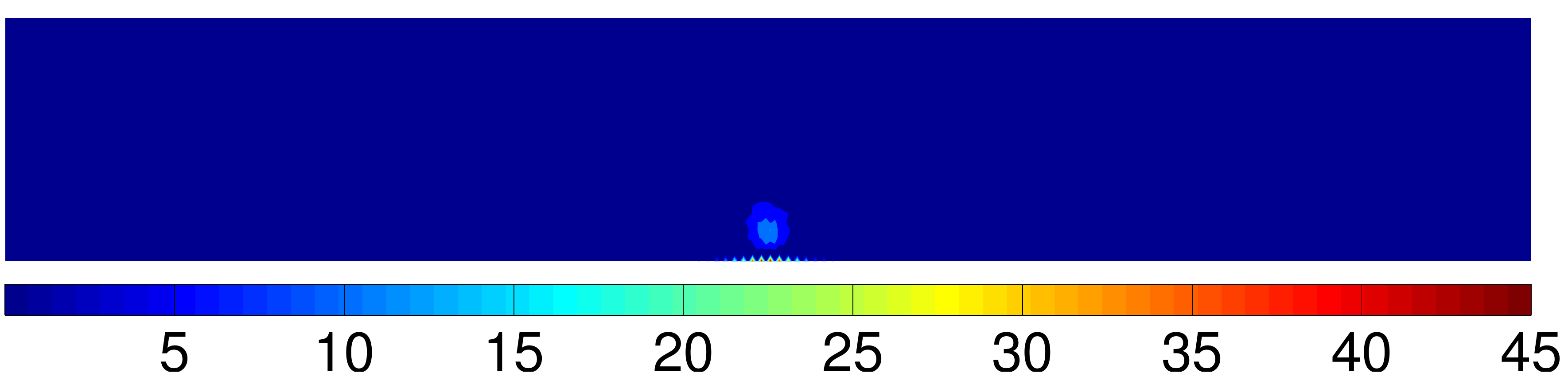}\put(0,0){}
\includegraphics[width=.138\textwidth] {AF_ar_2_spring_6}\put(0,0){}

\includegraphics[width=.138\textwidth] {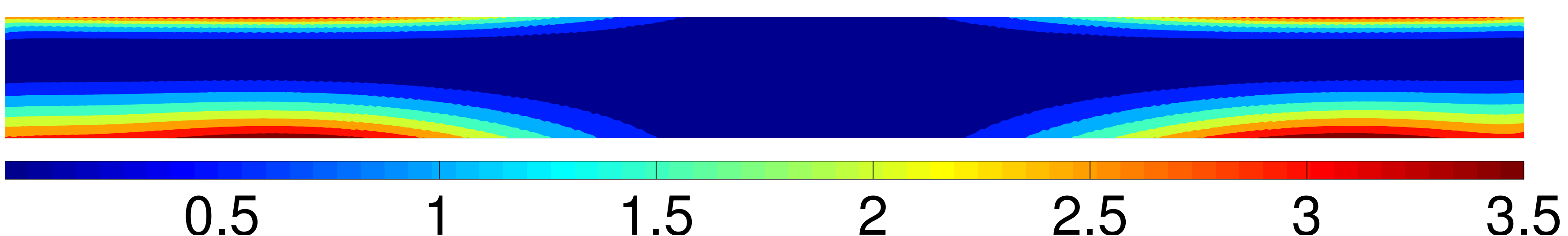}\put(0,0){}
\includegraphics[width=.138\textwidth] {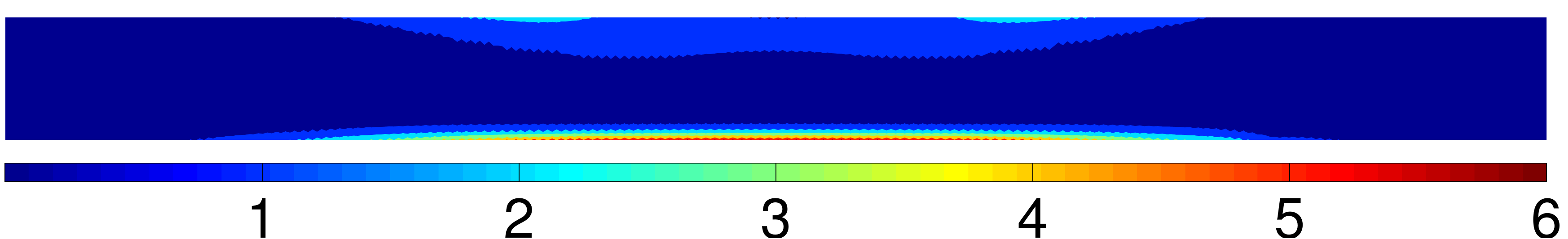}\put(0,0){}
\includegraphics[width=.138\textwidth] {AF_ar_4_spring_2}\put(0,0){}
\includegraphics[width=.138\textwidth] {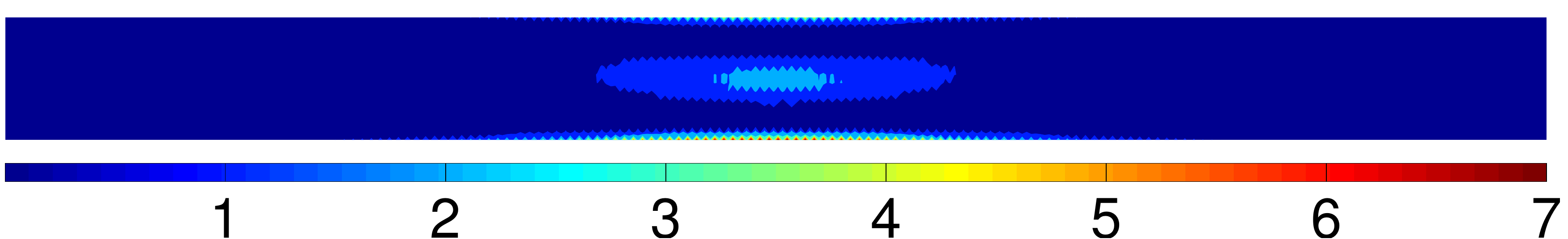}\put(0,0){}
\includegraphics[width=.138\textwidth] {AF_ar_4_spring_4}\put(0,0){}
\includegraphics[width=.138\textwidth] {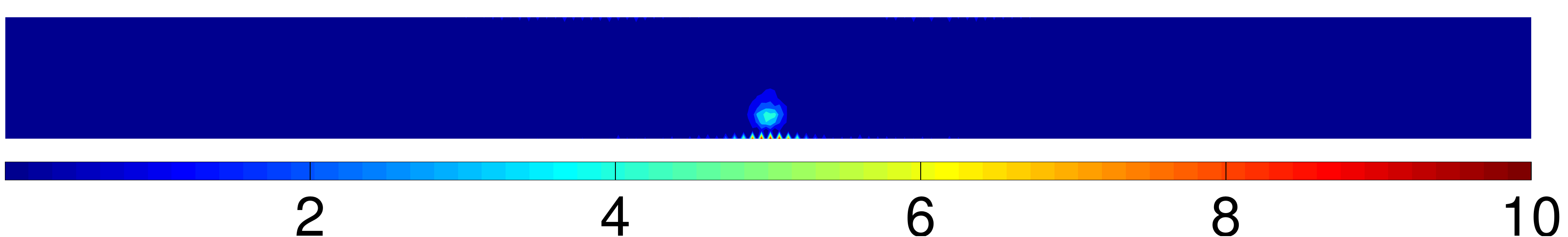}\put(0,0){}
\includegraphics[width=.138\textwidth] {AF_ar_4_spring_6}\put(0,0){}

\includegraphics[width=.138\textwidth] {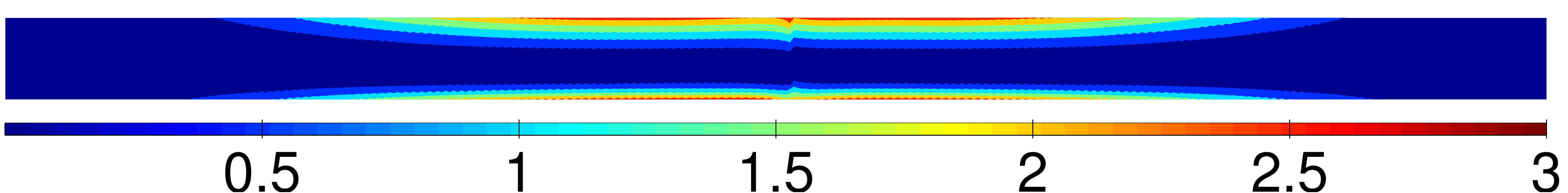}\put(0,0){}
\includegraphics[width=.138\textwidth] {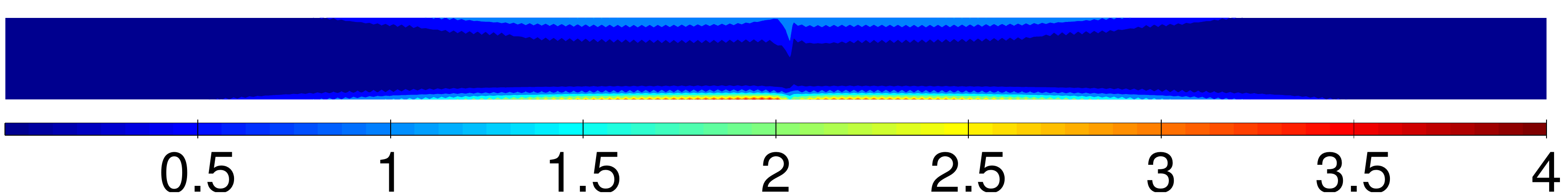}\put(0,0){}
\includegraphics[width=.138\textwidth] {AF_ar_6_spring_2}\put(0,0){}
\includegraphics[width=.138\textwidth] {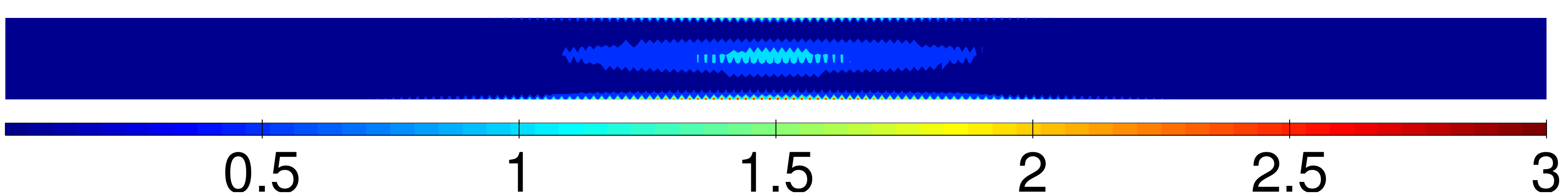}\put(0,0){}
\includegraphics[width=.138\textwidth] {AF_ar_6_spring_4}\put(0,0){}
\includegraphics[width=.138\textwidth] {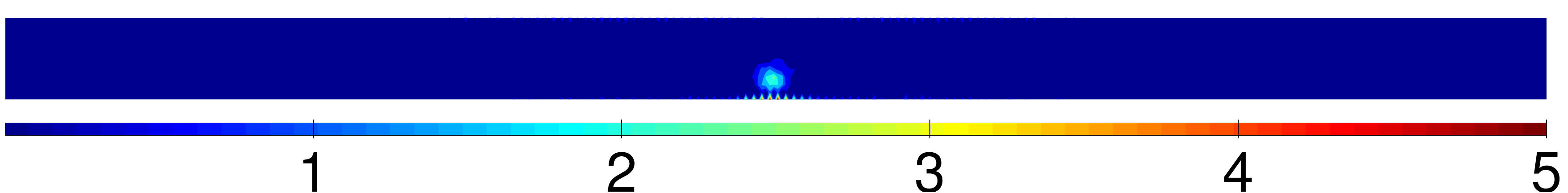}\put(0,0){}
\includegraphics[width=.138\textwidth] {AF_ar_6_spring_6}\put(0,0){}

\includegraphics[width=.138\textwidth] {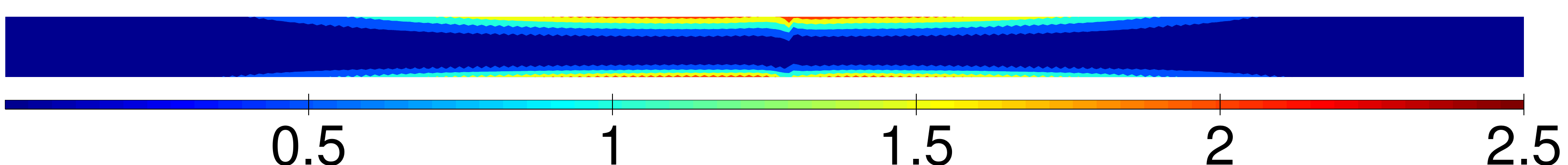}\put(0,0){}
\includegraphics[width=.138\textwidth] {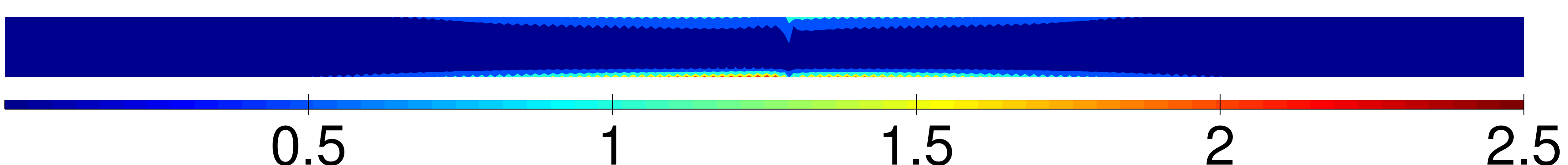}\put(0,0){}
\includegraphics[width=.138\textwidth] {AF_ar_8_spring_2}\put(0,0){}
\includegraphics[width=.138\textwidth] {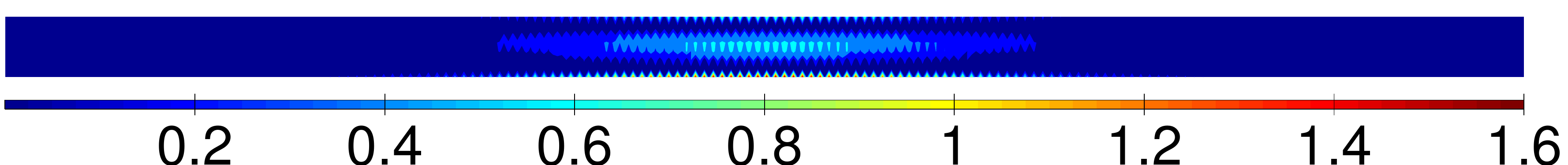}\put(0,0){}
\includegraphics[width=.138\textwidth] {AF_ar_8_spring_4}\put(0,0){}
\includegraphics[width=.138\textwidth] {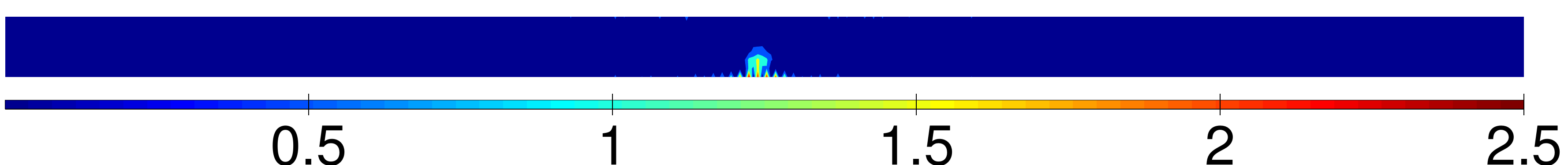}\put(0,0){}
\includegraphics[width=.138\textwidth] {AF_ar_8_spring_6}\put(0,0){}

\caption{Both mean and Gaussian curvature are more evenly distributed for increasing stretchability ${k}$: (a) Square ${H^2}$ of the mean curvature ${H}$, (b) magnitude $|{K}|$ of the Gaussian curvature, and (c) square ${\epsilon^2}$ of the strain ${\epsilon}$ of generalized M\"obius bands for different values of the aspect ratio ${a}$ and stretchability ${k}$. The aspect ratio of each contour plot is equal to the aspect ratio of the band it represents. Note the different scales adapted individually to capture the entire range of values.}
\label{fig:MeanCurvatureSqAll}
\end{figure*}
%%%%%%%%%%%%%%%

%%%%%%%%%%%%%%%%%%%%%%%%	  
\section{Model validation}
\label{sec:validation}
%%%%%%%%%%%%%%%%%%%%%%%%

To validate our model, we first investigate the influence of the discretization on the equilibrium shape and corresponding energy obtained after minimization. In view of the preceeding discussion, we expect that the model approaches a continuum limit as the number $N$ of mesh points is increased. We study the impact of mesh size. To that end, we consider the convergence of the discrete approximations of the curvature $\kappa$ and the torsion $\tau$ of the centerlines for bands made from effectively unstretchable materials, whereas for M\"obius bands made from stretchable materials we study the convergence of $\Psi_s$, $\Psi_b$, and $\Psi$. 

Since $\kappa$ and $\tau$ respectively depend on the directions of $\bft$ and $\bfn$, given a point spacing $r_0$ their convergence does not depend on a length scale. Instead it depends only on the number of points along the centerline. If we denote the arclength of the centerline by $s$, the arclength at the $i$-th point along the centerline by $s_{i}$, and the number of points along the centerline of the band by $N_l$, then it is most suitable to compare the discrete approximations of the curvature and torsion of the centerline for different values of $N_l$. These approximations are observed to converge to limits as $N_l$ is increased (Figure \ref{fig:convergence}(a--b)). Our results indicate that the shapes of bands made of effectively unstretchable materials remain unchanged on increasing $N_l$ from 173 to 347, demonstrating that a choice of $N_l$ in the range $173<N_l<347$ produces converged results and that further increasing $N_l$ has negligible impact on the shape of the band (Figure \ref{fig:convergence}(a--b)).

For bands made from stretchable materials, we compare $\Psi, \Psi_s,$ and $\Psi_b$ for increasing $N_l$ and different values of $k$. All energies show convergence to limiting values (Figure \ref{fig:convergence}(c--e)), which is consistent with the previously discussed convergence of the curvature and torsion (Figure \ref{fig:convergence}(a--b)) in the case of bands made from effectively unstretchable materials. Note that the value of $k$ for which energy due to stretching becomes negligible and the bands approach the unstretchable limit is not immediately evident. However, for $k = 10^{-6}$, $\Psi_s$ is approximately $1\%$ of $\Psi$, indicating that it is reasonable to conclude that the choice $k = 10^{-6}$ provides a very close approximation to unstretchability.

\section{Connection between lattice and continuum models}
%%%%%%%%%%%%%%%%%%%%%%%%

\subsection{Bending energy}

Sadowsky's~\cite{Hinz2013,Hinz2013a,Hinz2013b} work on M\"obius bands made from effectively unstretchable materials hinges on a continuum description in which a strip is treated as a surface $S$ endowed with bending-energy density proportional to the square of its mean curvature $H$. By the \emph{Theorema Egregium} of Gauss,\cite{Gauss1827} bending a surface without stretching or shrinking leaves its Gaussian curvature $K$ unchanged. Simultaneous bending and stretching of a surface will, however, generally alter its Gaussian curvature. To model stretchable bands, it is therefore natural to incorporate energetic dependence on Gaussian curvature. In keeping with a quadratic dependence on mean curvature, this amounts to adding a term linear in the Gaussian curvature and, granted that the band is free of spontaneous mean curvature, leads to the Canham--Helfrich bending-energy functional
\begin{equation}\label{CH01}
E_{\rm bend} = \int_{S}(2\mu H^2+\overline{\mu}K)\,\text{d}A,
\end{equation}
involving the bending-energy density 
\begin{equation}\label{eq:CHdens01}
\psi_{\rm bend} = 2\mu H^2 + \bar\mu K,
\end{equation}
where $\mu>0$ is the splay modulus and $\bar\mu<0$ is the saddle-splay modulus. Importantly, since $K=0$ for a developable surface, \eqref{CH01} reduces in the unstretchable limit to the functional considered by Sadowsky \cite{Hinz2013,Hinz2013a,Hinz2013b}.

Although connections between the Canham--Helfrich functional \eqref{CH01} and triangular lattice parameters have already been established by Seung and Nelson~\cite{Seung1988} and Schmidt and Fraternali \cite{Schmidt2011}, these rely on potentials involving the normals of neighboring triangular cells. Defining the normals to the cells of a triangulation of a nonorientable surface like a M\"obius band is problematic. Hence, we use other elementary geometric considerations to establish a connection between \eqref{CH01} and the bending-energy functional 
\begin{equation}
E_b=\sum_{i\in S_{\theta}}U_{\theta i}
=\frac{k_{\theta}}{2}\sum_{i\in S_{\theta}}(\theta(i)-\pi)^{2}
\label{eq:EB}
\end{equation}
of the discrete lattice, where $S_{\theta}$ is the set of all angular springs.

To determine the effective splay and saddle-splay moduli $\mu$ and $\bar\mu$ corresponding to the underlying triangular lattice, we first use the definition~\eqref{disCurv} of the discrete curvature $\kappa(i)$ in the expression for the bending energy $E_b$. Taking into account that the lattice is periodic lattice, this yields
\begin{equation}\label{eq:EBcurv01}
E_b = \half k_\theta r_0^2\sum_{i=1}^N\sum_{j=1}^3(\kappa^{(j)}(i))^2, 
\end{equation}
where $r_0$ is the distance between lattice points, $N$ is the number of points on the surface, and $k_\theta>0$ is the lattice modulus associated with bending ($E_b$ in \eqref{eq:EBcurv01} carries the same dimensions as $E_{\rm bend}$ in \eqref{CH01}), $\kappa^{(j)}(i)$ is the curvature of the $j$-direction lattice line at point $i$ for $j=1, 2, 3$. Further, letting the angle between the $1$-direction lattice line at point $i$ and the first principal direction at point $i$ be $\alpha(i)$, we can use the expressions~\eqref{eq:kappa_n02} and~\eqref{eq:kappa_g02} for $\kappa_n$ and $\kappa$ to write~\eqref{eq:EBcurv01} as
 \begin{multline}\label{eq:EBcurv02}
 \displaystyle E_b = \half k_\theta r_0^2\sum_{i=1}^N\left(\sum_{j=1}^3(\kappa_g^{(j)}(i))^2+(\kappa_1(i))^2(\cos^4(\alpha(i))+\cos^4(\alpha(i)+\frac{\pi}{3})+\cos^4(\alpha(i)+\frac{2\pi}{3}))\right.\\\left.+(\kappa_2(i))^2(\sin^4(\alpha(i))+\sin^4(\alpha(i)+\frac{\pi}{3})+\sin^4(\alpha(i)+\frac{2\pi}{3}))\right),
 \end{multline}
where the angle between lattice lines is required to remain equal to $\pi/3$. To enforce this constraint on the angle, we assume the distance between points remains constant. This assumption yields reasonable approximations for small deformations. Further, it implies that lattice lines are geodesics---i.e., $\kappa_g\equiv 0$ along lattice lines. We next use the identities
\begin{equation}
\left.
\begin{split}
\cos^4\vartheta+\cos^4(\vartheta+\frac{\pi}{3})+\cos^4(\vartheta+\frac{2\pi}{3})&=\frac 98,
\\
\sin^4\vartheta+\sin^4(\vartheta+\frac{\pi}{3})+\sin^4(\vartheta+\frac{2\pi}{3})&=\frac 98,
\\
\cos^2\vartheta+\cos^2(\vartheta+\frac{\pi}{3})+\cos^2(\vartheta+\frac{2\pi}{3})&=\frac 32,
\\
\sin^2\vartheta+\sin^2(\vartheta+\frac{\pi}{3})+\sin^2(\vartheta+\frac{2\pi}{3})&=\frac 32,
\end{split}
\mskip2mu
\right\}
\end{equation}
to reduce~\eqref{eq:EBcurv02} to
\begin{equation}
\label{eq:EBcurv03}
E_b = \frac{k_\theta r_0^2}{2}\sum_{i=1}^N\left(\frac 92 (H(i))^2-\frac 32 K\right),
\end{equation} 
where $H(i)$ and $K(i)$ are the respective discrete mean and Gaussian curvatures at the lattice point $i$. Finally, a Riemann sum argument yields
\begin{equation}
\label{eq:EBcurv04}
\displaystyle E_b \approx \sqrt{3}k_\theta\int_S(3H^2-K)\,\text{d}A
\end{equation}
for which the corresponding bending-energy density is
\begin{equation}\label{eq:EBdens02}
\psi_{\rm bend} = \sqrt{3}k_\theta(3H^2-K).
\end{equation}
Comparison with~\eqref{CH01} yields
\begin{equation}
\label{eq:moduli01}
\mu = \frac{3\sqrt{3}k_\theta}{2} \quad {\text{and}} \quad \bar \mu = -\sqrt{3}k_\theta.
\end{equation}

To validate the foregoing calculations, we present the example of a plane strip discretized with a lattice mapped isometrically to a helically wrapped strip (Figure \ref{triLat}). Relative to a rectangular Cartesian basis $\{\bfe_1,\bfe_2,\bfe_3\}$, 

Let $\{\bfe_1,\bfe_2,\bfe_3\}$ denote a rectangular Cartesian basis and consider a helix of radius $r>0$ and pitch $2\pi b>0$ parametrized by 
\begin{equation}
%\begin{aligned}\alpha(t):\;&\mathbb{R} \rightarrow \mathbb{R}^3\\&t\mapsto (r\cos t, r\sin t, bt)\end{aligned}
\bfalpha(t)=r\cos t\mskip2mu\bfe_1+r\sin t\mskip2mu\bfe_2+bt\mskip2mu\bfe_3
\label{helix}
\end{equation} 
for $t\in \mathbb{R}$. Eliminating $b$ between the relations

Solving for $b$ in the relation
\begin{equation}
\varphi=\arctan\left(\frac{b}{r}\right)
\qquad\text{and}\qquad
\kappa = \frac{r}{r^2 + b^2}
\label{lead}
\end{equation} 
defining the lead angle and curvature of the helix yields 
\begin{equation}
\kappa= \frac{\cos^2 \varphi}{r}.
\label{curvHel}
\end{equation}

If a plane strip approximated by a triangular lattice is wrapped helically around a cylinder, then at any point $i$, the curve in the $j$-direction is a helix. Further, letting $\varphi^{(j)}(i)$ be the lead angle of the helix through the point $i$ in the $j$-direction, it is evident that 
\begin{equation}\varphi^{(3)}(i) = \varphi^{(2)}(i)+\frac{\pi}{3} = \varphi^{(1)}(i)+\frac{2\pi}{3}.\label{lead2}\end{equation} 
Now substituting \eqref{lead2} into \eqref{curvHel} yields 
\begin{equation}
\left.
\begin{split}
\kappa^{(1)}(i)&=\frac{\cos^2(\varphi^{(1)}(i))}{r},
\\
\kappa^{(2)}(i)&=\frac{\cos^2(\varphi^{(2)}(i)+\frac{\pi}{3})}{r},
\\
\kappa^{(3)}(i)&=\frac{\cos^2(\varphi^{(3)}(i)+\frac{2\pi}{3})}{r}.
\end{split}
\mskip2mu\right\}
\label{curvLead}
\end{equation}
Finally, substituting \eqref{curvLead} into \eqref{eq:EBcurv01} yields 
\begin{equation}E_{B} = \frac 12 k_{\theta} \frac{r_0^2}{r^2}\sum_{i=1}^{N}\left(\cos^4(\varphi^{(1)}(i)) +\cos^4\left(\varphi^{(1)}(i)+\frac{\pi}{3}\right)+\cos^4\left(\varphi^{(1)}(i)+\frac{2\pi}{3}\right)\right).\label{nrg}\end{equation}
Reducing, as in the general case, results in
\begin{equation}E_{B}=\frac{3\sqrt{3}k_{\theta}A}{4r^2}.\label{EL}\end{equation}
Since $H^2\equiv 1/4r^2$ for a helically wrapped strip, comparison of~\eqref{EL} with the Canham--Helfrich bending energy functional~\eqref{CH01} confirms that
\begin{equation}
\mu = \frac{3\sqrt{3}k_\theta}{2}.
\end{equation}

Notice that \eqref{EL} is independent of $\varphi^{(j)}(i)$ and thus of the orientation of the triangular lattice; consistent with the discussion of Vigliotti et al.~\cite{Vigliotti2014}, this assures that the triangular lattice behaves isotropically in bending. Additionally, in the approximately unstretchable regime, where $K$ is very close to zero, \eqref{eq:EBcurv04} is dominated by the term involving $H^2$, demonstrating that, for small $k$, we approximately minimize the integral of $H^2$ over the surface. As reported by Sadowsky \cite{Hinz2013,Hinz2013a,Hinz2013b}, Mahadevan and Keller \cite{Mahadevan1993}, and Starostin and van der Heijden \cite{Starostin2007, Starostin2007a}, this result is consistent with analogous results from continuum theory. 

Finally, we can further compare \eqref{eq:EBdens02} to the bending-energy density
\begin{equation}
\label{flex}
D(2H^2 - (1-\nu)K)
\end{equation} 
for a thin plate derived by Keller and Merchant \cite{Keller1991} in which $D$ is the bending modulus and $\nu$ is the Poisson ratio. To achieve consistency between \eqref{eq:EBdens02} and \eqref{flex}, the material must have $\nu = 1/3$. In contrast, similar calculations using expressions for $\mu$ and $\bar{\mu}$ provided by Seung and Nelson~\cite{Seung1988} and Schmidt and Fraternali~\cite{Schmidt2011} respectively yield $\nu = 0$ and $\nu = -1/3$.

\subsection{Stretching energy}

Apart from energy due to bending, M\"obius bands made from stretchable materials store elastic energy due to in-plane stretching. Using the purely two-dimensional theory of linearized elasticity  and considering only neglecting contributions due to shear results in the area stretching-energy functional of the form (Deserno~\cite{Deserno})
\begin{equation}\label{eq:areFunc01}
E_{\rm stretch} = 2\mu_s \int_{S} \epsilon^2 \;\text{d}A,
\end{equation}
where $\mu_s>0$ is the two-dimensional bulk (area) modulus and $\epsilon\approx (J-1)/2 \approx (A/A_0 - 1)/2$ is an approximation for the local in-plane strain, with $J$ being the areal Jacobian determinant and $A$ the current area of an area element with reference area $A_0$. ($E_{\rm stretch}$ carries dimensions of energy.) In the present two-dimensional setting, $\mu_s$ carries dimensions of energy/length$^2$. In view of \eqref{eq:areFunc01}, the stretching-energy density is
\begin{equation}\label{eq:areDens01}
\psi_{\rm stretch} = 2\mu_s\epsilon^2.
\end{equation}

We now relate the energy of a discrete set of springs to the continuum expression \eqref{eq:areFunc01} with two primary objectives: (I) to determine the effective value of $\mu_s$ for an equilateral triangular lattice of linear springs, and (II) to verify the isotropy of such a lattice. The stretching energy $E_s$ of a collection of linear springs is given by
\begin{equation}\label{ELs}
E_s = \frac{k_{l}}{2} \sum_{i=1}^{N_{s}} [(r(i)-r_0)^2],
\end{equation}
where $k_l>0$ is the lattice modulus associated with stretching carrying the same dimensions as the two-dimensional bulk modulus $\mu_s$,  $N_s$ is the number of linear springs, $r(i)$ is the current length of the $i$-th spring, and $r_0$ is the equilibrium length of the springs. 
Notice that in the discrete setting, pure dilatation amounts to uniform stretching of the lattice. That is, each unit cell stretches (or contracts) into a larger (or smaller) equilateral triangle. The stretching energy of a unit cell in a periodic equilateral triangular lattice is
\begin{equation}\label{E_Ls_01}
E_{\rm UC} = \frac{k_l}{4}[(r(1)-r_0)^2+(r(2)-r_0)^2+(r(3)-r_0)^2],\end{equation}
where $r(1),$ $r(2),$ and $r(3)$ are the lengths of the sides of the unit cells. Notice that the additional factor $1/2$ in~\eqref{E_Ls_01} accounts for periodicity, namely that each spring is shared by neighboring unit cells. For pure dilatation (uniform stretching), $r(1)=r(2)=r(3)=r$ and \eqref{E_Ls_01} reduces to
\begin{equation}
E_{\rm UC}=\frac{3k_l}{4} (r-r_0)^2.\label{ESred}
\end{equation}
Using the area $A_0=\sqrt{3}r_0^2/4$ of a unit cell and the magnitude of the pure dilatational strain $\epsilon = (r-r_0)/r_0$ yields 
\begin{equation}\label{ESred_02}
E_s = N_{\rm UC}E_{\rm UC}=  \sqrt{3} k_l \int_S \epsilon^2\,\mathrm{d}A.
\end{equation}
Importantly,~\eqref{ESred_02} is the lattice counterpart of~\eqref{eq:areFunc01}, and comparison of terms yields the connection
\begin{equation}\label{k_l_mu_s}
\mu_s = \frac{\sqrt{3}k_l}{2}
\end{equation}
between the effective continuum modulus $\mu_s$ and the lattice modulus $k_l$. Notice that \eqref{k_l_mu_s} is consistent with a previously establish connection between lattice and continuum parameters due to Seung and Nelson \cite{Seung1988}. 

The relationship~\eqref{k_l_mu_s} is valid for periodic lattices. In the special case of one isolated unit cell without neighbors, the factor of $1/2$ in~\eqref{E_Ls_01} vanishes, and it correspondingly follows that
\begin{equation}\label{k_l_mu_s_1}
\mu_{s_1} = \sqrt{3}k_l.
\end{equation}
Further, within the continuum theory, the following approximations hold
\begin{equation}\label{eq:linElast_approx}
\trace (\epsilon\bfI)= 3\epsilon\approx J-1 \approx \frac{A}{A_0} -1,
\end{equation}
where $J$ is the areal Jacobian of the deformation and $\bfI$ is the two-dimensional identity tensor. In view of~\eqref{eq:linElast_approx}, the expression~\eqref{ESred_02} for $E_s$ can be written as
\begin{equation}\label{ESred_03}
E_s = N_{\rm UC}E_{\rm UC}=  \frac{ k_lN_{\rm UC}r_0^2}{12}\left(\frac{A}{A_0} -1\right)^2.
\end{equation}

\subsection{Numerical results}

%%%%%%%%%%%%%%%
\begin{figure*}[t!]
\begin{center}
%\hspace{-0.5cm}
\includegraphics[width=.345\linewidth] {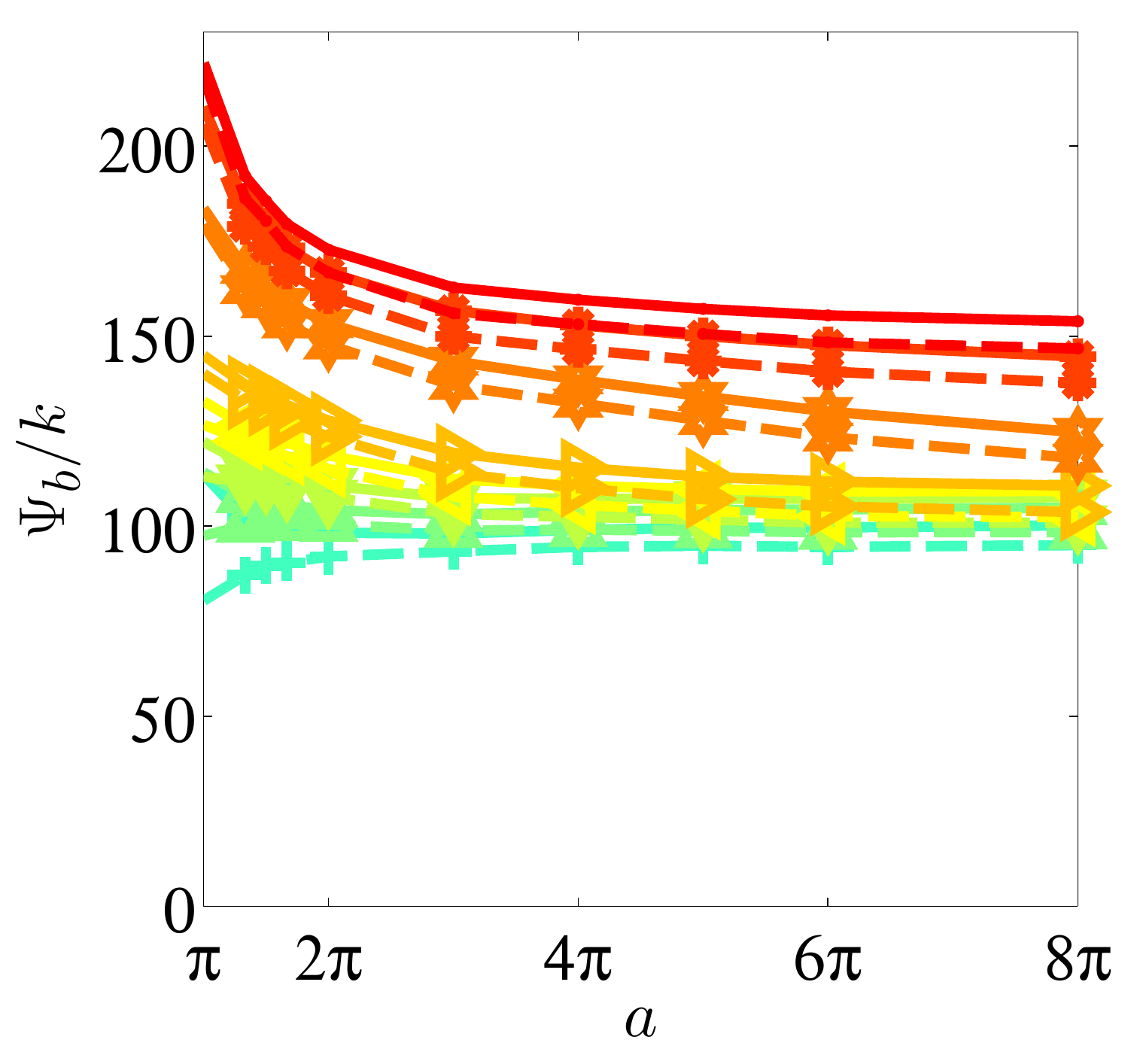}
\put(-140,148){(a)}
%\vspace{-0.3cm}
\includegraphics[width=.16\linewidth] {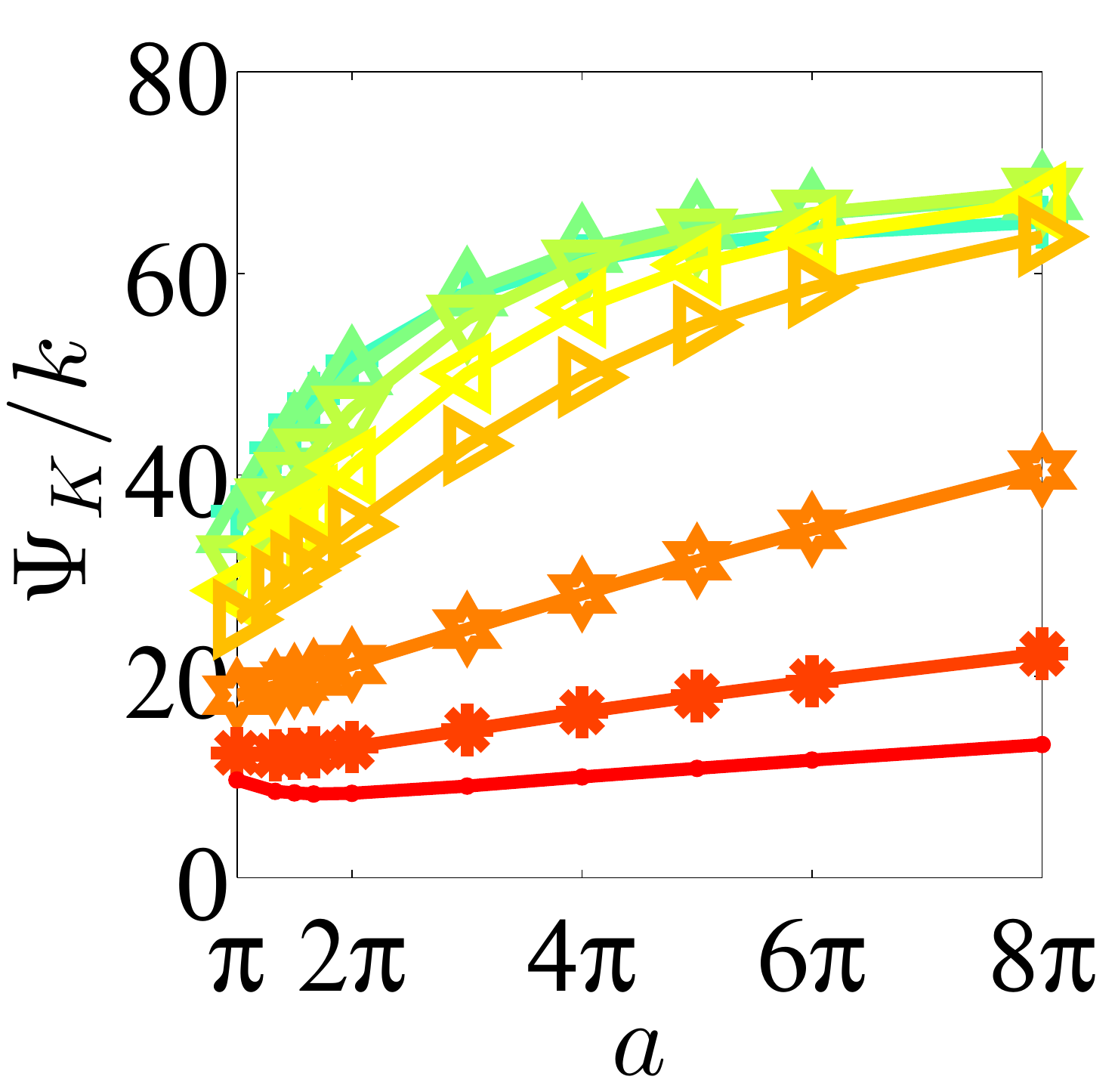}
\put(-87,83){\includegraphics[width=.17\linewidth] {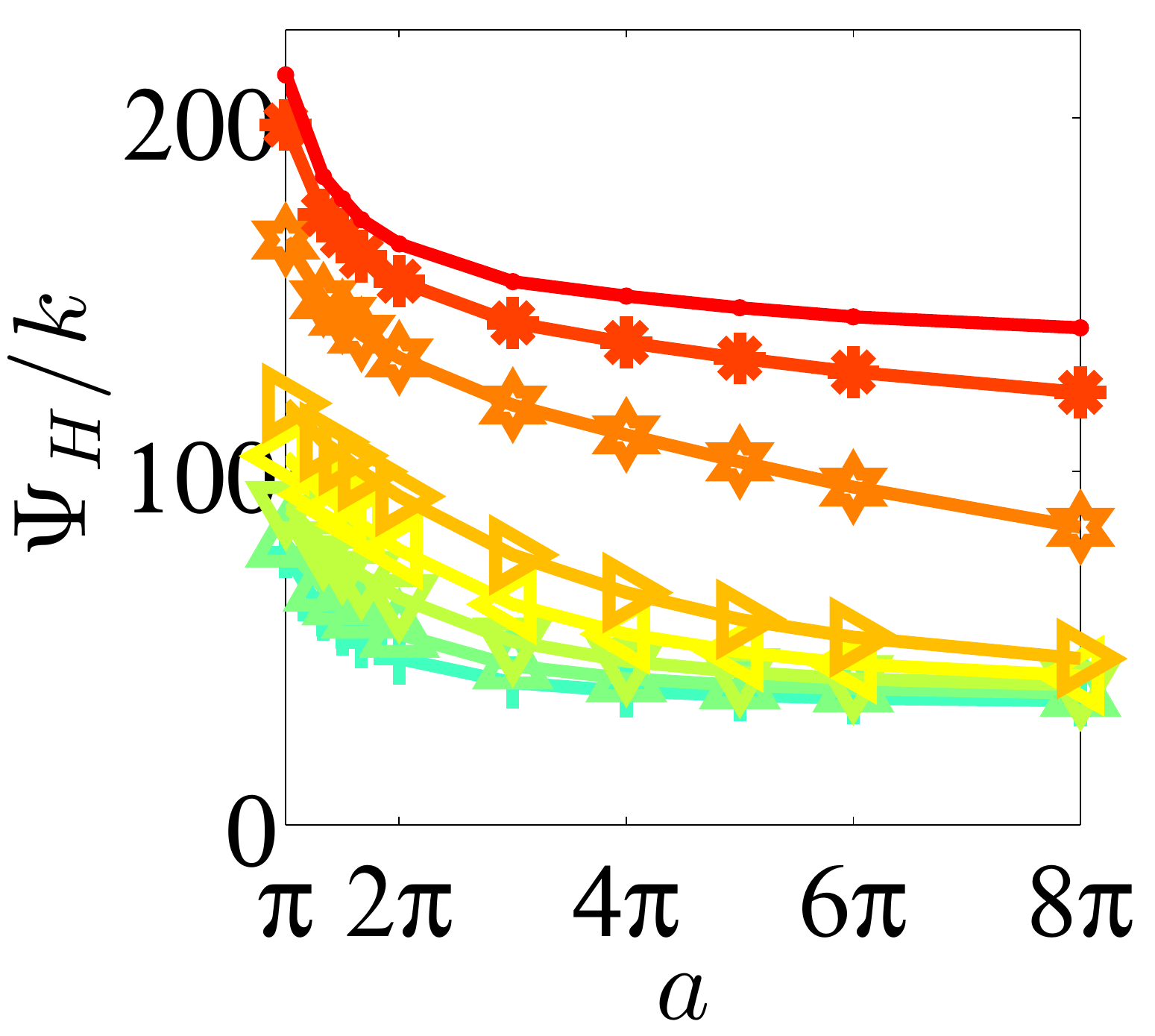}}
\put(-57,148){(b)}
\put(-57,65){(c)}
\includegraphics[width=.335\linewidth] {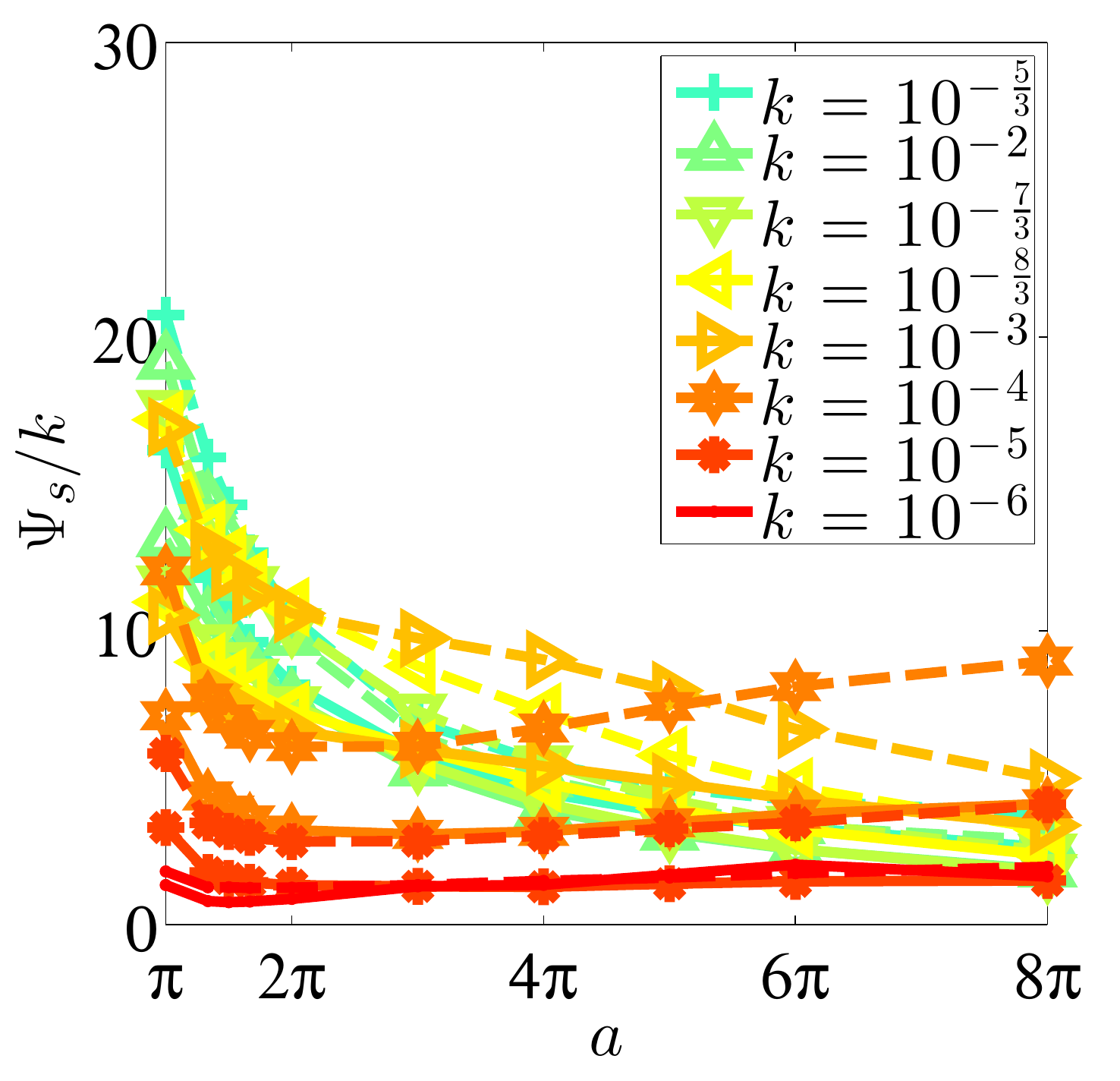}
\put(-142,148){(d)}
\end{center}
\caption{The dimensionless bending energy of the lattice model coincides with the bending energy of the Canham--Helfrich continuum functional using the theoretically determined splay modulus ${\mu}$ and saddle splay modulus ${\bar\mu}$: (a) Dimensionless bending energy of the lattice model (dashed lines) and that from the Canham--Helfrich functional (solid lines). Dimensionless energy contributions to the Canham--Helfrich energy functional from (b) ${H^2}$ and (c) ${K}$. (d) Dimensionless stretching energy due to dilatation (solid lines) is consistently of lower magnitude than that of the lattice model (dashed lines), but both exhibit the same trend. }
\label{fig:Lattice_vs_cont}
\end{figure*}

Consistent with the linearized theory elasticity, the derivations leading to \eqref{eq:moduli01} and \eqref{k_l_mu_s} are predicated on assuming that the strains and rotations are infinitesimal. We therefore expect deviations from the continuum theory for sufficiently large values of the stretchability $k$. For $k\lesssim10^{-5/3}$, we find that the dimensionless bending energy $\Psi_b$ computed from a suitably normalized version of \eqref{eq:EBcurv03} (Figure \ref{fig:Lattice_vs_cont}(a) solid lines) and its counterpart for the discrete lattice \eqref{eq:EB} (Figure \ref{fig:Lattice_vs_cont}(a) dashed lines) coincide. Importantly, the dimensionless energies agree for both small and large contributions associated with $K$ (Figure \ref{fig:Lattice_vs_cont}(b--c)), which confirms the expressions \eqref{eq:moduli01} for the continuum parameters. For $k\gtrsim10^{-5/3}$, we observe deviations from the continuum theory for small aspect ratios. This occurs since, as bands approach a self-intersecting shape, the dimensionless continuum functional becomes orders of magnitude larger than the dimensionless lattice energy due to divergent values of $H^2$ and $K$ near points of self-intersection (not shown).

From Figure \ref{fig:Lattice_vs_cont}(d), we observe that the dimensionless stretching energy due to dilatation is consistently of lower magnitude than that of the lattice model, both energy functionals exhibit the same trend. This indicates that for thin, stretchable materials capable of supporting shear strain, energy contributions due to in-plane shearing are not negligible when a strip of material is twisted into a M\"obius band. As Deserno~\cite{Deserno} explains, the Canham--Helfrich theory neglects energy contributions due to shear and describes simple surfaces, like spheres, very well, but might not be suitable to model topologically complex objects such as M\"obius bands.

%%%%%%%%%%%%%%%%%%%%%%%%%%%%%%%%
\subsection{Numerical validation of the stretching energy}
\label{sec:numVal}
%%%%%%%%%%%%%%%%%%%%%%%%%%%%%%%%

\begin{figure*}[t]
\begin{center}
\includegraphics[width=.4\linewidth, trim=0cm 0cm 0cm 0cm, clip=true] {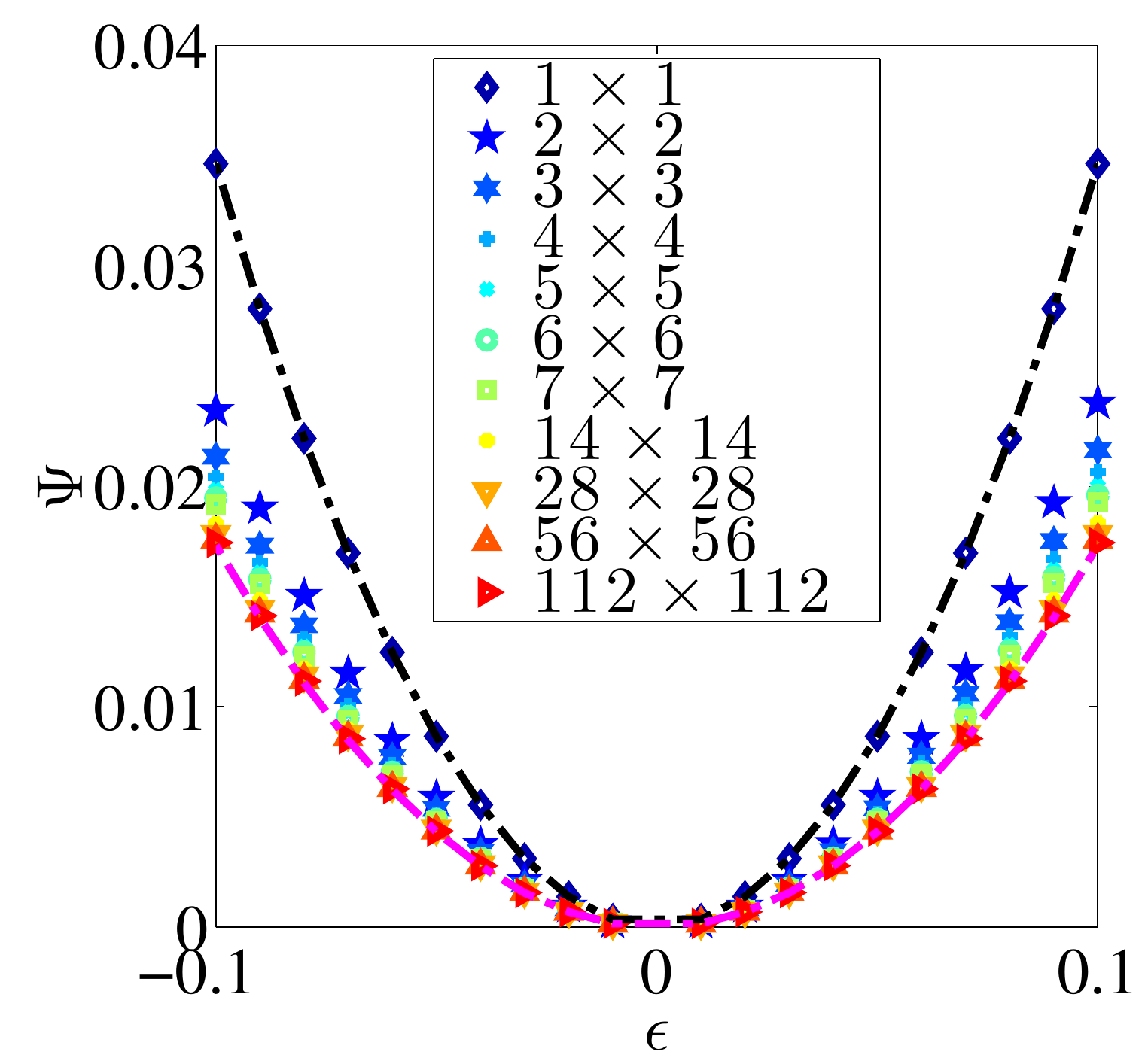}
\put(-40,150){$2\sqrt{3}\epsilon^2$}
\put(-38,40){$\sqrt{3}\epsilon^2$}
\put(-200,200){(a)}
%\quad
\includegraphics[width=.4\linewidth] {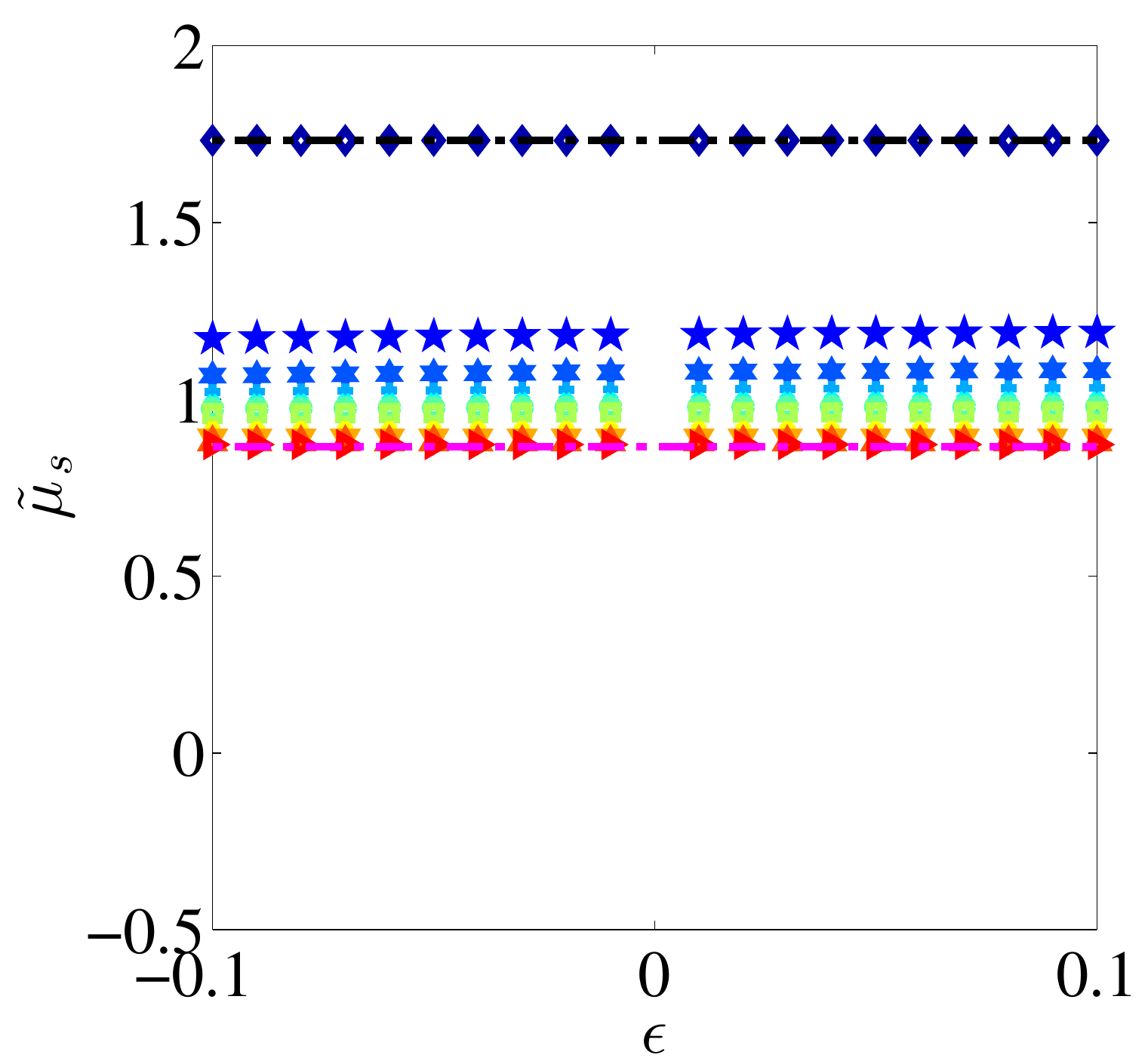}
\put(-100,35){\includegraphics[width=.15\linewidth] {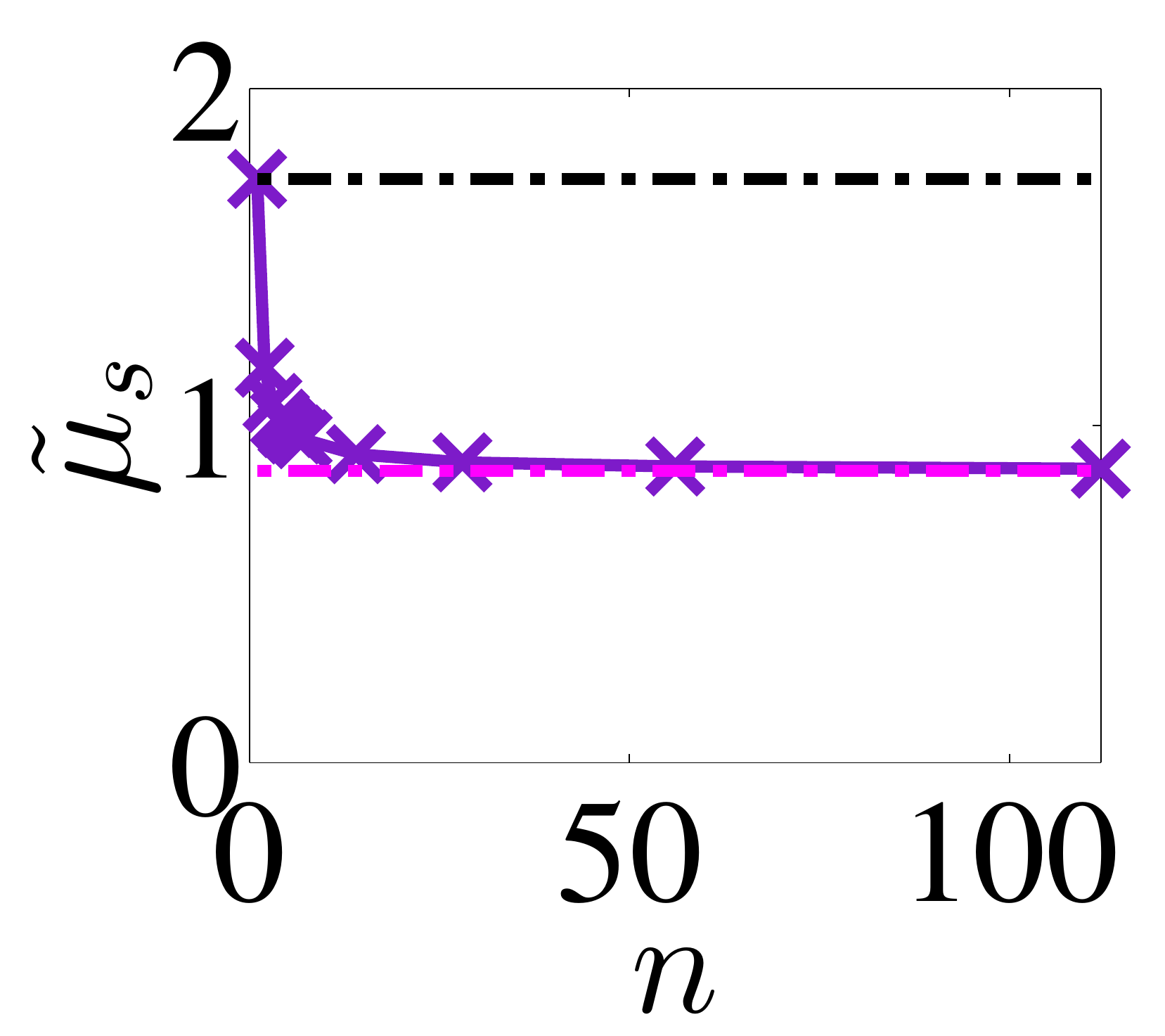}}
\put(-150,148){$\sqrt{3}$}
\put(-150,95){$\sqrt{3}/2$}
\put(-200,200){(b)}
\put(-53,81){$\sqrt{3}$}
\put(-53,60){$\sqrt{3}/2$}
\put(-80,60){(b.i)}
\end{center}
\caption{The lattice behaves linearly elastically, and the effective bulk modulus $\tilde\mu_s$ converges towards the theoretical prediction: (a) Dimensionless energy ${\Psi}_{\color{green}s}$ as a function of strain ${\epsilon}$ along with the theoretical predictions ${\Psi=\sqrt{3} \epsilon^2}$ and ${\Psi_1=2\sqrt{3} \epsilon^2}$. (b) Dimensionless effective bulk modulus ${\tilde \mu_s}$ as a function of strain ${\epsilon}$ with the theoretical predictions ${\tilde \mu_s=\sqrt{3}/2}$ and ${\tilde \mu_{s_1}=\sqrt{3}}$. (b.i) Convergence of ${\tilde\mu_s}$ for increasing number of segments $n$. }
\label{fig:numValBulk01}
\end{figure*}
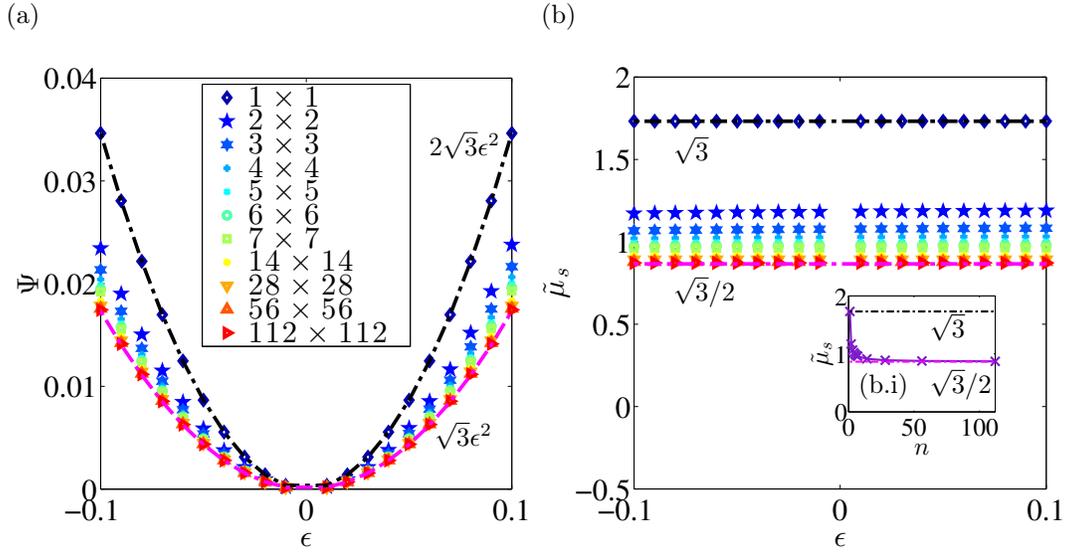

For pure dilatation and assuming periodic boundaries, the total energy $E_s$ of a collection of springs~\eqref{ELs} and the energy of a unit cell are related through
\begin{equation}\label{eq:lammps_UC}
E_{s} = N_{\rm UC} E_{\rm UC},
\end{equation}
where $N_{\rm UC}$ is the number of unit cells. In view of~\eqref{eq:lammps_UC} and \eqref{eq:areFunc01}, the effective stretching modulus can be extracted from the value of \eqref{ELs} determined by simulation through
\begin{equation}\label{eq:lammps_mu_s}
\mu_s = \frac{E_s}{2\epsilon^2 N_{\rm UC} A_0}.
\end{equation}
By analogy to the various scalings introduced in the main text, using $k_l$ and the reference area $A = N_{\rm UC}A_0$ to nondimensionalize~\eqref{eq:lammps_UC} and~\eqref{eq:lammps_mu_s} yields the dimensionless counterparts
\begin{equation}\label{eq:lammps_Psi_nd}
\Psi_s= \frac{E_s}{A k_l} = \frac{E_s}{N_{\rm UC} k_lA_0}
\end{equation}
and
\begin{equation}\label{eq:lammps_mu_s_nd}
\tilde \mu_s = \frac{\mu_s}{k_l} = \frac{E_s}{2\epsilon^2 N_{\rm UC} A_0 k_l}.
\end{equation}
In view of~\eqref{ESred_02}, the following identifications hold: $E_s = \sqrt{3}\int_S \epsilon^2\,\mathrm{d}A$, $\tilde \mu_s = \sqrt{3}/2$, and $\tilde \mu_{s_1} = \sqrt{3}$. To quantify the parameter range in which the plane lattice subject to in-plane pure dilatation behaves linearly elastically, we measure~\eqref{eq:lammps_mu_s_nd} for different imposed $\epsilon$ and different sizes of the lattice (or, equivalently, different $N_{\rm UC}$). The lattice sizes range from $1\ctimes1$ to $112\ctimes112$ triangular unit cells (SI Figure \ref{fig:numValBulk01}), and the strains range from $\epsilon = -0.1$ to $\epsilon = 0.1$. The energy is found to converge to the theoretically determined value, with little difference between results obtained for the $14\ctimes14$ lattice and for the $112\ctimes112$ lattice.
%
%%%%%%%%%%%%%%%%%%%%%%%%%%%%%%%%
%%%%%%%%%%%%%%%%%%%%%%%%%%%%%%%%

\footnotesize

\providecommand*{\mcitethebibliography}{\thebibliography}
\csname @ifundefined\endcsname{endmcitethebibliography}
{\let\endmcitethebibliography\endthebibliography}{}

%\setlength{\bibsep}{4pt}
%\bibliography{soft_matter_refs}		% expects file "myrefs.bib"
%\bibliographystyle{rsc}	         % (uses file "plain.bst")
%\bibliographystyle{rspublicnat}	         % (uses file "plain.bst")
\normalsize

\end{document}